\documentclass[nohyper,notoc]{article}
\usepackage{axodraw}
\usepackage{epsfig}
\usepackage{cite}

\usepackage{tabularx} 
\usepackage{rotating} 
\usepackage{epsfig}

\begin{document}

\catcode`\@=11
\@addtoreset{equation}{section}
\@addtoreset{figure}{section}

\newcommand{\beq}{\begin{equation}}
\newcommand{\eeq}{\end{equation}}
\newcommand{\bea}{\begin{eqnarray}}
\newcommand{\eea}{\end{eqnarray}}
\def \lsim{\mathrel{\vcenter
     {\hbox{$<$}\nointerlineskip\hbox{$\sim$}}}}
\def \gsim{\mathrel{\vcenter
     {\hbox{$>$}\nointerlineskip\hbox{$\sim$}}}}
\def\gappeq{\mathrel{\rlap {\raise.5ex\hbox{$>$}}
{\lower.5ex\hbox{$\sim$}}}}
\def\lappeq{\mathrel{\rlap{\raise.5ex\hbox{$<$}}
{\lower.5ex\hbox{$\sim$}}}}
\def\simlt{\stackrel{<}{{}_\sim}}
\def\simgt{\stackrel{>}{{}_\sim}}

\def\mnu{[m_{\nu}]_{ij}}

\newcommand{\eqn}[1]{eq.~(\ref{#1})}
\newcommand{\eqns}[2]{eqs.~(\ref{#1}) and (\ref{#2})}
\newcommand{\Eqn}[1]{Eq.~(\ref{#1})}
\newcommand{\Eqns}[2]{Eqs.~(\ref{#1}) and (\ref{#2})}

\def\CPv{{\begin{picture}(13,0)(0,0)\put(0,0){\rm
CP}\put(0,0){\line(2,1){15}}\end{picture}}~}

\def\BLv{{\begin{picture}(13,0)(0,0)\put(0,0){\rm
B+L}\put(2,-1){\line(2,1){17}}\end{picture}}~}
\def\pslash{p \! \! \! /~}

\def\subBLv{{\begin{picture}(12,0)(0,0)\put(0,0){$\scriptscriptstyle \rm
B{\scriptscriptstyle+}L$}\put(-.7,-.1){\line(4,1){14}}\end{picture}}~}

\def\pslash{p \! \! \! /~}

\renewcommand{\theequation}{\thesection.\arabic{equation}}
\renewcommand{\thefigure}{\thesection.\arabic{figure}}

\vspace*{2cm}
 \begin{center}
 {\Large{\bf Leptogenesis}}
 \end{center}
 \vspace*{1 cm}

\begin{center}
\vskip 25pt
{\bf Sacha Davidson $^{*,}$\footnote{E-mail address:
s.davidson@ipnl.in2p3.fr},
 Enrico Nardi $^{\dagger,}$\footnote{E-mail address:
enrico.nardi@lnf.infn.it }, and Yosef Nir
$^{\ddagger,}$}\footnote{E-mail address: 
yosef.nir@weizmann.ac.il; The
Amos de-Shalit chair of theoretical physics}

\vskip 10pt
$^*${\it  IPN de Lyon, Universit\'e  Lyon 1, IN2P3/CNRS,  4 rue Enrico Fermi, Villeurbanne,  69622 cedex France
}\\
$^\dagger${\it
INFN, Laboratori Nazionali di Frascati, C.P. 13,
      100044 Frascati, Italy, {\rm \&} \\
    Instituto de F\'{i}sica, Universidad de Antioquia,
    A.A.1226, Medell\'{i}n, Colombia
} \\
$^\ddagger${\it Department of Particle Physics,
  Weizmann Institute of Science, Rehovot 76100, Israel}
\vskip 20pt
{\bf Abstract}
\end{center}
Leptogenesis is a class of scenarios where the baryon asymmetry
of the Universe is produced from a lepton asymmetry
 generated in the decays of a heavy sterile neutrino.
 We explain the motivation for
leptogenesis. We review the basic mechanism, and describe subclasses
of models. We then focus on recent developments in the understanding of
leptogenesis: finite temperature effects, spectator processes, and
in particular the significance of flavour physics.

\newpage

\newpage
\tableofcontents
\newpage

\section{The Baryon Asymmetry of the Universe}
\label{intro}

\subsection{Observations}
\label{observations}
Observations indicate that 
the number of baryons (protons and neutrons) in the Universe  is unequal
 to the number of antibaryons (antiprotons and antineutrons). 
To the best of our understanding, all the structures that we
see in the Universe -- stars, galaxies, and clusters -- consist of
matter (baryons and electrons) and there is no antimatter (antibaryons
and positrons) in  appreciable quantities. Since various
considerations suggest that the Universe has started from a state with
equal numbers of baryons and antibaryons, the observed baryon
asymmetry must have been generated dynamically, a scenario that is
known by the name of {\it baryogenesis}.

One may wonder why we think that the baryon asymmetry has been
dynamically generated, rather than being
 an initial condition. There are at least
two reasons for that. First, if a baryon asymmetry had been an initial
condition, it would have been
a highly fine-tuned one. For every 6,000,000 antiquarks, there
should have been 6,000,001 quarks. Such a fine-tuned condition seems
very implausible. Second, and perhaps more important, we have
excellent reasons, based on observed features of the cosmic microwave
background radiation, to think that inflation took place during
the history of the Universe.  Any primordial baryon asymmetry
would have been exponentially diluted away 
by the required amount of  inflation.

The baryon asymmetry of the Universe poses a puzzle in particle
physics. The Standard Model (SM) of particle interactions contains 
all the ingredients that are necessary to dynamically
generate such an asymmetry in an initially baryon-symmetric
Universe. Yet, it fails to explain an asymmetry as large as the one
observed (see {\it e.g.} \cite{Dolgov:1991fr}). 
New physics is called for. The new physics must \cite{Sakharov:1967dj}, first,
distinguish matter from antimatter in a more pronounced way than do  the
weak interactions of the SM. Second, it should 
depart from thermal equilibrium during the history of the Universe.

The baryon asymmetry of the Universe  can be defined in two
equivalent ways:
\bea
\label{etaB}
\eta&\equiv& \frac{n_B- n_{\bar{B}}}{n_{\gamma }} {\Big |}_{0}
= (6.21 \pm 0.16)  \times 10^{-10}, \\
Y_{\Delta B} &\equiv&  \frac{n_B- n_{\bar{B}}}{s} {\Big |}_{0}
= ( 8.75 \pm 0.23)  \times 10^{-11}
\label{YB}
\eea
where $n_B$, $n_{\bar B}$, $n_\gamma$ and $s$  are the number
densities of, respectively, baryons, antibaryons, photons and entropy,
 a subscript $0$ implies ``at present time'', and the numerical value
is from combined microwave background  and large scale structure
data (WMAP 5 year data, Baryon Acoustic Oscillations and Type Ia Supernovae) 
\cite{WMAP5cosmo}. It is convenient to
calculate $Y_{\Delta B}$, the baryon asymmetry
relative to the entropy density $s$, because $s =g_*(2\pi^2/45)T^3$
is conserved during the expansion of the Universe ($g_*$ is
the number of degrees of freedom in the plasma, and $T$ is
the temperature; see the discussion
and definitions below eqn (\ref{defYX})). The two definitions
(\ref{etaB}) and (\ref{YB}) are related through $Y_{\Delta
  B}=(n_{\gamma 0}/s_0)\eta\simeq\eta/7.04$. 
A third, related way to
express the asymmetry is in terms of the baryonic fraction of the
critical energy density,
\beq
\Omega_B\equiv\rho_B/\rho_{\rm crit}.
\eeq
The relation to $\eta$ is given by
\beq
\eta=2.74\times10^{-8}\ \Omega_B\ h^2,
\eeq
where $h\equiv H_0/100\ {\rm km\ s}^{-1}\ {\rm Mpc}^{-1}=0.701\pm0.013$
\cite{WMAP5param}
is the present Hubble parameter.

The value of baryon asymmetry of the Universe is inferred from
observations in two independent ways. The first way is via big bang
nucleosynthesis \cite{Yao:2006px,Steigman:2005uz,Cyburt:2004yc}.
This chapter in cosmology predicts the
abundances of the light elements, D, $^3{\rm He}$, $^4{\rm He}$, and
$^7${\rm Li}.  These predictions depend on essentially a single
parameter which is $\eta$. The abundances of D and $^3${\rm He} are
very sensitive to $\eta$. The reason is that they are crucial in the
synthesis of $^4${\rm He} via the two body reactions
D($p,\gamma$){}$^3${\rm He} and $^3${\rm He}(D,$p$){}$^4${\rm He}. The
rate of these reactions is proportional to the number densities of the
incoming nuclei which, in turn, depend on $\eta$: $n(\rm D)\propto
\eta$ and $n({}^3{\rm He})\propto\eta^2$. Thus, the larger $\eta$, the
later these $^4$He-producing processes will stop (that is, become
slower than the expansion rate of the Universe), and consequently the
smaller the freeze-out abundances of D and of $^3${\rm He} will be.
The abundance of $^4$He is less sensitive to $\eta$. Larger values of
$\eta$ mean that the relative number of photons, and in particular
photons with energy higher than the binding energies of the relevant
nuclei, is smaller, and thus the abundances of D, $^3$He and $^3$H
build up earlier. Consequently, $^4$He synthesis starts earlier, with
a larger neutron-to-proton ratio, which results in a higher $^4$He
abundance.  The dependence of the $^7$Li-abundance on $\eta$ is more
complicated, as two production processes with opposite
$\eta$-dependencies play a role.

The primordial abundances of the four light elements can be inferred
from various observations. The fact that there is a range of $\eta$
which is consistent with all four abundances gives a strong support to
the standard hot big bang cosmology.  This range is given (at 95\% CL)
by \cite{Yao:2006px}
\beq\label{nbbbn}
4.7\times10^{-10}\leq\eta\leq6.5\times10^{-10},\ \ \
0.017\leq\Omega_B h^2\leq0.024.
\eeq

The second way to determine $\Omega_B$ is from measurements of the
cosmic microwave background (CMB) anisotropies (for a pedagogical
review, see \cite{Hu:2001bc,Dodelson:2003ft}). The CMB spectrum
corresponds to an excellent approximation to a blackbody radiation
with a nearly constant temperature $T$. The basic observable is the
temperature fluctuation $\Theta(\hat{\bf n})=\Delta T/T$ ($\hat{\bf
  n}$ denotes the direction in the sky). The analysis is simplest in
Fourier space, where we denote the wavenumber by $k$.

The crucial time for the CMB is that of recombination, when the
temperature dropped low enough that protons and electrons could form
neutral hydrogen. This happened at redshift $z_{\rm rec}\sim1000$.
Before this time, the cosmological plasma can be described, to a good
approximation, as a photon-baryon fluid. The main features of the CMB
follow from the basic equations of fluid mechanics applied to perfect
photon-baryon fluid, neglecting dynamical effects of gravity and the
baryons:
\beq\label{perflu}
\ddot\Theta+c_s^2k^2\Theta=0,\ \ \
c_s\equiv\sqrt{\dot p/\dot\rho}=\sqrt{1/3},
\eeq
where $c_s$ is the sound speed
in the dynamically baryon-free fluid ($\rho$ and $p$ are the photon
energy density and pressure). These features in the anisotropy
spectrum are: the existence of peaks and troughs, the spacing between
adjacent peaks, and the location of the first peak. The modifications
due to gravity and baryons can be understood from adding their effects
to eqn (\ref{perflu}),
\beq
\ddot\Theta+c_s^2k^2\Theta=F,\ \ \
c_s=\frac{1}{\sqrt{3(1+3\rho_B/4\rho_\gamma)}},
\eeq
where $F$ is the forcing term due to gravity, and $\rho_B$ is the
baryon energy density. The physical effect of the baryons is to
provide extra gravity which enhances the compression into potential
wells. The consequence is enhancement of the compressional phases
which translates into enhancement of the odd peaks in the
spectrum. Thus, a measurement of the odd/even peak disparity constrains
the baryon energy density. A fit to the most recent observations
(WMAP5 data only, 
assuming  a $\Lambda$CDM model with  a scale-free power
spectrum for the primordial density fluctuations) 
gives (at 2 $\sigma$) \cite{WMAP5param}
\beq\label{omecmb}
0.02149 \leq\Omega_B h^2\leq 0.02397.
\eeq

The impressive consistency between the nucleosynthesis (\ref{nbbbn})
and CMB (\ref{omecmb}) constraints on the baryon density of the
Universe is another triumph of the hot big-bang cosmology. A
consistent theory of baryogenesis should
explain  $n_B\approx 10^{-10} s$.

\subsection{ Ingredients and Mechanisms}
\label{ingredients}
The three ingredients  required to dynamically generate a baryon
asymmetry  were given  by Sakharov in Ref. \cite{Sakharov:1967dj}:
\begin{enumerate}
\item  Baryon number violation:
  This condition is required in order to evolve from an initial state
  with $Y_{\Delta B}=0$ to a state with $Y_{\Delta  B}\neq0$.
\item C and CP violation:
  If either C or CP were conserved, then processes involving baryons
  would proceed at precisely the same rate as the C- or CP-conjugate
  processes involving antibaryons, with the overall effects that no
  baryon asymmetry is generated.
\item Out of equilibrium dynamics:
In chemical equilibrium, there are no asymmetries in quantum numbers
that are not conserved (such as $B$, by the first condition).
\end{enumerate}

These ingredients are all present in the Standard Model. However,
no  SM mechanism  generating a large enough baryon asymmetry 
has been found. 
\begin{enumerate}
\item Baryon number is violated in the Standard Model, and the
resulting baryon number violating processes are fast in the early
Universe \cite{Kuzmin:1985mm}. The violation is due to the triangle
anomaly, and leads to processes that involve nine left-handed quarks
(three of each generation) and three left-handed leptons (one from
each generation). A selection rule is obeyed,\footnote{This selection
rule implies that the sphaleron processes do not mediate proton decay.}
\beq
\Delta B=\Delta L = \pm3.
\eeq
At zero temperature, the amplitude of the baryon number violating
processes is proportional to $e^{-8 \pi^2/g^2}$
\cite{'tHooft:1976up}, which is too small to have any observable
effect. At high temperatures, however, these transitions become
unsuppressed \cite{Kuzmin:1985mm}.
\item The weak interactions of the SM violate C maximally and violate CP via
the Kobayashi-Maskawa mechanism \cite{Kobayashi:1973fv}. This CP
violation can be parameterized by the Jarlskog invariant
\cite{Jarlskog:1985ht} which, when appropriately normalized, is of
order $10^{-20}$. Since there are practically no kinematic
enhancement factors in the thermal bath
\cite{Gavela:1994ds,Gavela:1994dt,Huet:1994jb}, it is impossible to
generate $Y_{\Delta B}\sim10^{-10}$ with such a small amount of CP
violation. Consequently, baryogenesis implies that there must exist
new sources of CP violation, beyond the Kobayashi-Maskawa phase of the
Standard Model.
\item Within the Standard Model, departure from thermal equilibrium occurs
at the electroweak phase transition
\cite{Rubakov:1996vz,Trodden:1998ym}. Here, the non-equilibrium
condition is provided by the interactions of particles with the bubble
wall, as it sweeps through the plasma. The experimental lower bound on
the Higgs mass implies, however, that this transition is not strongly
first order, as required for successful baryogenesis. Thus, a
different kind of departure from thermal equilibrium is required from
new physics or, alternatively, a modification to the electroweak phase
transition.
\end{enumerate}
This shows that baryogenesis requires  new physics that
extends the SM in at least two ways: It must introduce new sources of
CP violation and it must either provide a departure from thermal
equilibrium in addition to the  electroweak phase transition (EWPT) 
or modify the EWPT itself.
Some possible new physics mechanisms for baryogenesis are the
following:

{\bf GUT baryogenesis}  \cite{Ignatiev:1978uf,Yoshimura:1978ex,Toussaint:1978br,Dimopoulos:1978kv,Ellis:1978xg,Weinberg:1979bt,Yoshimura:1979gy,Barr:1979ye,Nanopoulos:1979gx,Yildiz:1979gx} generates the baryon asymmetry  in the
out-of-equilibrium decays of heavy bosons in Grand Unified Theories
(GUTs).  The  Boltzmann Equations (BE) for the bosons and the  baryon asymmetry
are studied, for instance, in \cite{Fry:1980ph,Fry:1980bc,Kolb:1979qa}.
BE are also required for
leptogenesis, and  in this review, we will  follow closely the analysis
of Kolb and Wolfram\cite{Kolb:1979qa}. 
The  GUT baryogenesis  scenario  has difficulties with 
the non-observation of proton decay, which puts a lower bound on the mass of
the decaying boson, and therefore on the reheat temperature after
inflation \footnote{Bosons with $m > T_{reheat}$ could nonetheless
be produced   during ``preheating''\cite{Kolb:1996jt,Kolb:1998he}, 
which is a stage between the end of inflation and
the filling of the Universe with a thermal bath.}. 
Simple inflation models do not
give such a  high reheat  temperature, which
in addition,   might regenerate 
unwanted
relics. Furthermore, in the simplest GUTs, $B+L$ is violated but $B-L$
is not. Consequently, the $B+L$ violating SM sphalerons, which are in
equilibrium at $T \lsim 10^{12}$ GeV,  would destroy this asymmetry
\footnote{A solution to this problem \cite{Fukugita:2002hu}, in
the seesaw model, could be to
have fact $L$ violation (due to the righthanded
neutrinos) at $T> 10^{12}$ GeV. This would
destroy the $L$ component of the asymmetry,
and the remaining $B$ component would
survive the sphalerons.}.

{\bf Leptogenesis} was invented by Fukugita and
Yanagida in Ref. \cite{Fukugita:1986hr}.
 New particles -- singlet neutrinos -- are
introduced via the seesaw mechanism
\cite{Minkowski:1977sc,Yanagida:1979as,Glashow,GellMann:1980vs,Mohapatra:1980yp}.
Their Yukawa couplings provide the necessary new source of
CP violation. The rate of these Yukawa interactions can be slow enough
(that is slower than $H$, the expansion rate of the Universe, at the
time that the asymmetry is generated) that departure from thermal
equilibrium occurs. Lepton number violation comes from the
Majorana masses of these new particles, and the Standard Model
sphaleron processes still play a crucial role in partially converting
the lepton asymmetry into a baryon asymmetry \cite{Khlebnikov:1988sr}.
This review focuses on the simplest and theoretically best motivated  
realization of leptogenesis:  thermal leptogenesis 
with hierarchical  singlet neutrinos.

{\bf Electroweak baryogenesis} \cite{Rubakov:1996vz,Riotto:1999yt,Cline:2006ts}
 is the name for a class of models where the departure from
thermal equilibrium is provided by the electroweak phase transition.
In principle, the SM belongs to this class, but the phase transition
is not strongly first order \cite{Kajantie:1995kf} 
and the CP violation is too small\cite{Gavela:1994ds,Gavela:1994dt}. Thus,
viable models of electroweak baryogenesis need a modification of the
scalar potential such that the nature of the EWPT changes, and new
sources of CP violation. One example\cite{Losada:1996ju} is the 2HDM (two Higgs doublet
model), where the Higgs potential has more parameters and, unlike the
SM potential, violates CP. Another interesting example is the MSSM
(minimal supersymmetric SM), where a light stop modifies the Higgs
potential in the required way\cite{Carena:1996wj,Delepine:1996vn}
 and where there are new,
flavour-diagonal, CP violating phases. Electroweak baryogenesis and, in
particular, MSSM baryogenesis, might soon be subject to experimental
tests at the CERN LHC.

{\bf The Affleck-Dine mechanism} \cite{Affleck:1984fy,Dine:1995kz}.
The asymmetry arises in a classical scalar field, which later decays
to particles. In a SUSY model, this field could be some combination of
squark, Higgs and slepton fields. The field starts with a large
expectation value, and rolls towards the origin in its scalar
potential. At the initial large distances from the origin, there can
be contributions to the potential from baryon or lepton number
violating interactions (mediated, for instance, by heavy particles).
These impart a net asymmetry to the rolling field.  This generic
mechanism could produce an asymmetry in any combination of $B$ and
$L$.

Other, more exotic scenarios, are described in
Ref. \cite{Dolgov:1991fr}.

\subsection{Thermal leptogenesis: advantages and alternatives}
\label{alternatives}
There are many reasons to think that the Standard Model is only a low
energy effective theory, and that there is new physics at a higher
energy scale. Among these reasons, one can list the experimental
evidence for neutrino masses, the fine-tuning problem of the Higgs
mass, the dark matter puzzle, and gauge coupling unification.

It would be a particularly interesting situation if a new physics
model, motivated by one of the reasons mentioned above, would also
provide a viable mechanism for baryogenesis. Leptogenesis
\cite{Fukugita:1986hr} is such a scenario, because it is almost
unavoidable when one invokes the seesaw mechanism
\cite{Minkowski:1977sc,Yanagida:1979as,Glashow,GellMann:1980vs,Mohapatra:1980yp}
to account for the neutrino masses and to explain their
unique lightness. Indeed, the seesaw mechanism requires that lepton
number is violated, provides in general new CP violating phases in the
neutrino Yukawa interactions, and for a large part of the parameter
space predicts that new, heavy singlet neutrinos decay out of
equilibrium. Thus, all three Sakharov conditions are naturally
fulfilled in this scenario. The question of whether leptogenesis is
 the source of the observed baryon asymmetry is then, within the
seesaw framework, a quantitative one.

In the context of the seesaw extension of the SM, there are several
ways to produce a baryon asymmetry. They all have in common the
introduction of singlet neutrinos $N_i$ with masses $M_i$ that are
usually \footnote{ An exception is the 
$\nu$MSM \cite{Asaka:2005an}.}
heavier than the electroweak breaking scale, $M_i\gg v_u$. They may
differ, however, in the cosmological scenario, and in the values of
the seesaw parameters. (The number of seesaw parameters is much larger
than the number of measured, light neutrino parameters.) A popular
possibility, which this review focuses on, is ``thermal leptogenesis''
with hierarchical masses, $M_1 \ll M_{j>1}$ \cite{Fukugita:1986hr}.
The $N_1$ particles are produced by scattering in the thermal bath,
so that their number density can be calculated from the seesaw
parameters and the reheat temperature of the Universe.

Thermal leptogenesis has been studied in detail by many people, and
there have been numerous  clear and pedagogical reviews. 
Early analyses, focusing on hierarchical
singlet neutrinos,  include
\cite{Luty:1992un,Gherghetta:1993kn,Plumacher:1996kc,Plumacher:1997ru} 
(see the thesis \cite{Plumacher:1998ex} for details, 
in particular of supersymmetric leptogenesis with
superfields). The
importance of including the wave function
renormalisation of the decaying  singlet,
in calculating the CP asymmetry, was  recognised
in \cite{Covi:1996wh}. Various reviews
\cite{Buchmuller:1999cu,Buchmuller:2000as}
were written at this stage, 
a pedagogical presentation 
that introduces BE,
and discusses models and supersymmetry
can be found in \cite{Buchmuller:2000as}.
 More detailed calculations,
but which do not include flavour effects, are presented in
\cite{Buchmuller:2004nz,Giudice:2003jh}:
thermal effects were included in
\cite{Giudice:2003jh}, and \cite{Buchmuller:2004nz}
gives many useful analytic approximations.
 The importance of
flavour effects was emphasized in
\cite{Pilaftsis:2005rv} for resonant  leptogenesis
with  degenerate $N_i$, and  in
\cite{Abada:2006fw,Nardi:2006fx,Abada:2006ea}
for hierarchical singlets. 
An earlier  ``flavoured''
analysis is \cite{Barbieri:1999ma}, and flavoured  BE are 
presented in \cite{Endoh:2003mz}.
The aim of this review is to pedagogically
introduce flavour effects, which can change
the final baryon asymmetry by factors of few or more.
Some previous  reviews of flavour
effects can be found  in the TASI lectures
of \cite{Chen:2007fv}, and in the conference proceedings
\cite{Davidson:2007xu,Nardi:2007fs,Nardi:2007cf}.
They are also mentioned in Refs. 
\cite{Strumia:2006qk,Strumia:2006db}.

A potential drawback of thermal leptogenesis, for hierarchical masses
$M_i$'s, is the lower bound on $M_1$\cite{Hamaguchi:2001gw,Davidson:2002qv}
 (discussed in Section \ref{esteff}),
which gives a lower bound on the reheat temperature. This bound
might be in conflict with an upper bound on the reheat temperature
that applies in supersymmetric models with a ``gravitino problem''
\cite{Weinberg:1982zq,Khlopov:1984pf,Ellis:1984eq,Kawasaki:1994af,Kohri:2005wn,Rychkov:2007uq,Kawasaki:2008qe}.
The lower bound on $M_1$  can be avoided with
quasi-degenerate $M_i$'s
\cite{Flanz:1994yx,Flanz:1996fb,Covi:1996wh,Pilaftsis:1998pd,Pilaftsis:2005rv,Pilaftsis:2003gt}, where the CP violation can be enhanced in $N_i-N_j$ mixing.
This scenario is discussed in Section \ref{lessh}.

Other leptogenesis scenarios, some of which work at lower reheat
temperatures, include the following:
\begin{itemize}
\item ``Soft leptogenesis''
  \cite{Grossman:2003jv,D'Ambrosio:2003wy,Grossman:2004dz}, which
can work even in a one-generation SUSY seesaw. Here the source of both
lepton number violation and CP violation is a set of soft
supersymmetry breaking terms.
\item The  Affleck Dine mechanism
  \cite{Affleck:1984fy,Dine:1995kz,Hamaguchi:2002vc} where the scalar
  condensate carries lepton number. For a detailed discussion,
see, for instance, \cite{Hamaguchi:2002vc}.
\item The $N_i$'s could be produced non-thermally, for instance in inflaton decay
\cite{Lazarides:1991wu,Asaka:1999yd}, or  in preheating
\cite{Boubekeur:2002gv,Giudice:1999fb}.
\item The singlet sneutrino $\tilde{N}$ could be the inflaton, which
  then produces a lepton asymmetry in its decays  \cite{Ellis:2003sq}.
\item Models with a ({\it e.g.} flavour) symmetry  that breaks
below  the leptogenesis
    scale, and  involve additional heavy states that carry lepton number
    \cite{AristizabalSierra:2007ur}.
\end{itemize}

\subsection{Reading this review}
\label{reading}

This review contains chapters at different levels of detail,
which might not all be equally interesting to all readers.
Analytic  formulae, which can be used to 
estimate the baryon asymmetry, can be found in Appendix
\ref{recipes}.  Appendix   \ref{notn} is a dictionary
of  the notation used in this review. 
A reader interested in a qualitative understanding 
and wishing to avoid Boltzmann Equations (BE), 
can read Section  \ref{models}, perhaps Section \ref{BLV},
Sections \ref{toy}, \ref{ssCP} and  \ref{esteff},
and browse Sections  \ref{sec:flavor}  and  \ref{variations}.

For a more quantitative understanding
of thermal leptogenesis,   some  acquaintance 
with the BE  is required.
In addition to the Sections listed above,
readers may wish to review  preliminary 
definitions in  Appendix \ref{notn},  read
 the introduction to simple BE of
Sections \ref{sec:BE} and  \ref{simplestBE},
and browse Sections  
 \ref{sec:thermal} and \ref{sec:spec}, and Appendix  \ref{chempot}.

{\it Aficionados} of  leptogenesis may also
be interested in Section  \ref{sec:1to3-2to3},
in which  more complete BE are derived, and  Appendix
\ref{apenrho} that describes the toy model
which motivates our  claims about flavour effects.

The layout of the review is the following: 
In Section \ref{models} we briefly review the seesaw mechanism and
describe various useful parameterizations thereof.
Section \ref{BLV} introduces 
the non-perturbative $B+L$ violating interactions
which are a crucial ingredient for baryogenesis through leptogenesis.
 Section \ref{toy}
gives  an overview of leptogenesis, using rough estimates
to motivate the qualitative  behaviour.  
Flavour dependent quantities (CP asymmetries and efficiencies) 
are introduced already in this section: flavour blind equations  
 in section \ref{toy} 
 would be  inapplicable later in the review. 
The calculation of the CP
asymmetry and bounds thereon are presented in Section \ref{sec:CP}. Subsections
\ref{comments} and \ref{secCPscat} discuss the implications of CPT and
unitarity, 
relevant to a reader who wishes to derive the BE. 
Simple   BE 
that describe interactions  mediated by the neutrino
Yukawa coupling  are derived in Section \ref{simplestBE}, using definitions
from Section \ref{sec:BE}. (Formulae for the
number  densities  and definitions of the rates, calculated
at zero temperature, can be found in  Appendix  \ref{kinetic}.)
In Section \ref{sec:1to3-2to3}
 the most
relevant scattering interactions, and  a large set of ${\cal
  O}(1)$ effects, are included. 
 The next three sections refine the analysis
by implementing various ingredients that are omitted in the basic
calculation. Section \ref{sec:thermal} is an overview of  finite temperature
effects. Section \ref{sec:spec} analyzes the impact of spectator
processes. (These are often
described  by chemical equilibrium  conditions,
which are reviewed  in Appendix  \ref{chempot}.)
 Section \ref{sec:flavor} is devoted to a detailed discussion
of flavour effects. In section \ref{variations}, we present some
variations on the simplest framework: non-hierarchical singlets 
(subsection \ref{lessh}), soft leptogenesis (subsection \ref{soft}),
Dirac leptogenesis (subsection \ref{dirac}), leptogenesis with scalar
triplets (subsection \ref{triplet}), and with fermion triplets (subsection 
\ref{sec:triplet}). Our conclusions and some prospects for future 
developments are summarized in Section \ref{conclusions}.
We collect all the symbols that are used in this review into a table
in Appendix \ref{notn}, where we also give the equation number where
they are defined. The basic ingredients and consequences of kinetic
equilibrium are reviewed in Appendix \ref{kinetic}, while those of
chemical equilibrium are reviewed in Appendix \ref{chempot}. The
BE used in the main text are not covariant under transformations of the
lepton flavour basis. Appendix \ref{apenrho} discusses, via a toy
model, covariant equations for a ``density matrix'', and explains how 
a flavour basis for the BE is singled out by the dynamics.
  Approximate solutions to simple BE are given in Appendix \ref{recipes}.

\newpage
\section{Seesaw Models}
\label{models}

Measurements of fluxes of solar, atmospheric, reactor and accelerator
neutrinos give clear evidence that (at least two of) the observed
weakly interacting neutrinos have small masses. Specifically, two
neutrino mass-squared differences have been established,
\beq
\Delta m^2_{21}=(7.9\pm0.3)\times10^{-5}\ \mbox{eV}^2 \equiv m_{\rm
  sol}^2,\ \ \
|\Delta m^2_{32}|=(2.6\pm0.2)\times10^{-3}\ \mbox{eV}^2 \equiv m_{\rm
  atm}^2.
\eeq
If these are  ``Majorana'' masses, then they violate lepton number
and  correspond to a mass matrix of the following form:
\beq
{\cal L}_{m_\nu}=\frac{1}{2} \overline{\nu^c}_\alpha [m]_{\alpha
  \beta} \nu_\beta + {\rm h.c.}.
\label{mlight}
\eeq
The mass terms $[m]_{\alpha\beta}$ break the $SU(2)_L$ gauge symmetry
as a triplet. Consequently, they cannot come from renormalizable
Yukawa couplings with the Standard Model Higgs doublet. (This stands
in contrast to the charged fermion masses which break $SU(2)_L$ as
doublets and are generated by the renormalizable Yukawa couplings.)
Instead, they are likely to arise from the dimension
five operator $(\ell_\alpha \phi) (\ell_\beta \phi)$. Seesaw models
are full high energy models which induce this effective low energy
operator at tree level. Seesaw models are attractive because they
naturally reproduce the small masses of the doublet
neutrinos. Explicitly, if the exchanged particle has a mass $M$ which,
by assumption, is much heavier than the electroweak breaking scale
$v_u$, then the light neutrino mass scale is $v_u^2/M$, which is much
smaller than $v_u$, the charged fermion mass scale.

There are three types of seesaw models, which differ by the properties
of the exchanged heavy particles:
\begin{itemize}
\item {\bf Type I}: $SU(3)\times SU(2)\times U(1)$-singlet fermions;
\item {\bf Type II}: $SU(2)$-triplet scalars;
\item {\bf Type III}: $SU(2)$-triplet fermions.
\end{itemize}
We now describe these three types of seesaw models in more detail,
with particular emphasis on the type I seesaw, which is well motivated
by various extensions of the SM. We also comment on the supersymmetric
seesaw framework.

\subsection{Singlet fermions (Type I)}
\label{type1}
In the type I seesaw
\cite{Minkowski:1977sc,Yanagida:1979as,GellMann:1980vs,Mohapatra:1980yp},
two or three singlet fermions $N_i$ (sometimes referred to as
``right-handed neutrinos'') are added the Standard Model. They are
assumed to have large Majorana masses. The leptonic part of the
Lagrangian can be written in the mass basis of the charged leptons and
of the singlet fermions as follows:\footnote{The Yukawa indices
  are ordered left-right, and the definition of $\lambda$ is aimed to
  reproduce the Lagrangian that corresponds to the superpotential of
  eqn (\ref{superpot}).}
\bea
{\cal L}& =&+ [{\bf h}]^*_{ \beta}(
{\overline{\ell}}_\beta { \phi^{c*}}) \, e_{R \beta} - [{\bf
  \lambda}]^*_{\alpha k} ( {\overline{\ell}}_\alpha \phi^* ) \, N_k -
\frac{1}{2}{\overline{N_{j}}}{M_j}{N_{j}^c}+{\rm h.c.},
\label{L}
\eea
where $\phi$ is the Higgs field, with vev $v_u \simeq 174$ GeV, and
the parentheses indicate anti-symmetric SU(2) contraction:
\beq
(\overline{\ell}_\alpha  { \, \phi^*})=
\left( \overline{\nu_L}_{ \alpha}, \overline{e_L}_{ \alpha} \right)
\left(\begin{array}{cc} 0 & -1 \\ 1 & 0 \end{array} \right)
\left(\begin{array}{c} \phi^- \\ \phi^{*0} \end{array} \right).
\label{contraction}
\eeq
Twenty-one parameters are required  to fully determine the Lagrangian
of eqn (\ref{L}). To see this, notice that it is always possible to choose
 a basis for the
$N_i$'s where the mass matrix $M$ is diagonal  $M =D_M$, with 
three positive and real
eigenvalues. Similarly, one can choose a basis for the
$\ell_\alpha$ and $e_{R\beta}$ such that the charged lepton Yukawa
matrix ${\bf h}$ is diagonal and real
(in particular, ${\bf h} {\bf h}^\dagger = D_{{\bf h}}^2$)  
 giving  other three parameters.
Then, in this basis,  
the neutrino Yukawa matrix $\lambda$ is  a generic
complex matrix, from which three phases can be removed by phase
redefinitions on the $\ell_\alpha$, leaving   
9 moduli and 6 phases as physical parameters. 
Therefore there are in total 21 real parameters 
for the lepton sector. See \cite{Santamaria:1993ah} for a more elegant
counting, in particular of the phases.

If the effective mass matrix $m$ of the light neutrinos is normalized
as in eqn (\ref{mlight}) then, in the charged lepton mass basis, it is
given by
\beq
[m]_{\alpha \beta} = [{\mathbf \lambda}]_{\alpha k}M_k^{-1}
[{\bf  \lambda}]_{\beta k} v_u^2.
\label{mass}
\eeq
One has to take care of using a consistent set of definitions of the
mass and the Higgs vev \cite{GonzalezGarcia:2002dz}. In particular,
note the factor of $1/2$ in eqn (\ref{mlight}) and our use of
$v_u=174\>$GeV, both of which are important in the relation
(\ref{mass}). These choices are also important for the proof of
eqn (\ref{tildemin}), so $v_u = 174$ GeV should be used in the
definition of $\tilde{m}$ and $m_*$ of eqn (\ref{tildem}).

The leptonic mixing matrix $U$ is extracted by diagonalizing $[m]$:
\beq\label{defU}
[m] = U^* D_m U^\dagger,
\eeq
where $D_m={\rm diag} \{m_1, m_2, m_3 \}$. The matrix $U$ is
$3\times3$ and unitary and therefore depends, in general, on six
phases and three mixing angles. Three of the phases can be removed by
redefining the phases of the charged lepton doublet fields. Then, the
matrix $U$ can be conveniently parametrized as
\bea
U= \hat{U}\cdot {\rm diag}(1 ,e^{i\alpha} ,e^{i\beta}) ~~~,
\label{UV}
\eea
where $\alpha$ and $\beta$ are termed ``Majorana'' phases (if
the neutrinos were Dirac particles, these two phases could be removed
by redefining the phases of the neutrino fields), and $\hat{U}$ can be
parametrized in a fashion similar to the CKM matrix:
\beq
 \label{Vdef}
\hat{U}= \left[ \begin{array}{ccc}
c_{13}c_{12} & c_{13}s_{12} & s_{13}e^{-i\delta} \\
-c_{23}s_{12}-s_{23}s_{13}c_{12}e^{i\delta} &
 c_{23}c_{12}-s_{23}s_{13}s_{12}e^{i\delta} & s_{23}c_{13} \\
s_{23}s_{12}-c_{23}s_{13}c_{12}e^{i\delta} &
 -s_{23}c_{12}-c_{23}s_{13}s_{12}e^{i\delta} &
  c_{23}c_{13}
\end{array} \right]  ~~~,
\eeq
where $c_{ij}\equiv\cos\theta_{ij}$ and
$s_{ij}\equiv\sin\theta_{ij}$. There are various other possible phase
conventions for $U$ (see {\it e.g} the first appendix of
\cite{King:2002nf} for a list and translation rules).

In the leptonic sector of the SM augmented with the Majorana neutrino
mass matrix of eqn (\ref{mass}), there are twelve physical parameters:
The 3 charged lepton masses $m_e, m_\mu, m_\tau$, the 3 neutrinos
masses $m_1, m_2, m_3$, and the 3 angles and 3 phases of the mixing matrix
$U$. Seven of these parameters are measured ($m_e,m_\mu,m_\tau,\Delta
m^2_{21},|\Delta m^2_{32}|,s_{12},s_{23}$). There is an upper bound on
the mixing angle $s_{13}$. The mass of the lightest neutrino and three
phases of $U$ are unknown.

In addition, there are 9 unknown parameters in the high scale
theory. They are, however, relevant to leptogenesis.

\subsubsection{Parametrizing the seesaw}
\label{sec:seepar}
 
The  usual {\bf ``top-down parametrization''} of the theory,
which applies at energy scales $\Lambda \gappeq M_i$, is given in
eqn (\ref{L}).  To relate various parametrizations of the seesaw, 
it is useful to
diagonalize $\lambda$, which can be done with a bi-unitary
transformation:
\beq
\lambda = V_L^\dagger D_{\lambda} V_R
\label{2.lambda}
\eeq
Thus in the top-down approach, the lepton sector can be described by
the nine eigenvalues of $D_M , D_{\lambda}$ and $D_{{\bf h}}$, and the
six angles and six phases of $V_L$ and $V_R$. In this parametrization,
the inputs are masses and coupling constants of the propagating
particles at energies $\Lambda$,  so it makes ``physical'' sense.

Alternatively, the (type I) seesaw Lagrangian of eqn (\ref{L}) can be
described with inputs from the left-handed sector
\cite{Davidson:2001zk}.  This is referred to as a {\bf ``bottom-up
  parametrization''}, because the left-handed ($SU(2)_L$-doublet)
particles have masses $\lappeq$ the weak scale. $D_{{\bf h}}$, $U$ and
$D_{m}$, can be taken as a subset of the inputs. To identify the
remainder, consider the $\ell$ basis where $m$ is diagonal,
so as to emphasize the parallel between this parametrization and the
previous one (this is similar to the $N$ basis being chosen to
diagonalize $M$). If one knows $\lambda \lambda^\dagger \equiv
W_L^\dagger D_{\lambda}^2 W_L$ in the $D_{m}$ basis, then the $N$
masses and mixing angles can be calculated:
\beq
 M^{-1}=  D_\lambda^{-1} W_L D_m W_L^T D_\lambda^{-1}
  =   V_R D_M^{-1} V_R^T
\eeq
In this parametrization, there are three possible basis choices for
the $\ell$ vector space: the charged lepton mass eigenstate basis
($D_{{\bf h}}$), the neutrino mass eigenstate basis ($D_m$), and the
basis where the $\lambda$ is diagonal. The first two choices are
physical, that is, $U$ rotates between these two bases. $D_{{\bf
 h}}$, $D_m$ and $U$ contain the 12 possibly measurable parameters
of the SM seesaw. The remaining 9 parameters can be taken to be
$D_{\lambda}$ and $V_L$ (or $W = V_L U^*$). If supersymmetry (SUSY) is
realized in nature, these parameters  may contribute to the Renormalization
Group (RG) running of the slepton mass matrix \cite{Borzumati:1986qx}.

The {\bf Casas-Ibarra  parametrization} \cite{Casas:2001sr} is very
convenient for calculations. It uses the three diagonal matrices
$D_M$, $D_m$ and $D_{{\bf h}{\bf h}^\dagger}$, the unitary matrix $U$
defined in eqn (\ref{defU}) and a complex orthogonal matrix $R$. In
the mass basis for the charged leptons and for the singlet fermions,
which is used in the Lagrangian of eqn (\ref{L}), $R$ is given by
\beq
R = D_m^{-1/2}U^T \lambda D_{M}^{-1/2} v_u.
\label{eq:R}
\eeq
$R$ can be written  as $R =\hat{R}~{\rm diag} \{\pm 1,\pm 1,\pm 1\}$
where  the $\pm1$ are related to the CP parities of the $N_i$, and
$\hat{R}$ is  an orthogonal complex matrix
\beq
\hat{R} =  \left[ \begin{array}{ccc}
c_{13}c_{12} & c_{13}s_{12} & s_{13} \\
-c_{23}s_{12}-s_{23}s_{13}c_{12} &
 c_{23}c_{12}-s_{23}s_{13}s_{12} & s_{23}c_{13} \\
s_{23}s_{12}-c_{23}s_{13}c_{12} &
 -s_{23}c_{12}-c_{23}s_{13}s_{12} &
  c_{23}c_{13}
\end{array} \right]  ~~~,
\label{R}
\eeq
where $c_{ij}=\cos z_{ij}$,  $s_{ij}=\sin z_{ij}$, with $z_{ij}$  complex
angles.

An alternative parametrization \cite{Pascoli:2003rq}, that can be
useful for quasi-degenerate $N$ or $\nu$ masses, is given by
\beq
R = O \times \exp\{iA\},
\eeq
where $O$ is a real orthogonal matrix, and $A$ a real anti-symmetric
matrix.

In summary, the lepton sector of the seesaw extension of the SM can be
parametrized with $D_{{\bf h}}$,  the real eigenvalues of two more
matrices, and the transformations among the bases where the matrices
are diagonal.  The  matrices-to-be-diagonalized can be chosen in
various ways:
\begin{enumerate}
\item `` top-down'' -- input the $N$ sector: ${ D_M}$,
${ D_{ \lambda^\dagger\lambda}},$ and ${ V_R}$ and $V_L$.
\item `` bottom-up'' -- input the $\nu_L$ sector:  ${ D_m}$,
${ D_{\lambda \lambda^\dagger}},$ and ${ V_L}$ and $U$.
\item ``intermediate'' -- the Casas-Ibarra parametrization:
${ D_M}$, ${ D_m}$, and $U$ and $R$.
\end{enumerate}

\subsubsection{The two right-handed neutrino model}
\label{sec2rhn}
The minimal seesaw models that are viable have only two $N_i$'s
\cite{Frampton:2002qc}. Such models, known as ``two right-handed
neutrino'' (2RHN) models, have a strong predictive power. The 2RHN
model can be thought of as the limit of the three generation model,
where $N_3$ decouples from the theory because it is either very heavy
or has very small couplings, that is,
$(|\lambda_{3\alpha}|^2M_3^{-1})/(|\lambda_{i\alpha}|^2M_i^{-1})\to0$
for $i=1,2$. In the 2RHN models, $M$ is a $2\times2$ matrix, and
$\lambda$ is a $3\times2$ matrix. It was originally introduced and
studied with a particular texture
\cite{Frampton:2002qc,Raidal:2002xf}, and later studied in general
(see {\it e.g.} \cite{Guo:2006qa} for a review).
Here we follow the parametrization of the general model used in
\cite{Endoh:2002wm}.

The Lagrangian of the lepton sector is of the form of eqn (\ref{L}),
but with the sum over the singlet indices restricted to two
generations ($\alpha,\beta=e,\mu,\tau$; $j,k=1,2$). The model has
fourteen independent parameters, which can be classified as follows: 5
masses ($m_e,m_\mu,m_\tau,M_1,M_2$), and 9 real parameters in
$\lambda$ (three phases can be removed from the six complex elements
of $\lambda$ by phase redefinitions on the doublets).

In this model, the lightest neutrino is massless, $m_1=0$, and so its
associated phase vanishes, $\alpha=0$. This situation implies that
there are ten parameters that can be determined by low energy physics,
instead of the usual twelve.

In the mass eigenstate bases of charged leptons and singlet fermions,
the $3 \times 2 $ matrix $\lambda$  can be written, analogously to eqn
(\ref{2.lambda}), as follows:
\beq
\left[V_L\right]^\dagger\left[
\begin{array}{ll}
0 & 0 \\ \lambda_2 & 0 \\ 0 & \lambda_3
\end{array} \right]
\left[V_R\right]
\eeq
where $[V_R]$ is a $2\times 2$ unitary matrix with one angle and one
phase, $[V_L]$ is a $3\times3$ matrix which can be written as $U^*
W^\dagger$, with 4 of the six off-diagonal elements of $W$ vanishing.

As mentioned above, the main prediction of this model is that one of the
doublet neutrinos is massless. This excludes the possibility of three
quasi-degenerate light neutrinos, and leaves only two other options:
\begin{enumerate}
\item Normal hierarchy, with $m_1=0$, $m_2=m_{\rm sol}$, $m_3=m_{\rm atm}$.
\item Inverted hierarchy, with $m_3=0$, $m_{1,2}\approx m_{\rm atm}$,
  $m_2-m_1=m_{\rm sol}^2/(2m_{\rm atm})$.
\end{enumerate}

\subsection{Triplet scalars (Type II)}
\label{seesawii}
One can generate neutrino masses by the tree level exchange of
$SU(2)$-triplet scalars
\cite{Mohapatra:1980yp,Magg:1980ut,Schechter:1980gr,%
Wetterich:1981bx,Lazarides:1980nt}.
These $SU(2)$-triplets should be color-singlets and carry hypercharge
$Y=+1$ (in the normalization where the lepton doublets have $Y=-1/2$).
In the minimal model, there is a single such scalar, which we denote
by $T$. The relevant new terms in the Lagrangian are
\beq\label{ltrisca}
{\cal L}_T=-M_T^2|T|^2+\frac12\left([\lambda_L]_{\alpha\beta}\,\ell_\alpha
\ell_\beta T +M_T\lambda_\phi\,\phi\phi T^*+{\rm h.c.}\right).
\eeq
Here, $M_T$ is a real mass parameter, $\lambda_L$ is a symmetric
$3\times3$ matrix of dimensionless, complex Yukawa couplings, and
$\lambda_\phi$ is a dimensionless complex coupling.

The exchange of scalar triplets generates an effective dimension-5
operator $\ell\ell\phi\phi$ which, in turn, leads to neutrino
masses. The triplet contribution to the neutrino masses, $m_{\rm II}$, is
\beq
[m_{\rm II}]_{\alpha\beta}=[\lambda_L]_{\alpha\beta}\frac{\lambda_\phi
  v_u^2}{M_T}.
\eeq

This model for neutrino masses has eleven parameters beyond those of
the Standard Model: 8 real and 3 imaginary ones. Of these, $6+3$ can
in principle be determined from the light neutrino parameters, while 2
($M_T$ and $|\lambda_\phi|$) are related to the full high energy theory.

The model involves lepton number violation because the co-existence of
$\lambda_L$ and $\lambda_\phi$ does not allow a consistent way of
assigning a lepton charge to $T$, and new sources of CP violation via
the phases in $\lambda_L$ and $\lambda_\phi$.

The supersymmetric triplet model \cite{Hambye:2000ui} must include,
for anomaly cancellation, two triplet superfields, $T$ and $\bar T$,
of opposite hypercharge. Only one of these couples to the leptons. The
relevant superpotential terms are
\beq
W_T=M_T T\bar T+\frac{1}{\sqrt2}\left([\lambda_{L}]_{\alpha\beta}\,L_\alpha
TL_\beta+\lambda_{H_d}H_d TH_d+\lambda_{H_u}H_u\bar TH_u\right),
\eeq
leading to
\beq
[m_{\rm II}]_{\alpha\beta}=[\lambda_L]_{\alpha\beta}\frac{\lambda_{H_u}
  v_u^2}{M_T}.
\eeq

\subsection{Triplet fermions (Type III)}
\label{seesawiii}
One can generate neutrino masses by the tree level exchange of
$SU(2)$-triplet fermions $T_i^a$ \cite{Foot:1988aq,Ma:1998dn,Ma:2002pf}
($i$ denotes a heavy mass eigenstate while $a$ is an SU(2)
index). These $SU(2)$-triplets should be color-singlets
and carry hypercharge $0$. The relevant Lagrangian terms have a form
that is similar to the singlet-fermion case (\ref{L}), but the
contractions of the SU(2) indices is different, so we here show it
explicitly:
\beq
{\cal L}_{T^a}=[\lambda_T]_{\alpha
  k}\tau^a_{\rho\sigma}\ell_\alpha^\rho \phi^\sigma T_k^a-\frac12
  M_iT_i^aT_i^a+{\rm h.c.}.
\eeq
Here, $M_i$ are real mass parameters, while $\lambda_T$ is a
$3\times3$ matrix of dimensionless, complex Yukawa couplings.

The exchange of fermion triplets generates an effective dimension-5
operator $\ell\ell\phi\phi$ which, in turn, leads to neutrino
masses. The triplet contribution to the neutrino masses, $m_{\rm III}$, is
\beq
[m_{\rm III}]_{\alpha\beta}=[\lambda_T]_{\alpha k}M_k^{-1}\lambda_{\beta k}
v_u^2.
\eeq

As in the standard seesaw model, this model for neutrino masses has
eighteen parameters beyond those of the Standard Model: 12 real and 6
imaginary ones.

The model involves lepton number violation because the co-existence of
$\lambda_T$ and $M_k$ does not allow a consistent way of
assigning a lepton charge to $T_k$, and new sources of CP violation via
the phases in $\lambda_T$.

\subsection{Supersymmetry and singlet fermions}
\label{sec:susy}
One of the motivations of supersymmetry (SUSY) is to cancel
quadratically divergent contributions to the Higgs mass. In the seesaw
extension of the SM, a large mass scale $M_i$ is present. This
results in an additional set of corrections of $O(\lambda^2
M^2)$.\footnote{The requirement of no excessive fine tuning in the
  cancellation of these contributions has been used to set the bound
  $M_i\lsim 10^7\,$GeV~\cite{Casas:2004gh}.}  Cancelling the
contributions from the large seesaw scale motivates the supersymmetric
version of the model.

The superpotential for the leptonic sector of the type I seesaw  is
\beq
W=\frac{1}{2}N^c_i M_iN^c_i +  (L_\alpha H_u) [\lambda]_{ \alpha i} N^c_i -
 (L_\alpha H_d)[{\mathbf h}]_{ \alpha}  E^c_\alpha ,
\label{superpot}
\eeq
where $L_\alpha$ and $E^c_\alpha$ are respectively the $SU(2)$ doublets and
singlets lepton superfields, and $H_u$ and $H_d$ are the Higgs
superfields. (To resemble the SM notation, the scalar components of
the Higgs superfields will be denoted as $\phi_{u}$ and $\phi_{d}$.)
The $SU(2)$ contractions in parentheses are anti-symmetric,
as in eqn (\ref{contraction}).
Both Higgs bosons, $\phi_u$ and $\phi_d$, have vacuum expectation
values (vevs): $\langle \phi_i \rangle \equiv v_i$. Their ratio is
defined as
\beq
\tan\beta\equiv \frac{v_u}{v_d}.
\label{tgbeta}
\eeq
When including ``flavour effects'' in supersymmetric leptogenesis, the
value of $\tan \beta$ is relevant, because the Yukawa coupling
${\bf h}_{e,\mu ,\tau}\propto m_{e, \mu , \tau}/\cos\beta$.

To agree with experimental constraints (while keeping supersymmetry as
a solution to the $m_H^2$-fine-tuning problem), it is important to add
soft SUSY-breaking terms:
\beq\label{eq:sosu}
\frac{1}{2}
[\widetilde{m}^2_L ]_{\alpha \beta}\tilde{L}^*_\alpha \tilde{L}_\beta
+\ldots + \frac{1}{2}\tilde{N}^c_i [BM]_{ij}\tilde{N}^c_j +
[A\lambda]_{ \alpha i} ( \tilde{L}_\alpha \phi_u )  \tilde{N}^c_i
 +
[A{\mathbf h}]_{\beta \alpha } ( \tilde{L}_\alpha \phi_d ) \tilde{E}^c_\alpha  +
{\rm h.c.},
\eeq
where the ... represent soft masses-squared for all the scalars. In
the thermal leptogenesis scenario that this review concentrates on,
the soft SUSY breaking terms give $O(m^2_{\rm SUSY}/M^2)$ corrections,
and can be neglected. In other mechanisms, such as soft
leptogenesis \cite{Grossman:2003jv,D'Ambrosio:2003wy,Grossman:2004dz}
and Affleck-Dine leptogenesis
\cite{Affleck:1984fy,Hamaguchi:2002vc,Dine:1995kz}, the soft
parameters play a central role.

The interesting feature of the SUSY seesaw, is that the neutrino
Yukawa couplings may contribute to the RG running
of the soft slepton mass matrix, and induce flavour-changing
mass-squared terms. Consider, for example, the Type I seesaw, with
universal soft masses at some high scale $\Lambda$:
$[\tilde{m}^2_L]_{\alpha \beta} = m_0^2 \delta_{\alpha \beta}$,
$[A \lambda]_{\alpha i} = A_0 [\lambda]_{\alpha i}$.
Then  at the electroweak scale, in the flavour basis,
the RG contributions to the off-diagonal elements of the
mass-squared matrix can be estimated at leading log as follows:
\beq
[\tilde{m}^2_L]_{\alpha \beta} = -\frac{3 m_0^2 + A_0^2}{16 \pi^2}
\sum_i
\lambda_{\alpha i} \log \frac{M^2_i}{\Lambda^2} \lambda^*_{\beta i}
\label{LFV}
\eeq
In general, the soft mass matrix is unknown. One can argue, however,
that the flavour-changing mass-squared matrix elements (off-diagonal in
the flavour basis) are at least of order eqn (\ref{LFV}), because we
do not expect fine-tuned cancellations among different
contributions. The flavour off-diagonal terms induce processes like
$\mu\to e\gamma$, $\tau\to\mu\gamma$ and $\tau\to e\gamma$
\cite{Borzumati:1986qx}, at rates of order \cite{Hisano:1995cp}
\beq
\frac{\Gamma(\ell_\alpha\to\ell_\beta \gamma)}
{\Gamma(\ell_\alpha\to\ell_\beta \nu \bar{\nu})}=\frac{\alpha^3}{G^2_F}
\frac{|[\tilde{m}^2_L]_{\alpha \beta}|^2}{{m}_{\rm susy}^8}
(1+\tan^2\beta),
\eeq
where $m_{\rm susy}$ is a typical slepton mass. For reasonable values
of the unknown parameters, one obtains predictions
\cite{Hisano:1995cp,Lavignac:2001vp,Casas:2001sr,Petcov:2005jh}   that
are in the range of current or near-future experiments
\cite{Yao:2006,Hayasaka:2007vc,Aubert:2005wa}.

The neutrino Yukawa couplings make real and imaginary contributions to
the soft parameters. These phases give contributions to lepton
electric dipole moments \cite{Ellis:2001yza,Masina:2003wt,Farzan:2004qu}
which are orders of magnitude below current bounds, but possibly
accessible to future experiments.

The seesaw contribution to RG running of the slepton masses is
relevant to leptogenesis, because the slepton mass matrix could be an
additional low energy footprint of the seesaw  model. As discussed
in Section \ref{type1}, it is possible to reconstruct the Yukawa
matrix $\lambda$ and the masses $M_i$, from the light neutrino mass
matrix $m_\nu$ of eqn (\ref{mass}), and the Yukawa combination
$\lambda \lambda^\dagger$ that enters the RG equations. So it is interesting
to study correlations between low energy observables and a large
enough baryon asymmetry at low enough $T_{\rm reheat}$. Early
(unflavoured) studies can be found in
refs. \cite{Davidson:2003cq,Akhmedov:2003dg}, and  in many other
works: \cite{Petcov:2006pc} (degenerate $N_i$),
\cite{Petcov:2005jh} (hierarchical $N_i$), \cite{Pascoli:2003rq}
(degenerate light neutrinos), and \cite{Rossi:2002zb} (type~II).
Recent flavoured analyses can be found in refs.  
\cite{Antusch:2006gy,Antusch:2006cw}
(hierarchical $N_i$) and \cite{Cirigliano:2006nu,Branco:2006hz}
(degenerate $N_i$).

\newpage

\section{Anomalous $B+L$ Violation}
\label{BLV}
The aim of this section is to give a qualitative
introduction\footnote{This Section is based on a lecture given by V.
  Rubakov at the Lake Louise Winter Institute, 2008.}  to the
non-perturbative baryon number violating interactions that play a
crucial role in leptogenesis. A similar discussion can be found in
Ref. \cite{Chen:2007fv}, while more details and references can be
found, for instance, in Section 2 of Ref. \cite{Rubakov:1996vz}, and
in Refs. \cite{Polyakov,VARbook,Vainshtein:1981wh,Coleman:1978ae}.

{}From a theoretical perspective, the baryon number $B$ and the three
lepton flavour numbers $L_\alpha$ are conserved in the renormalisable
Lagrangian of the Standard Model. Furthermore, experimentally, the
proton has not been observed to decay: $\tau_p\gsim10^{33}$ years
\cite{Shiozawa:1998si,Kobayashi:2005pe}.  (For a review of proton
decay, see \cite{Nath:2006ut}.)  However, due to the chiral anomaly,
there are non-perturbative gauge field configurations
\cite{'tHooft:1976up,'tHooft:1976fv,Callan:1976je} which can act as
sources for $B+L_e+L_\mu+L_\tau$. (Note that $B-L_e-L_\mu-L_\tau$ is
conserved.)  In the early Universe, at temperatures above the
electroweak phase transition (EWPT), such configurations occur
frequently \cite{Kuzmin:1985mm,Linde:1977mm,Dimopoulos:1978kv}, and
lead to rapid $B+L$ violation. These configuations are commonly
referred to as ``sphalerons''
\cite{Klinkhamer:1984di,Arnold:1987mh,Arnold:1987zg}.

\subsection{The chiral anomaly}

For a pedagogical introduction to the chiral  anomaly, see
for instance Ref. \cite{Polyakov}.

Consider the Lagrangian for a massless Dirac fermion $\psi$ with
$U(1)$ gauge interactions:
\beq
{\cal L} =  \overline{\psi}  
\gamma^\mu (\partial_\mu  - i  A_\mu) 
\psi - \frac{1}{4e^2} F_{\mu \nu} F^{\mu \nu}.
\label{BLtoyU1}
\eeq 
It is invariant under the local symmetry:
\beq
\psi(x) \rightarrow e^{i  \theta(x)} \psi(x) ~, ~~~~
A_\mu(x)  \rightarrow A_\mu(x)  +   \partial_\mu \theta(x). 
\label{BLV:gauge}
\eeq
It is also invariant under a global ``chiral'' symmetry: 
\beq
\psi (x) \rightarrow e^{i \gamma_5 \phi} \psi(x).   
\eeq
The associated current,
\beq
j^\mu_5 =  \overline{\psi} \gamma_5 \gamma^\mu  \psi, 
\eeq 
is conserved at tree level, but not in the quantum theory. This can be
related to the regularization of loops---renormalization introduces a
scale, and the scale breaks the chiral symmetry, as would a fermion
mass (see, for instance, chapter 13 of Ref. \cite{Polyakov}).  Indeed, at
one loop, one finds
\beq
\partial_\mu j^\mu_5 =  \frac{1}{16 \pi^2} 
\widetilde{F}_{\mu \nu} F^{\mu \nu} =   \frac{
\epsilon_{\rho \sigma \mu \nu}}{16 \pi^2} F^{\rho \sigma} F^{\mu \nu} ~.
\label{BLV:div}
\eeq 
The right-hand side can be written as a total divergence involving
gauge fields, and is related to their topology: it counts the
``winding number'', or Chern-Simons number, of the field
configuration.  (An instructive 1+1 dimensional model, where the
topology is easy to visualize, can be found in Ref. \cite{Coleman:1978ae}.)
In four dimensions, the space-time integral of the right-hand side of
eqn (\ref{BLV:div}) vanishes for an Abelian gauge field, but can be
non-zero for non-Abelian fields.

In the context of leptogenesis, we are looking for an anomaly in the
$B+L$ current. Within the four-dimensional SM, it arises due to the
$SU(2)$ gauge interactions,
which are chiral and non-Abelian.  We neglect other interactions in
the following (see Ref. \cite{Rubakov:1996vz} for a discussion of the
effects of Yukawa and $SU(3)_C\times U(1)_Y$ interactions).
The fermions that are relevant to our discussion are the three
generations of quark and lepton doublets: $ \{ \psi_L^i \} = \{
q_L^{a, \beta} , \ell_L^\alpha \} $, where $\alpha, \beta$ are
generation indices, $a,b$ are colour indices, and $A,B$ are 
$SU(2)$ indices.  The Lagrangian terms for the $SU(2)$ gauge
interactions read
\beq {\cal L} = \sum_i \overline{\psi_L}^i \gamma^\mu (\partial_\mu -
i \frac{g}{2} \sigma^A W^A_\mu) \psi_L^i.
\label{BLtoy}
\eeq 
It has twelve global $U(1)$  symmetries (one for each  field):
\beq
\psi_L^i(x) \rightarrow e^{i  \beta} \psi(x)_L^i  ~~~.
\eeq
The chiral currents   associated  to these transformations,
\beq
j^i_\mu =  \overline{\psi_L}^i  \gamma_\mu  \psi_L^i,
\eeq 
are   conserved at tree level, but are ``anomalous''
in the quantum theory:
\beq
\partial^\mu j_\mu^{~i} = \frac{1}{64 \pi^2}
{F}_{\mu \nu}^A \widetilde{  F}^{\mu \nu A}.  
\eeq 

\begin{figure}[ht]
\hskip4cm
 \includegraphics[width=8cm,height=6cm,angle=0]{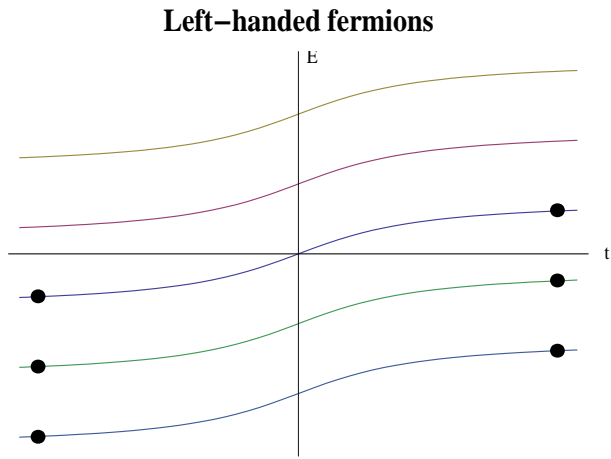}
 \caption[]{
Evolution with time  of the energy eigenstates of chiral fermions
in a gauge field background  with $\tilde{F} F \neq 0$. }
\label{figdeVAR}
\end{figure}

Let us define $Q^i(t)=\int j_0^{~i} d^3 x$,~ $\Delta
Q^{\>i}=Q^{\>i}(+\infty)-Q^{\>i}(-\infty)$, and let us suppose for the moment
that there exist field configurations such that
\beq
\Delta  Q^{~i} =   \frac{1}{64 \pi^2} \int d^4 x
 { F}_{\mu \nu}^A \widetilde{  F}^{\mu \nu A}  
\label{BLvabove}
\eeq
is a non-zero integer. This implies that fermions will be created,
even though there is no perturbative interaction in the Lagrangian
that generates them. One way \cite{Christ:1980ku} to understand
``where they come from'' is to think in the Dirac sea picture, and
place the chiral fermions $\{ \psi_L^i \}$ in an external gauge field
for which the right hand side of eqn (\ref{BLvabove}) is non-zero.  In
the ground state at $t \rightarrow - \infty$, all the negative energy
states are filled, and all the positive energy states are empty. As
the fermions are massless, there is no mass gap at $E = 0$.  At any
given $t$, one can solve for the eigenvalues of the fermion
Hamiltonian.  See, for instance, Ref. \cite{VARbook} for a discussion.
One finds that the levels move as a function of $t$: negative energy
states from the sea acquire positive energy, and empty positive energy
states could become empty sea states. 
In the case of the chiral $SU(2)$ of the SM, one finds that, for each
species of doublet, what was a filled left-handed state in the sea at
$t \rightarrow - \infty$, becomes a particle at $t \rightarrow +
\infty$.  See figure \ref{figdeVAR}.  This ``level-crossing'' occurs
for each type of fermion, so the gauge field configuration centered at
$t = 0$ in figure \ref{figdeVAR}, is a source for nine quarks and
three leptons.

\subsection{$B+L$ violating rates}
At zero temperature, gauge field configurations that give non-zero 
$\int d^4x \widetilde{F}{F}$ correspond to tunneling
configurations, and are called instantons \cite{Belavin:1975fg} (for
reviews, see {\it e.g} Refs. \cite{Vainshtein:1981wh,Coleman:1978ae}).
They change fermion number by an integer $N$, so the instanton
action is large:
$$\left| \frac{1}{4g^2} \int d^4x F_{\mu \nu}^A  F^{\mu \nu A} \right|  \geq 
\left| \frac{1}{4g^2}\int d^4x F_{\mu\nu}^A \widetilde{F}^{\mu\nu A}
\right|    \geq \frac{64 \pi^2 N}{4 g^2} ~~.
$$
The first inequality follows from the Schwartz inequality (see
\cite{Coleman:1978ae}). Consequently, the associated rate is highly
suppressed, 
$$\Gamma \propto  e^{- 
{\rm (instanton~ Action)} } \sim e^{- 4 \pi/\alpha_W}~~~,$$ 
and the mediated $B+L$ violation is unobservably small. Moreover,
the instantons do not threaten the stability of the proton
\cite{'tHooft:1976up}, because  an instanton
acts as a source for  three leptons (one  from each generation),
and nine quarks (all colours and generations), so it induces  
$\Delta B = \Delta L = 3$ processes.  Notice that the three quantum
numbers $B/3-L_\alpha$ are not anomalous, 
so they are conserved in the SM.

If the ground state of the gauge fields is pictured as a periodic
potential, with minima labeled by integers, then the instantons
correspond to  vacuum fluctuations that  tunnel between minima.
With this analogy, one can imagine that at finite temperature,
a thermal fluctuation of the  field could climb {\it over} the
barrier. The sphaleron  
\cite{Klinkhamer:1984di,Arnold:1987mh,Arnold:1987zg}
is such a configuration, in the presence of the Higgs vacuum
expectation value. The $B+L$ violating rate mediated
by sphalerons is Boltzmann suppressed:
$$
\Gamma_{\rm sph} \propto e^{-E_{\rm sph}/T},
$$
where $E_{\rm sph} = 2Bm_W/\alpha_W $ is the height of the barrier at
$T=0$, and $1.5\lsim B\lsim 2.75$ is a parameter that depends on the
Higgs mass. 

For leptogenesis, we are interested in the $B+L$ violating rate at
temperatures far above the EWPT. The large $B+L$ violating gauge field
configurations occur frequently at $T\gg m_W$
\cite{Arnold:1998cy,Arnold:1996dy,Bodeker:1998hm,Bodeker:1999gx,Moore:2000mx}.
The rate can be estimated as (see \cite{Burnier:2005hp} for a recent
discussion) 
\beq
\label{eq:BLVrate}
\Gamma_{\subBLv}  \simeq \, 250\,  \alpha_W^5\, T ~~~.
\eeq
This implies that, for temperatures below  $10^{12}$ GeV and above  
the EWPT,  $B+L$ violating rates  are in equilibrium
\cite{Bento:2003jv,Burnier:2005hp}.

\newpage
\section{A Toy Model of Thermal Leptogenesis}
\label{toy}
In this section, the baryon asymmetry produced by thermal leptogenesis
is {\it estimated}. The goal is to provide a basic understanding and
useful formulae, while avoiding, at this stage, the details. The CP
asymmetry in the singlet fermion decay is introduced as a parameter,
and discussed in more detail in Section \ref{sec:CP}.  The dynamics is
estimated by dimensional analysis; Boltzmann equations appear in
Section \ref{real}.

In this section, we make the following simplifying assumptions:
\begin{enumerate}
\item The lepton asymmetry is produced in a single flavour $\alpha$;
\item The $N$-masses are hierarchical: $M_1 \sim 10^{9} ~{\rm GeV} \ll
  M_2, M_3$. (The kinematics is  simpler in an effective theory of a
  propagating $N_1$ and effective dimension-five operators induced by
  $N_2$ and $N_3$).
\item  Thermal production of $N_1$ and negligible production of
  $N_2, N_3$.
\end{enumerate}
When any of these three assumptions does not hold, there are
interesting modifications of the simplest scenario that we describe in
this section. The effects of flavour are discussed in section
\ref{sec:flavor}, the consequences of non-hierarchical $M_i$'s
(and the possible effects of $N_{2,3}$)  are
summarised  in section \ref{lessh}, and other leptogenesis 
mechanisms were mentioned in  section \ref{alternatives}.

The basic idea is the following. Scattering processes produce a
population of $N_1$'s at temperatures $T \sim M_1$. Then this $N_1$
population decays away because, when the temperature drops below
$M_1$, the equilibrium number density is exponentially suppressed
$\propto e^{-M_1/T}$. If the $N_1$ interactions are CP violating,
asymmetries in all the lepton flavours can be produced. If the relevant
interactions are out-of-equilibrium, the asymmetries may survive. They
can then be reprocessed into a baryon asymmetry by the SM $B+L$
violating processes that have been  discussed in Section~\ref{BLV}.

A unique feature of thermal leptogenesis, which distinguishes it from
other ``out-of-equilibrium decay'' scenarios of baryogenesis, is that
the same coupling constant controls the production and later
disappearance of the population of $N$'s. As demonstrated in the
seminal works \cite{Buchmuller:1996pa,Plumacher:1996kc}, a sufficient
number density of $N$'s can be produced via their Yukawa coupling
$\lambda$. The CP asymmetry in the processes that produce the $N$
population is closely related to the CP asymmetry in the $N$ decays,
which wipe out the $N$ population. In particular, in our toy model of
hierarchical $N_i$'s, the CP asymmetry in the scattering interactions
by which the $N_1$ population is produced is equal in magnitude but
opposite in sign to the CP asymmetry in $N_1$ decays. At first sight,
this suggests that the final lepton asymmetry is zero.\footnote{Notice
  that the potential cancellation is between the CP asymmetry in
  processes with $N$ and $\ell_\alpha$ in the final state, such as
  $X\to N\ell_\alpha$ scattering, and the asymmetry in processes with
  $N$ in the initial state and $\ell_\alpha$ in the final state, such
  as $N\to\phi\ell_\alpha$. Only processes with $\ell_\alpha$ in the
  final state can generate an asymmetry. In particular, there is no
  cancellation between the asymmetry produced in decays and inverse
  decays. It is intuitive, and straightforward to verify (see Section
  \ref{real}), that interactions with the lepton in the initial
  state can wash-out the asymmetry, but not produce it.}  A non-zero
asymmetry survives, however, because the initial anti-asymmetry made
with the $N$ population is depleted by scattering, decays, and inverse
decays. This depletion is called {\bf washout}, and is critical to
thermal leptogenesis. The importance of flavour effects in leptogenesis
is a consequence of the crucial role played by washout: the initial
state of washout interactions contains a lepton, so it is important to
know which leptons are distinguishable.

Our aim here is to estimate the asymmetry-to-entropy ratio by
considering the Sakharov conditions. Each condition gives a
suppression factor. The baryon asymmetry can be approximated as
\begin{equation}
Y_{\Delta B}\simeq\frac{135\zeta(3)}{4\pi^4g_*}
\sum_{\alpha}\epsilon_{\alpha \alpha}\times\eta_\alpha\times C.
\label{approx}
\end{equation}
The first factor is the equilibrium $N_1$ number density divided by
the entropy density at $T \gg M_1$, of ${\cal O}(4\times10^{-3})$ when
the number of relativistic degrees of freedom  $g_*$ is taken
$\simeq106$, as in the SM. An equilibrium number density of $N_1$'s is
the maximum that can arise via thermal $N_1$ production. It is
produced when $\lambda_{\alpha 1}$ is large. A smaller $N_1$ density is
parameterized in $\eta_\alpha$. As concerns the other factors in
(\ref{approx}), we note the following:
\begin{enumerate}
\item $\epsilon_{\alpha \alpha}$ is the CP asymmetry in $N_1$ decay.
For every $1/\epsilon_{\alpha\alpha}$ $N_1$ decays, there is one
more $\ell_\alpha$ than there are $\bar{\ell}_ \alpha$'s.
\item $\eta_\alpha$ is the efficiency factor. Inverse decays, other
``washout'' processes, and inefficiency in $N_1$ production, reduce
the asymmetry by $0<\eta_\alpha<1$. In particular, $\eta_\alpha=0$ is
the limit of $N_1$ interactions in perfect equilibrium, so no
asymmetry is created.
\item  $C$ describes further reduction of the asymmetry due to fast
processes which redistribute the asymmetry that was produced in lepton
doublets among other particle species. These include gauge, third
generation Yukawa, and  $B+L$ violating non-perturbative processes.
As we discuss in sections \ref{sec:spec} and \ref{sec:flavormix}, $C$
is a matrix in flavour space, but for simplicity we approximate it here
as a single number.
\end{enumerate}
Formulae for $\epsilon_{\alpha \alpha}$ and $\eta_\alpha$  can
be located in this review, by consulting  the table in section
\ref{notn}. 
Our aim is now to estimate $\epsilon_{\alpha \alpha}$, $\eta_\alpha$
and $C$.

\subsection{CP violation ($\epsilon_{\alpha\alpha}$)}
\label{ssCP3a}
To produce a net asymmetry in lepton flavour $\alpha$, the $N_1$ must
have $L_\alpha$-violating interactions (see section \ref{B}), and
different decay rates  to final states with particles or
anti-particles. The asymmetry in lepton flavour $\alpha$
produced in the decay of $N_1$ is defined by
\bea
 \!  \! \! \epsilon_{\alpha \alpha} & \equiv& \frac{    \!
 \Gamma(N_1  \! \rightarrow
\!  \phi \ell_\alpha)
- \Gamma({N}_1  \!\rightarrow  \! \bar{\phi} \bar{\ell}_\alpha) \!}{
\Gamma(N_1  \! \rightarrow  \! \phi \ell) \! +  \!
 \Gamma({N}_1 \rightarrow  \! \bar{\phi} \bar{\ell})}
\label{epsaa1}
\eea
where $\bar{\ell}$ denotes the anti-particle of $\ell$. $N_1$
is a Majorana fermion, so $\overline{N_1}   =  N_1$
\footnote{ In supersymmetry, $N_1$ represents the
chiral superfield, so one may wish to distinguish
$N_1$ from $\overline{N_1}$. In that case, the $\epsilon_{\alpha\alpha}$
arise in corrections to a ``D-term'', and can be defined by
replacing 
$ \Gamma({N}_1  \rightarrow   \phi \ell_\alpha)
\rightarrow 
 \Gamma(\bar{N}_1  \rightarrow   \phi \ell_\alpha)
$ in eqn \ref{epsaa1}.}.
The asymmetry $\epsilon_{\alpha \alpha}$ is normalized to the total
decay rate, so that the Boltzmann Equations are linear in flavour
space. When we include additional lepton generations, we
find that the CP asymmetry is a diagonal element of a matrix,
so we give it a double flavour index already.

By definition, $|\epsilon_{\alpha\alpha}|\leq1$. Usually, it is much
smaller than that. It is a function of the parameters of the
Lagrangian (\ref{L}). This dependence is evaluated in section
\ref{sec:CP}. The requirement that it is large enough to
account for the observed baryon asymmetry (roughly speaking,
$|\epsilon_{\alpha\alpha}|>10^{-7}$) gives interesting constraints on
these parameters.

\subsection{Out-of-equilibrium dynamics ($\eta_\alpha$)}
\label{ssec:TEv}
The non-equilibrium which is necessary for thermal leptogenesis is
provided by the expansion of the Universe: interaction rates which are
of order, or slower, than the Hubble expansion rate $H$ are not fast
enough to equilibrate particle distributions.
 Interactions can be
classified as much faster than $H$, much slower, or of the same order.
For the purposes of making analytic estimates (and writing Boltzmann
codes), it is convenient to have a single scale problem.
The timescale
of leptogenesis is $H^{-1}$, so we neglect interactions that are much
slower than $H$. Interactions whose rates are faster than $H$ are
resummed into thermal masses, and impose chemical and kinetic
equilibrium conditions on the distributions of particles whose
interactions are fast.

For the initial conditions, we assume that after inflation the
Universe reheats to a thermal bath at temperature $T_{\rm reheat}$
which is composed of particles with gauge interactions.  A thermal
number density of $N_1$ ($n_{N_1} \simeq n_\gamma$) is produced if
$T_{\rm reheat}\gappeq M_{1}/5$ \cite{Buchmuller:2004nz,Giudice:2003jh}, 
and if the
production timescale for $N_1$'s, $1/\Gamma_{\rm prod}$, is shorter
than the age of the Universe $\sim 1/H$. The $N_1$ can be produced by
inverse decays $\phi\ell_\alpha\rightarrow N_1$ and, most effectively,
by $2\to2$ scatterings involving the top quark or electroweak gauge
bosons.  We here neglect the gauge interactions (although $g_2 >h_t$
at $10^{10}$ GeV) because it is simpler and formally consistent, and
because the $O(\alpha$) corrections do not give important new effects.
$N_1$ can be produced by $s$ or $t$ channel exchange of a Higgs : $q_L
t_R \to\phi \to \ell_\alpha N$ or $\ell_\alpha t_R \to \phi \to q_L
N$.  So the production rate can be estimated (by dimensional analysis in
zero temperature field theory) as
\beq
\Gamma_{\rm prod}\sim \sum_\alpha\frac{h_t^2|\lambda_{\alpha 1}|^2}{4
  \pi} T  ~.
\eeq
If $\Gamma_{\rm prod} > H$ then, since $h_t \sim 1$,
  the  $N_1$ total  decay  is also ``in equilibrium'':
\beq\label{GammaD}
\Gamma_D>H(T=M_1),
\eeq
where
\beq\label{eq:Gaa}
\Gamma_{D}=\sum_\alpha \Gamma_{\alpha\alpha}=\sum_\alpha
\Gamma(N_1 \rightarrow \phi \ell_\alpha,
\bar{\phi} \bar{\ell}_\alpha) =
\frac{ [\lambda^\dagger \lambda]_{11} M_1}{8 \pi},
\eeq
and
\beq
H(T=M_1) = 1.66 g_*^{1/2}  \left. \frac{T^2}{m_{\rm pl}}\right|_{T = M_1}.
\eeq
Here $g_*$ is the number of relativistic degrees of freedom in the
thermal bath (see eqn \ref{g*}). Within the SM, $g_* = 106.75$,

It is useful to introduce two dimensionful parameters
\cite{Buchmuller:1996pa}  $\tilde m$ and $m_*$, which are of the order
of the light neutrino masses, and which represent, respectively, the
decay rate $\Gamma_D$ and expansion rate $H(T=M_1)$:
\bea
\tilde{m} &\equiv&  \sum \tilde{m}_{\alpha\alpha}
\equiv \sum_\alpha \frac{\lambda^*_{\alpha 1} \lambda_{\alpha 1}v_u^2}{M_1} =
8 \pi\frac{v_u^2}{M_1^2}\Gamma_D,\nonumber\\
  m_{*}&\equiv& 8\pi  \frac{v_u^{2}}{M_1^2} \left. H \right|_{T = M_1}\simeq
1.1 \times 10^{-3}\ \mbox{eV}.
\label{tildem}
\eea
It can be shown \cite{Fujii:2002jw} that
\beq
\tilde{m} >m_{\rm min},
\label{tildemin}
\eeq
where $m_{\rm min}$ is the smallest light neutrino mass (this is relevant
for degenerate light neutrino masses), and that ``usually''
$\tilde{m}\gsim m_{\rm sol}$ \cite{Buchmuller:2003gz} (see also
\cite{Engelhard:2007kf}). The $\Gamma_{D}>H$ condition reads, in the
language of $\tilde m$ and $m_*$, simply as
\beq
\tilde{m} > m_*.
\eeq
If indeed $\tilde m\gsim m_{\rm sol}$, then this condition is
satisfied. This range of parameters is referred to as ``strong
washout''. The converse case, $\tilde{m}<m_*$, is referred to as ``weak
washout''.

In the {\bf strong washout scenario},  at $T\sim M_1$, a thermal number
density of $N_1$ is obtained ($n_{N_1}\sim n_\gamma$), and the total
lepton asymmetry $Y_L \simeq 0$ (any asymmetry made with the $N_1$ is
washed out). As the temperature drops and the $N_1$'s start to decay,
the inverse decays $\ell_\alpha\phi\to N_1$, which can wash out the
asymmetry, may initially be fast compared to $H$. Suppose that this is indeed
the case for flavour $\alpha$. Then the asymmetry  in lepton  flavour
$\alpha$ will survive once the partial inverse decays from flavour
$\alpha$ are ``out of equilibrium'':
\beq
\Gamma_{ID} (\phi\ell_\alpha\to N_1)\simeq
\frac12 \Gamma_{\alpha \alpha} e^{-M_1/T}
<  H = 1.66 \sqrt{g_*} \frac{ T^2}{m_{\rm pl}}
\label{10} 
\eeq
where  the partial decay rate $\Gamma_{\alpha \alpha}$ is defined in
eqn (\ref{eq:Gaa}), and $\Gamma_{ID}\simeq e^{-M_1/T} \Gamma_D$.
At  temperature $T_\alpha$ where eqn (\ref{10}) is satisfied, the
remaining $N_1$ density is Boltzmann suppressed, $\propto
e^{-M_1/T_\alpha}$. Below $T_\alpha$, the $N_1$'s decay `` out of
equilibrium'', and contribute to the lepton flavour asymmetry. So the
efficiency factor $\eta_\alpha$ for flavour $\alpha$ can be estimated
as
\beq
\eta_\alpha \simeq \frac{ n_{N_1}(T_\alpha)}{n_{N_1}(T\gg M_1)}
\simeq e^{-M_1/T_\alpha}\simeq \frac{m_*}{\tilde{m}_{\alpha \alpha}}
~~~~~~~~(\tilde{m} > m_*,\ \tilde{m}_{\alpha\alpha} > m_*),
\label{etas}
\eeq
where ${m_*}/{\tilde{m}_{\alpha \alpha}}= {H(T=M_1)}/{\Gamma_{\alpha
    \alpha}}$. This approximation applies for $\tilde{m}_{\alpha
  \alpha} > m_* \simeq 10^{-3}$eV.

Now consider an {\bf intermediate} case, where $\tilde{m} > m_*$
(strong washout), but $\tilde{m}_{\alpha \alpha}<m_*$. In this case,
the $N_1$ number density reaches its equilibrium value, because it has
large couplings to other flavours, but the coupling $\lambda_{\alpha1}$
to the flavour we are interested in is small. The (anti-)asymmetry in
flavour $\alpha$ is of order $-\epsilon_{\alpha \alpha}n_\gamma$.  As
the population of $N_1$ decays at $T \lsim M_1$, a lepton asymmetry
$\sim\epsilon_{\alpha\alpha}n_\gamma$ is produced. Consequently, at
lowest order in $\tilde{m}_{\alpha \alpha}$, the lepton asymmetry
vanishes.\footnote{This cancellation is discussed in more detail
in ref. \cite{Buchmuller:2004nz}, when they consider production by
inverse decays. This cancellation is absent in their later discussion
when production by scattering is included, because CP violation in
scattering was neglected.} A small part of the asymmetry (in flavour
$\alpha$) made during $N_1$ production is, however, washed out before
$N_1$ decay. This fraction can be estimated as $\sim -(
\tilde{m}_{\alpha \alpha}/m_* )\epsilon_{\alpha\alpha}n_\gamma$,
yielding an efficiency factor\footnote{This estimate assumes
the momentum distribution $f(p)$ is thermal (see Appendix
\ref{kinetic}). This is a usual assumption in leptogenesis,
where Boltzmann Equations for the total number densities
are solved. Differences that arise when the BE are solved mode by mode
were studied in \cite{Basboll:2006yx}.}
\beq
\eta_\alpha \sim \frac{\tilde{m}_{\alpha \alpha} }{ m_{*}}
~~~~~~~~(\tilde{m} > m_*,\ \tilde{m}_{\alpha\alpha} < m_*).
\label{etaw1}
\eeq

In the {\bf weak washout scenario}, not only $\tilde{m}_{\alpha
\alpha}< m_*$, as above, but also the total decay rate is small,
$\tilde{m}<m_*$. In this case, the $N_1$ number density does not
reach the equilibrium number density $\sim n_\gamma$. Production is
most efficient at $T \sim M_1$, when the age of the Universe satisfies
$2\tau_U=1/H$, so $n_{N_1}\sim\Gamma_{\rm prod}\tau_U n_\gamma
\sim (\tilde{m}/m_*)n_\gamma$. The cancellation, at lowest order, of
the lepton anti-asymmetry and asymmetry is as in the intermediate case
above, so that the efficiency factor can be estimated as
\beq
\eta_\alpha \sim \frac{\tilde{m}_{\alpha \alpha} \tilde{m}}{ m_{*}^{2}}
~~~~~~~~(\tilde{m} < m_*,\ \tilde{m}_{\alpha\alpha} < m_*).
\label{etaw}
\eeq

\subsection{Lepton and $B+L$ violation ($C$)}
\label{B}
The interactions of $N_1$ violate $L$ because lepton number cannot be
consistently assigned to $N_1$ in the presence of $\lambda$ and $M$.
If $L(N_1)=1$, then $\lambda_{\alpha1}$ respects $L$ but $M_1$
violates it by two units. If $L(N_1)=0$, then $M_1$ respects $L$ but
$\lambda_{\alpha1}$ violates it by one unit. The $N_1$ decay, which
depends on both $M_1$ and $\lambda_{\alpha1}$, does not conserve
$L$. The heavy mass eigenstate is its own anti-particle, so it can
decay to both $\ell \phi$ and $\bar{\ell} \phi^*$. If there is an
asymmetry in the rates, a net lepton asymmetry will be produced.

The baryon number violation is provided by $B+L$ changing SM non-perturbative
processes~\cite{Rubakov:1996vz} (see Section~\ref{BLV}). Their approximate rate is
given in eq.~(\ref{eq:BLVrate}) and is faster than the Hubble expansion $H$ in
the thermal plasma for $T\lsim 10^{12}$ GeV.  The asymmetry in lepton flavour
$\alpha$, produced in the $N_1$ decay, contributes to the density of
$B/3-L_\alpha$, which in conserved by the SM interactions. In equilibrium,
this excess of ${B-L}$ implies (for the SM) a baryon excess
\cite{Khlebnikov:1988sr}: \beq\label{prefactor} Y_{\Delta B} \simeq
\frac{12}{37} \sum_\alpha Y_{\Delta_\alpha} \eeq where $Y_{\Delta_\alpha}$ is
the asymmetry in $B/3 - L_\alpha$, divided by the entropy density.  The value
of $C = 12/37$ applies in the SM (see eqn \ref{eq:app6CSM}).  In the MSSM, it
is 10/31 (see eqn \ref{eq:app6CMSSM}).

So far, the focus has been on the neutrino Yukawa interactions, which
produce an asymmetry in the lepton doublets $\ell_\alpha$. The SM
interactions, which redistribute the asymmetries among other
particles, are included in Section \ref{sec:spec}. As discussed in
section \ref{sec:flavormix}, these interactions usually give $O(1)$
effects,  which are parameterized with the $A$-matrix
\cite{Barbieri:1999ma} that is derived in section \ref{chempot}.
One effect that can be explained already at this stage is the
following. When the charged lepton Yukawa coupling $h_\alpha$
is in chemical equilibrium, that is, when interaction rates such as
$\Gamma$(gauge boson$+e_R\leftrightarrow\ell+\phi$) are fast compared
to $H$, the lepton asymmetry in flavour $\alpha$ is shared between
$e_{R \alpha}$ and $\ell_\alpha$. But only the part that remains in
$\ell_\alpha$ is washed out by the neutrino Yukawa interactions, so
there is a mild  reduction in washout.

\subsection{Putting it all together}
An estimate for the baryon asymmetry can be obtained from eqn
(\ref{approx}), with the prefactor $C$  from eqn (\ref{prefactor}):
$$ Y_{\Delta B}  \sim  10^{-3}   \epsilon_{\alpha \alpha} \eta_\alpha, $$
where the CP asymmetry $\epsilon_{\alpha \alpha}$ is taken from eqn
(\ref{epsaatheff}) or (\ref{flavour-CPasym}), and the efficiency
factor is taken from eqn (\ref{etas}), (\ref{etaw1}) or (\ref{etaw}).
To obtain more accurate estimates, as can be found in Appendix
\ref{recipes}, the dynamics should be calculated via the Boltzmann
Equations, introduced in Section \ref{real}.

We can anticipate the flavour issues, which are discussed in Section
\ref{sec:flavor}, by supposing that there are CP asymmetries in all
flavours. Then, in the strong washout case for all flavours, we obtain
\beq
 Y_ {\Delta B}  \sim 10^{-3}  { \sum_\alpha} \epsilon_{\alpha \alpha} \eta_\alpha
\sim  10^{-3}  \, m_*  \, { \sum_\alpha}
\frac{ \epsilon_{ \alpha \alpha} } {
 \tilde{m}_{\alpha \alpha}}
~~~ ~~~{\rm (flavoured,\ strong~ washout)}
\label{matter}
\eeq
where the  flavours summed over are presumably the charged lepton mass
eigenstates. Alternatively, one might choose
$\ell_\alpha=\hat{\ell}_{N_1}$, the direction in flavour space into which
$N_1$ decays (see eqn \ref{toshow}). Then  $ \epsilon_{11}$ is the total
CP asymmetry $\epsilon$ and $\Gamma(N \rightarrow \phi \ell_1)$
is the total decay rate $\Gamma_D$. One obtains
\beq
 Y_{\Delta B}  \sim   10^{-3} \, m_*   \, \frac{\epsilon}{\tilde{m}}
~ ~~~~~~({\rm single ~ flavour,\ strong~ washout}),
\label{um}
\eeq
which is simpler but different. In 
 Appendix 
\ref{sec:appenrhosum} (see also Section \ref{sec:flavor}) 
we discuss why and when the charged lepton mass
basis is the relevant one.

\newpage
\section{CP Violation}
\label{sec:CP}

%
  In section
\ref{ssCP},  we review how to calculate a CP asymmetry
in $N_1$  decays,    including
the vertex and wavefunction\cite{Liu:1993tg}  contributions. 
In Section  \ref{esteff}, we calculate $\epsilon_{\alpha \alpha}$
for  hierarchical $N_i$,  and  give the formulae for less hierarchical
$N_i$, as  calculated
in \cite{Covi:1996wh}. 
 It was shown in \cite{Hamaguchi:2001gw,Davidson:2002qv}, that for
hierarchical $N_i$, there is   an
upper bound on the CP asymmetry  proportional to $M_1$.
 Section \ref{esteff} 
contains a derivation of 
this bound, which   gives a strong constraint on models, and a discussion
of various loopholes.  
Sections \ref{comments}--\ref{secCPscat} discuss
general constraints from S-matrix unitarity and CPT invariance, which
have implications for the generation of a cosmological asymmetry from
decays.  In particular, we explain a subtlety in the analysis: the
contribution of an on-shell $N_1$ to scattering rates should be
subtracted, as it is already included in the decays and inverse
decays.

\subsection{CP violation in $N_1$ decays}
\label{ssCP}
The CP asymmetry in lepton flavour $\alpha$, produced in the decay
of $N_1$, is defined in eqn (\ref{epsaa1}):
\bea
 \!  \! \! \epsilon_{\alpha \alpha} & \equiv& \frac{    \!
 \Gamma(N_1  \! \rightarrow
\!  \phi \ell_\alpha)
- \Gamma({N}_1  \!\rightarrow  \! \bar{\phi} \bar{\ell}_\alpha) \!}{
\Gamma(N_1  \! \rightarrow  \! \phi \ell) \! +  \!
 \Gamma({N}_1 \rightarrow  \! \bar{\phi} \bar{\ell})}
\label{epsaa}
\eea

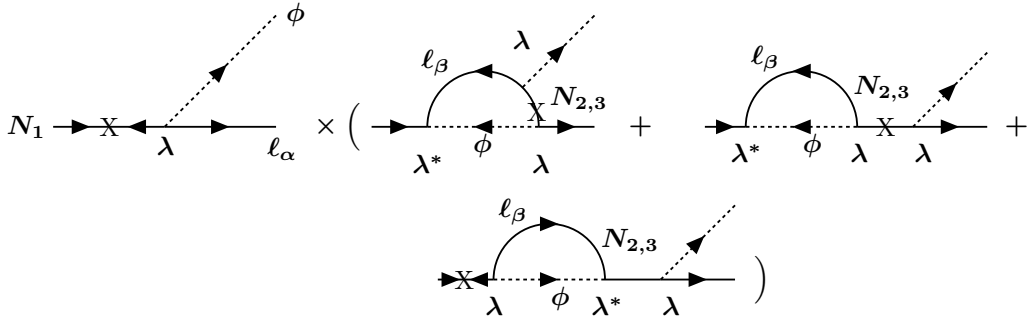
\begin{figure}[ht]
\unitlength0.5mm
\SetScale{1.4}
\begin{boldmath}
%
\hspace{1cm}
\begin{picture}(60,40)(0,0)
\ArrowLine(0,0)(15,0)
\ArrowLine(30,0)(15,0)
\Text(15,0)[c]{X}
\ArrowLine(30,0)(60,0)
\DashArrowLine(30,0)(60,30){1}
\Text(30,-5)[c]{{$\lambda$}}
\Text(-2,0)[r]{{$N_1$}}
\Text(57,-6)[l]{{$\ell_\alpha$}}
\Text(62,30)[l]{{$\phi$}}
\Text(63,0)[l]{{$~~\times {\Big (}$}}
\end{picture}
\hspace{1cm}
\begin{picture}(60,40)(0,0)
\ArrowLine(0,0)(15,0)
\DashArrowLine(45,0)(15,0){1}
\ArrowLine(45,0)(60,0)
\ArrowArc(30,0)(15,0,180)
\DashArrowLine(40.6,10.6)(60,30){1}
\Text(15,-10)[c]{{$\lambda^*$}}
\Text(45,-10)[c]{{$\lambda$}}
\Text(40,23)[c]{{$\lambda$}}
\Text(55,7)[c]{{$N_{2,3}$}}
\Text(44,4)[c]{X}
\Text(27,-5)[l]{{$\phi$}}
\Text(17,17)[c]{{$\ell_\beta$}}
\end{picture}
\hspace{1cm}
\begin{picture}(60,40)(0,0)
\Text(-10,0)[r]{{$+$}}
\ArrowLine(4,0)(15,0)
\DashArrowLine(45,0)(15,0){1}
\Line(45,0)(60,0)
\ArrowLine(60,0)(80,0)
\ArrowArc(30,0)(15,0,180)
\DashArrowLine(60,0)(80,20){1}
\Text(15,-7)[c]{{$\lambda^*$}}
\Text(45,-7)[c]{{$\lambda$}}
\Text(62,-7)[c]{{$\lambda$}}
\Text(51,10)[c]{{$N_{2,3}$}}
\Text(52,0)[c]{X}
\Text(30,-5)[l]{{$\phi$}}
\Text(20,18)[c]{{$\ell_\beta$}}
\Text(90,0)[r]{{$+$}}
\end{picture}
\newline
\begin{picture}(60,40)(0,0)
\Text(30,0)[c]{{$~~~~~~~~~~~~~~~~~~~~$}}
\end{picture}
\begin{picture}(60,40)(0,0)
\Text(30,0)[c]{{$~~~~~~~~~~~~~~~~~~$}}
\end{picture}
\begin{picture}(60,40)(0,0)
\ArrowLine(0,0)(7,0)
\Text(7,0)[c]{X}
\ArrowLine(15,0)(7,0)
\DashArrowLine(15,0)(45,0){1}
\Line(60,0)(45,0)
\ArrowLine(60,0)(80,0)
\ArrowArcn(30,0)(15,180,0)
\DashArrowLine(60,0)(80,20){1}
\Text(15,-7)[c]{{$\lambda$}}
\Text(45,-7)[c]{{$\lambda^*$}}
\Text(62,-7)[c]{{$\lambda$}}
\Text(51,10)[c]{{$N_{2,3}$}}
\Text(30,-5)[l]{{$\phi$}}
\Text(20,18)[c]{{$\ell_\beta$}}
\Text(85,0)[c]{{$~{ \Big ) }$}}
\end{picture}
\vspace{.3cm}
\end{boldmath}
\caption{The diagrams contributing to  the CP asymmetry
$\epsilon_{\alpha \alpha}$. The flavour of the internal lepton
$\ell_\beta$ is summed. The internal $\ell_\beta$ and Higgs $\phi$ are
on-shell. The X represents a Majorana mass insertion. Line direction
is ``left-handedness'', assigning to scalars the handedness of their
SUSY partners. The loop diagrams on the first line are lepton flavour
and lepton number violating. The last diagram is lepton flavour
changing but  ``lepton number conserving'', in the sense
that it makes no contribution to the total
CP asymmetry $\epsilon$.  It is suppressed by an
additional factor $M_1/M_{2,3}$ [see eqn (\ref{flavour-CPasym})].
\label{fig1}} 
\end{figure}

 The CP asymmetry
$\epsilon_{\alpha \alpha}$ arises from the  interference of tree-level
(subscript 0) and one-loop (subscript 1) amplitudes.
This is discussed further  in section \ref{comments}.
 As noted
in \cite{Liu:1993tg}, it is important to include all
the one loop diagrams,  including the  wavefunction corrections.  The tree and loop
matrix elements can each be separated into a coupling constant
part $c$ and an amplitude part ${\cal A}$:
\beq
{\cal M}={\cal M}_{0}+ {\cal M}_{1} = c_0 {\cal A}_0 + c_1 {\cal A}_1 \, .
\label{cetA}
\eeq
For instance, in the tree level decay of figure (\ref{fig1}),
\beq
c_0 =   \lambda^*_{\alpha 1} ~~~~
{\cal A}_0 (N \rightarrow \phi^\dagger\overline{\ell_\alpha}) =
\bar{u}_{\ell_\alpha} P_R u_N ~~~~.
\eeq
The matrix element for the CP conjugate process is
\beq
\overline{{\cal M}} = c_0^{*}\overline{{\cal A}}_0 + c_1^{ *}
\overline{{\cal A}}_1,
\eeq
where\footnote{In the CP conjugate amplitude, $\overline{{\cal A}}$,
  the $u_\ell$ spinors are replaced by $v_\ell$ spinors. Since,
  however, $\bar{u}_\ell u_\ell=\pslash= \bar{v}_\ell v_\ell$, the
  $|$magnitude$|^2$ is the same.} $|\overline{{\cal A}}_i|^2=|{{\cal
    A}}_i|^2$.  Thus the CP asymmetry can be written
\bea
{\epsilon}_{\alpha \alpha}&=&
\frac{\int \left| c_0 {\cal A}_0 + c_1 {\cal A}_1 \right|^2
\tilde{\delta}~d\Pi_{\ell, \phi}
- \int \left| c_0^{*} {\cal A}_0 + c_1^{ *} {\cal A}_1 \right|^2
\tilde{\delta}~d\Pi_{\ell,\phi}}
{2\sum_\beta \int \left| c_0 {\cal A}_0 \right|^2 \tilde{\delta}~
  d\Pi_{\ell \phi}} \nonumber \\
 &=&\frac{{\rm Im}\{c_0c_1^{*}\}}{\sum_\alpha|c_0|^2}
 \frac{2\int{\rm Im}\{  {\cal A}_0 {\cal A}_1^{*} \}  \tilde{\delta}~d\Pi_{\ell, \phi} }
{ \int |{\cal A}_0|^2 \tilde{\delta}~d\Pi_{\ell, \phi} }\, ,
\label{hateps}
\eea
where
\beq
\tilde{\delta} =(2 \pi)^4  \delta^4( P_i - P_f )~,  ~~~~
 d \Pi_{\ell,\phi} = d \Pi_{\ell} d \Pi_{\phi} =
\frac{d^3 p_\phi}{2E_\phi(2\pi)^3}  \frac{d^3 p_\ell}{2E_\ell(2\pi)^3} ~,
\label{tddPi}
\eeq
and $P_i$, $P_f$ are, respectively, the incoming four-momentum (in
this case $P_N$) and the outgoing four-momentum (in this case
$p_\phi+p_\ell$). The loop amplitude has an imaginary part when there
are branch cuts corresponding to intermediate on-shell particles (see
Cutkosky Rules in \cite{Itzykson:1980rh}, or eqn (\ref{CPloop})),
which can arise in the loops of figure \ref{fig1} when the $\phi$ and
$\ell_\beta$ are on-shell:
\beq
2  {\rm Im} \{  {\cal A}_0 {\cal A}_1^{*} \}   =
  {\cal A}_0 (N \rightarrow \phi \ell_{\alpha})
 \sum_\beta
\int  {\cal A}_0^{*} ( N \rightarrow \bar{\ell}'_{\beta}  \bar{\phi}' )
 ~ \tilde{\delta}'~d\Pi_{\ell'_\beta, \phi'}
 {\cal A}_0^{*} ( \bar{\ell}'_{\beta} \bar{\phi}'
\rightarrow \phi \ell_{\alpha} )  .
\label{ImAA}
\eeq
Here $\phi'$ and $\ell'_{\beta}$ are the (assumed massless)
intermediate on-shell particles, and $d\Pi_{\ell'_\beta, \phi'}$ is
the integration over their phase space.

\subsection{$\epsilon_{\alpha \alpha}$ and the lower bound on $M_1$}
\label{esteff}
In the limit $M_2,M_3 \gg M_1$, the effects of  $N_2,N_3$  can be
represented by an effective dimension-5 operator. In the diagram of
\ref{fig1}, this corresponds to shrinking the heavy propagator in the
loop to a point. For calculating $\epsilon_{\alpha \alpha}$, the
Feynman rule for the dimension-5 operator can be taken  $\propto
[m]/v_u^2$. (There is  a contribution to $[m]$ from $N_1$ exchange,
which is not present in the dimension-5 operator that is obtained by
integrating out $N_2$ and $N_3$. But the $N_1$-mediated part of  $[m]$
makes no contribution to the imaginary part for
$\epsilon_{\alpha\alpha}$.) Then, we obtain for the relevant coupling
constants
$$c_0 =   \lambda^*_{ \alpha 1} ~~~~~
c_1  = 3   \sum_\beta   \lambda_{ \beta 1}[ m^*]_{\beta \alpha} /v^2.$$
The  factor of three comes from careful book-keeping of weak $SU(2)_L$
indices; The dimension-5 operator is
\beq
\frac{[m]_{\alpha \beta}}{2
  v_u^2}(\nu_L^\alpha\phi_0-e_L^\alpha\phi^+)
(\nu_L^\beta \phi_0 - e_L^\beta \phi^+) + {\rm h.c.}.
\label{eq:3opeff}
\eeq
This leads to a Feynman rule $2(\delta_{\rho}^{\alpha}
\delta_{\sigma}^{\beta}+\delta_{\rho}^{\beta} \delta_{\sigma}^{\alpha}
)[m]_{\alpha \beta}/(2 v_u^2)$ for the vertex $\nu_L^\rho \phi_0
\nu_L^\sigma \phi_0$ or  $e_L^\rho \phi^+ e_L^\sigma \phi^+$, but to a
Feynman rule $-[m]_{\rho \sigma}/ v_u^2$ for $\nu_L^\rho \phi_0
e_L^\sigma \phi^+$ or $\nu_L^\sigma \phi_0  e_L^\rho \phi^+$.
Summing over all possible lepton/Higgs combinations in the loop gives
the factor of three. It can also be seen in the theory with
propagating $N_{2,3}$: The charged and the neutral components of the
intermediate $\phi'$ and $\ell'_{\beta}$ contribute in the $N_1$
wave-function correction, giving a factor of 2, but only the charged
or the neutral $\phi'$ and $\ell'_{\beta}$  appear in the vertex
correction.

To obtain the amplitude ratio, in the case of hierarchical $N$'s, we
take  ${\cal A}_0^{*}(\bar{\ell}'_{\beta} \bar{\phi}'
\rightarrow \phi \ell_{\alpha} ) =
\bar{v}_\ell^\alpha P_L u_\ell^\beta$, and after
spin sums we obtain:\footnote{The various 2's for the initial and
final state averages and sums  are discussed around eqn
(\ref{2bdps}). They cancel in the ratio and can be ignored here.}
\beq
\epsilon_{\alpha \alpha} = \frac{3 M_1}
{16 \pi v_u^2 [\lambda^\dagger \lambda]_{11}}
{\rm Im~} \{ [\lambda]_{\alpha 1} [m^*\lambda]_{\alpha 1} \}
\label{epsaatheff}
\eeq
where there is no sum on $\alpha$ in this equation.

The upper bound
\beq
|\epsilon_{\alpha \alpha}|\leq\frac{3M_1 m_{{\rm max}}}{16 \pi v_u^2}
\sqrt{B^{N_1}_{\phi \ell_\alpha} +
B^{N_1}_{ \overline{\phi} \;
    \overline{\ell}_\alpha}}
\label{epsaabd}
\eeq
where $B^{N_1}_{xy}\equiv\Gamma(N_1\to xy)/\Gamma_D$ and
$m_{{\rm max}}$  is the largest light neutrino mass,
can be derived by defining the unit vector
\beq
\label{toshow}
\hat{\ell}_{N_1}=\frac{\lambda_{\alpha 1}}{\sqrt{\sum_\beta
    |\lambda_{\beta 1}|^2}}
\eeq
and using $ |m\cdot\hat{\ell}_{N_1}^*|\leq|m_{{\rm max}}
\hat{\ell}_{N_1}|$.

The upper bound on $|\epsilon_{\alpha \alpha}|$ can be used to obtain
a lower bound on $M_1$ and the reheat temperature of the Universe for
thermal leptogenesis with hierarchical $N_i$. The first calculations
\cite{Hamaguchi:2001gw,Davidson:2002qv} used  the total asymmetry
\footnote{Notice that \cite{Davidson:2002qv} used $\epsilon$ for
the MSSM, which is twice as big, see eqn (\ref{g}). The bound
given in \cite{Davidson:2002qv}  therefore  had $8 \pi$ rather
than $16 \pi$ in the denominator.}
$\epsilon = \sum_\alpha \epsilon_{\alpha \alpha}$; the differences are
discussed around eqn (\ref{di2}). The estimate for the baryon
asymmetry, eqn (\ref{approx}), combined with eqn (\ref{prefactor}),
will match the observed asymmetry of eqn (\ref{YB}) for
$\sum_\alpha\epsilon_{\alpha \alpha}\eta_\alpha\sim10^{-7}$. The
efficiency factor, $0< \eta_{\alpha}<1$, is usually $\lsim0.1$
(see section \ref{real}), which implies
$$
\epsilon_{\alpha \alpha} \gsim 10^{-6} ~~~.
$$
Taking $m_{{\rm max}}=m_{\rm atm}$ in the upper bound of eqn
(\ref{epsaabd}), we find
\beq
M_1 \gsim 10^{9}  ~{\rm GeV} .
\label{eq:approxM1bd}
\eeq
A more precise bound can be obtained numerically. The CP asymmetries
$\epsilon_{\alpha\alpha}$ can be smaller when $\eta_\alpha$ is larger,
that is, when there is less washout, which occurs when $\Gamma_D\sim
H(T=M_1)$. For this range of parameters,  analytic approximations
are not so reliable (see the plots of section \ref{sec:Treheat}).

The bound of eqn (\ref{eq:approxM1bd}) is restrictive, so it is worth
repeating the list of assumptions that lead to it and to enumerate
some mechanisms that evade it:
\begin{enumerate}
\item The bound applies for non-degenerate heavy neutrinos. The CP
asymmetry can be much larger for quasi-degenerate $N_i$, with
$M_1-M_2\sim\Gamma_D$ (see section \ref{resonant}).
\item The bound actually applies only for strongly hierarchical heavy
neutrinos. To prove the upper bound on $\epsilon$, for arbitrary
$\lambda$ compatible with the observed $[m]$, requires $M_{2,3}>100
M_1$ \cite{Hambye:2003rt,Davidson:2003yk}. For a milder hierarchy, it
is possible to tune $\lambda$ such that a dimension-seven operator
gives a contribution to $\epsilon$ that can be as high as
$M_1^3/(M_2^2M_3)$ and thus possibly exceed the bound
\cite{Raidal:2004vt}, but no contribution to $[m]$. If such tuning is
neglected, the bound is ``usually'' good for
$M_{2,3}>10M_1$.\footnote{This can be seen from
  \cite{Davidson:2003yk}, where the procedure of scanning of the
  parameter space is described. There is no similar information in
  \cite{Hambye:2003rt}, where contrary claims are made.}
\item The bound can be evaded by adding particles and interactions
\cite{Hambye:2001eu}. Some  of the possibilities are the following:
\begin{itemize}
\item In multi-Higgs models,  the vev(s) of the scalars 
that appear in the light neutrino mass matrix may be unknown.
The CP asymmetries $\epsilon_{\alpha \alpha}$ increase
as these vevs become smaller.

For instance, writing $v_u=174\, \sin\beta$ GeV 
in a model with two Higgs doublets,  the
bound becomes  $M_1\gsim\sin^2\beta\times10^{9}$ GeV, which  can be
significantly weaker if $\sin\beta\ll 1$ \cite{Atwood:2005bf}. It is
interesting to note, in this regard, that this is not the case in the
supersymmetric Standard Model,\footnote{It is desirable, within the
supersymmetric framework, to avoid the bound (\ref{eq:approxM1bd})
because it may be in conflict with upper bounds on $T_{\rm reheat}$
from the gravitino problem.} in spite of its being a two-Higgs
model. The reason is that in this framework $\tan\beta>1$. If there
were extra Higgs doublets, or a non-analytic term $L H_d^* N$ with
large $\tan \beta$,  then thermal leptogenesis could be successful for
lower $M_1$ and $T_{\rm reheat}$ values.

In  ``inverse seesaw'' models\cite{Mohapatra:1986bd}, 
which contain additional singlet
fermions and   scalars,  the light neutrino masses are
proportional to the unknown vev(s) of the additional scalar(s).
For sufficiently small vev(s), hierarchical leptogenesis
can work down to the TeV scale \cite{Kang:2006sn,Hirsch:2006ft,Romao:2007jr}.

\item A minimal extension that works down to the TeV, is to
add a singlet\cite{Frigerio:2006gx}.

\item Including a fourth lepton generation allows  thermal
leptogenesis at $T\sim$ TeV
\cite{Abada:2003rh,Abada:2004wn,Abada:2005rt}.

\item
Consider the case that the number of heavy singlet neutrinos
is larger than three. Their contribution to the effective operator
$(\ell \phi)(\ell \phi)$ has no effect on the bounds on the
flavoured asymmetries $\epsilon_{\alpha \alpha}$. In contrast,
the contribution of many singlets  in
 weak washout can enhance the final baryon
asymmetry, allowing thermal leptogenesis at $T_{\rm reheat}$ that
is a factor  $\sim 30$ lower than in the case of three singlet
neutrinos \cite{Eisele:2007ws}. In the case where leptogenesis is
unflavoured, extra singlets weaken the upper bound on $|\epsilon|$
from $3M_1(m_3-m_1)/(16 \pi v^2)$  to $3M_1m_3/(16 \pi v^2)$
\cite{Eisele:2007ws,Ellis:2007wz}.

\end{itemize}

\item For degenerate light neutrinos, $m_{{\rm max}} > m_{\rm atm}$,
so the individual flavour asymmetries can be larger. However, the
efficiency factor is smaller, so for $m_{{\rm max}} \lsim$ eV (a
conservative interpretation of the cosmological bound
\cite{Cirelli:2006kt,Hannestad:2006mi,Seljak:2006bg})
the lower bound on $M_1$ is similar to eqn (\ref{eq:approxM1bd})
\cite{Abada:2006fw,DeSimone:2006,Blanchet:2006ch}.
\end{enumerate}

One can go beyond the effective theory and incorporate the $N_{2,3}$ 
states as dynamical degrees of freedom.
For a not-too degenerate $N_i$ spectrum, $M_i-M_j\gg\Gamma_D$
(the case of  $M_i-M_j\sim\Gamma_D$ is discussed in Section
\ref{resonant}), one obtains
\begin{eqnarray}
\epsilon_{\alpha\alpha}&=& \frac{1}{(8\pi)}\frac{1}{[\lambda^{\dagger}
  \lambda ]_{11}}\sum_j{\rm Im}\, \left\{(\lambda^*_{\alpha1})
  (\lambda^{\dagger}\lambda)_{1j}\lambda_{ \alpha j} \right\}
g\left(x_j\right)\, \nonumber \\
&&+\frac{1}{(8\pi)}\frac{1}{[\lambda^{\dagger}\lambda ]_{11}}
\sum_j{\rm Im}\, \left\{(\lambda^*_{\alpha1})
  (\lambda^{\dagger}\lambda)_{j1}\lambda_{ \alpha j} \right\}
\frac{1}{1 - x_j},
\label{flavour-CPasym}
\end{eqnarray}
where
$$x_j\equiv M_j^2/M_1^2,$$
and, within the SM \cite{Covi:1996wh},
\begin{equation}
g(x)=\sqrt{x}\left[ \frac{1}{1-x} + 1 -(1+x)\ln
\left(\frac{1+x}{x} \right)
 \right] \stackrel{x\gg 1}{\longrightarrow} - \frac{3}{2
\sqrt{x}} - \frac{5}{6x^{3/2}}+... \, .
\label{gSM}
\end{equation}
In the MSSM, $N_1$ decays to a slepton + Higgsino, as well as to
lepton + Higgs. The sum of the asymmetries to leptons and to sleptons
is about twice larger than the SM asymmetry \cite{Covi:1996wh}:
\beq
g(x) =- \sqrt{x} \left( \frac{2}{x -1} + \ln \left[ 1 + 1/x \right] \right)
 \stackrel{x\gg 1}{\longrightarrow} - \frac{3}{\sqrt{x}} - \frac{3}{2x^{3/2}}+... ~~.
\label{g}
\eeq

The first line of eqn (\ref{flavour-CPasym}) corresponds to the diagrams on
the first row of figure~\ref{fig1} while the second line
\cite{Covi:1996wh,Endoh:2003mz,Nardi:2006fx,AristizabalSierra:2007ur}
corresponds to the diagram of the second row. 
This contribution
violates the single lepton flavours but conserves the total lepton number, and
thus it vanishes when summed over $\alpha$:
\beq
\epsilon\equiv\sum_\alpha \epsilon_{\alpha\alpha}  =
\frac{1}{(8\pi)}\frac{1}{  [\lambda^{\dagger} \lambda ]_{11}}
\sum_j {\rm Im}\, \left\{ [(\lambda^{\dagger}\lambda)_{1j}]^2\right\}
g\left(x_j\right)\ .
\label{sumalpha}
\eeq

As discussed in Sections \ref{esteff} and \ref{sec:Treheat}, the upper
bound on the flavoured CP asymmetries $\epsilon_{\alpha \alpha}$ can
be used to obtain a lower bound on the reheat temperature.  Here we
discuss the upper bound on the total CP asymmetry
$\epsilon=\sum_\alpha\epsilon_{\alpha \alpha}$ of eqn
(\ref{sumalpha}) \cite{Davidson:2002qv,Hamaguchi:2001gw}:
\beq
\left|\epsilon\right| <
\frac{3}{16 \pi} \frac{(m_{\rm max} - m_{\rm min}) M_1}{v_u^2}
\times \beta(\tilde{m}, m_{\rm max}, m_{\rm min})
\label{di2}
\eeq
where $m_{\rm max}$ ($m_{\rm min}$)  is the largest (smallest) light
neutrino mass, and  $\beta \sim 1$ can be found  in \cite{Hambye:2003rt}.

The interesting feature of the bound (\ref{di2}), is that it {\it
 decreases} for degenerate light neutrinos.  This was used to obtain
an upper bound on the light neutrino mass scale from unflavoured
leptogenesis \cite{Buchmuller:2002jk,Buchmuller:2003gz} (discussed in
Section \ref{sec:upperbd}), and explains the interest in the form of the
function $\beta$.  However, the maximum CP asymmetry in a given
flavour is unsuppressed for degenerate light neutrinos
\cite{Abada:2006fw}, so flavoured leptogenesis can be tuned to work
for degenerate light neutrinos, as discussed in Section \ref{sec:upperbd}.

\subsection{Implications of CPT and unitarity for CP violation in decays
}
\label{comments}
S-matrix unitarity and CPT invariance give useful constraints on CP
violation (see {\it e.g.}
\cite{Kolb:1979qa,Weinberg:1979bt,Weinberg:1995mt}).  This section is
a brief review of some relevant results for CP violation in decays. CP
transforms a particle $\ell_\alpha$ into its antiparticle which we
represent as $\overline{\ell_\alpha}$.

Useful relations, between matrix elements and their CP conjugates, can
be obtained from the unitarity of the S-matrix ${\bf S = 1 + }i {\bf T}$:
\beq
{\bf 1 } = {\bf S}{\bf S}^\dagger =
 ({\bf 1 + }i {\bf T})({\bf 1 - }i {\bf T}^\dagger)
\label{S}
\eeq
which implies that $i{\bf T}_{ab}= i{\bf T}_{ba}^*-[{\bf T} {\bf
  T}^\dagger]_{ab}$. Assuming that the transition matrix ${\bf T}$
can be perturbatively expanded in some coupling constant $\lambda$, it
follows from
\beq
\left| {\bf T}_{ab} \right|^2 - \left| {\bf T}_{ba} \right|^2
= - 2 ~{\rm Im} \left\{  [{\bf T} {\bf T}^\dagger]_{ab}{\bf T}_{ba}^*
\right\} + \left|  [{\bf T} {\bf T}^\dagger]_{ab} \right|^2
\label{CPloop}
\eeq
that  CP violation in a tree process, such as $N_j$ decay, can first arise
in the loop corrections. Notice that the unstable $N_1$ is being
treated as an asymptotic state (the unitary S-matrix is defined
between asymptotic states); this approximation requires some care, as
discussed around eqn (\ref{eqvtau}).

CPT, which should be a symmetry of Quantum Field Theories, implies
\beq
\left| {\cal M}( a  \rightarrow  b  ) \right|^2 =
\left| {\cal M}(  \overline{b}   \rightarrow  \overline{a}  ) \right|^2
\label{CPT}
\eeq
where $i{\cal M}(a\to  b)(2\pi)^4\delta^4( \sum_i^n p_i-\sum_f^m q_f)$
is the $i{\bf T}_{ba}$ matrix element from an initial state of
particles $\{a_1(p_1), ...a_n(p_n)\}$ to a final state of
particles $\{ b_1(q_1), ...b_m(q_m) \}$. In particular, for a Majorana
fermion $N_1$, which is its own antiparticle,
\beq
\left| {\cal M}( N \rightarrow \ell_\alpha \phi ) \right|^2 =
\left| {\cal M}( \overline{\phi}  \overline{\ell}_\alpha
\rightarrow N ) \right|^2
\label{CPTN}
\eeq

Following many textbooks (for instance, section 3.6 of
\cite{Weinberg:1995mt}), one can show from unitarity and
CPT that the total decay rate of a particle $X$ and its antiparticle
$\overline{X}$ are the same. The unitarity condition
$ \sum_{ \{ b \} } \langle X |{\bf  S} | b\rangle
\langle b| {\bf S}^\dagger | X \rangle = 1$ implies
\beq
\sum_{ \{ b \} }\left| {\cal M}( X \rightarrow b ) \right|^2
 = \sum_{ \{ b \} }\left| {\cal M}( b \rightarrow X ) \right|^2,
\eeq
where the sum is over all accessible states $b$. Combined with the CPT
condition of eqn (\ref{CPT}) (with $a=X$), one obtains, as anticipated,
\beq
\sum_{ \{ b \} }
\left| {\cal M}( X \rightarrow b ) \right|^2
 = \sum_{ \{ b \} }
\left| {\cal M}( \overline{X} \rightarrow b ) \right|^2.
\label{CPTdecay}
\eeq
(Notice that $\{\overline{b}\}=\{b\}$.) It is nonetheless possible to
have a CP asymmetry in a partial decay rate. In the case of $N_1$,
which decays to $\phi \ell_\alpha$ and
$\overline{\phi}\overline{\ell}_\alpha$, the asymmetry of eqn
(\ref{epsaa}) can be non-zero.

$N_1$ can be approximated as an asymptotic state, for unitarity
purposes, if its lifetime is long compared to the S-matrix
timescale. This timescale can be identified as $1/\sqrt{s_{\rm kin}}$
(where $s_{\rm kin}$ is a Lorentz invariant measure of the center of
mass energy of the process, for instance the Mandelstam variable $s$
for $2\to2$ scattering), because, in calculating (for instance) a
decay rate, one squares the S-matrix element, using
\beq\label{eqvtau}
\left|\delta^4\left(\sum_i^n p_i-\sum_f^m q_f\right)\right|^2=
  \delta^4\left(\sum_i^n p_i - \sum_f^m q_f\right) V\tau,
\eeq
where $V$ and $\tau$ are the volume of the box and the time interval in
which the interaction takes place. For finite $\tau$, there must be an
uncertainty in the energy conservation $\delta$-function of order
$1/\tau$, so the $\delta(E)$ makes sense for $\sqrt{s_{\rm kin}} \gg
1/\tau$. Consequently, unitarity is satisfied for matrix elements with
$N_1$ in the initial or final state when
$\Gamma_D\ll\sqrt{s_{\rm kin}}\sim M_1$ (the narrow width approximation).

One must take care, however, to subtract from scattering rates into
true asymptotic states the contribution of on-shell $N_1$'s  (to avoid
double counting). This is usually done in the narrow width
approximation (see {\it e.g.} ref. \cite{Giudice:2003jh} for a clear
discussion). Then one can check the result by verifying the
CPT and unitarity constraints. Here we do the converse: use CPT and
unitarity to guess what should be subtracted (see section 3.8 of
\cite{Weinberg:1995mt} for a more complete analysis).

To see how this works, consider the process $\phi{\ell}_\alpha \to$
anything, at $O(\lambda^4)$ in the rate. The possible final states are
$N_1$, with one-loop corrections, and
$\overline{\phi}\overline{\ell}_\beta$ or $\phi{\ell}_\beta$, at tree
level. Then one can  make the following three observations:
\begin{enumerate}
\item Unitarity  and CPT  imply  (see eqn (\ref{CPTdecay}) with the
initial state $X =\phi{\ell}_\alpha$) that there can be no CP
asymmetry in this total rate:
\beq
\left| {\cal M}( \phi {\ell}_\alpha \to {\rm anything}) \right|^2
= \left| {\cal M}( \overline{\phi}\,
\overline{\ell}_\alpha \to {\rm anything}) \right|^2  ~.
\label{exunit}
\eeq
\item For leptogenesis to work, we need
$\epsilon_{\alpha\alpha}\neq0$ which, by CPT, implies that there is a
CP asymmetry in the partial rates to $N_1$:
\beq
\left| {\cal M}( \overline{\phi}\,
\overline{\ell}_\alpha \rightarrow N_1 ) \right|^2 -
\left| {\cal M}( \phi {\ell}_\alpha \rightarrow N_1 ) \right|^2
\neq 0.
\label{CPTabove}
\eeq
\item From the unitarity constraint (\ref{CPloop}), there should be
no CP asymmetry to cancel (\ref{CPTabove}) in the tree-level
scatterings.
\end{enumerate}

The apparent contradiction arises because on-shell $s$-channel
exchange of $N_1$ is included in the scattering, so we have counted it
twice. This on-shell part, also referred to as ``Real Intermediate
State'', should therefore be subtracted from the scattering:
\bea
|{\cal M}(\phi \ell_\alpha \rightarrow {\rm anything})|^2  & = &
|{\cal M}(\phi \ell_\alpha \rightarrow N)|^2 +  \sum_\beta \Big(
|{\cal M}(\phi \ell_\alpha \rightarrow \bar{\phi} \bar{\ell}_\beta )|^2 -
|{\cal M}^{\rm os}(\phi \ell_\alpha \rightarrow \bar{\phi} \bar{\ell}_\beta )|^2
\nonumber  \\
& &+
|{\cal M}(\phi \ell_\alpha \rightarrow \phi \ell_\beta )|^2 -
|{\cal M}^{\rm os}(\phi \ell_\alpha \rightarrow \phi \ell_\beta )|^2 \Big),
\eea
where ${\cal M}^{\rm os}$ stands for the on-shell contribution to the
amplitude. It is simple to check that the asymmetry (\ref{exunit})
vanishes as required if the subtracted matrix element squared,
denoted by ${\cal M}^\prime$,
\bea
|{\cal M}'(\phi \ell_\alpha \rightarrow \bar{\phi} \bar{\ell}_\beta )|^2
&\equiv&
|{\cal M}(\phi \ell_\alpha \rightarrow \bar{\phi} \bar{\ell}_\beta )|^2 -
|{\cal M}^{\rm os}(\phi \ell_\alpha \rightarrow \bar{\phi}
 \bar{\ell}_\beta )|^2,
\eea
is taken as follows:
\bea
|{\cal M}'(\phi \ell_\alpha \rightarrow \bar{\phi} \bar{\ell}_\beta
)|^2&=
&|{\cal M}(\phi \ell_\alpha \rightarrow \bar{\phi} \bar{\ell}_\beta
)|^2 -  \left| {\cal M}( \phi \ell_\alpha \rightarrow N) \right|^2
B^{N_1}_{\bar{\phi} \overline{\ell}_\beta} .
\label{RISM}~~~~~~
\eea
where $B^{N_1}_{\bar{\phi} \overline{\ell}_\beta} $ is the branching
ratio for $N \rightarrow \overline{\phi} \overline{\ell}_\beta $.
This is the subtracted matrix-element-squared one obtains in the narrow
width approximation. It will be useful for writing
Boltzmann Equations for the lepton asymmetry.

\subsection{CP violation in scattering}
\label{secCPscat}
Scattering processes are relevant for the production of the $N_1$
population, because decay and inverse decay rates are suppressed by a
time dilation factor $\propto M_1/T$. The $N_1 = \bar{N}_1$ particles
can be produced by $s$-channel $\phi$-exchange in $q t^c \to N
\ell_\alpha$ and  $\bar{q} \bar{t}^c \to N \bar{\ell}_\alpha$, and by
$t$-channel $\phi$-exchange in $q\bar{\ell}_\alpha \to N \bar{t}^c$,
$t^c \bar{\ell}_\alpha  \to N \bar{q} $, $\ell_\alpha\bar{t}^c \to N
q$ and $\bar{q} \ell_\alpha  \to N t^c$.

In this section, we explicitly calculate the CP asymmetry in
scattering processes, for the case of hierarchical $N_j$, and show
that it is the same as in decays and inverse decays
\cite{Abada:2006ea}.  This result was found in refs.
\cite{Pilaftsis:2003gt,Pilaftsis:2005rv,Anisimov:2005hr} for the case
of resonant leptogenesis. To introduce CP violation in scattering into
the Boltzmann Equations, one must correctly include all processes of
order $h_t^2 \lambda^4$ with the on-shell intermediate state $N_1$s
subtracted out \cite{Nardi:2007jp}. This is done in section
\ref{sec:1to3-2to3}, following the analysis of \cite{Nardi:2007jp}.

For simplicity, we work at zero temperature, in the limit of
hierarchical singlet fermions. This means we follow the framework of
subsection \ref{esteff}, that is, we calculate in the  effective field
theory with particle content of the SM $+N_1$, where the effects of
the heavier $N_2$ and $N_3$ appear in the dimension-five operator of
eqn (\ref{eq:3opeff}).

We define the CP asymmetries in $\Delta L=1$ scattering (mediated
by $s$- and $t$-channel Higgs boson exchange) as
\bea
\hat{\epsilon}^s_{\alpha \alpha}&=&\frac{
\sigma (t^c q \rightarrow N\ell_{\alpha})
- \bar{\sigma} (\bar{q} \overline{t^c} \rightarrow N\bar{\ell}_{\alpha})}
{ \sigma + \bar{\sigma}},\\
\hat{\epsilon}^t_{\alpha \alpha}&=&\frac{
\sigma (q  N \rightarrow  \overline{t^c}\ell_{\alpha})
- \bar{\sigma} (  \bar{q} N  \rightarrow   t^c \bar{\ell}_{\alpha})}
{ \sigma + \bar{\sigma}}\nonumber\\
&=& \frac{
\sigma (q  \bar{\ell}_{\alpha} \rightarrow  \overline{t^c}N )
- \bar{\sigma} (  \bar{q}\ell_{\alpha}  \rightarrow   t^cN )}
{ \sigma + \bar{\sigma}},
\label{?t}
\eea
where the cross-sections in the denominator are summed over flavour.
The initial state density factors cancel in the ratio, so the
cross-sections $\sigma$, $\bar{\sigma}$ can be replaced by the matrix
elements squared $|{\cal M}|^2$, integrated over final state phase
space $\int d \Pi$. Separating the tree and loop matrix elements into
a coupling constant part $c$ and an amplitude part ${\cal A}$, as in
eqn (\ref{cetA}), the CP asymmetry can be written as in eqn
(\ref{hateps}). The loop amplitude has an imaginary part when there
are branch cuts corresponding to intermediate on-shell particles,
which can arise here in a bubble on the $N$ line at the $N \phi
\ell_\alpha$ vertex, {\it e.g.} for $s$-channel Higgs exchange:
\beq
{\rm Im}\{  {\cal A}_0 (t^c q \rightarrow N\ell_{\alpha})
{\cal A}^{*}_1 (t^c q \rightarrow N\ell_{\alpha})   \}   =
{\cal A}_0 (t^c q \rightarrow N\ell_{\alpha})
\int  {\cal A}^{*}_0 ( t^c q \rightarrow \ell_{\alpha} \ell'_{\beta} \phi' )
d\Pi' {\cal A}^{*}_0 ( \ell'_{\beta} \phi' \rightarrow N )\, .
\eeq
Here $\phi'$ and $\ell'_{\beta}$ are the (assumed massless)
intermediate on-shell particles, and $d \Pi'$ is the integration over
their phase space.

In the scattering process, $c_0=h_t\lambda^*_{1 \alpha}$ and
$c_1=3h_t\lambda_{1 \beta}[ m^*]_{\beta \alpha} /v^2$,
where $h_t$ is the top Yukawa coupling. The complex coupling constant
combination in the scattering processes is clearly the same as in
$\epsilon_{\alpha \alpha}$ for decays discussed in section \ref{ssCP}.
To obtain the amplitude ratio [the  second ratio in eqn
(\ref{hateps})], we take, for instance, ${\cal
  A}_0(N\to\bar{\phi}\bar{\ell^\alpha}) = \bar{u}_\ell P_L u_N$. After
performing straightforward spin sums, we find that it is the same for
scattering and for $N$ decay,  so
\beq
\hat{\epsilon}^s_{\alpha\alpha}=\hat{\epsilon}^t_{\alpha\alpha}=
\epsilon_{\alpha\alpha}.
\eeq

CPT and unitarity are realized in the scattering process
$(q t^c \rightarrow N \ell_\alpha)$ in a similar way to inverse
decays. They should hold order by order in perturbation theory, so we
work at order $\lambda^2 \lambda^2_\alpha h_t^2$, and define
\bea
|{\cal M}(q t^c \rightarrow N \ell_\alpha)|^2  & = &
|{\cal M}_s|^2 (1 + \epsilon_{\alpha \alpha})\, ,
\label{scat} 
\eea
where $|{\cal M}_s|^2 \propto \lambda^2_\alpha h_t^2$,
and $|{\cal M}_s|^2 \epsilon_{\alpha \alpha}
 \propto \lambda^2 \lambda^2_\alpha h_t^2$.
At order $ \lambda^2  \lambda^2_\alpha h_t^2$,  we should also include
various $ 2 \rightarrow 3$  tree diagrams without $N_1$ in the final state.
Following the inverse decay discussion, one can write
\bea
|{\cal M}(q t^c \rightarrow X \ell_\alpha)|^2  & = &
|{\cal M}(q t^c \rightarrow N \ell_\alpha)|^2 +
\sum_\beta \left[ |{\cal M} (q t^c \rightarrow \ell_\beta \phi \ell_\alpha)|^2
-|{\cal M}^{\rm os} (q t^c \rightarrow \ell_\beta \phi \ell_\alpha)|^2  \right]
\nonumber \\
&&+   \sum_\beta \left[ |{\cal M} (q t^c \rightarrow
\overline{\ell}_\beta \overline{\phi} \ell_\alpha)|^2
-|{\cal M}^{\rm os} (q t^c \rightarrow \overline{\ell}_\beta
\overline{\phi} \ell_\alpha)|^2  \right]
\nonumber \\
& = &  |{\cal M}_s|^2 (1 + \epsilon_{\alpha \alpha}) +
|{\cal M}(q t^c \rightarrow  \ell \phi \ell_\alpha)|^2
 - |{\cal M}_s|^2
(1 + \epsilon_{\alpha \alpha})\frac{(1 + \epsilon)}{2} \nonumber \\
&&+
  |{\cal M}(q t^c \rightarrow   \overline{\ell}  \overline{\phi} \ell_\alpha)|^2
 - |{\cal M}_s|^2
(1 + \epsilon_{\alpha \alpha})\frac{(1 - \epsilon)}{2} \nonumber \\
& = & \sum_\beta \left[ |{\cal M}(q t^c \rightarrow \ell_\beta \phi \ell_\alpha)|^2
+  |{\cal M}(q t^c \rightarrow \overline{\ell}_\beta
\overline{\phi} \ell_\alpha)|^2\right]\, ,
\label{ttNl}
\eea
where, in the narrow width approximation,
\beq
|{\cal M}^{\rm os}(qt^c\to\overline{\ell}_\beta\bar{\phi}\ell_\alpha)|^2=
|{\cal M}(q t^c \rightarrow N \ell_\alpha)|^2 \times
 B^ N_{\bar{\phi} \overline{\ell}_\beta}
\eeq

In eqn (\ref{ttNl}), the CP asymmetry $\epsilon_{\alpha \alpha}$
has disappeared in the final result, so if  we repeat the calculation
for the CP conjugate initial state, $\bar{q} \overline{t^c}$,
we should obtain the same result, verifying that a CP asymmetry in  $q
t^c \rightarrow N \ell_\alpha$ is consistent with CPT and
unitarity. Furthermore, the final result of eqn (\ref{ttNl}) is
reassuring, because the unstable state $N$ has disappeared. There is
no CP violation in the total rate for $q t^c \to$ asymptotic (stable)
final states, but CP violation in the partial rate to the unstable
$N$ is possible. This can be relevant to the final value of the baryon
symmetry when some of the lepton flavours are weakly washed out.

\newpage
\section{Boltzmann Equations}
\label{real}

%
The lepton (and baryon) asymmetry produced
via  leptogenesis,  can be computed by solving the Boltzmann equations (BE).
These
describe the out-of-equilibrium dynamics of the processes involving
the heavy singlet fermions. The aim of this chapter is to derive the
basic Boltzmann equations, restricted to the non-supersymmetric
framework, and to the case where the processes that generate a lepton
asymmetry involve just the lightest singlet fermion $N_1$.
Modifications from supersymmetry (see {\it e.g.}
\cite{Plumacher:1998bv}) are described in Section \ref{sec:susylep}.
Possible contributions from heavier singlet neutrinos $N_{2,3}$
\cite{Barbieri:1999ma,DiBari:2005st,Engelhard:2006yg} are considered
in Section \ref{N2}.

To derive the BE one has to consider a large set of processes, as well
as abundances and density asymmetries of many types of particles, and
the use of notations in extended form can result in rather cumbersome
equations. Thus, we start in Section \ref{sec:BE} by introducing some
 compact notation. To make the navigation through the details in
the following subsections easier, we also present the general
structure of the final equations.

Simple Boltzmann equations, taking into account decays, inverse
decays, and $2\leftrightarrow2$ scatterings mediated by $N_1$
exchange, are derived in Section \ref{simplestBE} and are given 
in eqns (\ref{BE-N.DID})  and (\ref{mess3}). In Section
\ref{sec:1to3-2to3} we include scattering processes with $N_1$ on an
external leg, and we also discuss $1\leftrightarrow3$ and
$2\leftrightarrow3$ processes, since the CP asymmetries of their
off-shell parts must be taken into account for consistency.
The corresponding BE for the evolution of the $N_1$ density is
 given in  eqn (\ref{eq:finalBEYN}), and  for  the relevant flavour
asymmetry  in  eqns
(\ref{eq:finalBEYDeltai}),  (\ref{eq:finalBEYDeltai-s})    and 
(\ref{eq:finalBEYDeltai-w}).

We emphasize that to reach quantitatively accurate results, one has to
take into account (i) some relevant interactions that do not involve
the Majorana neutrinos, and (ii) flavour effects. These tasks are taken
in the following sections.

\subsection{Notation}
\label{sec:BE}
In this subsection we introduce our notation. A brief introduction
to particle number densities and rates in
the early Universe can be found on 
Appendix \ref{kinetic}.  We denote the thermally
averaged rate for an initial state $A$ to go into the final state $B$,
summed over initial and final spin and gauge degrees of freedom, as
(see eqn \ref{eq:intden})
\begin{equation}
\label{eq:AtoB}
\gamma^A_B\equiv \gamma(A\to B) ~ .
\end{equation}
The difference between the rates of CP-conjugate processes is written as
\begin{equation}
\label{eq:DeltaAtoB}
\Delta \gamma^A_B\equiv \gamma^A_B-\gamma^{\bar A}_{\bar B}.
\end{equation}
We denote by  $n_a$ the number density for the particle $a$, by
$n^{\rm eq}_a$ its equilibrium density, and by $s$ the entropy density
(see Appendix \ref{kinetic}). We define:
\begin{eqnarray}\label{eq:Ydef}
Y_a\equiv \frac{n_a}{s}, \qquad
y_a\equiv\frac{Y_a}{Y_a^{\rm eq}}, \qquad
\Delta y_a\equiv y_a-y_{\bar a}.
\end{eqnarray}
Thus, we write all particle densities ($Y_a$) normalized to the
entropy density. To simplify the expressions, we rescale the
densities $Y_a$ by the equilibrium density of the corresponding
particle ($Y_a^{\rm eq}=n^{\rm eq}_a/s$). We denote the asymmetries of
the rescaled densities by $\Delta y_a$.

The difference between a process and its time reversed, weighted by
the   densities of the initial state particles, is defined as
\begin{equation}\label{eq:tire}
[A\leftrightarrow B]\equiv(\prod_{i=1}^{n}y_{a_i})\gamma^A_B
-(\prod_{j=1}^{m}y_{b_j}) \gamma_A^B,
\end{equation}
where the state $A$ contains the particles $a_1,\ldots,a_n$ while the
state $B$ contains the particles $b_1,\ldots,b_m$. We consider only
processes in which at most one intermediate state heavy neutrino $N_1$
can be on-shell. In these cases, a primed notation $\gamma{}'^A_B$  (and
$[A\leftrightarrow B]'$) refers to rates with the resonant intermediate
state (RIS) subtracted. In other words, for the process $A\to B$, we
distinguish the on-shell piece ${\gamma^{\rm os}}^A_B$ from the
off-shell piece $\gamma'{}^A_B$:
\begin{equation}\label{eq:gammap}
\gamma'{}^A_B \equiv \gamma^A_B - {\gamma^{\rm os}}^A_B.
\end{equation}
In the simple case where only $2\leftrightarrow 2$ scatterings are
considered, the on-shell part is just
\begin{equation}
{\gamma^{\rm os}}^A_B\equiv \gamma^A_{N_1} B^{N_1}_B,
\label{eq:gamos}
\end{equation}
where $B^{N_1}_B$ is the branching ratio for $N_1$ decays into the
final state $B$. To include processes of higher order in the
couplings, eqn (\ref{eq:gamos}) needs to be generalized
\cite{Nardi:2007jp}.

We introduce from the start a set of BE that allow a proper treatment
of flavour effects. To do that, we write down the BE for the evolution
of the density of the heavy singlet fermions $N$ and of the asymmetry
for a generic lepton flavour $\alpha$. Below $T\sim10^{12}\,$GeV
($T\sim 10^{9}\,$GeV), reactions mediated by the $\tau$ ($\mu$) Yukawa
couplings become faster than the Universe expansion rate, possibly
resolving the flavour composition of these lepton doublets
(see Appendix \ref{sec:appenrhosum}).  As
discussed in Appendix \ref{apenrho}, if the charged lepton Yukawa
interactions are fast compared to both $H$ and $\Gamma_{ID}$, the
equations of motion for the lepton asymmetry reduce to the BE in the
flavour basis.  Assuming that the flavour basis does not change during
leptogenesis, we can work with simple projections onto flavour of all
the relevant quantitites. However, we still adopt a double-index
notation for most of the flavour-dependent quantities, as a reminder
that they correspond to diagonal elements of matrices in flavour space.
For example, we denote the density of leptons of flavour $\alpha$ by
\begin{equation}\label{eq:La}
 Y_{L}^{\alpha\alpha}\equiv  Y_{\ell_\alpha}+Y_{e_\alpha},
\end{equation}
where $Y_{\ell_\alpha}$ is the density of the two gauge degrees of
freedom in $\ell_\alpha$. 
The inclusion of the density $Y_{e_\alpha}$ for the right-handed charged
lepton $e_\alpha$ is required when (some of) the $L$-conserving charged lepton
Yukawa interactions become fast compared to the Universe expansion rate, since
in this case they  transfer part of the asymmetry to the right handed 
degrees of freedom (see section~\ref{sec:spec}). 

We define the asymmetry in the densities of leptons and antileptons
of flavour $\alpha$ as
\begin{equation}\label{eq:deltaalpha0}
Y_{\Delta L_\alpha}\equiv Y_{L}^{\alpha\alpha}-Y_{\bar
  L}^{\alpha\alpha}.
\end{equation}
When the rates of charged  lepton
Yukawa interactions are negligible, the lepton asymmetry is  
stored only in the lepton doublets, and one simply has 
$Y_{\Delta L_\alpha} = Y_{\ell_\alpha}-Y_{\bar\ell_\alpha}$.

As we explain below, the following asymmetries are particularly useful
in the context of leptogenesis:
\begin{equation}\label{eq:deltaalpha}
Y_{\Delta_\alpha}\equiv \frac{Y_{\Delta B}}{3}- Y_{\Delta L_\alpha},
\end{equation}
where $Y_{\Delta B}$ is the baryon asymmetry to entropy ratio.
For time derivative, we use
\beq\label{eq:dotY}
\dot Y \equiv \frac{sH_1}{z}\frac{{\rm d}Y}{{\rm d}z},
\eeq
where
\beq
z\equiv M_1/T,\ \ \ H_1\equiv H(T=M_1).
\eeq

We split the contributions to the evolution equation for
$Y_L^{\alpha\alpha}$ into three parts:
\begin{equation}\label{eq:YLi}
\dot  Y_{L}^{\alpha\alpha} =\left(\dot  Y_{L}^{\alpha\alpha}\right)_I+
\left(\dot  Y_{L}^{\alpha\alpha}\right)_{II}  +
\left(\dot  Y_{L}^{\alpha\alpha}\right)_{\rm sphal}.
\end{equation}
\begin{enumerate}
\item $(\dot Y_{L}^{\alpha\alpha})_I $ includes
  contributions of ${\cal O}(\lambda^2)$ and of ${\cal
    O}(\lambda^4)$. It is evaluated in section \ref{simplestBE}.
\item $(\dot Y_{L}^{\alpha\alpha})_{II} $ includes contributions of
  ${\cal O}(\lambda^2 h_t^2,\lambda^4 h_t^2)$ and ${\cal O}
  (\lambda^2 g^2,\lambda^4g^2)$, with $g$ a generic gauge coupling
  constant. It is evaluated in section \ref{sec:1to3-2to3}.
\item $(\dot Y_{L}^{\alpha\alpha})_{\rm sphal}$ represents
  the change in the lepton densities due to electroweak sphalerons.
\end{enumerate}

As concerns the sphaleron effects, although their precise rates are
hard to estimate, it is known that below $T\simeq 10^{12}\,$GeV they
are a source of rapid baryon number violation, but they leave $B-L$
unchanged. More precisely, sphalerons generate the same change in
the baryon and lepton number of each generation,
\begin{equation}\label{eq:Lsphal}
(\dot{Y}_{\Delta L_\alpha})_{\rm sphal}=\frac{1}{3}(\dot{Y}_{\Delta
  B})_{\rm sphal},
\end{equation}
leaving unchanged the charge densities of eqn (\ref{eq:deltaalpha}).
Hence, it is convenient to write an equation directly for these
quantities.  By subtracting from eqn (\ref{eq:YLi}) the analogous
equation for the density of antileptons $\dot{Y}_{\bar
  L}^{\alpha\alpha}$ and by subtracting again the result from the
equation that describes the evolution of the baryon asymmetry,
$(\dot{Y}_{\Delta B})/3= (\dot{Y}_{\Delta B})_{\rm sphal}/3$, one
obtains the evolution equations
\begin{equation}\label{eq:YDeltai}
\dot Y_{\Delta_\alpha} =-\left(\dot Y_{\Delta L_\alpha}\right)_I-
\left(\dot Y_{\Delta L_\alpha}\right)_{II}
\end{equation}
that do not depend on the sphaleron rates.

The heavy fermions $N_1$ are treated as quasi-stable particles on the
time scale of the Universe expansion.  This is justified by the fact
that leptogenesis requires that the $N_1$ lifetime is of the order of
the expansion time $H_1^{-1}$. The final baryon asymmetry depends on
the density of the neutrinos $N_1$ as a function of time, so a
Boltzmann equation for $Y_{N_1}$ is needed. It is convenient to split
also this equation into two parts:
\begin{equation}
  \label{eq:YN}
\dot Y_{N_1} =\left(\dot Y_{N_1}\right)_I+\left(\dot Y_{N_1}\right)_{II}.
\end{equation}
The term $(\dot Y_{N_1})_I$ includes the contributions of terms up
to ${\cal O}(\lambda^2)$ and is evaluated in section \ref{simplestBE}.
The term $(\dot Y_{N_1})_{II}$ includes the contributions up to
${\cal O}(\lambda^2 h_t^2)$ or ${\cal O}(\lambda^2 g^2)$ and is
evaluated in section \ref{sec:1to3-2to3}.

\subsection{The ${\cal O}(\lambda^2)$ and ${\cal O}(\lambda^4)$ terms}
\label{simplestBE}
This section aims to obtain the basic BE which depend only on the
neutrino Yukawa coupling $\lambda$, including only the terms $(\dot
Y_{\Delta L_\alpha})_I$ of eqn (\ref{eq:YDeltai}) and $(\dot
Y_{N_1})_I$ of eqn (\ref{eq:YN}).  The SM gauge interactions are
assumed to be fast, which ensures kinetic equilibrium for the particle
distributions. All other SM interactions, including sphalerons, are
neglected. The processes of $N_1$ decay and inverse decay and
two-to-two scattering mediated by the $N_i$'s are included. However,
the latter is important mainly to subtract real intermediate states,
but its effects are small in the temperature range $T < 10^{12}$ GeV
and in a first approximation can be neglected (see the appendix of
\cite{Abada:2006fw} for a brief discussion, and
\cite{Buchmuller:2004nz} for a detailed one).

The Boltzman equation for $Y_{N_1}$, including only decays and inverse
decays, is given by
\bea\label{BE-N.DID}
\left(\dot Y_{N_1} \right)_I &\!\!\!\!& =
\sum_\beta\Big\{[\ell_\beta \phi \leftrightarrow N_1] +
[\bar \ell_\beta \bar \phi  \leftrightarrow N_1]\Big\}
 \nonumber \\ 
&\!\!\!\!&= - \sum_\beta
\left[y_{N_1} \left(\gamma^{N_1}_{\phi\ell_\beta}+
\gamma^{N_1}_{\overline{\phi}\overline{\ell}_\beta}\right) -
y_\phi y_{\ell_\beta}
\gamma^{\phi\ell_\beta}_{N_1}
- y_{\bar{\phi}} y_{\overline{\ell}_\beta}
\gamma^{\overline{\phi}\overline{\ell}_\beta}_{N_1}  \right]
 \simeq - \left(y_{N_1} -1 \right) \gamma_{{N_1}\rightarrow 2},
\eea
where in the last expression we have approximated the $\phi$ and
${\ell_\beta}$ number densities with their equilibrium densities,
neglecting small corrections $\propto (\gamma^{\overline{\phi}
  \overline{\ell}_\beta}_{N_1} -\gamma^{\phi \ell_\beta}_{N_1}
)(y_{\overline{\ell}_\beta} - y_{\ell_\beta})$ that are second order
in the small quantity $\epsilon_{\beta\beta}$. We use
\beq
\label{eq:N1to2}
\gamma_{{N_1}\rightarrow 2}=
\sum_\beta(\gamma^{N_1}_{\ell_\beta \phi} + \gamma^{N_1}_{\overline{\ell}_\beta
  \overline{\phi}})
\eeq
for the thermally averaged two body $N_1$ decay rate.

To obtain the BE for the lepton asymmetry we should work to order
$\lambda^4$, because $\epsilon_{\alpha \alpha}={\cal O}(\lambda^4)$.
At this order, the doublet leptons participate in the following three
types of processes:
\begin{enumerate}
\item[\it (i)] $1\leftrightarrow 2$
processes: the (CP violating) decays $N_1 \to \ell_\alpha \phi$, and
  inverse decays $\ell_\alpha \phi \to N_1$;
\item[\it (ii)] 2$\leftrightarrow$2 scatterings mediated by $s$-channel
  $N_1$ exchange: $\ell_\alpha\phi\leftrightarrow\ell_\beta\phi$ with
  $\beta\neq \alpha$, and $\ell_\alpha
  \phi\leftrightarrow\bar\ell_\beta\bar\phi$;
\item[\it (iii)] $2\leftrightarrow 2$ scatterings mediated by $t$- and
  $u$-channel $N_1$ exchange:
  $\bar\phi\bar\phi\leftrightarrow\ell_\alpha\ell_\beta$ and
  $\phi\bar\phi\leftrightarrow\ell_\alpha\bar\ell_\beta$. (Neglecting
  provisionally thermal effects, no RIS can appear in this case, and
  hence there are no on-shell contributions to be subtracted.)
\end{enumerate}
Accordingly, at this order the evolution equation for the
density of the lepton flavour $\alpha$ reads
\beq
\label{eq:YLi-I}
\left(\dot{Y} _{L}^{\alpha\alpha}\right)_I =
\left(\dot{Y} _{L}^{\alpha\alpha}\right)_{1\leftrightarrow 2} +
\left(\dot{Y} _{L}^{\alpha\alpha}\right)_{2\leftrightarrow 2}^{N_s}+
\left(\dot{Y} _{L}^{\alpha\alpha}\right)_{2\leftrightarrow 2}^{N_t}\,,
\eeq
where
\begin{eqnarray}\label{eq:YLi1to2}
\left(\dot{Y} _{L}^{\alpha\alpha}\right)_{1\leftrightarrow 2} &=&
[N_1 \leftrightarrow  \ell_\alpha \phi], \\  [4pt]
\left(\dot{Y} _{L}^{\alpha\alpha}\right)_{2\leftrightarrow  2}^{N_s} &=&
\sum_\beta [\bar \ell_\beta \bar \phi \leftrightarrow\ell_\alpha\phi ]'+
\sum_{\beta\neq\alpha}[\ell_\beta \phi\leftrightarrow\ell_\alpha\phi]',
\label{eq:YLi2to2sub}\\
\left(\dot{Y} _{L}^{\alpha\alpha}\right)_{2\leftrightarrow 2}^{N_t} &=&
\sum_\beta \left\{[\phi \bar\phi\leftrightarrow\ell_\alpha\bar\ell_\beta]+
(1+\delta_{\alpha\beta})
[\bar\phi\bar\phi\leftrightarrow\ell_\alpha\ell_\beta ]\right\} .
\label{eq:YLi2to2Nt}
\end{eqnarray}
We first focus on {\it (i)} of eqn (\ref{eq:YLi1to2}) and {\it (ii)}
of eqn (\ref{eq:YLi2to2sub}) that give rise to the source term for the
asymmetry.  The processes {\it (iii)} of eqn (\ref{eq:YLi2to2Nt})
contribute only to the washout (at ${\cal O}(\lambda^4)$) and will be
added at the end of the section. As discussed in section
\ref{comments}, some care is required in combining $(i)$ and $(ii)$:
In the $2\leftrightarrow2$ scattering, the contribution from on-shell
$s$-channel $N_1$-exchange is already accounted for by the decays and
inverse decays. This contribution should be removed, to avoid
double-counting, by using a subtracted $|{\cal M}|^2$ (see eqn
\ref{RISM}), as indicated by the primed notation in eqn.
(\ref{eq:YLi2to2sub}). That is, we include only the off-shell part of
the scatterings rate density
\begin{equation}\label{eq:2-2}
 \gamma'{}^{\ell_\alpha\phi}_{\bar\ell_\beta \bar \phi }
= \gamma^{ \ell_\alpha\phi}_{\overline{\ell}_\beta\overline{\phi}}
-\gamma^{\ell_\alpha\phi}_{N_1}\
B^{N_1}_{\overline{\ell}_\beta\overline{\phi}}
\end{equation}
that has a CP asymmetry of the same order as the CP asymmetries of
decays and inverse decays.  The scatterings $\ell_\alpha
\phi\leftrightarrow \ell_\beta\, \phi$ (with $\beta\neq\alpha$) are
treated in a similar way.

The equations for the comoving number densities of the lepton doublet,
and of the anti-lepton doublet, are therefore
\bea
\label{eq:Yellalpha}
\left(\dot Y_{\ell_\alpha}\right)_I&=&[N_1\leftrightarrow\ell_\alpha\phi]+
\sum_\beta [\bar\ell_\beta\bar\phi\leftrightarrow\ell_\alpha\phi]'+
\sum_{\beta} [\ell_\beta\phi\leftrightarrow\ell_\alpha\phi]',\\
\label{eq:Ybarellalpha}
\left(\dot Y_{\overline{\ell}_\alpha}\right)_I&=&
[N_1 \leftrightarrow  \overline{\ell}_\alpha \overline{\phi}]+
\sum_\beta [\ell_\beta \phi \leftrightarrow \overline{\ell}_\alpha
 \overline{\phi} ]'+
\sum_{\beta} [ \overline{\ell}_\beta  \overline{\phi}
\leftrightarrow   \overline{\ell}_\alpha   \overline{\phi}]'.
\eea
Note that the sum in the last two terms of the two equations has been
extended to include the contributions from $\beta=\alpha$ which
cancel in the difference between the process and its time reversed.
Eqns (\ref{eq:Yellalpha}) and (\ref{eq:Ybarellalpha}) can  be written
more explicitely as
\bea
\left(\dot Y_{\ell_\alpha}\right)_I&=&
y_{N_1}\gamma^{N_1}_{\ell_\alpha\phi} +
\sum_\beta\left(y_\phi y_{\ell_\beta}
{\gamma'}^{\ell_\beta\phi}_{\ell_\alpha\phi}
+ y_{\overline{\phi}} y_{\overline{\ell}_\beta}{\gamma'}^{
\overline{\ell}_\beta\overline{\phi}}_{\ell_\alpha\phi} \right)
-y_\phi y_{\ell_\alpha}
\sum_\beta \left({\gamma'}^{\ell_\alpha\phi}_{\ell_\beta\phi}+
{\gamma'}^{\ell_\alpha\phi}_{\overline{\ell}_\beta\overline{\phi}}\right),
\label{mess1} \\  \label{mess2}
\left(\dot Y_{\overline{\ell}_\alpha}\right)_I&=&
y_{N_1}\gamma^{N_1}_{\overline{\ell}_\alpha\overline{\phi}} +
\sum_\beta\left(y_{\overline{\phi}} y_{\overline{\ell}_\beta}
{\gamma'}^{\overline{\ell}_\beta\overline{\phi}}_{\overline{\ell}_\alpha
\overline{\phi}}+ y_{\phi} y_{\ell_\beta}{\gamma'}^{
\ell_\beta\phi}_{\overline\ell_\alpha\overline{\phi}} \right)
-y_{\overline{\phi}}y_{\overline{\ell}_\alpha} \sum_\beta \left(
{\gamma'}^{\overline{\ell}_\alpha\overline{\phi}}_{\ell_\beta\phi}+
{\gamma'}^{ \overline{\ell}_\alpha\overline{\phi}}_
{\overline{\ell}_\beta\overline{\phi}}\right).
\eea
Using eqns (\ref{eq:gammap}) and (\ref{eq:gamos}) and CPT, we rewrite
eqn (\ref{mess1}) as follows:
\bea\nonumber
\left(\dot Y_{\ell_\alpha}\right)_I&=&\gamma^{N_1}_{\ell_\alpha\phi}
\left(y_{N_1}-\sum_\beta
\left[y_\phi y_{\ell_\beta} B^{N_1}_{\overline{\ell}_\beta\overline{\phi}}
+ y_{\overline{\phi}} y_{\overline{\ell}_\beta}
B^{N_1}_{\ell_\beta \phi}\right]\right)
-y_\phi y_{\ell_\alpha}
\gamma^{\ell_\alpha\phi}_{N_1}\left[1-
\sum_\beta(B^{N_1}_{\ell_\beta \phi}+ B^{N_1}_{\overline{\ell}_\beta\overline{\phi}})\right]
\\
&-&y_\phi y_{\ell_\alpha}\sum_\beta \left[
{\gamma}^{\ell_\alpha\phi}_{\ell_\beta\phi}+
{\gamma}^{\ell_\alpha\phi}_{\overline{\ell}_\beta\overline{\phi}}\right]
+\sum_\beta\left[y_\phi y_{\ell_\beta}
\gamma^{\ell_\beta \phi}_{\ell_\alpha \phi}
+ y_{\overline{\phi}} y_{\overline{\ell}_\beta}
\gamma^{\overline{\ell}_\beta\overline{\phi}}_{\ell_\alpha \phi}\right],
\label{eqn86}
\eea
and similarly for the analogous terms in eqn. (\ref{mess2}).
Some comments are in order with regard to eqn~(\ref{eqn86}):
\begin{enumerate}
\item At the order in $\lambda$ we are working in this section
  $\sum_\beta(B^{N_1}_{\ell_\beta \phi}+
  B^{N_1}_{\overline{\ell}_\beta\overline{\phi}})=1$. Consequently,
  the second term in the first line vanishes.
\item As concerns the second line of eqn. (\ref{eqn86}), the first
  term is proportional to $\gamma(\ell_\alpha\phi\to{\rm anything})$,
  while the second term is the sum of $\gamma({\rm
    anything}\to\phi\ell_{\alpha})$ and terms that are second
  order in the asymmetry.  We know from eqn (\ref{CPTdecay}) that
  there can be no CP asymmetry in differences of the form
  $\gamma(X\leftrightarrow {\rm anything}) - \gamma(
  \bar{X}\leftrightarrow{\rm anything})$. Consequently, these terms do
  not contribute to the CP asymmetry.
\item The source term ($\propto\epsilon_{\alpha \alpha}$) in the BE
  arises from the first term of (\ref{eqn86}) and from its analog in
  $Y_{\bar{\ell}_\alpha}$.  We can set the sum of the bracketed
  branching ratios to $1$, because corrections to this approximation
  are second order in the asymmetry. We thus obtain the correct
  behaviour: no asymmetry can be generated in thermal equilibrium
  ($y_{N_1}=1$).
\item The last term in  eqn. (\ref{eqn86}) does not contribute
to the washout of $y_{\ell_\alpha}$. Nevertheless  this term should
not be dropped since, as it will become clear in
section \ref{sec:flavormix}, it does induce washouts for the
charge density $Y_{\Delta_\alpha}$ in eqn. (\ref{eq:deltaalpha}).
\end{enumerate}

By subtracting (\ref{mess2}) from (\ref{mess1}), and including the
contribution of the $t$ and $u$-channel processes {\it (iii)} of
eqn.(\ref{eq:YLi2to2Nt}), we obtain the complete BE at ${\cal
  O}(\lambda^4)$:
\bea
\nonumber
\left(\dot Y_{\Delta L_\alpha}\right)_I &=& (y_{N_1}-1)
\Delta\gamma^{N_1}_{\ell_\alpha \phi}
-(\Delta y_{\ell_\alpha}+\Delta y_\phi)
 \sum_\beta \left(
\gamma^{\ell_\alpha\phi}_{\ell_\beta\phi}
+
\gamma^{\ell_\alpha\phi}_{\overline{\ell}_\beta\overline{\phi}}\right)
\\
&+& \sum_\beta(\Delta y_{\ell_\beta}+\Delta y_\phi)
\left(\gamma^{\ell_\beta\phi}_{\ell_\alpha\phi}
- \gamma^{\ell_\beta\phi}_{\overline{\ell}_\alpha\overline{\phi}}\right)
+\left(\dot Y_{\Delta L_\alpha}\right)^{w,N_t}_{2\leftrightarrow2}.
\label{mess3}
\eea
The term
\beq\label{eq:2to2Ntw}
\left(\dot{Y}_{\Delta L_\alpha}\right)_{2\leftrightarrow 2}^{w,N_t} = -
\sum_{\beta}\left[ (1+\delta_{\alpha\beta})
(\Delta y_{\ell_\alpha}+\Delta y_{\ell_\beta}+2\Delta y_\phi)
\gamma^{\ell_\alpha\ell_\beta}_{\bar \phi \bar \phi}
 + (\Delta y_{\ell_\alpha}-\Delta y_{\ell_\beta})
\gamma^{\ell_\alpha\bar\ell_\beta}_{\bar \phi  \phi}\right]
\eeq
is straightforwardly obtained by subtracting from eqn
(\ref{eq:YLi2to2Nt}) the analogous eqaution for
$\overline{\ell}_\alpha$. In eqns (\ref{mess3}) and (\ref{eq:2to2Ntw})
we approximate the washout rates with their tree level values, and we
linearize in the asymmetry densities $\Delta y$.

Note that there are no resonant contributions to the washout from the
second line of eqn (\ref{mess3}) since the on-shell parts contained in
$\gamma^{\ell_\beta\phi}_{\ell_\alpha\phi}$ and
$\gamma^{\ell_\beta\phi}_{\overline{\ell}_\alpha\overline{\phi}}$
cancel in the difference.  The washout term in the first line of
eqn. (\ref{mess3}) can be written as the sum of resonant $({\cal
  O}(\lambda^2))$ and non-resonant $({\cal O}(\lambda^4))$ parts. By
dropping all the subleading non-resonant terms, we obtain an
approximate expression, valid at ${\cal O}(\lambda^2)$:
\bea
\label{mess3a}
\nonumber
\left(\dot Y_{\Delta L_\alpha}\right)_I &\simeq&
(y_{N_1}-1)
\Delta\gamma^{N_1}_{\ell_\alpha\phi}-
(\Delta
y_{\ell_\alpha}+\Delta y_\phi)
\gamma^{N_1}_{\ell_\alpha\phi}
\\
&\simeq& \left[(y_{N_1}-1) \epsilon_{\alpha\alpha} 
- \frac{1}{2} (\Delta y_{\ell_\alpha}+\Delta y_\phi)
\frac{\tilde m_{\alpha\alpha}}{\tilde m} \right]\gamma_{N_1\to2},
\eea
where $\epsilon_{\alpha\alpha}$ is defined in 
eqn (\ref{epsaa1}),  
 $m_{\alpha\alpha}/\tilde m=
\Gamma(N_1\to \ell_\alpha\phi,\overline\ell_\alpha\overline\phi)/\Gamma_D$, 
and $\gamma_{N_1\to2}$ is defined in eqn~(\ref{eq:N1to2}). 
Using eqns~(\ref{eq:dotY}), (\ref{nMB}) and (\ref{2bdps}), and
provisionally neglecting the contribution to the washout of the
Higgs asymmetry $\Delta y_\phi$ (see section \ref{sec:spec}),
eqn. (\ref{mess3a}) can be written in a more explicit form:
\beq\label{mess3b}
\frac{d}{dz} \left( Y_{\Delta L_\alpha }\right)_I= \frac{z}{H_1}\left(
(Y_{N_1}-Y^{\rm eq}_{N_1} )\frac{K_1(z)}{K_2(z)}\epsilon_{\alpha\alpha}
-\frac{g_{N_1}}{4g_{\ell}}z^2 K_1(z)Y_{\Delta L_\alpha} 
\frac{\tilde m_{\alpha\alpha}}{\tilde m}
\right)\Gamma_D. 
\eeq
%

%
\begin{figure}[t!!]
\begin{center}
\begin{picture}(400,400)(-20,0)
\SetOffset(0,0)
%
\Text(-20,350)[]{$(a)$}
\ArrowLine(0,380)(25,350)
\SetWidth{1.5}
\Line(0,320)(25,350)
\SetWidth{.5}
\Text(10,385)[]{$\ell$}
\Text(10,320)[]{$N$}
\DashLine(25,350)(75,350){2}
\Text(50,360)[]{$\phi$}
\ArrowLine(75,350)(100,380)
\ArrowLine(75,350)(100,320)
\Text(107,385)[]{$q_3$}
\Text(107,320)[]{$\bar t$}
\ArrowLine(150,380)(200,380)
\SetWidth{1.5}
\Line(200,380)(250,380)
\SetWidth{.5}
\Text(160,390)[]{$\ell$}
\Text(240,390)[]{$N$}
\DashLine(200,380)(200,320){2}
\Text(208,350)[]{$\phi$}
\ArrowLine(150,320)(200,320)
\ArrowLine(200,320)(250,320)
\Text(160,330)[]{$\bar q_3$}
\Text(240,330)[]{$\bar t$}
\ArrowLine(300,380)(350,380)
\SetWidth{1.5}
\Line(350,380)(400,380)
\SetWidth{.5}
\Text(310,390)[]{$\ell$}
\Text(390,390)[]{$N$}
\DashLine(350,380)(350,320){2}
\Text(358,350)[]{$\phi$}
\ArrowLine(300,320)(350,320)
\ArrowLine(350,320)(400,320)
\Text(310,330)[]{$t$}
\Text(390,330)[]{$q_3$}
%
\Text(-20,250)[]{$(b)$}
\ArrowLine(0,280)(25,250)
\SetWidth{1.5}
\Line(0,220)(25,250)
\SetWidth{.5}
\Text(10,285)[]{$\ell$}
\Text(10,220)[]{$N$}
\DashLine(25,250)(75,250){2}
\Text(50,260)[]{$\phi$}
\Photon(75,250)(100,280){-2}{5}
\DashArrowLine(75,250)(100,220){2}
\Text(107,220)[]{$\bar \phi$}
\Text(107,285)[]{$A$}
\ArrowLine(150,280)(200,280)
\Photon(200,280)(250,280){-2}{5}
\Text(160,290)[]{$\ell$}
\Text(240,290)[]{$A$}
\ArrowLine(200,280)(200,220)
\Text(208,250)[]{$ \ell$}
\SetWidth{1.5}
\Line(150,220)(200,220)
\SetWidth{.5}
\DashArrowLine(200,220)(250,220){2}
\Text(160,230)[]{$N$}
\Text(240,230)[]{$\bar \phi$}
\ArrowLine(300,280)(350,280)
\Photon(350,280)(400,280) {-2}{5}
\Text(310,290)[]{$\ell$}
\Text(390,290)[]{$A$}
\ArrowLine(350,280)(350,220)
\Text(358,250)[]{$\ell$}
\DashArrowLine(300,220)(350,220){2}
\SetWidth{1.5}
\Line(350,220)(400,220)
\SetWidth{.5}
\Text(310,230)[]{$\phi$}
\Text(390,230)[]{$N$}
%
\Text(-20,150)[]{$(c)$}
\ArrowLine(0,180)(25,150)
\Photon(0,120)(25,150){-2}{5}
\Text(10,185)[]{$\ell$}
\Text(10,120)[]{$A$}
\ArrowLine(25,150)(75,150)
\Text(50,160)[]{$\ell$}
\DashArrowLine(75,150)(100,120){2}
\SetWidth{1.5}
\Line(75,150)(100,180)
\SetWidth{.5}
\Text(107,120)[]{$\bar \phi$}
\Text(107,185)[]{$N$}
\ArrowLine(150,180)(200,180)
\SetWidth{1.5}
\Line(200,180)(250,180)
\SetWidth{.5}
\Text(160,190)[]{$\ell$}
\Text(240,190)[]{$N$}
\DashLine(200,180)(200,120){2}
\Text(208,150)[]{$\phi$}
\Photon(150,120)(200,120){-2}{5}
\DashArrowLine(200,120)(250,120){2}
\Text(160,130)[]{$A$}
\Text(240,130)[]{$\bar \phi$}
\ArrowLine(300,180)(350,180)
\SetWidth{1.5}
\Line(350,180)(400,180)
\SetWidth{.5}
\Text(310,190)[]{$\ell$}
\Text(390,190)[]{$N$}
\DashLine(350,180)(350,120){2}
\Text(358,150)[]{$\phi$}
\Photon(350,120)(400,120){-2}{5}
\DashArrowLine(300,120)(350,120){2}
\Text(310,130)[]{$ \phi $}
\Text(390,130)[]{$A$}
%
\Text(-20,50)[]{$(d)$}
\ArrowLine(0,80)(20,50)
\DashArrowLine(0,20)(20,50){2}
\Text(10,85)[]{$\ell$}
\Text(10,20)[]{$\phi$}
\SetWidth{1.5}
\Line(20,50)(55,50)
\SetWidth{0.5}
\Text(40,60)[]{$N$}
\ArrowLine(55,50)(75,20)
\DashArrowLine(55,50)(75,80){2}
\Text(82,20)[]{$\bar\ell$}
\Text(82,85)[]{$\bar \phi$}
\ArrowLine(112,80)(148,80)
\DashArrowLine(148,80)(185,80){2}
\Text(118,90)[]{$\ell$}
\Text(180,90)[]{$\bar \phi$}
\SetWidth{1.5}
\Line(148,80)(148,20)
\SetWidth{0.5}
\Text(157,50)[]{$N$}
\DashArrowLine(112,20)(148,20){2}
\ArrowLine(148,20)(185,20)
\Text(118,30)[]{$\phi$}
\Text(180,30)[]{$\bar\ell$}
\ArrowLine(216,80)(254,80)
\DashArrowLine(254,80)(290,80) {2}
\Text(222,90)[]{$\ell$}
\Text(284,90)[]{$\bar \phi$}
\SetWidth{1.5}
\Line(254,80)(254,20)
\SetWidth{0.5}
\Text(262,50)[]{$N$}
\ArrowLine(216,20)(254,20)
\DashArrowLine(254,20)(290,20){2}
\Text(222,30)[]{$\ell$}
\Text(284,30)[]{$\bar \phi$}
\ArrowLine(320,80)(353,80)
\DashLine(353,80)(377,50) {2}
\DashArrowLine(377,50)(400,20) {2}
\Text(325,90)[]{$\ell$}
\Text(395,90)[]{$\bar \phi$}
\SetWidth{1.5}
\Line(353,80)(353,20)
\SetWidth{0.5}
\Text(362,50)[]{$N$}
\ArrowLine(320,20)(353,20)
\DashLine(353,20)(377,50){2}
\DashArrowLine(377,50)(400,80){2}
\Text(325,30)[]{$\ell$}
\Text(386,25)[]{$\bar \phi$}
\end{picture}
\end{center}
\caption{Diagrams for  various $2\leftrightarrow 2$
  scattering processes:
 $(a)$ scatterings with
  the top-quarks, $(b),\,(c)$ scatterings with the gauge bosons ($A=B,W_i$
  with $i=1,2,3$), $(d)$ $\Delta L=2$ scatterings mediated by $N_1$.
}
\label{fig:scatterings}
\end{figure}
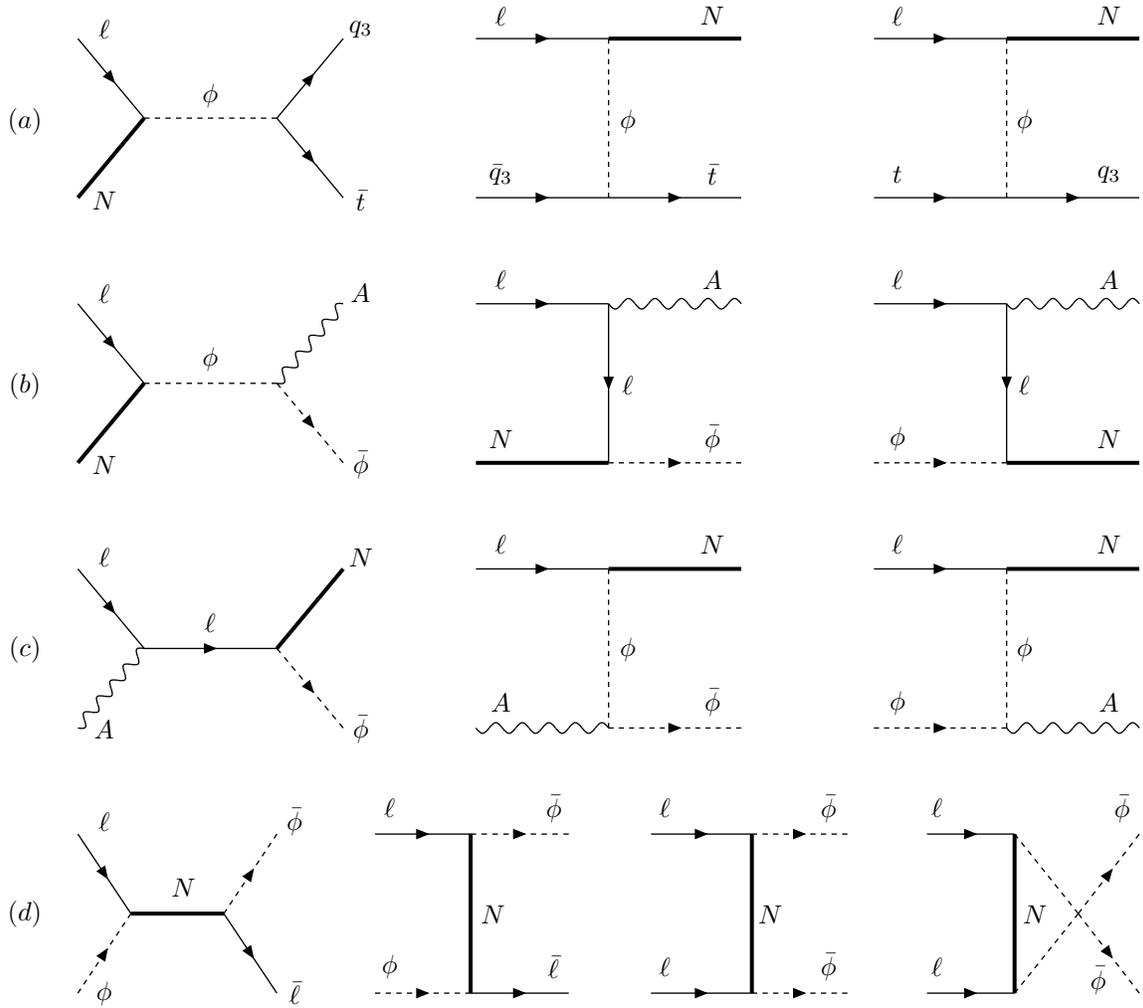

\subsection{The ${\cal O}(h_t^2\lambda^2)$ and ${\cal
    O}(h_t^2\lambda^4)$ terms}
\label{sec:1to3-2to3}
In this section, we include processes involving the top Yukawa
coupling $h_t$. Processes involving gauge bosons can be included in a
similar way and we add them in our final expressions.

We denote the left-handed third-generation quark doublet by $q_3$, and
the $SU(2)$-singlet top by $t$. The inclusion of $1\leftrightarrow3$
decays and inverse decays such as $N_1\leftrightarrow\ell_\alpha\bar
q_3t$, and of $N_1\ell_\alpha\leftrightarrow q_3\bar t$ scatterings
mediated by Higgs exchange, follows lines analogous to those presented
in the previous section. For the ${\cal O}(h_t^2\lambda^2)$
contributions to the evolution of the $N_1$ density, we obtain:
\begin{equation}
\left(\dot{Y}_{N_1}\right)_{II}=
-(y_{N_1}-1)\left[\gamma_{N\rightarrow 3} +\gamma^{2\leftrightarrow
    2}_{\rm top}\right].
\label{ypn}
\end{equation}
Here,
\begin{equation}
  \gamma_{N\rightarrow 3}\equiv \sum_\beta (
  \gamma^{N_1}_{\ell_\beta \bar q_3 t}
  +\gamma^{N_1}_{\bar \ell_\beta q_3\bar t}),
\end{equation}
is the contribution from decays into three-body final states, while
\begin{equation}
\gamma^{2\leftrightarrow 2}_{\rm top} =
\sum_\beta \left( \gamma^{{N_1}\ell_\beta }_{q_3\bar t}+
\gamma^{{N_1}\bar\ell_\beta }_{\bar q_3 t} +
\gamma^{{N_1}q_3}_{\ell_\beta  t}+
\gamma^{{N_1}\bar q_3}_{\bar\ell_\beta \bar t}+\gamma^{{N_1}t}_{\bar\ell_\beta q_3}+
\gamma^{{N_1}\bar t}_{\ell_\beta  \bar q_3}
\right),
\label{eq:gammatop}
\end{equation}
is the contribution from Higgs mediated scatterings: the first two
terms correspond to $s$-channel Higgs exchange, while the other four
(that are all equal at leading order) correspond to $t$- and
$u$-channel Higgs exchange (see fig~\ref{fig:scatterings}$(a)$).

Regarding the evolution of the lepton asymmetries, the derivation of
the BE is more subtle. Once we include the CP violating asymmetries
in $1 \leftrightarrow 3$ (inverse) decays, like ${N_1}\leftrightarrow
\ell_\alpha \bar q_3 t$, and in $2\leftrightarrow 2$ scatterings, like
${N_1} \ell_\alpha \leftrightarrow q_3 \bar t$, we must include also
the asymmetries of various off-shell $2\leftrightarrow 3$ scatterings,
which contribute to the source term at the same order in the
couplings. Accordingly, we write the term $(\dot
Y_{L}^{\alpha\alpha})_{II}$ of eqn (\ref{eq:YLi}) as follows:
\begin{equation}
\left(\dot{Y}_{L}^{\alpha\alpha}\right)_{II} =
\left(\dot{Y}_{L}^{\alpha\alpha}\right)_
{\stackrel{\scriptstyle1\leftrightarrow  3}
  {\scriptstyle 2 \leftrightarrow  2}}
+\left(\dot{Y}_{L}^{\alpha\alpha}\right)_{2\leftrightarrow 3}^{\rm sub} +
\left(\dot{Y}_{L}^{\alpha\alpha}\right)_{2\leftrightarrow 3}^{N_t},
\end{equation}
where
\begin{eqnarray}
\label{eq:1to3}
\left(\dot{Y}_{L}^{\alpha\alpha}\right)_{\stackrel{\scriptstyle1 \leftrightarrow
    3}{\scriptstyle 2 \leftrightarrow  2}}
\!\!\! &=& \!\!\! [{N_1}\leftrightarrow {\ell_\alpha  \bar q_3 t}]
+ [{q_3\bar  t}\leftrightarrow {{N_1}\ell_\alpha }]+
[{N_1}\bar t\leftrightarrow \bar q_3\ell_\alpha ]
+[{N_1} q_3\leftrightarrow  t\ell_\alpha ]\,;
\\[6pt]\nonumber
\left(\dot{Y}_{L}^{\alpha\alpha}\right)_{2\leftrightarrow 3}^{\rm sub}
\!\!\!&=&\!\!\!
\sum_{\beta\neq \alpha}\left\{
[\ell_\beta  \phi \leftrightarrow \bar q_3 t\ell_\alpha ]'+
[\ell_\beta  \phi q_3\leftrightarrow  t\ell_\alpha ]'+
[\ell_\beta  \phi \bar t\leftrightarrow \bar q_3 \ell_\alpha ]'
\right.
\\[-8pt]\nonumber
&& \hspace{1.cm}
+\left.
 [\ell_\beta  \bar q_3 t \leftrightarrow  \ell_\alpha  \phi ]' +
[\ell_\beta  \bar q_3  \leftrightarrow  \ell_\alpha  \phi  \bar t]' +
[\ell_\beta  t \leftrightarrow  \ell_\alpha  \phi  q_3]'
         \right\}  \qquad\\ [4pt] \nonumber
&& \hspace{-1.4cm} + \sum_\beta \left\{
[\bar \ell_\beta \bar \phi \leftrightarrow \ell_\alpha  \bar q_3t]'+
[\bar\ell_\beta  \bar \phi  q_3\leftrightarrow  t\ell_\alpha ]'+
[\bar \ell_\beta  \bar \phi \bar t\leftrightarrow \bar q_3 \ell_\alpha ]'+
  [\bar\ell_\beta q_3 \bar t\leftrightarrow \ell_\alpha  \phi ]'
\right.
\\[-2pt]\nonumber
 && \hspace{-1. cm}
+ \left.
[\bar\ell_\beta q_3 \leftrightarrow \ell_\alpha  \phi t]'+
[\bar\ell_\beta  \bar t\leftrightarrow \ell_\alpha  \phi \bar q_3]'+
[\bar q_3  t\leftrightarrow \ell_\alpha  \phi \bar \ell_\beta ]'+
[q_3 \bar t\leftrightarrow \ell_\alpha  \bar \phi \bar \ell_\beta ]'
 \right\}
 \\[6pt]
\label{eq:Lsub2to3}
 && \hspace{-1.2cm}
+ \sum_\beta (1+\delta_{\alpha\beta})
[q_3 \bar t\leftrightarrow \ell_\alpha  \phi \ell_\beta ]'\,;
\\[2pt]  \nonumber
\left(\dot{Y}_{L}^{\alpha\alpha}\right)_{2\leftrightarrow 3}^{N_t}
\!\!\!\!&=& \!\!\!\!
 \sum_{\beta\neq \alpha}\left\{
[\ell_\beta  q_3 \bar t\leftrightarrow \ell_\alpha  \bar \phi ] +
[\ell_\beta \bar \phi \leftrightarrow \ell_\alpha  q_3\bar t] +
[\ell_\beta \bar \phi  t\leftrightarrow \ell_\alpha  q_3]+
[\ell_\beta \bar \phi \bar q_3\leftrightarrow \ell_\alpha  \bar t] \right\}
 \\[-2pt]  \nonumber
&& \hspace{-1.3cm} +\sum_{\beta}\left\{
[q_3\bar t  \phi \leftrightarrow \bar \ell_\beta  \ell_\alpha ] +
[\bar q_3 t \bar \phi \leftrightarrow \bar \ell_\beta  \ell_\alpha ] +
[\bar q_3 \bar \phi \leftrightarrow \bar \ell_\beta  \ell_\alpha \bar t ] +
[t \bar \phi \leftrightarrow \bar \ell_\beta  \ell_\alpha \bar q_3 ]
\right\}
\\[-2pt]
&&\hspace{-1.3cm} + \sum_\beta   (1+\delta_{\alpha\beta})\left\{
[\bar t \bar \phi  \leftrightarrow \ell_\alpha  \bar q_3\ell_\beta ]+
[q_3 \bar t\bar \phi \leftrightarrow \ell_\alpha  \ell_\beta ]+
[q_3 \bar \phi \leftrightarrow \ell_\alpha  t\ell_\beta ]
\right\}\,.
\label{eq:2to3Nt}
\end{eqnarray}
As in the previous section, the asymmetries in the off-shell
$2\leftrightarrow3$ rates in eqn (\ref{eq:Lsub2to3}) can be estimated
by relating them to the asymmetries of the corresponding on-shell
parts.  However, for $2\leftrightarrow 3$ scatterings, the definition
of the on-shell part is more subtle, because after a real ${N_1}$ is
produced in a collision, it has a certain probability to scatter
before decaying. Consider, for example, the process ${\ell_\beta\phi
  q_3}\to{t\ell_\alpha}$. The contribution to this process from the
exchange of an on-shell ${N_1}$ corresponds to the production process
$\ell_\beta  \phi  \to {N_1}$, followed by the scattering
${N_1}+q_3\to t\ell_\alpha$ that is mediated by a Higgs in the $t$
channel. Processes of this kind can generally be written as $AX\to Y$
where $A$ denotes a possible state to which ${N_1}$ can decay.
The corresponding on-shell rate is then
\begin{equation}\label{eq:os2b}
\gamma^{\rm os}{}^{AX}_Y = \gamma^A_{N_1}P^{{N_1}X}_Y,\\
\end{equation}
where $P^{{N_1}X}_Y$ is the probability that ${N_1}$ scatters with $X$
to produce $Y$. Processes in which the on-shell ${N_1}$ can disappear
only by decaying (as for example
$\ell_\beta\phi\leftrightarrow\ell_\alpha\bar q_3 t$ or $\ell_\beta
t\leftrightarrow\ell_\alpha\phi q_3$) can generally be written as $A
\to B$ or as $X\to B Y$, where both $A$ and $B$ denote possible final
states for ${N_1}$ decays. The corresponding on-shell rates are
\begin{eqnarray}
\label{eq:os2a}
\gamma^{\rm os}{}^A_B &=& \gamma^A_{N_1}P^{N_1}_B,\\
\label{eq:os2c}
\gamma^{\rm os}{}^{X}_{BY} &=& \gamma^X_{{N_1}Y}P^{N_1}_B.
\end{eqnarray}
Note that because of the fact that in the dense plasma ${N_1}$ can
be scattered inelastically before decaying, as described by eqn
(\ref{eq:os2b}), the quantities $P^{N_1}_B$ in eqns (\ref{eq:os2a})
and (\ref{eq:os2c}) differ from the usual notion of branching ratios
at zero temperature. In particular, scattering rates should also be
included in normalizing properly the decay probabilities. The
quantities $P^a_b$ then denote the general probabilities that ${N_1}$
contained in state $a$ ends up producing a state $b$. In the case
under discussion, we have, for example,
\begin{eqnarray}
 \nonumber
 &&
P^{N_1}_{\ell_\alpha  \phi }=\frac{\gamma^{N_1}_{\ell_\alpha  \phi
 }}{\gamma_{\rm all}},
\quad\
P^{N_1}_{\ell_\alpha  \bar q_3 t}=\frac{\gamma^{N_1}_{\ell_\alpha
 \bar q_3 t}}{\gamma_{\rm all}},
 \\ \label{eq:PN}
 &&
P^{{N_1}\ell_\alpha }_{q_3\bar  t}=
 \frac{ \gamma^{{N_1}\ell_\alpha }_{q_3\bar  t}}{\gamma_{\rm all}},  \quad
P^{{N_1}q_3}_{t \ell_\alpha }=
\frac{ \gamma^{{N_1} q_3}_{t\ell_\alpha }}{\gamma_{\rm all}},  \quad
P^{{N_1}\bar t}_{\bar q_3 \ell_\alpha }=
 \frac{\gamma^{{N_1} \bar t}_{\bar q_3\ell_\alpha }}{\gamma_{\rm all}},
\end{eqnarray}
with similar definitions for the probabilities of the CP conjugate
processes. The probabilities are normalized in terms of the sum  of
all the rates:
\begin{equation} \label{eq:PNtot}
  \gamma_{\rm all}=\sum_\beta(\gamma^{N_1}_{\ell_\alpha\phi}
  +\gamma^{N_1}_{\bar\ell_\alpha  \bar \phi }
  +\gamma^{N_1}_{\ell_\alpha  \bar q_3t}
  +\gamma^{N_1}_{\bar\ell_\alpha  q_3 \bar t}
  +\gamma^{{N_1}\ell_\alpha }_{q_3 \bar t}
  +\gamma^{{N_1}\bar\ell_\alpha }_{\bar q_3 t}
  +\gamma^{{N_1}\bar q_3}_{\bar \ell_\alpha \bar t}
  +\gamma^{{N_1} q_3}_{\ell_\alpha  t}
  +\gamma^{{N_1}\bar t}_{\ell_\alpha \bar q_3}
  +\gamma^{{N_1} t}_{\bar \ell_\alpha   q_3}).
\end{equation}
To the order in the Yukawa couplings that we are considering, the
unitarity condition for the sum of the branching ratios of ${N_1}$
into all possible final states, $\sum_Y B^{N_1}_Y=1$, is then
generalized to $\sum_{X,Y} P^{{N_1}X}_Y=1$. In other words, the
probabilities for all the possible ways through which ${N_1}$ can
disappear add up to unity.

To include the new sources of CP asymmetries, we now need to subtract
from eqns (\ref{eq:1to3},\ref{eq:Lsub2to3}) the
analogous equations for $Y_{\bar L}^{\alpha\alpha}$. For the source
term, we obtain:
\begin{equation}
\label{eq:YDLi-II}
\left(\dot{Y}_{\Delta L_\alpha}\right)^s_{II}=
\left(\dot{Y}_{\Delta L_\alpha}\right)^s_
{\stackrel{\scriptstyle1 \leftrightarrow  3}{\scriptstyle 2 \leftrightarrow  2}}
+\left(\dot{Y}_{\Delta L_\alpha}\right)_{2\leftrightarrow 3}^{s,\>\rm sub},
\end{equation}
where we neglect the CP asymmetries of the $2\leftrightarrow 3$
processes with ${N_1}$ exchanged in the $t$-channel
eqn.~(\ref{eq:2to3Nt}) that are of higher order in the couplings.  For
the first term in the r.h.s.  of eqn (\ref{eq:YDLi-II}) we have
\begin{equation}
\label{eq:2to2s}
\left(\dot{Y} _{\Delta L_\alpha}\right)^s_{\stackrel{\scriptstyle1
    \leftrightarrow  3}
{\scriptstyle 2 \leftrightarrow 2}} =
(y_{N_1}+1)\left[\Delta\gamma^{N_1}_{\ell_\alpha \bar q_3 t}
  +\Delta\gamma^{{N_1}\bar t}_{\ell_\alpha \bar
    q_3}+\Delta\gamma^{{N_1}q_3}_{\ell_\alpha t}
  -\Delta\gamma^{{N_1}\ell_\alpha }_{ q_3\bar t}\right].
\end{equation}
Eliminating the subtracted rates by writing their CP asymmetries as
minus the CP asymmetries of the on-shell rates, and keeping terms up
to ${\cal O}(\lambda^4h_t^2)$, we obtain for the second term in eqn
(\ref{eq:YDLi-II})
\begin{equation}
\label{eq:sub2to3s}
\left(\dot{Y}_{\Delta L_\alpha}\right)_{2\leftrightarrow 3}^{s,\>\rm
  sub} =  - 2\, \Delta\gamma^{N_1}_{\ell_\alpha  \phi }
\left[ 1 - \sum_\beta  \left(P^{N_1}_{\ell_\beta  \phi }
    +P^{N_1}_{\bar \ell_\beta \bar \phi }\right)\right]
- 2 \left[\Delta\gamma^{N_1}_{\ell_\alpha \bar q_3 t}
  +\Delta\gamma^{{N_1}\bar t}_{\ell_\alpha \bar q_3}
  +\Delta\gamma^{{N_1}q_3}_{\ell_\alpha t}
  -\Delta\gamma^{{N_1}\ell_\alpha }_{ q_3\bar t}
\right]\,.
\end{equation}
Note that, at ${\cal O}(\lambda^2 h_t^2)$, the sum of the branching
ratios in eqn (\ref{eqn86}) of the previous section is not unity,
so the first term in that equation does not have the correct
thermodynamic behaviour $\propto (y_{N_1}-1)$, and the second term does
not vanish.  By keeping track carefully of the relevant higher order
terms, the source term arising from the difference between
(\ref{eqn86}) and the analog equation for $Y_{\bar{\ell}_\alpha}$
reads
\begin{equation}
\label{eq:sub2to2sb}
\left(\dot{Y} _{\Delta L_i}\right)^s_I =
\left(y_{N_1}+1- 2\sum_\beta  \left[P^{N_1}_{\ell_\beta  \phi }+P^{N_1}_{\bar \ell_\beta
    \bar \phi }\right]\right)\Delta\gamma^{N_1}_{\ell_\alpha  \phi },
\end{equation}
where $\sum_\beta(P^{N_1}_{\ell_\beta\phi} +P^{N_1}_{\bar
    \ell_\beta \bar \phi })<1$.  However, the first term in eqn
(\ref{eq:sub2to3s}) combines with eqn (\ref{eq:sub2to2sb}) to yield
for the source term involving $\Delta\gamma^{N_1}_{\ell_\alpha \phi}$
the correct behavior $\propto(y_{N_1}-1)$.  Summing up eqns
(\ref{eq:2to2s}), (\ref{eq:sub2to3s}) and (\ref{eq:sub2to2sb}), we
obtain the final expression for the source term that holds at ${\cal
  O}(\lambda^4 h_t^2)$:
\begin{equation}
\label{eq:sourcetot}
\left(\dot Y_{\Delta L_\alpha}\right)^s_{I+II}=(y_{N_1}-1)\left[
\Delta\gamma^{{N_1}}_{\ell_\alpha \phi }+
\Delta\gamma^{{N_1}}_{\ell_\alpha \bar q_3 t}
+\Delta\gamma^{{N_1}\bar t}_{\ell_\alpha \bar q_3}
+\Delta\gamma^{{N_1}q_3}_{\ell_\alpha t}
-\Delta\gamma^{{N_1}\ell_\alpha }_{q_3\bar t}
\right].
\end{equation}

Regarding the washouts, we neglect the contributions from eqns
(\ref{eq:Lsub2to3}) and (\ref{eq:2to3Nt}) since they both involve
non-resonant $2\leftrightarrow 2$ scatterings that are of higher order
in the couplings. We retain only the ${\cal O}(\lambda^2h_t^2)$
contributions of eqn (\ref{eq:1to3}).  Subtracting from eqn
(\ref{eq:1to3}) the analogous equations for $\overline{\ell}_\alpha$,
we obtain the relevant washout term:
\begin{equation}
  \label{eq:eq:YLi-w-II}
\left(\dot{Y} _{\Delta L_\alpha}\right)^w_{II} \simeq
 \left(\dot{Y}_{\Delta L_\alpha}\right)^w_{\stackrel
{\scriptstyle1 \leftrightarrow 3}{\scriptstyle 2 \leftrightarrow  2}}
\end{equation}
where
\begin{eqnarray}
 \nonumber
&& \hspace{-2cm}
\left(\dot{Y}_{\Delta L_\alpha}\right)^w_{\stackrel
{\scriptstyle1 \leftrightarrow 3}{\scriptstyle 2 \leftrightarrow  2}}
 =
\left[(\Delta y_{q_3}-\Delta y_t-\Delta  y_{\ell_\alpha })
\gamma^{N_1}_{\ell_\alpha  \bar q_3 t}+
(\Delta y_{q_3}-\Delta y_t-y_{N_1}\Delta y_{\ell_\alpha })
\gamma^{q_3\bar t}_{{N_1}\ell_\alpha }
\right.
 \\ [5pt]
\label{eq:1to3w}
&& \hspace{-.0cm}+\left.
(\Delta y_{q_3}-y_{N_1} \Delta y_t-\Delta y_{\ell_\alpha })
\gamma^{{N_1}\bar t}_{\bar q_3 \ell_\alpha }
+(y_{N_1}\Delta y_{q_3}-\Delta y_t-\Delta y_{\ell_\alpha })
\gamma^{{N_1} q_3}_{t \ell_\alpha }
\right]\,.
\end{eqnarray}
Note that while the contributions from the RIS-subtracted
$2\leftrightarrow3$ scatterings in eqn (\ref{eq:Lsub2to3}) must be
taken into account to obtain the correct form of the source term,
neglecting them in the washouts, as we do in eqn (\ref{eq:1to3w}),
does not have a significant effect on the numerical results.

The inclusion of ${N_1}$ decays into three body final states is
qualitatively required if we want to take into account all processes
of the same order in the couplings, and to incorporate consistently
$2\leftrightarrow3$ scatterings where the on-shell piece involves a $1
\to 3$ decay (for example, $\ell_\beta\phi\to {N_1}\to \ell_\alpha
\bar q_3 t$). The quantitative impact is, however, rather small: while
for zero temperature this decay has a large enhancement related to
small momentum-values for the Higgs propagating off-shell, for the
relevant temperatures, the finite value of the thermal mass prevents
this enhancement, and the decay rate is below 6\% of the two-body
decay rate \cite{Nardi:2007jp}. We thus neglect the washout term of
$3\to 1$ inverse decays, and the contribution of the three-body decay
CP asymmetry $\Delta\gamma^{N_1}_{\ell_\alpha \bar q_3 t}$ to the
source term in eqn (\ref{eq:sourcetot}).

Following the same procedure outlined above, it is possible to include
in the BE other relevant processes, such as those involving the gauge
bosons~\cite{Pilaftsis:2003gt,Pilaftsis:2004xx,Giudice:2003jh}.  With
all the subdominant terms neglected and with the effects of the gauge
bosons included, the simplified expression of the BE for the evolution
of $Y_{N_1}$ reads:
\begin{eqnarray}
  \label{eq:finalBEYN}
\dot Y_{N_1} = -(y_{N_1}-1)\left[ \gamma_{N\rightarrow 2}
+ \gamma^{2\leftrightarrow 2}_{\rm top} +
  \gamma^{2\leftrightarrow 2}_A\right],
\end{eqnarray}
where $A=W_i$ or $B$ for $SU(2)$ and $U(1)$ bosons respectively. The
term involving the gauge bosons is
\begin{equation}
\label{eq:AsAtAu}
\gamma^{2\leftrightarrow 2}_A=
\sum_\beta \left(\gamma^{{N_1}\ell_\beta }_{A\bar \phi }
  +\gamma^{{N_1}\bar\ell_\beta }_{A  \phi }
  +\gamma^{{N_1}\bar \phi }_{A\ell_\beta }
  +\gamma^{{N_1} \phi }_{A  \bar \ell_\beta } +
  \gamma^{{N_1} A}_{\ell_\beta  \phi }
  +\gamma^{{N_1} A}_{\bar \ell_\beta   \bar \phi }\right).
\end{equation}
where a sum over the gauge boson degrees of freedom in all the rate
densities is understood. In eqn (\ref{eq:finalBEYN}) we neglect
three-body decays involving the gauge bosons like $\gamma^{N_1}_{A\phi
  \ell_\alpha }$ that are suppressed by phase space factors. We also
neglect the contributions to the washouts from gauge bosons
$2\leftrightarrow3$ processes.

We can finally write the simplified evolution equation for the
charge-densities $Y_{\Delta_\alpha}$ [defined in eqn
(\ref{eq:deltaalpha})] in terms of the source and the washout terms:
\begin{equation}
  \label{eq:finalBEYDeltai}
\dot Y_{\Delta_\alpha}=
 \left(\dot Y_{\Delta_\alpha}\right)^s +  \left(\dot Y_{\Delta_\alpha}\right)^w.
\end{equation}
The source term is given by
\begin{eqnarray}
\left(\dot Y_{\Delta_\alpha}\right)^s &=&-(y_{N_1}-1)
\left[\Delta\gamma^{N_1}_{\ell_\alpha \phi}
+\Delta\gamma^{{N_1}\bar t}_{\ell_\alpha \bar  q_3}
+\Delta\gamma^{{N_1}q_3}_{\ell_\alpha t}
-\Delta\gamma^{{N_1}\ell_\alpha }_{ q_3\bar t}
-\Delta\gamma^{{N_1}\ell_\alpha }_{A\bar \phi }
+\Delta\gamma^{{N_1}\bar \phi }_{A\ell_\alpha }
+\Delta\gamma^{{N_1} A}_{\ell_\alpha  \phi }\right]
\nonumber\\
&\simeq&  - (y_{N_1}-1) \left[ \gamma_{N\rightarrow 2}
+ \gamma^{2\leftrightarrow 2}_{\rm top}+
\gamma^{2\leftrightarrow 2}_A \right]\, \epsilon_{\alpha\alpha}.
\label{eq:finalBEYDeltai-s}
\end{eqnarray}
In the second line we use the approximate equality between the
scatterings and the decay asymmetries that was discussed in section
\ref{secCPscat}, for example,
\begin{equation}
\frac{\Delta\gamma^{q_3\bar t}_{N_1\ell_\alpha}}
{\gamma^{q_3\bar t}_{N_1\ell_\alpha}}\simeq
\frac{\Delta\gamma^{N_1}_{\ell_\alpha \phi}}
{\gamma^{N_1}_{\ell_\alpha \phi}}=
\epsilon_{\alpha\alpha}.
\end{equation}
The washout term is given by
\begin{eqnarray}
\left(\dot Y_{\Delta_\alpha}\right)^w &=&
 \sum_\beta \left[ (\Delta y_{\ell_\alpha }+\Delta y_\phi )
\left( {\gamma}^{\ell_\alpha \phi }_{\bar\ell_\beta \bar \phi  }
+ {\gamma}^{\ell_\alpha \phi }_{\ell_\beta \phi }\right)
+(\Delta y_{\ell_\beta }+\Delta y_\phi )\left(
{\gamma}^{\ell_\alpha \phi }_{\bar\ell_\beta \bar \phi  }-
{\gamma}^{\ell_\alpha \phi }_{\ell_\beta \phi }\right)\right]
\nonumber\\
&+&\sum_{\beta}\left[(1+\delta_{\alpha\beta})
(\Delta y_{\ell_\alpha }+\Delta y_{\ell_\beta }+2\Delta y_\phi )
\gamma^{\ell_\alpha \ell_\beta }_{\bar \phi  \bar \phi }
 + (\Delta y_{\ell_\alpha }-\Delta y_{\ell_\beta })
\gamma^{\ell_\alpha \bar\ell_\beta  }_{\bar \phi   \phi }\right]
\nonumber\\
&+&(y_{N_1}\Delta y_{\ell_\alpha }-\Delta y_{q_3}+\Delta y_t)
\gamma^{{N_1} \ell_\alpha }_{q_3\bar t}
+\big[2\,\Delta y_{\ell_\alpha }
-(y_{N_1}+1)(\Delta y_{q_3}-\Delta y_t)\big]
\gamma^{{N_1} q_3}_{t \ell_\alpha }
\nonumber\\
&-&(y_{N_1}\Delta y_{\ell_\alpha }+\Delta y_\phi )
\gamma^{{N_1} \ell_\alpha }_{A\bar \phi }
+(y_{N_1}\Delta y_\phi +\Delta y_{\ell_\alpha })
\gamma^{{N_1} \phi }_{A \bar \ell_\alpha }
+(\Delta y_\phi +\Delta y_{\ell_\alpha })
\gamma^{{N_1} A}_{\phi  \ell_\alpha  }.
 \label{eq:finalBEYDeltai-w}
\end{eqnarray}
In the third line we use the equality of the $t$- and $u$-channels
top-quark scatterings to set $\gamma^{{N_1}\bar t}_{\bar
  q_3\ell_\alpha }=\gamma^{{N_1} q_3}_{t\ell_\alpha }$.

Confronting eqn (\ref{eq:finalBEYDeltai-s}) with eqn
({\ref{eq:finalBEYN}), we learn that the BE for the evolution of the
  $\Delta_\alpha$ charge density can be written as follows:
\begin{equation}\label{eureka}
\dot Y_{\Delta_\alpha}= \dot Y_{N_1} \, \epsilon_{\alpha\alpha}
+  \left(\dot Y_{\Delta_\alpha}\right)^w.
\end{equation}
Two comments are in order:
\begin{enumerate}
\item 
  The washout term of eqn (\ref{eq:finalBEYDeltai-w}) depends on the
  density asymmetries of various particle types which evolve with
  time. In principle, we need to know their time evolution in order to
  solve eqn (\ref{eureka}). However, as discussed in Sections
  \ref{sec:equilibrium} and \ref{sec:flavormix}, and, in more detail,
  in Appendix~\ref{chempot}, the chemical equilibrium constraints from
  fast SM reactions always allow one to express all the relevant
  density asymmetries $\Delta y_a$ in terms of the $Y_{\Delta_\beta}$.
  Doing that, one obtains a closed system of differential equations
  involving only the flavoured charge densities.
\item Eqn (\ref{eureka}) shows that if washouts were neglected (and
  the value of $\epsilon_{\alpha\alpha}$ assumed independent of the
  temperature; see Section~\ref{sec:thermal}), the final value of
  $Y_{\Delta_\alpha}$ would be simply proportional to the initial
  value of $Y_{N_1}$.  Therefore, in case that the
  $\lambda$-interactions are the only source of populating the $N_1$
  degree of freedom, the final asymmetry would vanish if not for the
  presence of the washouts.
\end{enumerate}

\newpage
\section{Thermal Effects}
\label{sec:thermal}

At the high temperatures at which leptogenesis occurs, the light
particles involved in the processes relevant for the generation of an
initial lepton number asymmetry are in equilibrium with the hot
plasma. The thermal effects give corrections to several ingredients in
the analysis: (i) coupling constants, (ii) particle propagators
(leptons, quarks, gauge bosons and the Higgs) and (iii) CP-violating
asymmetries. These three effects are discussed in turn in the
following three subsections. A dedicated study of thermal corrections
to leptogenesis processes with a discussion of the leading numerical
effects can be found in \cite{Giudice:2003jh}.

\subsection{Coupling constants}
\label{sec:couplings}
A detailed study of gauge and Yukawa couplings renormalization in a
thermal plasma can be found in \cite{Kajantie:1995dw}. In practice, it
is a very good approximation to use the zero-temperature
renormalization group equations for the top-quark Yukawa coupling and
for the gauge couplings, with a renormalization scale $\Lambda\sim
2\pi T$ \cite{Giudice:2003jh}. The value $\Lambda > T$ is related to
the fact that the average energy of the colliding particles in the
plasma is larger than the temperature.

The renormalization effects for the neutrino couplings are also well
known \cite{Casas:1999tg,Antusch:2003kp}. In the non-supersymmetric
case, to a good approximation these effects can be described 
by a simple rescaling of the low energy neutrino mass matrix 
$m(\mu)=r\cdot m$, where $1.2\lsim r\lsim 1.3$ 
for $10^8\,$GeV\,$\lsim \mu\lsim 10^{16}\,$GeV~\cite{Giudice:2003jh}.
Therefore,  RG effects on neutrino couplings can be accounted for 
by increasing the values of the neutrino mass parameters (for example,
$\tilde m$) as measured at low energy by $\approx 20\%-30\%$
(depending on the leptogenesis scale).  In the
supersymmetric case one expects a milder enhancement, but
uncertainties related with the precise value of the top-Yukawa
coupling can be rather large (see fig.3 in
ref.~\cite{Giudice:2003jh}).

\subsection{Decays and scatterings}
\label{sec:thermaldecays}

In the thermal plasma, any particle with sizeable couplings to the
background states acquires a thermal mass that, modulo the
renormalization of the relevant couplings, is proportional to the
plasma temperature.  Consequently, decay and scattering rates are
modified. The diagrams corresponding to the relevant leptogenesis
processes for which these corrections should be estimated are given in
fig.~\ref{fig1} ($1\leftrightarrow 2$ processes) and in
fig.~\ref{fig:scatterings} (for the $2\leftrightarrow 2$ scatterings).

Thermal corrections to particle masses have been thoroughly studied
in both the standard model and the supersymmetric standard model
\cite{Comelli:1996vm,Elmfors:1993re,Cline:1993bd,Weldon:1989ys,%
  Weldon:1982bn,Klimov:1981ka}. The singlet neutrinos have no gauge
interactions, their Yukawa couplings are generally small and, during
the relevant era, their bare mass is of the order of the temperature
or larger. Consequently, to a good approximation, corrections to their
masses can be neglected. We thus need to account for the thermal
masses of only the lepton doublets, the third generation quarks, the
Higgs and the gauge bosons (and, in the supersymmetric case, also
their superpartners).  Explicit expressions for the thermal masses
that enter the relevant leptogenesis processes are collected in
appendix B of \cite{Giudice:2003jh}.  For the following qualitative
discussion, it is enough to keep in mind that, within the leptogenesis
temperature range, $m_\phi(T)\gsim m_{q_3,t}(T) \gg m_\ell(T)$. The
most important effects relate to four classes of leptogenesis
processes:

(i) {\it Decays and inverse decays.} Since thermal corrections to the
Higgs mass are particularly large ($m_\phi(T) \approx 0.4\, T$),
decays and inverse decays become kinematically forbidden in the
temperature range in which $m_\phi(T)-m_\ell(T)
<M_{N_1}<m_\phi(T)+m_\ell(T)$. For lower temperatures, the usual
processes $N_1\leftrightarrow \ell \phi$ can occur.  For higher
temperatures, the Higgs is heavy enough that it can decay:
$\phi\leftrightarrow \ell N_1$. A rough estimate of the kinematically
forbidden region yields $2 \lsim T/M_1\lsim 5$. The important point
is that these corrections are effective only at $T>M_1$.  In the
parameter region $\tilde m > 10^{-3}\,$eV, that is favored by the
measurements of the neutrino mass-squared differences, the $N_1$ number
density and its $L$-violating reactions attain thermal equilibrium at
$T \approx M_1$ and erase quite efficiently any memory of the specific
conditions at higher temperatures. Consequently, in the strong washout
regime, these thermal corrections have practically no effect on the
final value of the baryon asymmetry.

(ii) {\it $\Delta L=2$ scatterings.}  A comparison of the scattering
rates for $\ell\ell \leftrightarrow \bar\phi\,\bar\phi$ with and
without thermal corrections is given in fig.~\ref{fig:bdM1}$(a)$
(adapted from~\cite{munoz:2007}). The reaction densities are computed
for $\tilde m=0.06\,$eV and $M_1=10^{10}\,$GeV and are plotted as a
function of decreasing temperature (increasing $z=M_1/T$). It is
apparent that for this process thermal effects are sizeable only in
the high temperature range $z<1$. For $\ell\,\bar\phi \leftrightarrow
\bar\ell\,\phi$ scatterings, a new resonant contribution can appear at
high temperatures due to the fact that $N_1$ can go on-shell when
exchanged in the $u$-channel \cite{Giudice:2003jh}. As regards the
off-shell contributions to this process, they are affected by thermal
corrections in a way similar to $\ell\,\bar\phi \leftrightarrow
\bar\ell\,\phi$, that is mainly at $M_1/T< 1$. We conclude that, for
the $\Delta L=2$ rates, thermal corrections are sizeable only at high
temperatures. In the theoretically preferred regime, $\tilde m > m_*$,
the related effects can be neglected.

(iii) {\it $\Delta L = 1$ scatterings with top-quarks.}  Comparisons
between the thermally-corrected and -uncorrected rates of the $\Delta
L=1$ top-quark scattering $\gamma^{\rm top}_{\phi_{s}}\equiv \gamma
(q_3\,\bar t \leftrightarrow\ell\,N_1)$ with the Higgs exchanged in
the $s$-channel, and of the sum of the $t$ and $u$-channel scatterings
$\gamma^{\rm top}_{\phi_{t+u}}\equiv\gamma(q_3\, N_1 \leftrightarrow
\ell\, t) +\gamma(\bar t\, N_1 \leftrightarrow \ell\, \bar q_3)$, are
given in fig.~\ref{fig:bdM1}$(b)$ (adapted from \cite{munoz:2007}). In
the case of $\gamma^{\rm top}_{\phi_{s}}$, mild corrections are
present only at high temperatures. However, in contrast to the
previous cases, the most relevant corrections to $\gamma^{\rm
  top}_{\phi_{t+u}}$ appear at low temperatures, reducing the
scattering rates and suppressing the corresponding contributions to
the $\Delta L=1$ washout. This peculiar situation arises from the fact
that in the zero temperature limit there is a large logarithmic
enhancement $\sim \ln(M_{N_1}/m_\phi)$ from the quasi-massless Higgs
exchanged in the $t$- and $u$-channels.  This enhancement disappears
when the Higgs thermal mass $m_\phi(T)\sim T\sim M_{N_1}$ is included.

(iv) {\it $\Delta L = 1$ scatterings with the gauge bosons} (see
figs.~\ref{fig:scatterings}$(b)$ and~\ref{fig:scatterings}$(c)$).  The
inclusion of thermal masses is required to avoid IR divergences that
would arise when massless $\ell$ (and $\phi$) states are exchanged in
the $t$- and $u$-channels. A naive use of some cutoff for the phase
space integrals to control the IR divergences can yield incorrect
estimates of the gauge bosons scattering rates and would be
particularly problematic at low temperatures, where gauge bosons
scatterings dominate over top-quark scatterings.

\begin{figure}[t!]
\vskip+4mm
\hskip-4mm
{
\vspace{0truecm}
\leftline{\hspace{7.2truecm}$(a)$ \hspace{7.2truecm}$(b)$}
 \vspace{-1.2truecm}
}
 \includegraphics[width=8cm,height=7cm,angle=0]{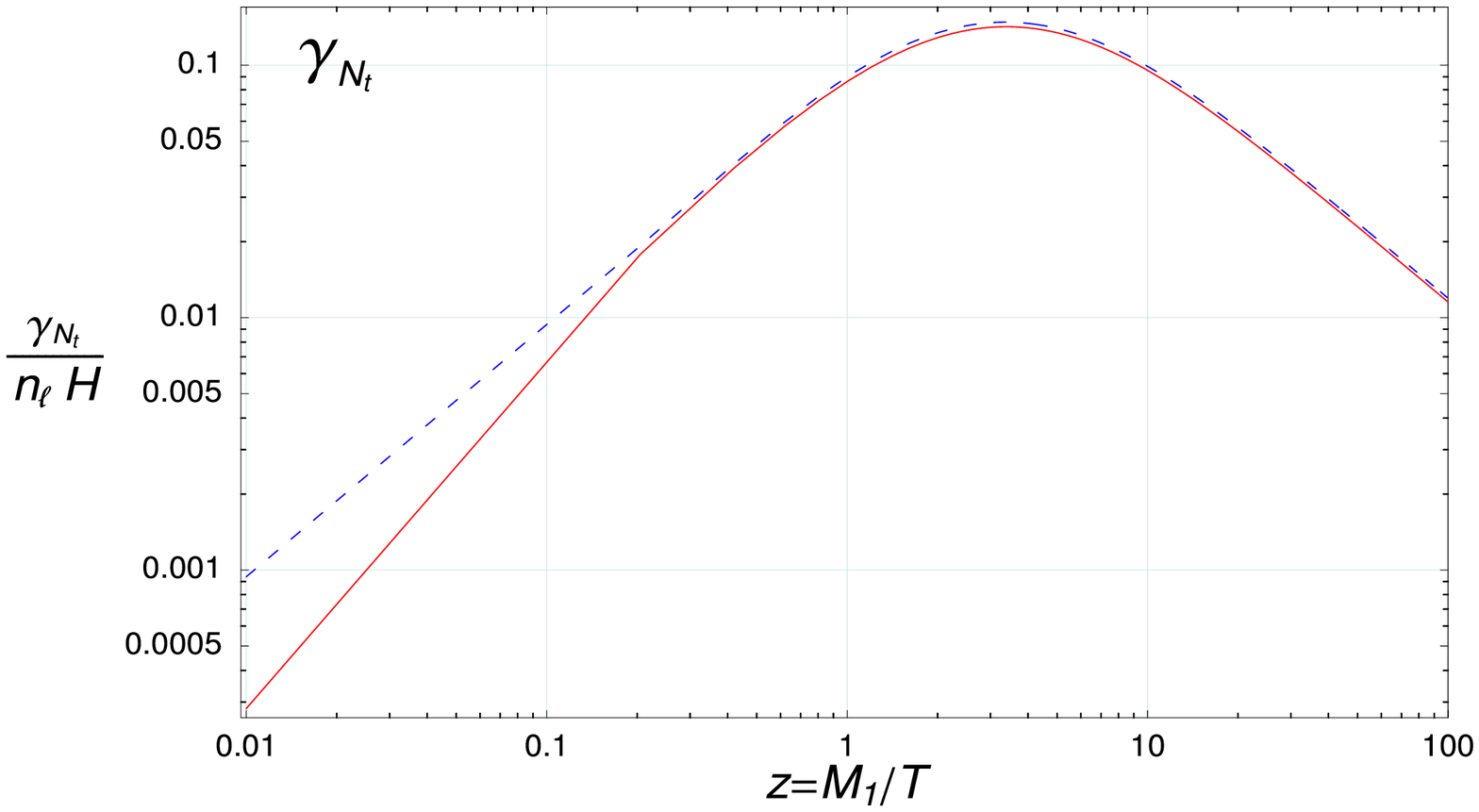}
 \includegraphics[width=8cm,height=7cm,angle=0]{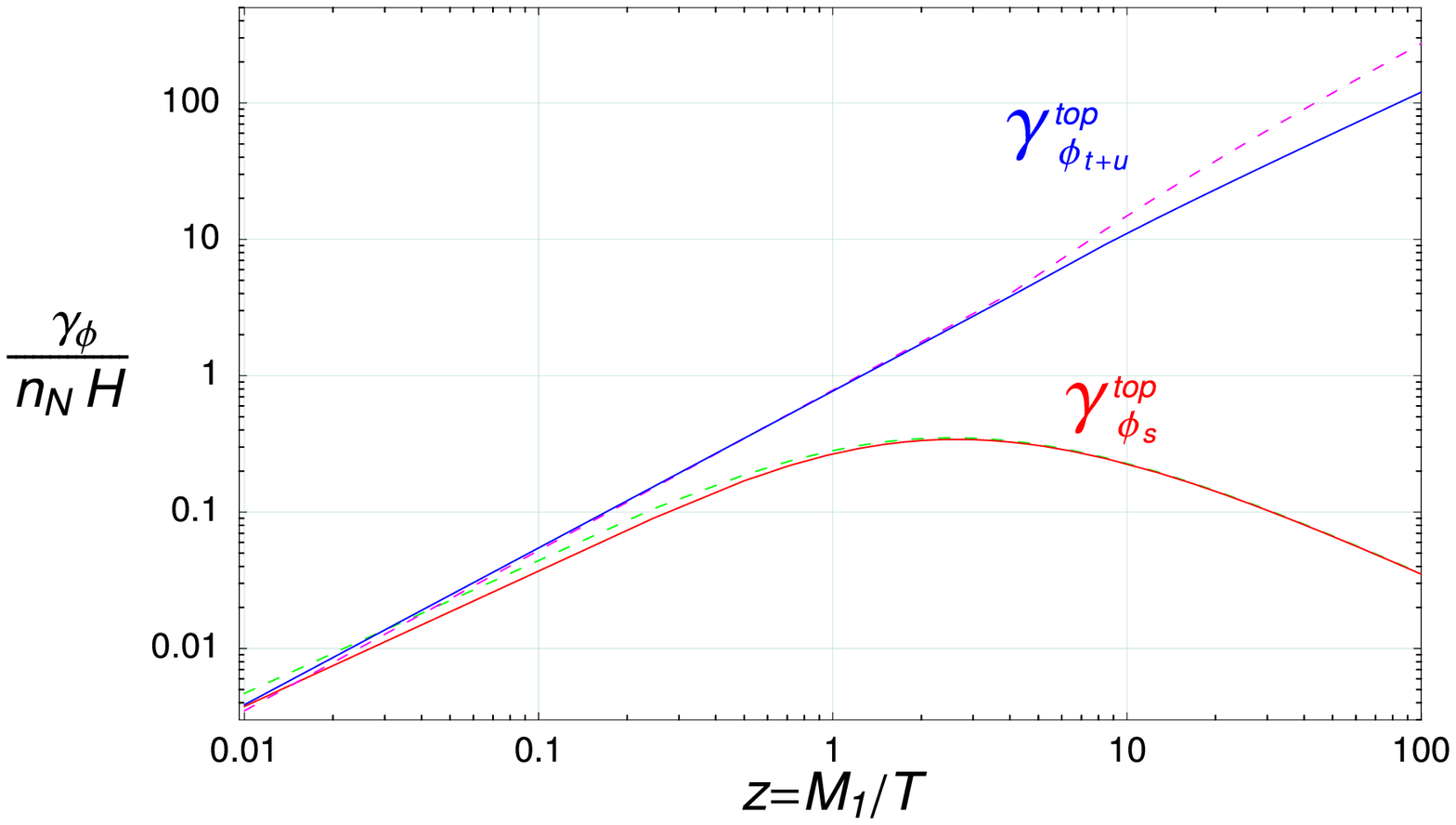}
\vskip-4mm
\hskip-4mm
\caption[]{
  Comparison between scattering rates in the Standard Model with
  (solid lines) and without (dashed lines) thermal corrections. We use
  $\tilde m=0.06\,$eV and $M_1=10^{10}\,$GeV.  $(a)$ $\gamma_{N_t}$:
  the rate density of the $t$-channel $N_1$-exchange scattering
  $\ell\,\ell\leftrightarrow\bar\phi\,\bar\phi$, normalized to $n_\ell
  H$. ($b$) $\gamma^{\rm top}_{\phi_s}$ and $\gamma^{\rm
  top}_{\phi_{t+u}}$: the rate densities of the Higgs exchange
  scaterrings in, respectively, the $s$-channel ($q_3\,\bar
  t\leftrightarrow\ell\,N$) and the $t$- and $u$-channels
  ($q_3\, N\leftrightarrow\ell\, t$ and $\bar t\, N \leftrightarrow
  \ell\, \bar q_3$), normalized to $n_{N_1} H$. (Figures adapted from
  ref.~\cite{munoz:2007}.)}
\label{fig:bdM1}
\end{figure}

\subsection{CP asymmetries}
\label{sec:thermalcp}
As discussed in section \ref{comments}, CP asymmetries arise from the
interference of tree level and one-loop amplitudes, when the relevant
couplings involved have complex phases, and the loop diagrams have an
absorptive part.  This last condition is satisfied whenever the loop
diagram can be cut in such a way that the particles in the cut lines
can be produced on shell. In the $N_1$ decay asymmetry at zero
temperature this is guaranteed by the fact that the decay products,
the Higgs $\phi$ and the lepton doublet $\ell$, coincide with the
states circulating in the loops. However, at the high temperatures at
which the $N_1$'s decay, the Higgs and the lepton doublets are in
equilibrium with the hot plasma, and their interactions with the
background particles modify the CP asymmetries and introduce a
dependence on the temperature: $\epsilon\to\epsilon(T)$. Thermal
corrections to CP asymmetries arise from various effects:
\begin{itemize}
\item[\it i)] The possibility of absorption and re-emission of the
  loop particles by the medium requires the use of finite temperature
  propagators for computing the absorptive parts of the Feynman
  diagrams.
\item[\it ii)] The stimulation of decays into bosons and the blocking
  of decays into fermions due to the dense background requires a
  proper modification of the density distributions of the final
  states.
\item[\it iii)] Thermal motion of the decay particles with respect to
  the background breaks the Lorentz symmetry and affects the
  evaluation of the CP asymmetries.
\item[\it iv)] Thermal masses should be included in the finite
  temperature resummed propagators, and also modify the fermion and
  boson dispersion relations.  Their inclusion yields the most
  significant modifications to the zero temperature results for the
  CP asymmetries.
\end{itemize}
The first three effects were investigated in \cite{Covi:1997dr}. A
rather complete account of thermal corrections to the decay CP
asymmetries, that includes also the effects of thermal masses, can be
found in \cite{Giudice:2003jh}. In principle, at finite temperature,
there are additional effects related to new cuts that involve the
heavy $N_{2,3}$ neutrino lines (see fig.~\ref{fig1}). These new cuts
appear because the heavy particles in the loops may absorb energy from
the plasma and go on-shell.  However, for hierarchical spectrum,
$M_{2,3}\gg M_1$, the related effects are suppressed to a negligible
level by a Boltzmann factor $\exp(-M_{2,3}/T)$ that, at the
temperatures relevant for the $N_1$ decays, is tiny.

\subsubsection{Propagators and statistical distributions}
\label{sec:stat}
The real time formalism of thermal field theory
\cite{Bellac:1996,Landsman:1986uw} can be used to compute the particle
propagators at finite temperature. In this formalism, ghost fields
dual to each of the physical fields have to be introduced, and
consequently the thermal propagators have $2\times2$ matrix
structures. For the one-loop computations of the absorbtive parts of
the Feynman diagrams, the relevant propagator components are just
those of the physical fields, that for fermions ($\ell$) and bosons
($\phi$) are:
\bea
\label{eq:thermalF}
S_\ell(p,m_\ell)&=&\left[\frac{i}{p^2-m_\ell^2+i0^+}-2\pi\,
n_\ell\,\delta(p^2-m_\ell^2)\right](p\!\!\!\slash +m_\ell),\\
\label{eq:thermalB}
D_\phi(p,m_\phi)&=& \left[\frac{i}{p^2-m_\phi^2+i0^+}+2\pi\,
  n_\phi\,\delta(p^2-m_\phi^2)\right].
\eea
The leading effects in {\it i)} are proportional to the factor
$-n_\ell+n_\phi-2 n_\ell n_\phi$, where $n_{\ell,\phi}=
[\exp(E_{\ell,\phi}/T\pm 1]^{-1}$.  This factor vanishes when the
thermal masses of the leptons and of the Higgs are neglected, because
the Bose-Einstein and Fermi-Dirac statistical distributions depend on
the same argument, $E_\ell=E_\phi=M_1/2$, and consequently the thermal
correction to the fermion propagator ($n_\ell$), the thermal piece of
the boson propagator ($n_\phi$) and the product of the two thermal
corrections ($n_\ell n_\phi$) cancel each other. This can be
interpreted as a complete compensation between stimulated emission and
Pauli blocking. As regards the effects in {\it ii)}, they lead to
overall factors that cancel between numerator and denominator in the
expression for the CP asymmetry $\epsilon$.

In the supersymmetric case, the situation is more subtle. First,
the singlet neutrino $N_1$ decays not only to the standard $\ell \phi$
final state, but also to their superpartners: $\tilde \ell\tilde
\phi$.  Given that both decay channels contribute to the imaginary
part of both decay modes, the same cancellation as in the previous
case occurs (when thermal masses are neglected). Second, a new source
of lepton asymmetry comes from the scalar neutrino $\tilde N_1$ that
decays both into final fermions $\ell\tilde \phi$, and into final
bosons $\tilde \ell \phi$.  Thermal effects modify the asymmetry in
each channel. One-loop diagrams with the $\tilde \ell\phi$ bosons
contribute to the CP-asymmetry for decays into fermions, while
one-loop diagrams with internal $\ell\tilde \phi$ fermions contribute
to the CP-asymmetry for decays into bosons. As a result, the finite
temperature propagator corrections to the partial asymmetries of the
single decay channels do not vanish~\cite{Covi:1997dr}.  When the
statistical functions that block and stimulate the final state
emission (corrections of type {\it ii)}) are taken into account, the
branching fractions into fermion and boson final states differ.  When
the CP-asymmetries of the two channels are summed up, the effects of
the two types of corrections {\it i)} and {\it ii)} compensate each
other, and the zero temperature result is again
reproduced~\cite{Covi:1997dr}.

\subsubsection{Particle motion}
We have seen that when a large hierarchy $M_{2,3}\gg M_1$ is assumed,
the particle thermal masses are neglected, and the decaying particle
is considered at rest in the thermal bath, there are no thermal
corrections to the zero temperature results for the CP-asymmetries.
However, since the decaying particle is moving with respect to the
background with velocity $\vec\beta$, due to their statistics, the
fermionic decay products are preferentially emitted in the direction
anti-parallel to the plasma velocity (for which the thermal
distribution is less occupied), while the bosonic ones are emitted
preferentially in the forward direction (for which stimulated emission
is more effective). This induces an angular dependence in the decay
distribution at order ${\cal O}(\beta)$.  In the total decay rate the
${\cal O}(\beta)$ anisotropy effects are integrated out, and only
${\cal O}(\beta^2)$ effects remain~\cite{Covi:1997dr}.  Therefore,
while the inclusion of the effects of the thermal motion of the
decaying particle do modify the zero temperature results, these
corrections are numerically small \cite{Covi:1997dr,Giudice:2003jh}
and generally negligible.

\subsubsection{Thermal masses}
When the finite values of the light particle thermal masses are taken
into account, the arguments of the Bose-Einstein and Fermi-Dirac
statistical distributions are different. It is a good approximation
\cite{Giudice:2003jh} to use for the particle energies
$E_{\ell,\phi}=M_1/2\mp(m^2_\phi-m^2_\ell)/2M_1$. Since now
$E_\ell\neq E_\phi$, the prefactor $n_\ell-n_\phi+2 n_\ell n_\phi$
that multiplies the thermal corrections does not vanish anymore, and
sizeable corrections become possible.  The most relevant effect is
that the CP-asymmetry vanishes when, as the temperature increases, the
sum of the light particles thermal masses approaches
$M_1$~\cite{Giudice:2003jh}. This is not surprising, since the
particles in the final state coincide with the particles in the loop,
and therefore when the decay becomes kinematically forbidden, also the
particles in the loop cannot go on the mass shell.  The same happens
in the supersymmetric case for the CP-asymmetry in $N_1$ decays.
However, this is not the case for the decays of the scalar neutrino
$\tilde N_1 \to \ell \tilde \phi$ and $\tilde N_1 \to \tilde \ell
\phi$ for which the particles running in the loop are different from
the particles in the final states.  Since thermal masses are larger
for the scalars than for the fermions, the $\tilde N_1$ CP-asymmetry
vanishes when the decay $\tilde N_1\to \ell\tilde \phi$ is still
kinematically allowed. The relevant analytical expressions for this
case and detailed numerical results can be found
in~\cite{Giudice:2003jh}.

When the temperature is large enough that the Higgs can decay (see
section~\ref{sec:thermaldecays}), there is a new source of lepton
number asymmetry associated with the decay processes $\phi \to \ell
N_1$.  The CP-asymmetry in Higgs decays $\epsilon_\phi$ can be up to
one order of magnitude larger than the CP-asymmetry in $N_1$
decays~\cite{Giudice:2003jh}. This is mainly due to a kinematical
suppression of the tree level decay rate appearing in the denominator
of $\epsilon_\phi$ that is roughly proportional to the thermal mass
difference $m^2_\phi -m^2_\ell$.  While this represents a dramatic
enhancement of the CP-asymmetry, $\epsilon_\phi$ is non-vanishing only
at temperatures $T \gsim T_\phi \sim 5 M_1$, when the kinematical
condition $m_\phi(T)> m_\ell(T)+M_1$ is satisfied.  Therefore, in the
strong washout regime, no trace of this effect survives. On the other
hand, rather large $\lambda$ couplings are required in order that
Higgs decays can occur before the phase space closes: the decay rate
can attain thermal equilibrium only when $\tilde m \gsim
(T_\phi/M_1)^2 m_*\gg m_*$, and therefore, in the weak washout regime
($\tilde m \lsim m_*$), these decays always remain strongly out of
equilibrium.  This means that only a small fraction of the Higgs
particles have actually time to decay, and the lepton-asymmetry
generated in this way is accordingly suppressed.

In summary, while the corrections to the CP-asymmetries can be
significant at $T\gsim M_1$ (and quite large at $T\gg M_1$ for Higgs
decays), in the low temperature regime, where the precise value of
$\epsilon$ plays a fundamental role in determining the final value of
the baryon asymmetry, there are almost no effects, and the zero
temperature results still give a reliable approximation.

Before concluding this section let us mention that effects similar to
the ones described above can be expected also for the CP-asymmetries
of scattering processes, like top-quark or gauge-boson scatterings.
These asymmetries were considered in
\cite{Pilaftsis:2003gt,Pilaftsis:2005rv,Abada:2006ea} in the
approximation in which they are proportional to the CP-asymmetry in
decays, and were further analyzed in \cite{Nardi:2007jp} by going
beyond this approximation, but still in the zero temperature limit.
An important difference is that while $2\leftrightarrow 2$ scatterings
are always kinematically allowed, the absorptive part of the one-loop
diagrams vanishes at the same thresholds when the decay CP-asymmetries
vanish.  This can have the peculiar effect of thermalizing the $N_1$
degree of freedom without producing an associated lepton asymmetry. To
our knowledge, a study of the effects of thermal corrections to
CP-asymmetries in scattering has not been carried out yet.

\newpage
\section{Spectator Processes}
\label{sec:spec}

\subsection{Introduction}
\label{sec:basics}
During leptogenesis, various processes can modify the densities of
particle species.  Some of these, such as the heavy neutrino decays or
various interactions that washout the lepton number, occur on a time
scale comparable to the expansion rate of the Universe, and hence
should be accounted for via appropriate Boltzmann equations. Other
processes can be very fast (depending on the temperature considered)
and their effect is to impose certain relations among the chemical
potentials of various particle species. These processes include the
gauge interactions, Yukawa interactions involving the heavier
fermions, and the electroweak and QCD non-perturbative `sphaleron'
processes.  They are called `spectator processes' because they do not
change lepton number directly. Instead, they affect it indirectly, by
changing the densities of the lepton doublets and of the Higgs on
which the rates of washout processes depend. The issue of spectator
processes and their effects on leptogenesis was first raised in ref.
\cite{Buchmuller:2001sr}.  A rather complete analysis of the numerical
relevance of each spectator processes can be found in ref.
\cite{Nardi:2005hs}. The main results of this section can be red off
from the last column in Table~\ref{tab:spectator-results},   
where the effects of various spectator processes are quantified as 
a percentage variation of the final asymmetry resulting from leptogenesis.

We work in the scenario in which the singlet neutrino masses are
hierarchical, $M_1\ll M_{2,3}$, and the lepton asymmetry is generated
mainly via the CP and lepton number violating decays of the lightest
singlet neutrino $N_1$. We restrict the discussion to the
non-supersymmetric case, since no qualitative new features appear in
the supersymmetric case.

Several spectator processes become relevant only in the temperature
regime in which lepton flavour effects are also important.  In
particular, reactions mediated by the $\tau$ ($\mu$) Yukawa couplings
become faster than the expansion rate of the Universe below $T\sim
10^{12}\,$GeV ($10^{9}\,$GeV). Then it becomes important to know the
flavour composition of the lepton asymmetry.
For simplicity, in our discussion of spectator processes we assume
that leptons and antileptons produced in the decays of $N_1$ are
aligned with or orthogonal to some specific lepton and antilepton
flavours $\ell_\alpha$ and $\bar \ell_\alpha$ (with
$\alpha=e,\,\mu,\,\tau$).
That is,  we assume that the only
non-vanishing element in the matrix
\beq\label{eq:Pab}
 P_{\alpha\beta}= \langle \ell_\alpha|
\ell_{N_1} \rangle \langle \ell_\beta| \ell_{N_1} \rangle^*.
\eeq
that characterizes the flavour composition of the lepton $\ell_{N_1}$
into which $N_1$ decays, is $P_{\alpha\alpha}=1$ while all the other
diagonal and non-diagonal entries are 0.  Accordingly, in the
following we replace the notation $\ell_{N_1}$ with $\ell_\alpha$,
that is appropriate to denote pure flavour states.

We divide the processes that generate and washout the
$B/3-L_\alpha$ asymmetry into three classes:
\begin{itemize} \itemsep=1pt
\item[\it i)] $N_1$ decays and inverse decays, $N_1 \leftrightarrow
  \ell_{\alpha}\, \phi$, and on-shell $N_1$-mediated $\Delta L=2$
  scatterings $\ell_{\alpha}\, \phi \leftrightarrow
  \overline{\ell}_{\alpha}\, \overline{\phi}$ (the on-shell part of
  the first diagram in  fig.~\ref{fig:scatterings}$(d)$);
\item[\it ii)] $\Delta L=1$ Higgs-mediated scattering processes involving
the top-Yukawa coupling, $\ell_{\alpha}\, N_1 \leftrightarrow q_3\, \bar t$,
$\ell_{\alpha}\, \bar q_3 \leftrightarrow N_1\, \bar t$ and
$\ell_{\alpha}\, t\leftrightarrow N_1\, q_3$
(fig.~\ref{fig:scatterings}$(a)$).
\item[\it iii)] $\Delta L=1$ scatterings with the gauge bosons, such a
  $\ell_\alpha N_1\to A\bar\phi$, $\ell_\alpha A\to N_1\bar\phi$ and
  $\ell_\alpha \phi \to A N_1$ with $A=W_i $ or $B$
  (figs.~\ref{fig:scatterings}$(b)$ and \ref{fig:scatterings}$(c)$).
\end{itemize}

Note that all the rates above depend (at tree level) on a single
combination of neutrino Yukawa couplings that can be parameterized, as
in eqn (\ref{tildem}), by $\tilde m_{11} \equiv \tilde
m=[\lambda\lambda^\dagger]_{11} v^2_u/M_1 $.  Other (subleading)
washout reactions couple to lepton states that are different from
$\ell_{N_1}$ and $\bar\ell_{N_1}$.  In particular, we refer here to
the $\Delta L=2$ scatterings which go through off-shell $s$-channel,
$\ell \phi \leftrightarrow \bar\ell\bar\phi$, and $t$-channel,
$\ell\ell\leftrightarrow\bar\phi\bar \phi$
(fig.~\ref{fig:scatterings}$(d)$). In the temperature regime $T<M_1$,
and including the contributions from $N_{2,3}$, the amplitude for
these processes is proportional to the light neutrino mass matrix,
$[m]_{\alpha\beta}$ of eqn (\ref{mass}). Consequently, the fastest
rate couples to the lepton doublet containing the heaviest light
neutrino state $\nu_3$. However, being of higher order in the
$\lambda$ couplings, these processes are generally negligible, with a
possible exception in the high temperature regimes, $M_1 >
10^{13}$~GeV, where some Yukawa couplings are of order unity.  Since
flavour effects, as well as the majority of the spectator processes,
become relevant only at $T < 10^{13}$~GeV, in the following we
consider $\ell_{\alpha}$ and $\bar \ell_{\alpha}$ as the only relevant
directions in flavour space.

By definition, washout processes are lepton number violating reactions
that tend to destroy any excess of lepton or of antileptons. In
general, they depend also on the abundances of other particle species.

The washout reactions that we consider are contained in the first
three lines in eqn (\ref{eq:finalBEYDeltai-w}).  Note that since we
are assuming alignment conditions, $\ell_\alpha=\ell_\beta$, some of
the terms vanish.  Washout reactions that involve the gauge bosons
(figs.~\ref{fig:scatterings}$(b)$, \ref{fig:scatterings}$(c)$) that
appear in the last line in eqn (\ref{eq:finalBEYDeltai-w}) are equally
important. However, since the chemical potential for gauge bosons
vanishes, no further density asymmetries are associated with these
reactions, and for simplicity we neglect them. As a result of our
approximations and simplifications, washout rates are controlled by
the following density asymmetries: $\Delta y_{\ell_\alpha}$ for the
lepton doublets, $\Delta y_\phi$ for the Higgs, $\Delta y_{q_3}$ for
the third generation quark doublet and $\Delta y_{t}$ for the
top-quark singlet.  The latter two quantities always appear in the
combination $\Delta y_{t}-\Delta y_{q_3}$. It is useful to recall that
$\Delta y_{\ell_\alpha}$, $\Delta y_\phi$ and $\Delta y_{q_3}$
represent the sum of the asymmetries in the two components of the
relevant $SU(2)$-doublet, and that (in the non-supersymmetric case)
the densities of the lepton and baryon number charges are
\begin{eqnarray}
\label{eq:Ltot}
Y_{\Delta L} &=& \sum_{\alpha=e,\mu,\tau}  Y_{\Delta L_\alpha}=
\sum_\alpha( \Delta y_{\ell_\alpha}+\Delta y_{e_\alpha})Y^{\rm eq},\\
Y_{\Delta B}&=&\frac{1}{3}\sum_{i=1,2,3}( \Delta y_{q_i}+
\Delta y_{u_i}+\Delta y_{d_i})Y^{\rm eq},
\end{eqnarray}
where $u_i$, $d_i$ denote the $SU(2)$-singlet quarks of the $i$-th generation
and $e_\alpha$ are the $SU(2)$-singlet leptons.  Similar relations hold also
for the densities of the flavoured charges $Y_{\Delta_\alpha}\equiv
Y_{\Delta B}/3-Y_{\Delta L_\alpha}$, that are individually conserved by
the electroweak sphalerons.

\subsection{Detailed analysis}
\label{sec:equilibrium}
Relations among the asymmetry-densities $\Delta y_{\ell_\alpha}$,
$\Delta y_\phi$ and $\Delta y_{t}-\Delta y_{q_3}$ are determined by
the chemical equilibrium conditions enforced by the reactions that are
faster than the expansion rate of the Universe. The dependence of the
expansion rate on temperature is different from that of various
particle interaction rates, and as the temperature drops down, more
and more interactions become faster than the expansion rate and `enter
into equilibrium'. Thus, the equilibrium conditions change with
temperature.

Since leptogenesis takes place at temperatures $T\sim M_1$, the
relevant constraints depend on the value of $M_1$. As discussed in
Appendix \ref{chempot}, the density of the charge $B/3-L_\alpha$
(where in this section $\alpha$ is the flavour direction into which
$N_1$ decays) is conserved by all standard model processes, but not by
interactions involving $N_1$. All the relevant asymmetry-densities can
be expressed as linear functions of the charge asymmetry
$Y_{\Delta_\alpha}\equiv \frac13Y_{\Delta B}-Y_{\Delta L_\alpha}$, and
the equilibrium conditions fix the coefficient of proportionality for
each temperature range:
\begin{equation}
\label{eq:cc}
\Delta y_{\ell_\alpha}=-c_\ell\> \frac{Y_{\Delta_\alpha}}{Y^{\rm eq}},
\qquad  \qquad
\Delta y_\phi = -c_\phi\> \frac{Y_{\Delta_\alpha}}{Y^{\rm eq}}.
\end{equation}
The two coefficient $c_\ell$ and $c_\phi$ encompass all the effects of
the relevant spectator processes (charged lepton and quark Yukawa
interactions, and the electroweak and QCD sphalerons). When
generalized to multiple generations, $c_\ell$ generalizes to a matrix
(that is the inverse of the $A$-matrix \cite{Barbieri:1999ma} that is
introduced at the end of section~\ref{sec:flavor}).

We distinguish between six relevant temperature ranges according to
the set of interactions that are in equilibrium. For each such
temperature regime we present the equilibrium conditions. We impose,
when relevant, various conditions of flavour alignment and calculate
the corresponding $c_\ell$ and $c_\phi$ defined in eqn (\ref{eq:cc}).
Note that $c_\ell$ and $c_\phi$ give a crude understanding of the
impact of the respective asymmetries: $c_\phi/c_\ell$ gives a rough
estimate of the relative contribution of the Higgs to the washout,
while $c_\ell + c_\phi$ gives a measure of the overall washout
strength.  The quantitative significance of the different spectator
processes can be red off from the last column in
Table~\ref{tab:spectator-results} (adapted from
ref.~\cite{Nardi:2005hs}) where for the different temperatures, and
assuming a strong washout regime with $\tilde m=0.06\,$eV, we give the
percentage variations with respect to the case when all spectator
processes are neglected ($c_\ell=1$, $c_\phi=0$).

\bigskip
\noindent
1) {\it Only gauge and top-Yukawa interactions in equilibrium}
($T> 10^{13}\,$GeV). \\[3pt]
Since in this regime the electroweak sphalerons are out of
equilibrium, no baryon asymmetry is generated during leptogenesis.
Moreover, since the charged lepton Yukawa interactions are negligible,
the lepton asymmetry is just in the left-handed degrees of freedom and
confined in the $\ell_\alpha$ doublet, yielding $Y_{\Delta L}= \Delta
y_{\ell_\alpha}\ Y^{\rm eq} =-Y_{\Delta_\alpha}$. As concerns $\Delta
y_\phi$, although initially equal asymmetries are produced by the
decay of the heavy neutrino in the lepton and in the Higgs doublets,
the Higgs asymmetry is partially transferred into a chiral asymmetry
for the top quarks ($\Delta y_t -\Delta y_{q_3}\neq 0$) implying
$\Delta y_{\ell_\alpha}\neq \Delta y_\phi$.
We see from the last column in Table~\ref{tab:spectator-results} that
the inclusion of the Higgs asymmetry yields a sizeable reduction
in the surviving asymmetry.

\bigskip
\noindent
2) {\it Strong sphalerons in equilibrium} ($T\sim 10^{13}\,$GeV). \\[3pt]
QCD sphalerons enter equilibrium at higher temperatures than the
corresponding electroweak processes because of their larger rate
($\Gamma_{\rm QCD}\sim 11(\alpha_s/\alpha_W)^5\Gamma_{\rm EW}$
\cite{Moore:1997im}). These processes are likely to be in equilibrium
already at temperatures $T_s\sim 10^{13}$~GeV
\cite{Moore:1997im,Bento:2003jv,Mohapatra:1991bz}) and yield the
constraint
\begin{equation}
\sum_i\left(2 \mu_{q_i}-\mu_{u_i}-\mu_{d_i} \right)=0\,.
\label{qcdsph}
\end{equation}
Direct comparison with the previous case allows us to estimate the
corresponding effects. The relation $Y_{\Delta L}= \Delta
y_{\ell_\alpha} Y^{\rm eq}=-Y_{\Delta_\alpha}$, implying $c_\ell=1$,
holds also for this case. However, switching on the QCD sphalerons
reduces the Higgs number asymmetry by a factor of $21/23$. This effect
yields a suppression of the washout that does not exceed the few
percent level.

\bigskip

%
\begin{table}[t!]
\begin{center}
  \renewcommand{\arraystretch}{1.4}
\begin{tabular}
{p{0cm}|>{\centering\small}p{1.4cm}>{\raggedleft\arraybackslash\small}p{2.3cm}>
{\raggedleft\small}c>{\raggedleft\small}c>{\raggedright\small}p{0.4cm}c}
\omit
 & \multicolumn{6}{c}{\bf Equilibrium processes, constraints,
  coefficients and effects on $Y_{\Delta_\alpha}$}
\\ \hline\hline
&&&&& \\[-14pt]
&{\small$T\,$(GeV)}& {\small Equilibrium}  & {\small Constraints} &
$\hbox{\large \it c}_\ell$ &
$\hbox{\large \it c}_\phi$ &
$\delta(|Y_{\Delta_\alpha}|)$ 
 \\[4pt]  \hline
& --  & \quad no spectators  &
-- 
                              & $1 $ & $0\, $ &
--
\\
& $\ \ \gg 10^{13}$ & \quad $h_t$, gauge  &
$B=\sum_i(2q_i+u_i+d_i)=0$\hspace{.1cm}
                              & $1 $ & $\frac{2}{3}\, $
& $-40\%$
\\
 &$\  \sim 10^{13}$ &  +\ QCD-Sph  & $\sum_i(2q_i-u_i-d_i)=0\hspace{1.cm} $
                              & $1 $ & $\frac{14}{23}\, $
&  $-37\%$
\\ [4pt]
\begin{rotate}{90}
$\ B=0$
\end{rotate}
&$10^{12\div 13}$ & + \quad  $h_b$,\ $h_\tau \quad $ &
$
\begin{array}{l}
\raise 3pt \hbox{$b=q_3-\phi,$}\\[-2pt]
 \tau=\ell_\tau-\phi
\end{array}
\left\{\begin{array}{l}
\!\!\! P_{\tau\tau}\!=\!0 \\[-4pt]
\!\!\! P_{\tau\tau}\!=\!1
\end{array}\right.$ &
$\begin{array}{l}\!
   1  \\[-2pt]
   \! \frac{3}{4}
\end{array}$ &
    $\hspace{-2mm}
\begin{array}{l}
\frac{3}{8} \\[-2pt]
\> \frac{1}{2}
\end{array}$
&
$ \begin{array}{l}
\> -27\% \\[-2pt]
\> -17\%
\end{array}$
\\[20pt] \hline
 &&&&& \\[-16pt]
&$10^{11\div 12}$  &  +\quad  EW-Sph &
          $\sum_{i,\alpha}(3q_i+\ell_\alpha)=0\ \
\left\{\begin{array}{l}
\!\!\! P_{\tau\tau}\!=\!0 \\[-2pt]
\!\!\! P_{\tau\tau}\!=\!1
\end{array}\right.$ &
    $\begin{array}{l}
\!\! \frac{98}{115}   \\[-2pt]
\!\! \frac{78}{115}
\end{array}$
&
    $\hspace{-2mm}
    \begin{array}{l}\,
  \frac{41}{115}   \\[-0pt]
  \, \frac{56}{115}
\end{array}$
  &
    $\hspace{-2mm}
    \begin{array}{l}\,
   -15\%   \\[-0pt]
  \,  -10\%
\end{array}$
\\[14pt]
\begin{rotate}{90}
$ B\neq 0$
\end{rotate}
&$10^{8\div 11}$ & +\ $h_c$, $h_s$, $h_\mu$
& $ \begin{array}{l}
c\!=\!q_2\!+\!\phi, \\[-3pt]
s\!=\!q_2\!-\!\phi,  \\[-3pt]
\mu\!=\!\ell_\mu\!-\!\phi
\end{array}
\ \left\{ \begin{array}{l}
\!\!\! P_{ee}=\!1 \\ [-2pt]
\!\!\! P_{\tau\tau}\!=\!1
\end{array}\right.\ \ $ &
$\begin{array}{l}
\!\!   \frac{151}{179}  \\[-2pt]
\!\!  \frac{344}{537}
\end{array}$ &
$\begin{array}{l}
\! \frac{37}{179} \\[-2pt]
\! \frac{52}{179}
\end{array}$
&
$\begin{array}{l}
\!\ -3\% \\[-2pt]
\! +12\%
\end{array}$
\\[20pt]
& $\ll 10^{8}$
& 
$ \begin{array}{l}
  \hbox{\rm \small
    all Yukawas}\ h_i,\ h_\alpha
 \end{array} $
&
$P_{ee}=\!1$
  &
  $\frac{442}{711}$
&
$\frac{16}{79}$
& $+27\%$
\\[12pt]
\hline \hline
\end{tabular}
\caption{\baselineskip 12pt
  The relevant quantities in the various temperature regimes. Chemical
  potentials are labeled here with the same notation used for the
  fields: $\mu_{q_i}\!=\!q_i$, $\mu_{\ell_\alpha}\!=\!\ell_\alpha$ for
  the $SU(2)$ doublets, $\mu_{u_i}\!=\!u_i$, $\mu_{d_i}\!=\!d_i$,
  $\mu_{e_i}\!=\!e_i$ for the singlets and $\mu_\phi=\phi$ for the
  Higgs.  The relevant reactions in equilibrium in each regime are
  given in the second column and the constraints imposed in the
  third. The conditions adopted for $P_{\alpha\alpha}$ are indicated
  (the appropriate constraints on the conserved quantities $
  \Delta_\beta =B/3 \!-\!  L_\beta$ with $\beta=e,\,\mu,\,\tau$ should
  also be imposed).  The values of the coefficients $c_\ell$ and
  $c_\phi$ are given in, respectively, the fourth and fifth
  column. For each regime (and assuming $\tilde m=0.06\,$eV in all the
  cases), the last column quantifies the percentage variation in the
  absolute value of $Y_{\Delta_\alpha}$ with respect to the case when
  all spectator processes are neglected (first line in the table).}
\label{tab:spectator-results}
\end{center}
\end{table}

\noindent
3) {\it Bottom- and tau-Yukawa interactions in equilibrium}
($10^{12}\ {\rm GeV}\lsim T\lsim 10^{13}\,$GeV). \\[3pt]
The asymmetries in the $SU(2)$-singlet $b$ and $e_\tau$ degrees of freedom
are populated. The corresponding chemical potentials obey the equilibrium
constraints
\beq
\mu_b=\mu_{q_3}-\mu_\phi,\ \ \ \mu_\tau=\mu_{\ell_\tau}-\mu_\phi.
\eeq
Possibly, $h_b$ and $h_\tau$ Yukawa interactions enter into
equilibrium at a similar temperature as the electroweak sphalerons
\cite{Bento:2003jv}.  However, in order to quantify separately the
impact of these two effects, we first consider the possibility of a
regime with only gauge, QCD sphalerons and the Yukawa interactions of
the whole third family in equilibrium.  As concerns the flavour
composition of the lepton asymmetry, we distinguish two alignment
cases: first, when the lepton asymmetry is produced in a direction
orthogonal to $\ell_\tau$ ($P_{\tau\tau}=0$) and second, when it is
produced in the $\ell_\tau$ channel ($P_{\tau\tau}=1$). When
$P_{\tau\tau}=0$, the lepton asymmetry is produced in one of the two
directions orthogonal to $\ell_\tau$ and therefore it does not `leak'
into the $SU(2)$ singlet degrees of freedom, implying that $c_\ell=1$
still holds. In the case where $P_{\tau\tau}=1$, the washout effects
are somewhat suppressed, since the lepton asymmetry is partially
shared with $e_\tau$ that does not contribute directly to the washout
processes.  Our results for these two cases suggest that the effect on
the final value of $Y_{\Delta_\alpha}$ associated to the $\tau$ Yukawa
interactions is of the order of 10\%.

\bigskip
\noindent
4) {\it Electroweak sphalerons in equilibrium }
($10^{11}\ {\rm GeV}\lsim T\lsim 10^{12}\,$GeV). \\[3pt]
The electroweak sphaleron processes take place at a rate per unit volume
$\Gamma_{EW}/V\propto T^4\alpha_W^5\log(1/\alpha_W)$
\cite{Arnold:1998cy,Bodeker:1998hm,Arnold:1996dy}, and are expected to be
in equilibrium from temperatures $\sim 10^{12}$~GeV, down to the
electroweak scale or below \cite{Bento:2003jv}. Electroweak sphalerons
equilibration implies
\begin{equation}
\label{ewsph}
\sum_i\left( 3\mu_{q_i}+\mu_{\ell_\alpha} \right)=0\,.
\end{equation}
As concerns lepton number, each electroweak sphaleron transition creates all
the doublets of the three generations, implying that individual lepton flavour
numbers are no longer conserved. As concerns baryon number, electroweak
sphalerons are the only source of $B$ violation, implying that baryon number
is equally distributed among the three families of quarks. In particular,
for the third generation, $B_3=B/3$ is distributed between the doublets
$q_3$ and the singlets $t$ and $b$.
In Table \ref{tab:spectator-results} we give the coefficients $c_\ell$ and
$c_\phi$ for the two aligned cases: (i) $P_{\tau\tau}=0$
implying $Y_{\Delta_\tau}=0$, and (ii) $P_{\tau\tau}=1$ implying
$Y_{\Delta_e}=Y_{\Delta_\mu}=0$. We see that in this case the transfer of part
of the lepton asymmetry to a single right-handed lepton ($e_\tau$) can have
a 5\% enhancing effect on the final value of $Y_{\Delta_\alpha}$.

\bigskip
\noindent
5) {\it Second generation Yukawa interactions in equilibrium}
($10^{8}\ {\rm GeV}\lsim T\lsim 10^{11}\,$GeV). \\[3pt]
In this regime, the $h_c$, $h_s$ and $h_\mu$ interactions enter into
equilibrium. We consider two cases of alignment: (i) $P_{ee}=1$
implying $Y_{\Delta_\mu}=Y_{\Delta_\tau}=0$, and (ii) $P_{ee}=0$.  To
ensure a pure states regime we further assume complete alignment with
one of the two flavours with Yukawa interactions in equilibrium, for
definiteness $P_{\tau\tau}=1$, and therefore
$Y_{\Delta_e}=Y_{\Delta_\mu}=0$.  The difference in $c_\ell$ between
the two aligned cases is larger than in the regimes 3 and 4, and
accordingly the difference in the corresponding values of
$Y_{\Delta_\alpha}$, of the order of 15\%, is somewhat larger than in
the cases in which just the third generation Yukawa couplings are in
equilibrium.

\bigskip
\noindent
6) {\it All Yukawa interactions in
  equilibrium} ($T\lsim 10^{8}\,$GeV). \\[3pt]
In this regime, since all quark Yukawa interactions are in equilibrium
(actually this only happens for $T<10^6$~GeV), the QCD sphaleron
condition becomes redundant. Hence ignoring the constraint of
eqn (\ref{qcdsph}), as is usually done in the literature, becomes fully
justified only within this regime. If, however, leptogenesis takes
place at $T> 10^8$~GeV, as occurs for hierarchical singlet $N$'s, the
constraint implied by the QCD sphalerons is non-trivial, even if the
associated numerical effects are not large.

Due to the symmetric situation of having all Yukawa interactions in
equilibrium we have just one possible flavour alignment (the other two
possibilities being trivially equivalent). We take for definiteness
$P_{ee}=1$, implying $Y_{\Delta_\mu}=Y_{\Delta_\tau}=0$.  In this case
$c_\ell$ is reduced by a factor of almost two with respect to the case
in which the spectator processes are neglected ($c_\ell =1$) and the
final value of $Y_{\Delta_\alpha}$ is correspondingly enhanced. The
reason for the reduction in $c_\ell$ can be traced mainly to the fact
that a sizable amount of $B$ asymmetry is being built up at the
expense of the $L$ asymmetry, and also a large fraction of the
asymmetry is being transferred to the right-handed degrees of freedom
at the same time when inverse decays and washout processes are active,
reducing the effective value of $\Delta y_{\ell_e}$ that contributes
to drive these processes. 

To summarize, we considered the possible impact of the spectator
processes.  Conditions of flavour alignment/orthogonality were imposed,
to ensure that these effects are cleanly disentangled from lepton
flavour effects (these are numerically more significant, but of a
different nature). A rough quantitative understanding of spectator
processes can be obtained by relying on the fact that the surviving
asymmetry is inversely proportional to the washout rate, as discussed
in section~\ref{toy}.  Hence, the final $Y_{\Delta_\alpha}$
asymmetries obtained in the relevant temperature regimes will be
inversely proportional to $c_\ell+c_\phi$. This suggests that
numerical corrections related to the proper inclusion of spectator
processes have at most ${\cal O}(1)$ effects~\cite{Nardi:2005hs}.
Inspecting the table, we learn that when the electroweak sphalerons
are not active and all Yukawa interactions (except those of the
top-quark) are negligible, the Higgs contribution enhances the washout
processes, leading to a smaller final $Y_{\Delta_\alpha}$ asymmetry.
As more and more spectator processes become fast (compared to the
expansion rate of the Universe), the general trend is towards reducing
the value of the washout coefficients and hence increasing the
$Y_{\Delta_\alpha}$ up to values that can be slightly larger then what
is obtained when all spectator processes are neglected.

\newpage
\section{Flavour Effects}
\label{sec:flavor}

The aim of this section is to discuss   
what are  flavour effects in leptogenesis, why
they arise, and when they matter.
It makes use of results from Sections \ref{real}, \ref{sec:spec}
and Appendices \ref{kinetic}, \ref{chempot} and  \ref{apenrho},  
but should be independently
readable. 
In  Section \ref{puzzle},  we  introduce the puzzle 
of flavour in leptogenesis calculations, which is
that the baryon asymmetry depends on
the  choice of lepton flavour basis. 
 We also
give heuristic rules for how to choose
 the correct basis. 
Sections \ref{sec:basicflavour} and \ref{sec:flavormix}
illustrate how flavour effects can  modify the final baryon
asymmetry. 
Spectator processes (including sphalerons)
are neglected in Section \ref{sec:basicflavour},
where we  focus on  how flavour dynamics can strongly
enhance the asymmetries in lepton doublets $Y_{\Delta \ell_\alpha}$.
  The sum of the  $Y_{\Delta \ell_\alpha}$
 is usually of the order of the final baryon asymmetry; this 
is discussed  in Section \ref{sec:flavormix}, where  spectator  
effects are included, and the relation between the 
charge densities $Y_{\Delta_\alpha}$ and the 
asymmetries for the various particle species is elucidated.
 Notice that it is important to distinguish the asymmetries in  lepton 
doublets $Y_{\Delta \ell_\alpha}$ 
from  the  $B/3 - L_\alpha$ charge densities $Y_{\Delta_\alpha}$,
even though they are similar in our notation.
In
Section~\ref{sec:phenoflavour} we analyze the main phenomenological
consequences of including the effects of the lepton flavours.

Historically,  leptogenesis calculations were performed in
the ``single flavour approximation'',  which consists
of studying Boltzmann Equations for the  $B-L$ asymmetry.
 Reference \cite{Barbieri:1999ma}  considered the BE for the
asymmetries in $B/3 - L_\alpha$,
but  did not emphasize  that the results were
significantly different from  the single flavour approximation. 
In subsequent years, various authors 
\cite{Nielsen:2001fy,Vives:2005ra,Abada:2004wn}
 noticed that flavour effects could
be used to enhance the baryon asymmetry in particular
models. The  flavoured BE are
given  in \cite{Endoh:2003mz}.   The importance of  studying  lepton
asymmetries flavour by flavour,  in order to obtain
 a ``reliable''  estimate of the baryon asymmetry,
was  recently highlighted  in 
\cite{Pilaftsis:2005rv,Abada:2006fw,Nardi:2006fx, Abada:2006ea}.

\subsection{The flavour puzzle}
\label{puzzle}
In the introduction to Section \ref{toy},  washout
interactions were presented as 
a critical ingredient for thermal leptogenesis, and
the importance of flavour effects was said to follow from the importance of
washout: the washout interactions have lepton doublets in the initial state,
so to compute the washout rates one needs to know which leptons are
distinguishable.  In Section \ref{real} we became acquainted with the
Boltzmann Equations, which describe the detailed evolution of asymmetry
production and of washout processes.  Upon solving the BE, one could expect
to get a correct description of washout, and discover if flavour
matters in leptogenesis.  Unfortunately, this is not the case.  The BE are not
covariant in lepton flavour space, and solving them in different bases gives
different answers.

A simple example can illustrate this problem. Suppose that $N_1$
decays  to $\ell _\mu$ and $\ell_\tau$  with equal branching ratios, 
so that 
$\hat{\ell}_{N_1} = (\hat\ell_\mu + \hat\ell_\tau)/\sqrt{2} $ 
(see eqn (\ref{toshow})). 
Furthermore, assume that the asymmetries in both flavours are the 
same: $\epsilon_{\mu \mu} = \epsilon_{\tau \tau}$.
Let us consider the simple BE eqn (\ref{mess3a}), 
and  neglect the Higgs asymmetry $\Delta y_\phi$ and all spectator 
processes, that are irrelevant for the present discussion. 
In this approximation the total lepton asymmetry coincides with the 
asymmetry stored in the lepton doublets: $Y_{\Delta L}=
Y^{\rm eq}\sum_\alpha \Delta y_{\ell_\alpha}\equiv Y^{\rm eq}\Delta y_{\ell} $.
To write down the BE   for the total lepton asymmetry 
we can proceed in two ways.
We can chose to work in the  basis of $\hat{\ell}_{N_1}$ and 
of two other flavours orthogonal 
to  $\hat{\ell}_{N_1}$ that, by assumption, 
are not produced in $N_1$ decays.
In this case the  only  CP asymmetry
is  $\epsilon_{\hat{\ell}_{N_1}}= \sum_\alpha \epsilon_{\alpha
  \alpha}\equiv\epsilon$,
and furthermore $\tilde m_{\hat{\ell}_{N_1}}=\tilde m$.
The  BE for the total lepton asymmetry then is 
\beq
\label{eq:noflavor}
\dot Y_{\Delta L } =
 \left[(y_{N_1}-1)\epsilon -
\frac{1}{2}\Delta
y_{\ell}\right]\gamma_{N_1\to2}\,.
\label{8.1.1}
\eeq
Alternatively, we can chose to work in the 
``flavour'' (charged lepton mass
eigenstate) basis, and in this case we  shall write 
two BE for the $\mu$ and $\tau$ asymmetries. 
By summing them up we  obtain
\bea
\label{eq:withflavor}
\dot Y_{\Delta L } = 
\dot Y_{\Delta L_\mu } + 
\dot Y_{\Delta L_\tau } 
 & =  &
 \left[(y_{N_1}-1)(\epsilon_{\mu\mu}+\epsilon_{\tau\tau}) -
\frac{1}{2}\Delta
y_{\ell_\mu}\frac{\tilde m_{\mu\mu}}{\tilde m}
- \frac{1}{2}\Delta
y_{\ell_\tau}\frac{\tilde m_{\tau\tau}}{\tilde m} \right]\gamma_{N_1\to2}
\nonumber \\
& = &  \left[(y_{N_1}-1)\epsilon -
\frac{1}{4}\Delta
y_{\ell} \right]\gamma_{N_1\to2},
\label{8.1.2}
\eea
where we have used $ \tilde m_{\mu\mu}/\tilde m =
\tilde m_{\tau\tau}/\tilde m =1/2$.
Eqns (\ref{8.1.1}) and (\ref{8.1.2}) differ by a factor of 1/2
in the washout, and therefore they  yield different
 baryon asymmetries. This simple example 
illustrates  what was (cryptically) mentioned
at the end of section \ref{toy}:  the 
rough estimates  for the  baryon asymmetry
given in eqns (\ref{matter}) and (\ref{um})
depended on the choice of the flavour basis.  

To obtain a reliable result, we either need a formalism that is flavour
covariant, or we need to guess which is the correct basis.  Here
we opt to guess, based on  physical intuition.
(A flavour covariant toy model, that motivates our guesses,
is discussed in Appendix \ref{apenrho}.) \smallskip

Two basic points provide guidance  in making the
correct guess:
\begin{enumerate}
\item  washout processes are  critical in  leptogenesis; 
\item   interactions whose  timescale  is very different from 
that of  leptogenesis, drop out of the BE.
\end{enumerate}
The second point is usual Effective Field Theory:
interactions which are strong should be resummed,
very weak interactions can be neglected.
 In particular,
interactions that are much faster than
the timescale of leptogenesis and of the Universe expansion rate, are 
``resummed''  into  thermal corrections (see Section \ref{sec:thermal})
 and into chemical equilibrium conditions (see Section \ref{sec:spec}). 
The second point relates to the first, because
to perform a correct calculation of 
 the  washout rates,   one 
 should  make the correct choice of basis
for the  lepton doublets which are in the 
initial state.   
The fast flavour-dependent interactions,  mentioned in point 2. are precisely 
the ones that  
can resolve the flavour basis ambiguity.

The lepton doublets of different flavours are distinguished in the Lagrangian
by their Yukawa couplings $h_\alpha$. 
During leptogenesis,  they will also be distinguishable, 
if the $h_\alpha$ mediated interactions are fast
compared to those of leptogenesis and to 
the Universe expansion rate.
Since the $h_{e,\mu,\tau}$  mediated interactions are of widely different strengths, 
they would, for instance, induce differences in the thermal masses of 
the different leptons. 
The answer to the ``flavour puzzle'' in the previous example is
that when (some of) the charged lepton Yukawa interactions are ``fast
enough'', the ``flavour'' (charged lepton mass eigenstate) basis is the
correct basis for the BE, and eqn. (\ref{eq:withflavor}) should be used. In
the opposite situation in which charged lepton Yukawa interactions are much
slower than the Universe expansion rate, leptogenesis has no knowledge of
lepton flavours, and the correct BE is eqn~(\ref{eq:noflavor}).

To be more quantitative, let us estimate the temperature below which 
lepton flavor effects cannot be neglected. 
The interaction rate for  a charged lepton  Yukawa coupling $h_\alpha$
 \cite{Cline:1993bd,Blanchet:2006be} can be estimated as 
\beq
\Gamma_{\alpha}  \simeq 5 \times 10^{-3} h_\alpha^2 T,
\eeq
(for details see Appendix
\ref{chempot}  around eqn (\ref{app7gam})).
The condition $\Gamma_{\tau(\mu)}\gsim H$ then implies 
that  the rate for  $h_{\tau(\mu)}$  mediated processes 
 becomes faster than the expansion rate of the Universe below
$T\sim10^{12}$ GeV ($10^9$ GeV),\footnote{In SUSY 
$h_\tau = m_\tau/(v_u \cos\beta )$, so the tau Yukawa is in
equilibrium for $T < (1+ \tan^2 \beta) \times 10^{12}$ GeV.}
while for
 $T\gg 10^{12}\,$GeV the charged lepton Yukawa couplings are irrelevant. 
Hence, above this temperature all flavour effects can be neglected, and 
the only ``basis-choosing''  
interactions  are those involving  the neutrino  Yukawa 
couplings $\lambda$.\footnote{$\Delta L = 2$ interactions mediated
by $N_i$ exchange could also become fast at large 
temperatures. Besides  giving additional contributions
to the washout, this also complicates the issue of the physical basis
for the lepton 
doublets. See {\it e.g.} \cite{Barbieri:1999ma}.} 

If leptogenesis occurs below $ T\sim10^{12}\,$GeV, the  $h_\tau$-related
reactions 
participate, along with the neutrino Yukawa, in determining a physical flavour
basis (via, for instance, the ``thermal mass matrix'').
The conditions under which $\ell_\tau$ is 
singled out as a distinct lepton are
discussed in Section \ref{sec:appenrhosum}; see e.g. eqn~(\ref{eq:app7dernier}).
For instance,
if at the time the lepton asymmetry starts to survive, the $h_\tau$
interactions are also faster than the $N_1$ inverse
decays~\cite{DeSimone:2006,Blanchet:2006ch}, then the relevant basis states
are $\ell_\tau$ and the component in $\ell_{N_1}$ that is orthogonal to
$\ell_\tau$, and the total lepton asymmetry is shared between these two
flavours.  This situation can have significant consequences.  In particular,
it can happen that the asymmetry in the $\tau$ flavour is larger in size than
(and even opposite in sign to) the total lepton asymmetry generated in $N_1$
decays (see eqn (\ref{epsaabd})).  If, in addition, the $\tau$ flavour is
weakly washed-out, a sizeable fraction of its asymmetry can survive until the
end of the leptogenesis era. 
An example  of two cases in which   flavour effects yield a large 
enhancement of the baryon asymmetry  is given in figure \ref{strongmixed}.


\begin{figure}[t!]
\vskip-2mm
\hskip-4mm
 \includegraphics[width=8cm,height=6cm,angle=0]{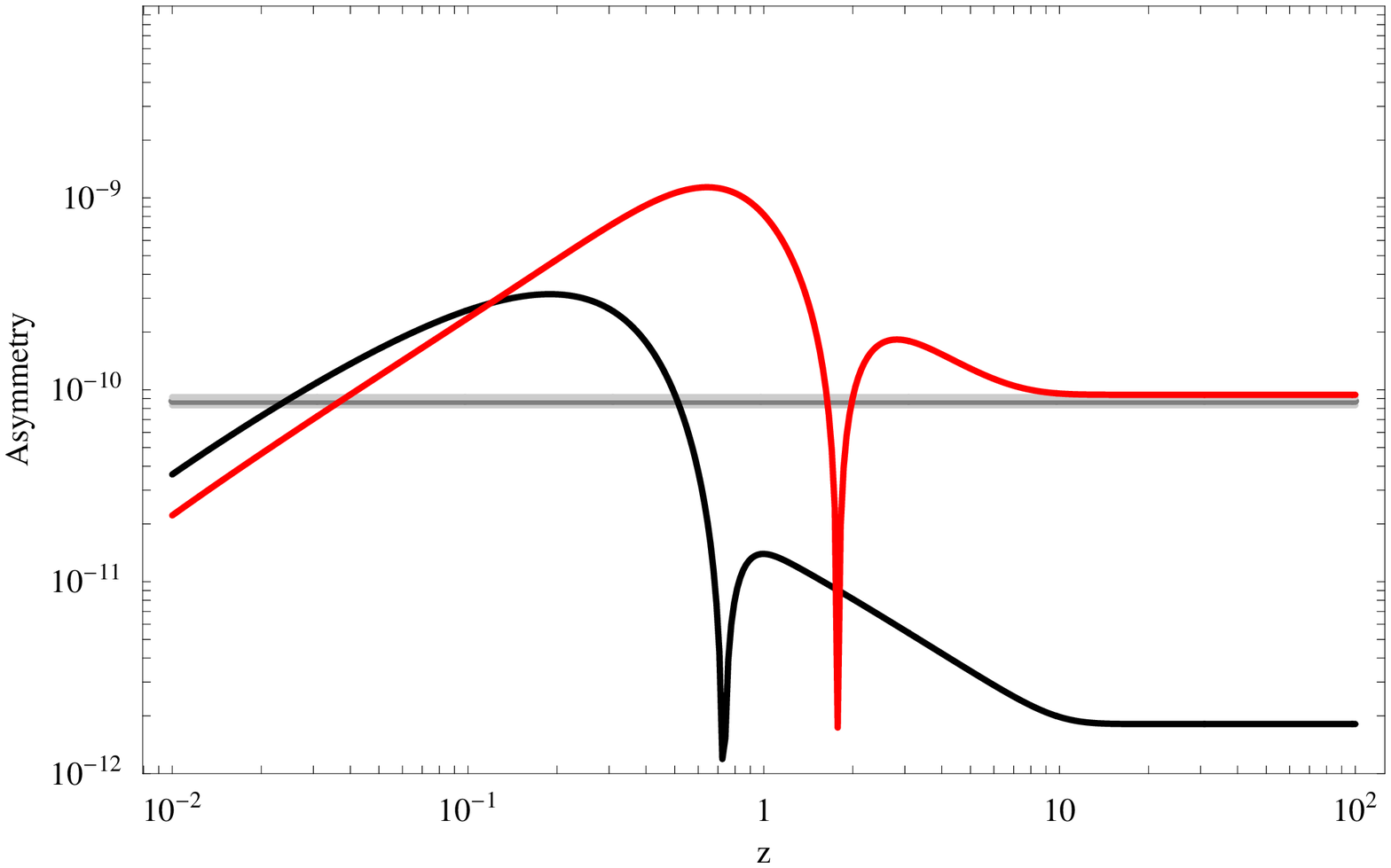}
 \includegraphics[width=8cm,height=6cm,angle=0]{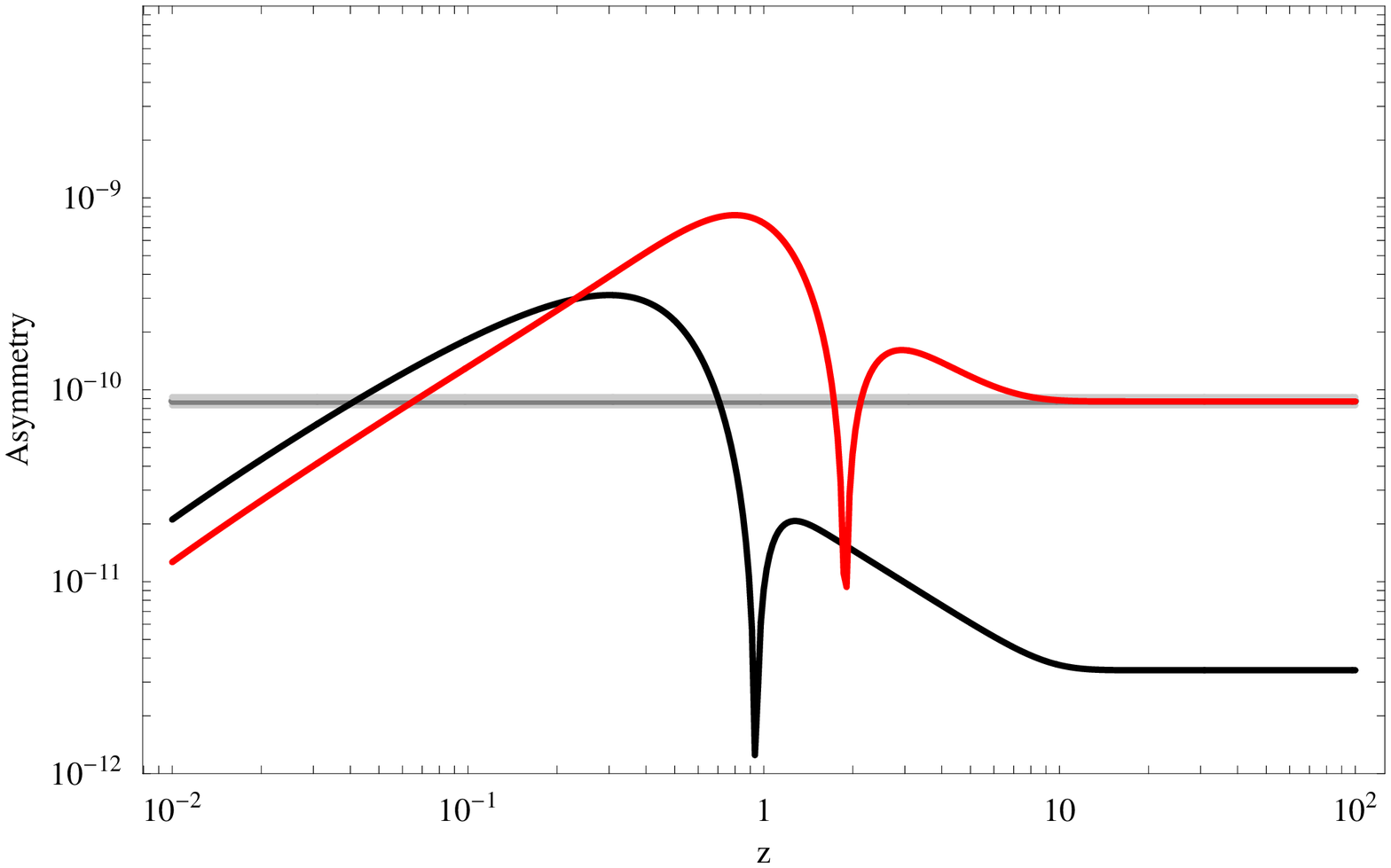}
 \caption[]{\small The total baryon asymmetry in the two flavour
   calculation (upper curves) and within the one-flavour approximation
   (lower curves) as a function of $z$, for two different sets of 
washout parameters. 
Left picture:
$\tilde m_{\tau\tau}/m_*=10$, 
$\tilde m_{\mu\mu}/m_*=30$,
$\tilde m_{ee}/m_*=30$. 
Right picture: 
$\tilde m_{\tau\tau}/m_*=10$, 
$\tilde m_{\mu\mu}/m_*=30$, 
$\tilde m_{ee}/m_*=10^{-2}$.
In both cases 
$\epsilon_{\tau \tau}=2.5\times 10^{-6}$, 
$\epsilon_{\mu\mu}=-2\times 10^{-6}$,
$\epsilon_{ee}=10^{-7}$ and $M_1=10^{10}$ GeV.
These plots are from \cite{Abada:2006ea}. }
\label{strongmixed}
\end{figure}

\subsection{Enhancement of the $B-L$ asymmetry}
\label{sec:basicflavour}

To simplify the analysis,  we focus in
this Section  on  the asymmetries in lepton doublets,
and neglect spectator and sphaleron processes.  
We make two assumptions to fix the flavour basis:
\begin{itemize}
\item[(i)] The rate of the $h_\tau$-interactions $\Gamma_{\tau}$ is
much larger than $H$ and $\Gamma_{ID}$, the rates of, respectively,
the expansion of the Universe, and the $N_1$ inverse decay;
\item[(ii)] During the entire period of leptogenesis, the rate of the
$h_\mu$-interactions, $\Gamma_{\mu}$, is either
$\ll H,\Gamma_{ID}$, or $\gg H, \Gamma_{ID}$,
so that the flavour basis does not change during leptogenesis.
\end{itemize}
When these two assumptions hold, it is a good approximation to work just
with the projections of  the $\ell_{N_1}$ densities onto (for
$\Gamma_\mu\ll H,\Gamma_{ID}$) the two-flavour basis $(\ell_o,\ell_\tau)$,
where $\ell_o$ is the direction of the component in $\ell_{N_1}$ that is
orthogonal to $\ell_\tau$, or (for $\Gamma_\mu\gg H,\Gamma_{ID}$)
the three flavour basis $(\ell_e,\ell_\mu,\ell_\tau)$.

The flavour projectors correspond to the diagonal entries of the
matrix $P$ in eqn (\ref{eq:Pab}) (and $\overline P$ for the CP
conjugate states) \cite{Barbieri:1999ma}, that is
\beq
\label{eq:gammaalpha}
\Gamma(N_1\to
\ell_\alpha\phi)= \Gamma(N_1\to\ell_{N_1}\phi)\,P_{\alpha\alpha}.
\eeq
The CP violating differences,
\beq
\label{eq:deltaP}
\Delta P_{\alpha\alpha}=P_{\alpha\alpha}-\bar P_{\alpha\alpha},
\eeq
are important quantities that account for the misalignment in flavour
space of the states $\ell_{N_1}$ and $\overline{\ell}_{N_1}$. At tree
level, $\Delta P_{\alpha\alpha}=0$ and the flavoured washout parameters
$\tilde m_{\alpha\alpha} $ 
can be simply written as
\beq
\label{eq:tildemalpha}
\tilde m_{\alpha\alpha}=\tilde m\, P_{\alpha\alpha}.
\eeq
At the loop level, however, the $\Delta P_{\alpha\alpha}$ do not vanish in general.
We use eqns (\ref{eq:gammaalpha},\ref{eq:deltaP},\ref{eq:tildemalpha}) to
rewrite the asymmetry in lepton flavour $\alpha$, eqn.(\ref{epsaa1}), as
\begin{equation}
  \label{eq:DeltaP}
  \epsilon_{\alpha\alpha}=
\epsilon \,\frac{  \tilde m_{\alpha\alpha}}{\tilde m}
+ \frac{1}{2}\Delta P_{\alpha\alpha}.
\end{equation}
The crucial observation is that, while washouts are dominated by the
inverse-decay rates, which are ${\cal O}(\lambda^2)$ and do
not involve the couplings to $N_{2,3}$, the CP-violating difference
$\Gamma(N_1\to \ell_\alpha\phi) -\Gamma(N_1\to
\bar\ell_\alpha\bar\phi)$ is ${\cal O}(\lambda^4)$ and does
involve also the couplings $\lambda_{\beta k}$ with $k=2,3$.  Therefore,
as shown by eqn (\ref{eq:DeltaP}), in general $\epsilon_{\alpha \alpha}$
is not proportional to $m_{\alpha\alpha}$. This implies that, for fixed
values of $\epsilon$ and $\tilde m$, one can have
$\epsilon_{\alpha\alpha}$ large together with $\tilde
m_{\alpha\alpha}$ small, and this can yield a strong enhancement of the
asymmetry $Y_{\Delta \ell_\alpha}$.
Taking into account that $\sum_\alpha P_{\alpha\alpha}=\sum_\alpha
\overline{P}_{\alpha\alpha}=1$, we obtain the following `sum rules':
\begin{equation}
  \label{eq:sums}
\sum_\alpha \tilde m _{\alpha\alpha} =\tilde m,
\qquad
 \sum_\alpha \epsilon_{\alpha\alpha}=\epsilon.
\end{equation}
The first term in eqn (\ref{eq:DeltaP}) corresponds simply to the
projection of the total asymmetry $\epsilon$ onto the flavour
$\alpha$.  The second contribution plays a more subtle role. It
represents a type of CP-violation that is specific of the flavour
CP-asymmetries and does not affect the value of $\epsilon$.

The Boltzmann equation for $Y_N$ is given in eqn (\ref{eq:finalBEYN}):
\beq\label{eq:YN5.6}
\dot Y_N = -(y_{N_1} -1)
[\gamma_{N\rightarrow 2} + \gamma^{2\leftrightarrow 2}_{\rm top} +
  \gamma^{2\leftrightarrow 2}_A ].
\eeq
As concerns the Boltzmann Equations (\ref{eureka}) for the $Y_{\Delta_\alpha}$,
we remarked at the end of Section~\ref{real} that, to write them in closed
form, one has to include all the spectator processes. This
induces a mixing between $Y_{\Delta_\alpha} $ and $Y_{\Delta_\beta} $
that complicates the expressions of the BE. For the sake of clarity, we
provisionally neglect these effects (we discuss them in Section
\ref{sec:flavormix}). This approximation allows us to consider, instead of
coupled equations for the $Y_{\Delta_\alpha}$'s,  independent equations
for each of the lepton doublet densities $Y_{\Delta \ell_\alpha}$'s:
\beq\label{eq:YDL5.6}
\dot Y_{\Delta \ell_\alpha} \simeq - \dot Y_N \epsilon_{\alpha\alpha}
+(\dot Y_{\Delta \ell_\alpha})^w.
\eeq
Note that since we neglect the redistribution of the asymmetry to singlet leptons,
we took $(\dot Y_{\Delta \ell_\alpha})^w=-(\dot Y_{\Delta_\alpha})^w$ when using
eqn (\ref{eq:finalBEYDeltai-w}).
The solutions are of the form of eqn (\ref{matter}):
\begin{equation}\label{YBflavor}
 Y_{\Delta \ell} \equiv  \sum_\alpha Y_{\Delta \ell_\alpha}
\sim  4 \times  10^{-3} \sum_\alpha \epsilon_{\alpha \alpha}
\eta_\alpha(\tilde m,\tilde m_{\alpha\alpha}) .
\end{equation}
where the summation is over
$\alpha = e, \mu, \tau$ for $T <  10^{9}$ GeV,
and $\alpha = o, \tau$ for  $ 10^{9}$ GeV $< T < 10^{12}$ GeV.
Approximate analytic  solutions (reviewed in Appendix \ref{recipes})
to eqns (\ref{eq:YN5.6}) and (\ref{eq:YDL5.6}) give
\beq\label{eq:etastrong}
\eta_\alpha \simeq \left\{ \begin{array}{cc}
\left[ \left(\frac{m_*}{2 \tilde{m}_{\alpha \alpha}}\right)^{-1.16} +
\left(\frac{\tilde{m}_{\alpha \alpha}}{2.1 m_*}\right)^{-1} \right]^{-1}
& \tilde{m} > m_* \cr
\frac{\tilde{m}}{2 m_*}\frac{\tilde{m}_{\alpha \alpha}}{m_*} & \tilde{m} < m_*.\cr
\end{array}\right.
\eeq

Eqns (\ref{YBflavor}) and (\ref{eq:etastrong}) 
allow us to estimate the lepton doublet asymmetry $Y_{\Delta\ell}$
in various washout regimes. In the case that all the flavours are in
the strong or mildly strong washout regime, $\tilde m_{\alpha\alpha}\geq
m_*$ for all $\alpha$, the effects of the $\Delta P$ term in
eqn (\ref{eq:DeltaP}) are most striking. The efficiency
for  $Y_{\Delta \ell_\alpha}$  is approximately given by
$\eta_\alpha \sim m_*/(2\tilde m_{\alpha\alpha})$.  Inserting
eqn (\ref{eq:DeltaP}) in eqn (\ref{YBflavor}), we obtain
\begin{equation}\label{eq:strongflavor}
\frac{ Y_{\Delta \ell}}{2 \times 10^{-3}}
\sim {\cal N}_f \frac{m_*}{\tilde m} \epsilon
+  \sum_\alpha \frac{m_*}{\tilde m_{\alpha\alpha}}
\frac{\Delta P_{\alpha\alpha} }{2}.
\end{equation}
Here ${\cal N}_f$ denotes the number of lepton flavours that are effectively
resolved by the fast charged lepton Yukawa interactions. 
The first term represents an enhancement by a factor ${\cal N}_f$ with respect
to the doublet asymmetry that would be obtained by neglecting flavour effects.
It is clearly independent of particular flavour structures and it would be the
only effect in the special cases where $\epsilon_{\alpha\alpha} \propto \tilde
m_{\alpha\alpha}$.  Its origin is simple to understand. On one hand, the asymmetry
in each flavour is reduced by a factor of the branching ratio, $\tilde
m_{\alpha\alpha}/\tilde m$. On the other hand, the efficiency factor is increased
by the same factor, because all the washout rates for $Y_{\Delta \ell_\alpha}$ are
reduced by it. The two factors cancel each other, and an asymmetry
$Y_{\Delta \ell_\alpha} \propto \epsilon m_*/\tilde m$ is
generated in each flavour. The sum over flavours yields the ${\cal N}_f$ enhancement.

In contrast, the second term depends crucially on the details of the flavour
structure. Let us analyze its possible effects in the ${\cal N}_f=2$ case, in
which $\Delta P_{\tau\tau}=-\Delta P_{oo}$ (with
$\Delta P_{oo}=\Delta P_{ee}+\Delta P_{\mu\mu}$). If we assume $\tilde m \gg m_*$,
then the first term in the r.h.s of eqn (\ref{eq:strongflavor}) is strongly
suppressed. However, flavour configurations are possible for which
$\tilde m_{\tau\tau} \sim m_*$ (or $\tilde m_{oo}\sim m_*$). Then ,the production
of an asymmetry $Y_{\Delta \ell_\tau}$ (or $Y_{\Delta \ell_o}$) driven by the
second term occurs under `optimal' conditions.
\begin{figure}[t!]
\hskip15mm
 \includegraphics[width=9cm,height=12cm,angle=270]{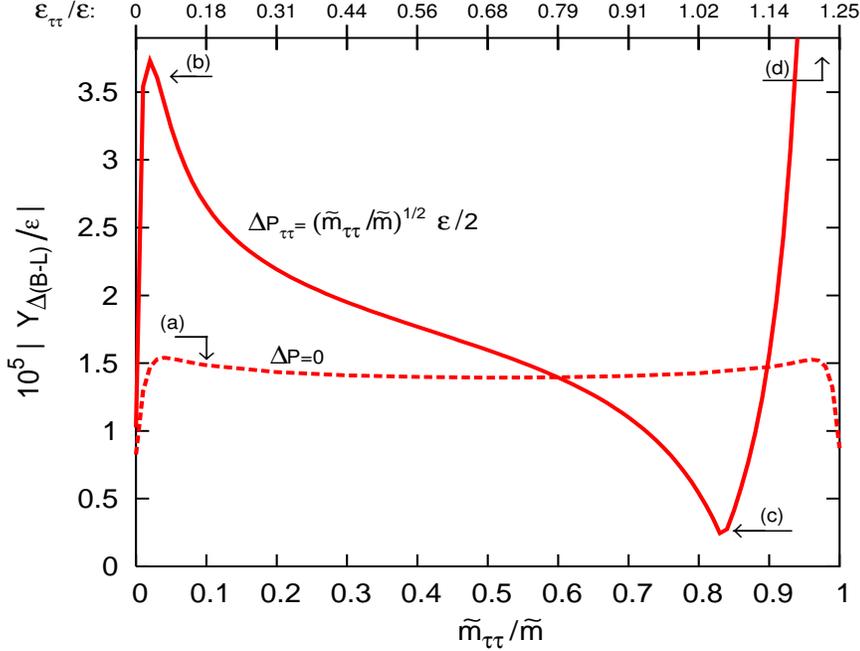}
 \caption[]{ The absolute value of the final $|Y_{\Delta(B-L)}|$ (in
   units of $10^{-5}|\epsilon|$) as a function of $\tilde
   m_{\tau\tau}$. We use $M_1=10^{10}$~GeV and $\tilde m=0.06$~eV.
   The dashed line corresponds to the case
   $\epsilon_{\alpha\alpha}\propto\tilde m_{\alpha\alpha}$ for which
   $\Delta P=0$.  The solid line gives an example of the results for
   $\Delta P \neq 0$. We take $\Delta P_{\tau\tau} /\epsilon \propto
   \sqrt{\tilde m_{\tau\tau}/\tilde m}$. The corresponding values of
   $\epsilon_{\tau\tau}/\epsilon$ are marked on the upper $x$-axis.
   The arrows with labels $(a)$, $(b)$, $(c)$ and $(d)$ correspond to
   the four situations discussed in the text. Note that around $(c)$
   $Y_{\Delta(B-L)}$ changes sign. (Figure adapted from
   ref.~\cite{Nardi:2006fx}).  }
\label{fig:kplot}
\end{figure}
This qualitative expectation is confirmed by numerical integration of the
flavour dependent Boltzmann equations.  Fig.~\ref{fig:kplot} (adapted from
ref.~\cite{Nardi:2006fx}) depicts a set of possible cases.  The absolute value
of the final $|Y_{\Delta(B-L)}|$ (in units of $10^{-5}|\epsilon|$) computed with
$M_1=10^{10}$~GeV and $\tilde m=0.06$~eV is shown as a function of $\tilde
m_{\tau\tau}$. Two flavour-aligned cases at $\tilde m_{\tau\tau}=0$ ({\it i.e.} no
$\tau$ component in $\ell_{N_1}$) and at $\tilde m_{\tau\tau}=\tilde m$ ({\it i.e.}
$\ell_{N_1} \equiv \ell_\tau$) coincide with the results that would have been
obtained by neglecting flavour. For $\tilde m_{\tau\tau}/\tilde m\neq0,\,1$
quite different results are obtained.  Case $(a)$ (dashed line) gives the
results obtained by setting $\Delta P=0$, that is by taking
$\epsilon_{\alpha\alpha}\propto\tilde m_{\alpha\alpha}$.  As predicted in our
qualitative analysis, the final $B-L$ asymmetry is practically independent of
the flavour structure, and gets enhanced by a factor $\sim {\cal N}_f=2$ with
respect to the flavour-aligned results.

For the more general cases where $\Delta P \neq 0$, the final doublet
lepton asymmetry strongly depends on the particular flavour structure.
To depict in a simple way the various possibilities, we adopt, as a
convenient ansatz, the relation $\Delta P_{\tau\tau} \propto \sqrt{\tilde
  m_{\tau\tau}/\tilde m} $.  This is based on the fact that, if we take the
tree-level $N_1\to\ell_\tau \phi$ decay amplitude to zero
($\lambda_{\tau1}\to0$) while keeping the total decay rate fixed
[$(\lambda^\dagger\lambda)_{11}=\>$const.], $\tilde m_{\tau\tau}$ vanishes as the
square of this amplitude ($\propto|\lambda_{\tau1}|^2$) while $\Delta
P_{\tau\tau}$ vanishes as the amplitude ($\propto\lambda_{\tau1}$).  We
extrapolate this proportionality to finite values of $\tilde m_{\tau\tau}$
because this allows to represent the $\Delta P_{\tau\tau}$ effects in a simple
two-dimensional plot of $ Y_{\Delta(B-L)}$ versus $\tilde m_{\tau\tau}$.

On the upper $x$-axis of Fig.~\ref{fig:kplot} we mark for reference
the relative value of the $\tau$ CP-asymmetry
$\epsilon_{\tau\tau}/\epsilon$ corresponding to the different values
of $\tilde m_{\tau\tau}$. The most peculiar features are the two
narrow regions marked with $(b)$ and $(d)$ where $|Y_{\Delta(B-L)}|$
is strongly enhanced. In $(b)$ this happens around the `optimal' value
$\tilde m_{\tau\tau}\sim m_*$ that yields a strong enhancement, even
though $\epsilon_{\tau\tau}/\epsilon$ is rather small. The region
marked with $(d)$ corresponds to $\tilde m_{oo}\sim m_*$ and to
$\epsilon_{oo}$ of the opposite sign with respect to $\epsilon$.  Here
the steep rise of $|\Delta Y_{\Delta(B-L)}|$ can reach values up to
one order of magnitude larger than the vertical scale of the
figure. (Note, however, that the ansatz $\Delta P_{\tau\tau} \propto
\sqrt{\tilde m_{\tau\tau}/\tilde m} $ does not yield the required
behavior $\Delta P=0$ at the boundary $\tilde m_{\tau\tau}/\tilde m
=1$ and therefore the continuous line in the plot should not be
extrapolated too close to this limit.) It is worth noticing that the
two peaks correspond to values of $Y_{\Delta(B-L)}/\epsilon$ of
opposite sign. In fact, the $B-L$ asymmetry changes sign around 
point $(c)$, and for $\tilde m_{\tau\tau}/\tilde m \gsim 0.85$ it is
of the opposite sign with respect to what one would obtain neglecting
flavour effects.  Similar results are expected in the temperature
regime below $T\sim 10^{9}\,$GeV, when both the $\tau$ and the $\mu$
Yukawa reactions are in equilibrium. In this case even larger
enhancements are possible because the dynamical production of the
asymmetry of two flavours can occur in an `optimal' regime.

In the weak washout regime for all the flavours, there can be no dynamical
enhancement of the asymmetry and flavour effects are less important.
The efficiencies for this regime are given in eqn (\ref{etaw})
(or (\ref{eq:etastrong})). Using also eqn (\ref{eq:DeltaP}), we obtain
\begin{equation}
  \label{eq:weakflavor}
-\frac{ Y_{\Delta \ell}}{ 2 \times 10^{-3}} \sim \frac{\tilde m^2}{m_*^2}
\left[\epsilon \sum_\alpha \frac{\tilde m_{\alpha\alpha}^2}{\tilde m^2}
+ \frac{1}{2} \sum_\alpha \frac{m_{\alpha\alpha}}{\tilde m}
\Delta P_{\alpha\alpha} \right].
\end{equation}
The first sum within square brackets is always $\leq 1$ so this
contribution by itself is suppressed with respect to the unflavoured
case.  The second sum is always less than or equal to the largest of the
$\Delta P_{\alpha\alpha}$'s.  Therefore, in this regime a necessary
condition to have a large enhancement of the resulting doublet lepton
asymmetry is that $\Delta P_{\alpha\alpha}\gg \epsilon$ for at least
one lepton flavour.

Finally, in the case where one flavour $\beta$ is weakly coupled to $N_1$,
$\tilde m_{\beta\beta}/m_* \ll 1$, but other flavours $\alpha$ are strongly
coupled, $\tilde m_{\alpha\alpha}/m_* \sim \tilde m/m_* \gg 1$, the
singlet neutrinos are brought into equilibrium by the
reactions involving the strongly coupled flavours, so that $n_{N_1}\sim
n_\gamma$. The efficiency factor $\eta_\beta$ `loses' the factor of $\tilde
m/m_*$, which corresponds to a suppressed $N_1$ abundance [see eq.~(\ref{etaw})].
Consequently, the asymmetry for the weakly coupled flavour $\beta$ is given by
\begin{equation}
  \label{eq:weakstrong}
 - \frac{Y_{\Delta \ell_\beta}}{2 \times 10^{-3}} \sim \frac{\tilde m}{m_*}
\left[\epsilon\frac{\tilde m_{\beta\beta}^2}{\tilde m^2}
+  \frac{m_{\beta\beta}}{\tilde m}
\frac{\Delta P_{\beta\beta}}{2} \right].
\end{equation}
The asymmetry for the strongly coupled $\alpha$ flavours is again given by
eqn (\ref{eq:strongflavor}) with ${\cal N}_f=1$ or 2.

\subsection{Mixing of lepton flavour dynamics}
\label{sec:flavormix}
In Section~\ref{real} we derive BE for the charge densities
$Y_{\Delta_\alpha}$, without taking into account the SM spectator processes.
For this reason, the dependence of the $\Delta y_a$'s on the
$Y_{\Delta_\alpha}$'s is left implicit. In Section~\ref{sec:spec} we analyze
the spectator processes, assuming that $N_1$ decays to pure
(flavour) states, that is, neglecting flavour effects. In this
section we combine these two issues, and we outline how to express the
washout term in eqn (\ref{eq:finalBEYDeltai-w}) in terms of the
$Y_{\Delta_\alpha}$ (see Appendix~\ref{chempot} for a more complete treatment).

We assume that asymmetries can be generated in two flavour directions, which we
take to be $\hat{\tau}$ and $\hat{o}$, where the latter is some linear
combination of $\hat{\mu}$ and $\hat{e}$.
Below  $T\sim 10^{12}\,$GeV, baryon and lepton numbers violating
electroweak sphaleron processes, and $h_\tau$-related
processes are fast. In this situation, the charge densities that
are (slowly) evolving because of the leptogenesis processes are
\begin{equation}
\label{eq:BmLalpha}
Y_{\Delta_\alpha} \equiv \frac{1}{3} Y_{\Delta B} - Y_{\Delta L_\alpha}=
\frac{1}{3}\sum_{i=1,2,3}(\Delta y_{q_i}+
\Delta y_{u_i}+\Delta y_{d_i})Y^{\rm eq}-
(\Delta y_{\ell_\alpha}+\Delta y_{e_\alpha})Y^{\rm eq}.
\end{equation}
Note that, in the temperature range of interest, all the
asymmetry-densities in the right-hand-side of this equation are generally
non-vanishing. In particular, $N_1$ decays generate a non-vanishing chemical
potential for the Higgs particles which, in turn (see eqn (\ref{tophiggs})),
induces a non-vanishing chemical potential to the top-quarks. The QCD sphaleron
condition of eqn (\ref{qcdsph}) further implies that this happens for all the lighter
quarks. The only exceptions are $\Delta y_{e_\mu}$ above $T\sim
10^9\,$GeV, and $\Delta y_{e_e}$ that is not populated during standard leptogenesis.

The asymmetry densities for the quarks and for the  $SU(2)$-singlet leptons
can  always be re-expressed in terms of the asymmetry densities of the $SU(2)$-doublet
leptons. This procedure makes use of the condition of hypercharge neutrality, eqn
(\ref{hyper}), that involves all three generation fermions. Consequently, once the
appropriate substitutions are implemented, one obtains that the densities of the
$\Delta_\alpha$ charges in eqn (\ref{eq:BmLalpha}) correspond to linear
combinations of the asymmetry densities of all the lepton doublets:
\begin{equation}
\label{eq:BmLalpha2}
 Y_{\Delta_\alpha} = - \left[A\right]^{-1}_{\alpha\beta}\, \Delta
y_{\ell_\beta}\, Y^{\rm eq}.
\end{equation}
The matrix $\left[A\right]^{-1}$ can then be inverted to give
$\Delta y_{\ell_\alpha}$ in terms of the charge
densities $Y_{\Delta_\alpha}$ (see Appendix~\ref{chempot}).  The
same is true also for the Higgs asymmetry. We can thus write:
\begin{equation}
\label{eq:CC}
\Delta y_{\ell_\alpha} = - \sum_\beta A_{\alpha\beta}\>
\frac{  \scriptstyle
  Y_{\Delta_\beta}}{ \scriptstyle Y^{\rm eq}},
\qquad \qquad
\Delta y_{\phi} = - \sum_\beta C^\phi_{\beta}\>
 \frac{\scriptstyle Y_{\Delta_\beta}}{\scriptstyle Y^{\rm eq}}.
\end{equation}
The matrix $A$ (introduced in ref. \cite{Barbieri:1999ma}) and the vector $C^\phi$
(introduced in ref. \cite{Nardi:2005hs}) generalize the coefficients $c_\ell$ and
$c_\phi$, defined in eqn (\ref{eq:cc}) for the case of pure flavour states, to the
general flavour case. Their numerical values are determined by the relevant set of
chemical equilibrium conditions, and therefore depend on temperature.

\begin{figure}[t!]
\hskip15mm
 \includegraphics[width=9cm,height=12cm,angle=270]{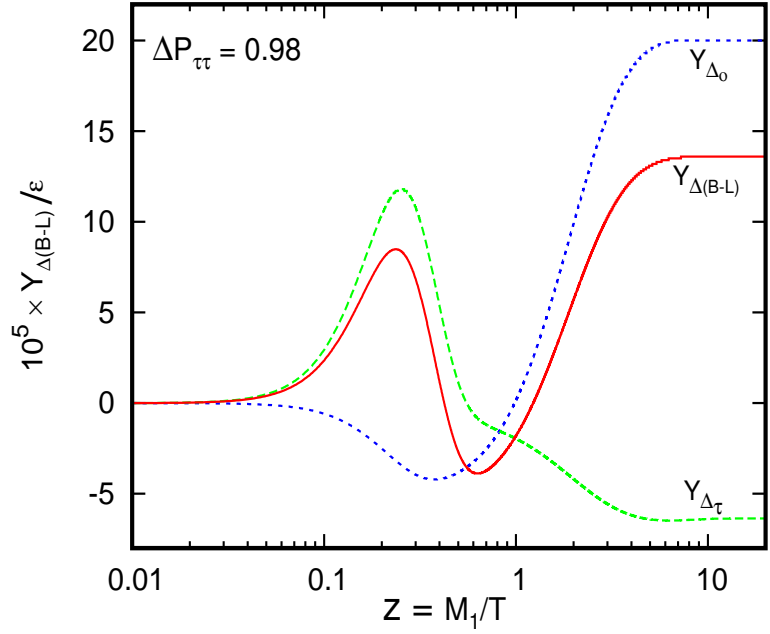}
\caption[]{ The evolution of the charges densities ${Y_{\Delta_\tau}}$
  (green dashed line), ${Y_{\Delta_{o}}}$ (blue dotted line)
  and $ Y_{\Delta(B-L)}$ (red solid line) in units of $10^{-5}\, \epsilon$ as
  a function of $z=M_1/T$. We use $M_1=10^{10}$~GeV and $\tilde
  m_1=0.06$~eV.  The $b$- and $\tau$-Yukawa reactions and the electroweak
  sphalerons are in equilibrium.  The figure refers to the point
  labeled $(d)$ in fig.~\ref{fig:kplot} that corresponds to $\Delta
  P_{\tau\tau}=0.98$ and $\epsilon_{\tau\tau}\approx 1.2\,\epsilon$.  (Figure adapted from ref.~\cite{Nardi:2006fx}).}
\label{fig:washout}
\end{figure}

As a consequence of eqns~(\ref{eq:CC}), the evolution of one
density $Y_{\Delta_\alpha}$ depends on the other charge densities, that is,
the BE for the lepton flavours form a system of coupled equations.  This
phenomenon is sometimes referred to as `electroweak sphaleron flavour mixing',
and was studied in refs. \cite{Nardi:2006fx,JosseMichaux:2007zj}.
One example of the possible effects is depicted in fig.~\ref{fig:washout}.
The plot gives the evolution of $Y_{\Delta_\tau}$, $Y_{\Delta_{o}}$
and $Y_{\Delta(B-L)}=Y_{\Delta_\tau}+Y_{\Delta_{o}} $ for the flavour
configuration corresponding to the point labeled with $(d)$ in
fig.~\ref{fig:kplot}.  $Y_{\Delta_\tau}$ is in the strong washout regime,
$\tilde m_{\tau\tau}\approx 0.06\,$eV, and the $\tau$ CP asymmetry is
relatively large, $\epsilon_{\tau\tau}\approx 1.2\,\epsilon$. For
$Y_{\Delta_{o}}$, the value $\tilde m_{oo} \approx
10^{-3}$eV is close to optimal, and the asymmetry is smaller in size and opposite
in sign to $\epsilon_{\tau\tau}$: $\epsilon_{oo}\approx -0.2\,\epsilon$. For the
relevant temperature, $T\lsim10^{12}$GeV, electroweak sphaleron, $h_b$ and $h_\tau$
reactions are in equilibrium, yielding the matrices of coefficients [in the
$(Y_{\Delta_{o}},Y_{\Delta_\tau})$ basis]:
\begin{equation}
A = \frac{1}{230}\pmatrix{
196 &  -24 \cr
-9 &  156 }\,,
\qquad \qquad \qquad
C^\phi = \frac{1}{115}(41,\> 56)\,.
\label{case4}
\end{equation}

As apparent from fig.~\ref{fig:washout}, $Y_{\Delta_{o}}$
gives the dominant contribution to the final $B-L$ asymmetry. A qualitative
explanation goes as follows. For $z\gsim 0.1$, a large asymmetry builds up
quickly in the $\tau$ flavour and contributes to the early washout of
$Y_{\Delta_{o}}$ (the sign of $Y_{\Delta_{\tau}}$ that is opposite to
that of $Y_{\Delta_{o}}$ combines with the negative sign of
$A_{o\tau}=-24/230$ yielding an increase of the washout rates).
Thus, the early (negative) asymmetry in $Y_{\Delta_{o}}$ remains
suppressed.  It follows that the opposite sign $o$ asymmetry generated from $N_1$
decays at $z\gsim 1$ is largely unbalanced.  As concerns the $\tau$
asymmetry, typically of strong washout regimes, it `freezes-out' at
$z > 1$  with an absolute value that is sizably smaller than the values
reached at earlier times.

Remarkably, as can be clearly seen in fig.~\ref{fig:washout}, the $B-L$ asymmetry
changes sign twice during its evolution. This peculiar behavior is qualitatively
different from what would be obtained by neglecting the sphaleron induced flavour
mixing, when there is a single change of sign.
Other interesting washout effects induced by sphaleron mixing have been
analyzed in ref.~\cite{Shindou:2007se}.

\subsection{Phenomenological consequences}
\label{sec:phenoflavour}
As discussed in Section \ref{sec:basicflavour}, starting from a specific set of
parameters within a given seesaw model, the inclusion of flavour effects
in the calculation of the baryon asymmetry can change the result by factors of
a few to orders of magnitude \cite{Nardi:2006fx,Abada:2006ea}. In this section, we
take a more ``bottom-up'' perspective of flavoured leptogenesis.
As discussed below eqn (\ref{mass}), there are (currently)
14 unknown parameters in the three generation type I seesaw model, of which
only a few are accessible to low energy experiments.
It is interesting, from a phenomenological perspective, to constrain
the unknown parameters by requiring that leptogenesis works successfully.
We will find that the inclusion of flavour effects does not change the
 resulting constraints in a significant way.

\subsubsection{The upper bound on the light neutrino mass scale}
\label{sec:upperbd}
In the ``single flavour'' calculation, successful thermal leptogenesis
requires a light $\nu$ mass scale $\lsim0.2$ eV
\cite{Buchmuller:2003gz,Buchmuller:2002rq}. The argument leading to
this bound does not apply in the flavoured calculation. Currently,
there is no  consensus in the literature on
the value of the bound in the flavoured case.  Analytic
 arguments (which we reproduce below) and
the numerical analysis of Refs \cite{Abada:2006fw,JosseMichaux:2007zj,DeSimone:2006},  suggest that flavoured leptogenesis
 can be tuned to work for $m_{\rm min} $ up to $ 0.5-1$  eV.
 This  is of the order of current bounds from Large Scale Structure data and
WMAP  \cite{Hannestad:2006mi,Seljak:2006bg,Cirelli:2006kt,Yao:2006}.
The leptogenesis bound is relaxed because there is  more CP violation
available in flavoured leptogenesis.  The upper limit of eqn
(\ref{di2}) on  the total CP asymmetry {\it decreases}
like $\Delta m_{\rm atm}^2/m_{\rm max}$, as the light neutrino
mass scale  $m_{\rm max}$ increases. There is
therefore an upper bound on $m_{\rm max}$ ($\simeq m_{\rm min}$
for degenerate neutrinos). However, the limit (\ref{di2}) does not
apply to the flavoured CP asymmetries, which can
increase with the light neutrino mass scale,
as shown in eqn (\ref{epsaabd}).

The limit $m_\nu \lsim 0.2$ eV\cite{Buchmuller:2003gz}, obtained in the
single flavour approximation, can be understood as follows. For such a
high mass scale, the light neutrinos are almost degenerate, with
$m_1 \simeq m_2 \simeq m_3$. Due to the lower bound $\tilde m_1\geq
m_{\rm min}$, leptogenesis takes place in the strong washout regime.
Taking the total CP asymmetry to be close to the upper bound of eqn
(\ref{di2}), the final baryon asymmetry can be approximated as
\beq
Y_{\Delta B} \sim 10^{-3}\epsilon \frac{
  m_*}{\tilde{m}_1} \propto \frac{M_1 m_*}{v_u^2} \frac{\Delta
  m^2_{\rm atm}}{m_{\rm min}^2}.
\eeq
As the light neutrino mass scale is increased, $M_1$ and, correspondingly, the
temperature of leptogenesis, must increase to compensate the $\Delta
m^2_{\rm atm}/ m_{\rm min}^2$ suppression. The leptogenesis temperature is,
however, bounded from above, by the requirement of having lepton number
violating processes out of equilibrium. For instance, the requirement that
the $\Delta L=2$ processes are out of equilibrium (we estimate
$\Gamma\sim\gamma/n^{\rm eq}$, with $\gamma$ from eqn \ref{scatapp4d}) gives
\beq
\frac{2  m_{\rm min}^2 T^3}{ \pi^2 v^4}
\lsim \frac{20 T^2}{M_{\rm pl}},
\label{DL2mbd}
\eeq
which implies  $M_1 \lsim 10^{10} ({\rm eV}/ m_{\rm min})^2$ GeV.
There is therefore an upper bound on the baryon asymmetry
which scales as $1/ m_{\rm min}^4$. A large enough baryon asymmetry is obtained for
$m_{\rm min}\lsim0.1$ eV.

Consider now the flavoured case. We define $c^2=P_{\tau\tau}$
and $s^2=P_{oo}$ (with $s^2 +c^2 = 1$), so that $\Gamma_{\tau \tau} =
c^2 \Gamma_{D}$ and $\Gamma_{oo} = s^2 \Gamma_{D}$.
The individual flavour asymmetries, $\epsilon_{\alpha \alpha}$,
satisfy the bound of eqn (\ref{epsaabd}), while
their sum satisfies the stricter bound of eqn (\ref{di2}).
So, if we can arrange  $- \epsilon_{oo} =  \epsilon_{\tau \tau} \sim
3 M_1 m_{\rm min}/(16 \pi v^2)$, then for strong  flavoured washout, and
for sufficiently slow $\Delta L = 2$ processes,
the final baryon asymmetry (from  {\it e.g.} eqn (\ref{matter})) is
\beq
\label{kk}
 Y_{\Delta B} \sim  10^{-3}
\frac{3 M_1 {m}_*}{16 \pi v_u^2} \frac{c^4 - s^4}{c^2s^2}
\sim 3\times 10^{-11}\frac{c^4 - s^4}{c^2s^2}\frac{M_1}{10^{10} {\rm GeV}}.
\label{big}
\eeq
To show that the $m_{\rm min}\lsim 0.2$ eV  bound does not
apply in the flavoured case,  we must demonstrate that the three
ingredients that entered the derivation of eqn (\ref{big}) -- (i) an
appropriate set of flavoured CP asymmetries, (ii) negligible $\Delta L = 2$
processes, and (iii) strong flavoured washout -- can be realized in a model.

(i) To obtain the desired decay rates and asymmetries we use
the Casas-Ibarra \cite{Casas:2001sr}  parametrization,
in terms of $m_i$, $M_j$, $U$ and the complex orthogonal matrix
$R\equiv  v D_m^{-1/2}U^T\lambda D_M^{-1/2}$ (see the discussion around
eqn~(\ref{R})).  This ensures that we obtain the correct low-energy
parameters.  We take $o=\mu$ for simplicity, and write $R$ as a rotation
through the complex angle $\theta = \varphi + i \eta$.
The flavour asymmetries, for maximal atmospheric mixing, are
\bea
\epsilon_{\mu \mu}&  =& \frac{3 M_1 m_{\rm min}}{16 \pi v^2}
\frac{{\rm Im} \{  U_{\mu k}^* R_{k1}U_{\mu p}  R_{p1}  \}}
{\sum_\alpha |\sum_j U^*_{\alpha j} R_{j1}|^2}
=- \frac{3 M_1  m_{\rm min}}{16 \pi v^2}\frac{\sinh \eta \cosh \eta}{\sinh^2 \eta +
\cosh^2 \eta} \cos 2 \varphi ~,
 \nonumber \\
\epsilon_{\tau \tau}&  =& \frac{3 M_1  m_{\rm min}}{16 \pi v^2}
\frac{{\rm Im} \{ R_{k1} U_{\tau k}^*  R_{p1} U_{\tau p} \}}
{\sum_\alpha |\sum_j R_{j1}U^*_{\alpha j}|^2}
=  + \frac{3 M_1  m_{\rm min}}{16 \pi v^2}\frac{\sinh \eta \cosh \eta}{\sinh^2 \eta +
\cosh^2 \eta} \cos 2 \varphi ~ \nonumber\\
&&\label{particular1}
\eea
The flavour dependent decay rates $\Gamma_{\mu \mu}$
and $\Gamma_{\tau \tau}$ are proportional to
\bea
|\lambda_{ \mu 1}|^2 & =& \frac{M_1  m_{\rm min}}{2v^2 }| R_{11} + R_{21}|^2  =
\frac{M_1  m_{\rm min}}{2v^2} (\sinh^2 \eta +
\cosh^2 \eta +   \sin 2 \varphi) ~,
\nonumber \\
|\lambda_{ \tau 1}|^2 & =&  \frac{M_1  m_{\rm min}}{2v^2}| R_{11} - R_{21} |^2
= \frac{M_1  m_{\rm min}}{2v^2} (\sinh^2 \eta +
\cosh^2 \eta -  \sin 2 \varphi).
\label{particular2}
\eea
We see that we can choose $\eta$ and $\varphi$
as required to obtain eqn (\ref{big}) with $c^4-s^4 \gsim s^2 c^2$.

(ii) Eqn (\ref{DL2mbd}) gives the condition for the $\Delta L = 2$ processes
to be out of equilibrium: $M_1\lsim10^{10} ({\rm eV}/m_{\rm min})^2$ GeV.
Combined with eqn (\ref{big}), this gives
\beq
Y_B \sim 3\times10^{-11}\ \frac{c^4 - s^4}{c^2s^2}
\left(\frac{{\rm eV}}{m_{\rm min}}\right)^2.
\eeq

(iii) In this strongly washed out area of parameter space,
$N_1$ decays  take place in the flavoured regime
if  $\Gamma_\tau \gg \Gamma_{ID}$
\cite{DeSimone:2006,Blanchet:2006ch} (see Appendix
\ref{sec:appenrhosum}).
Suppose $s^2 \ll c^2$. Then eqn (\ref{big}) is
a valid estimate, provided $\Gamma_\tau \gg \Gamma_{ID}$
when $s^2 \Gamma_{ID} = H$:
\beq
\frac{\Gamma_\tau}{H} \simeq  \frac{10^{12} {\rm GeV}}{T} \gg
\frac{1}{s^2} = \frac{1}{B^{N_1}_{\phi \ell_o}}
\eeq

We conclude that with careful tuning, thermal leptogenesis can work
for $m_{\rm min} \lsim$ eV. This can be seen in Fig.~\ref{fig:thermal}.
For the plot on the left, flavour effects are included, and
leptogenesis can work for values $10^{-4}\ {\rm eV}\lsim \tilde
m_1\lsim1\ {\rm eV}$. At the upper bound we have $\tilde m_1\simeq
m_1$, so the cosmological bound is saturated.


%
\begin{figure}
\vskip-2mm
\hskip-8mm
\includegraphics[width=17cm,height=7cm,angle=0]{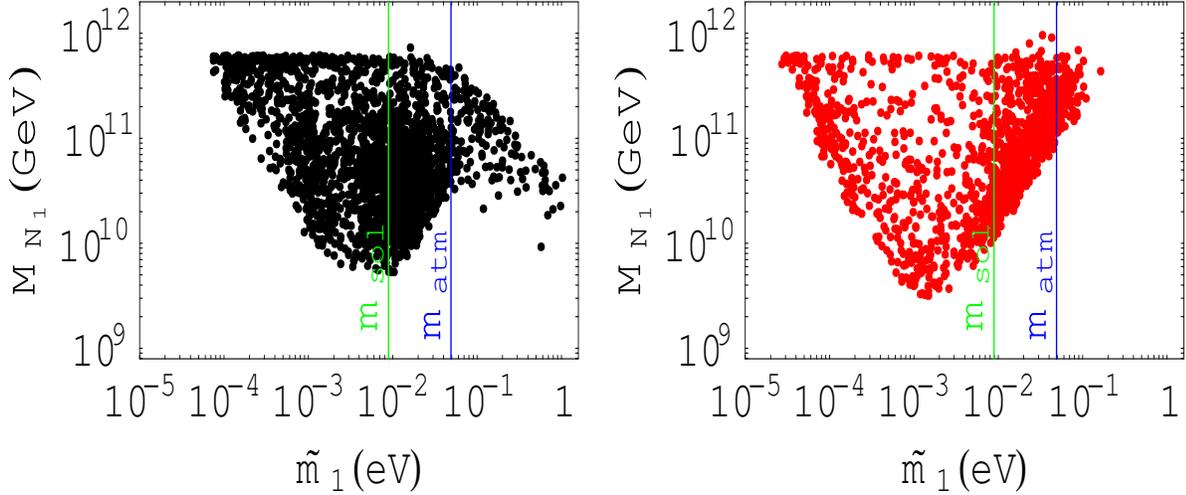}
\vskip-2mm
\hskip-4mm
\caption[]{Points in the $M_1-\tilde m_1$ plane where thermal leptogenesis
generates a sufficient baryon asymmetry (updated from
\cite{JosseMichaux:2007zj}). These plots clearly show the lower
bound on $M_1$  arising when the $N_i$  are hierarchical.
The plot on the right corresponds to a single flavour calculation.
The plot on the left includes flavour effects. The lower bound on
$M_1$ is similar in both cases. Leptogenesis works  for
larger values of $\tilde{m}_1$ in the flavoured plot, which
is obtained with two-flavour BE, and imposing the constraint
(\ref{eq:app7dernier}).  }
\label{fig:thermal}
\end{figure}

\subsubsection {Sensitivity of $Y_{\Delta B}$  to low energy phases}
An important, but disappointing, feature of ``single-flavour'' leptogenesis
is the lack of a model-independent relation between CP violation in the
leptogenesis processes and the (observable) phases of the lepton mixing
matrix $U$ (see eqn (\ref{UV})). With three singlet neutrinos $N_i$, thermal
leptogenesis can work with no CP violation in $U$
(see {\it e.g.} \cite{Rebelo:2002wj}) and, conversely,
leptogenesis can fail in spite of non-vanishing phases in $U$ \cite{Branco:2001pq}.
In some specific models, however, it is possible to establish a relation
\cite{Branco:2002kt,Branco:2001pq}.

In unflavoured leptogenesis, it is simple to see that the baryon asymmetry
is insensitive to phases of $U$ \cite{Branco:2001pq}. Using the Casas-Ibarra
parametrization of eqn (\ref{eq:R}) in eqn (\ref{flavour-CPasym}), $\epsilon$
can be written in a form where $U$ does not appear:
\beq
\epsilon   =-  \frac{ 1}{16 \pi v_u^2}
\frac{\sum_{j,k,p} {\rm Im} \{ ( R_{j1}^* m_j R_{jk})(R_{p1}^* m_p R_{pk})
  \}}{\sum_i m_i |  R_{i1}|^2}
 M_k~ g\left(\frac{M_k^2}{M_1^2}\right).
\eeq
Alternatively, in the top-down parametrization in terms of $V_R$, $V_L$ and
$D_\lambda$ (see Section \ref{sec:seepar}), $\epsilon$ depends on the phases in
$V_R$, while the $U$ phases {\it could } arise only from $V_L$.

The latter observation can be  exploited to relate the $U$ phases to flavoured
leptogenesis in models where there is no CP violation in $V_R$\footnote{Such
models may be difficult to construct \cite{Branco}.}
\cite{Nardi:2006fx,Abada:2006ea,Pascoli:2006ci,Molinaro:2007uv,Pascoli:2006ie,Branco:2006ce}.
However, for flavoured leptogenesis with three $N_i$'s, it is still
true that the baryon asymmetry  is not sensitive  to phases of $U$
\cite{Davidson:2007va}, in the sense that the baryon asymmetry can
be accounted for with the phases of $U$ having any value.
Conversely, if the phases of $U$ are measured, the baryon asymmetry is
still not constrained. This can be seen from the scatter plots of
ref. \cite{Davidson:2007va}, which show that, for any value of
the $U$ phases, the baryon asymmetry can be large enough.

In the two $N_i$ model (see section \ref{sec2rhn}), there are three phases.
Relations between $U$  phases and leptogenesis were obtained in
unflavoured leptogenesis \cite{Endoh:2002wm}. The relation between the
flavoured CP asymmetries $\epsilon_{\alpha \alpha}$ and the phases of $U$,
focusing on various texture models, was discussed in \cite{Fujihara:2005pv}.
Possible relations between low energy observables and the baryon asymmetry,
in the general case and for particular texture models, were discussed in
\cite{Abada:2006ea}.

In minimal left-right models with spontaneous CP violation,
there is a single phase which controls all CP violation
including that for leptogenesis \cite{Chen:2004ww,Chen:2006bv}. 

\subsubsection{The lower bound on $T_{\rm reheat}$}
\label{sec:Treheat}
In the space of leptogenesis parameters, there is an envelope
inside which  leptogenesis {\it can} work. In the single-flavour
calculation, the most important parameters are $M_1$, $\Gamma_D$
(or, equivalently $\tilde{m}$), $\epsilon$ and the light neutrino
mass scale \cite{Buchmuller:2002rq}. Including  flavour gives the
parameter space more dimensions ($M_1,\epsilon_{\alpha \alpha},
\Gamma_{\alpha \alpha}$...), but the envelope can still be projected
onto the $M_1-\tilde{m}$ plane by fixing $\epsilon$ to its maximum
value of eqn (\ref{di2}), which is a function of $M_1$.

For thermal leptogenesis to take place, an adequate number density of
$N_1$ must be produced by thermal scattering in the plasma, suggesting
$T_{\rm reheat} \gsim  M_1/5$ \cite{Buchmuller:2004nz,Giudice:2003jh}.
There is a lower bound on $M_1$, from requiring a large enough CP asymmetry.
The corresponding bound on $T_{\rm reheat}$ depends on the fine details of
reheating, the dynamics of which has been studied in refs. \cite{Giudice:2003jh,Buchmuller:2004nz}. To avoid these complications, the
lower bound on $M_1$ is usually quoted.

For non-degenerate light neutrinos, the lowest allowed value of $M_1$ occurs
at $\tilde{m} \sim m_*$, where analytic approximations are unreliable. It was
shown in \cite{Blanchet:2006be,Antusch:2006gy,JosseMichaux:2007zj}
that including flavour effects does not relax the bound with respect to
the unflavoured case \cite{Buchmuller:2002rq,Giudice:2003jh}, and we still have
\beq
M_1 > 2 \times 10^9 ~ {\rm GeV}, ~~~~~~{\rm non-degenerate~ } m_i.
\label{eq:4.8M1bd}
\eeq
This is shown in Fig.~\ref{fig:thermal} \cite{JosseMichaux:2007zj}.

\newpage
\section{Variations}
\label{variations}
The bulk of this review focuses on thermal leptogenesis, with
hierarchical singlet neutrinos, and where the CP asymmetry is
generated in the decays of the lightest singlet neutrino.  In this
chapter, we give brief overviews of various alternative scenarios of
leptogenesis. More detailed presentations can be found in the original
literature, to which we give appropriate references.

\subsection{Supersymmetric thermal leptogenesis}
\label{sec:susylep}
In the framework of supersymmetric seesaw models, new leptogenesis
mechanisms become possible, such as Affleck-Dine leptogenesis or soft
leptogenesis. In this subsection, we focus on the standard thermal
leptogenesis scenario with hierarchical singlet $N$'s, where the
presence of supersymmetry (SUSY) makes only small qualitative and
quantitative differences \cite{Campbell:1992hd,Plumacher:1997ru}. We
have often referred to the supersymmetric modifications in previous
chapters, so here we collect this information and summarize the
differences that arise when calculating the baryon asymmetry in the
supersymmetric type I seesaw. 

Neglecting small SUSY-breaking terms, a singlet neutrino $N_i$ and its
superpartner, the singlet sneutrino $\tilde N_i$, have equal masses,
equal decay rates $\Gamma_{N_i}=\Gamma_{\tilde N_i}$ and equal CP
asymmetries $\epsilon_{N_i}=\epsilon_{\tilde N_i}$. In this
approximation, estimating the $B-L$ asymmetry in the minimal
supersymmetric standard model (MSSM) with respect to the SM case (at
fixed values of the Yukawa couplings and $M_{N_i}$ masses) amounts to
a (careful) counting of a few numerical factors:
\begin{enumerate}
\item There are twice the number of states running in the loops, and
  thus the CP asymmetries $\epsilon_{N_i}$ and $\epsilon_{\tilde N_i}$
  are roughly twice the SM value of $\epsilon_{N_i}$ (see the loop
  function $g(x_j)$ in eqn (\ref{g})).
\item Including the asymmetry generated in $\tilde N_i$ decays gives
  another enhancement of a factor of two.
\item In the MSSM, the plasma is populated by  twice as many 
particles as there are in the SM. More precisely, we have $g^{\rm
  MSSM}_*/g_*^{\rm SM}=228.75/106.75\approx 2.14$. Thus the lepton
asymmetries to entropy ratios are reduced by $1/2$.
\item Due to the additional $\tilde\ell\tilde\phi$ final states, the
  $N_i$ decay rate is twice faster than in the SM.  (For the scalar
  neutrinos there are also two channels $\tilde N_i \to
  \tilde\ell\phi, \ell\tilde\phi$.) In the strong washout regime, the
  associated inverse-decay reactions double the washout rates,
  reducing the asymmetry by a factor of two. In the weak washout
  regime, the production of $N_i$ and $\tilde N_i$ is more efficient,
  and this increases the asymmetry by the same factor.
\item The expansion rate of the Universe is proportional to
  $\sqrt{g_*}$ and is therefore roughly a factor of $\sqrt{2}$ faster
  in SUSY ($m_*^{\rm MSSM}\approx \sqrt{2} m_*^{\rm SM}$). This
  reduces the time during which the strong washout processes can erase
  the asymmetry, yielding a $\sqrt{2}$ enhancement.  For weak
  washouts, it reduces the time for $N_i$ and $\tilde N_i$ production,
  yielding a $\sqrt{2}$ suppression of the final asymmetry.
\item Finally, in both the SM and the MSSM cases, the asymmetry in baryons
  $Y_{\Delta B}$ is of order 1/3 of the asymmetry in $B-L$. The exact
  relation differs only slightly: $28/79$ for the SM {\it vs.} $8/23$
  for the MSSM (see eqns
  (\ref{eq:app6CSM}) and (\ref{eq:app6CMSSM}), respectively).
\end {enumerate}
Putting all these factors together, we estimate: 
\beq
\frac{\left(Y_{\Delta B}\right)^{\rm MSSM}}{\left(Y_{\Delta
    B}\right)^{\rm SM}}\Big|_{M_i,\lambda_\alpha,h_i} 
\approx 
\left\{
\matrix{ \phantom{\Big|} \ \> 
 \sqrt{2}\qquad  
\mbox{(strong washout)},\cr  
2\,\sqrt{2} \qquad 
\mbox{(weak washout)}.} 
\right. 
\eeq
Thus, in the MSSM, in spite of the doubling of the particle spectrum
and of the large number of new processes involving
superpartners~\cite{Plumacher:1998bv}, one does not expect
major numerical changes with respect to the non-supersymmetric case.

Supersymmetric thermal leptogenesis has a potential gravitino problem
\cite{Khlopov:1984pf,Ellis:1984eq,Kohri:2005wn,Rychkov:2007uq}.  After
inflation, the universe thermalizes to a reheat temperature $T_{\rm
  reheat}$. Gravitinos are produced by thermal scattering in the bath,
and the rate is higher at higher temperatures. The gravitinos are
long-lived; if there are lighter SUSY particles (the gravitino is not
the LSP), the decay rate can be estimated as \cite{Kohri:2005wn}
\beq
\Gamma \propto \frac{m^3_{\rm grav}}{m^2_{\rm pl}}\simeq
\left(\frac{m_{\rm grav}}{20 TeV}\right)^3 {\rm sec}^{-1}.
\eeq
If too many gravitinos decay during or after Big Bang Nucleosynthesis
(at $t\sim$ seconds), the resulting energetic showers in the thermal
bath destroy the agreement between predicted and observed light
element abundances\cite{Weinberg:1982zq,Kawasaki:1994af,Kawasaki:2008qe}.
 There are several possible solutions to the
gravitino problem:
\begin{enumerate}
\item The reheat temperature is low enough, $T_{\rm reheat}\lsim10^6-10^{10}$
 GeV \cite{Kawasaki:2008qe,Giudice:2008gu} that the gravitino density is
 small. The  precise bound depends on the gravitino mass and SUSY spectrum.
\item The gravitinos decay before BBN; sufficiently
heavy  gravitinos
can arise in anomaly-mediated scenarios \cite{Ibe:2004tg}.
\item Late time entropy production can dilute the gravitino
  abundance, but it also dilutes the $B-L$ asymmetry
\cite{Buchmuller:2006tt}.
\item The gravitino is the LSP (and the dark matter), in which case
the bound on $T_{\rm reheat}$ is less restrictive
\cite{Feng:2004mt,Kanzaki:2006hm}. The gravitino can be the LSP in
gauge-mediated scenarios.
\end{enumerate}

As discussed in Section \ref{esteff}, there is a lower bound on the
reheat temperature $T_{\rm reheat} \gsim 10^9$ GeV, to produce a big
enough baryon asymmetry by thermal leptogenesis (in the three
generation type I seesaw with hierarchical $N_i$ masses). This is
difficult to reconcile with the first solution described above
\cite{Bolz:1998ek}, which is arguably the most plausible one in
gravity mediated scenarios. A large enough baryon asymmetry can be
produced at lower $T_{\rm reheat}$, if $M_2 - M_1 \lsim \Gamma_D$, the
$N_1$ decay width.

\subsection{Less hierarchical $N$'s}
\label{lessh}
It is crucial for thermal  leptogenesis that there are more than one singlet
neutrino. In particular, the CP asymmetry that is produced in $N_1$
decays comes from interference between the tree diagram and the loop
diagrams involving the heavier $N_i$ as virtual particles in the loop.
In the conventional picture, however, one assumes a strong hierarchy,
$M_{i>1}\gg M_1$. Then, an upper bound on the CP asymmetry in $N_1$
decays applies (see eqn (\ref{epsaabd})),which 
provides much of the predictive power of the conventional leptogenesis
scenario.

However, as mentioned in section \ref{esteff}, when the singlet
neutrino masses are not very strongly hierarchical, a term that is
higher order in $M_1/M_{i>1}$ can become important. Specifically,
there is a contribution to the CP asymmetry that is proportional to
$M_1^3/(M_2^2M_3)$ and does not vanish when the light neutrino masses
are degenerate \cite{Hambye:2003rt,Davidson:2003yk}.

In addition to this effect of the heavier singlet neutrino masses on
the CP asymmetry in the $N_1$ decays, there are also interesting
effects related to the decays of the heavier singlet neutrinos
$N_{2,3,\ldots}$ themselves. These effects are discussed in the
following subsection.

\subsubsection{$N_2$ effects}
\label{N2}
When analyzing leptogenesis, the  effects of $N_2$ and $N_3$
are often neglected. This is reasonable if $T_{\rm reheat}< M_{2,3}$.
However,   if $T_{\rm reheat}> M_{2,3}$,  one should not assume that the
$L$-violating interactions of $N_1$ would washout any lepton asymmetry
generated at temperatures $T\gg M_1$ and, in particular, the asymmetry
generated in the decays of $N_{2,3}$. Under various (rather
generic) circumstances, the lepton asymmetry generated in $N_{2,3}$
decays survives the $N_1$ leptogenesis phase. Thus, it is quite
possible that the lepton asymmetry relevant for baryogenesis
originates mainly (or, at least, in a non-negligible part) from
$N_{2,3}$ decays.

The possibility that $N_2$ leptogenesis can successfully explain the
baryon asymmetry of the Universe has been shown in two limiting cases:
\begin{enumerate}
\item The ``$N_1$-decoupling'' scenario, in which the Yukawa couplings
  of $N_1$ are simply too weak to washout the $N_2$-generated
  asymmetry \cite{Vives:2005ra,DiBari:2005st,Blanchet:2006dq}.
\item The ``strong $N_1$-coupling'' scenario, where $N_1$-related
  decoherence effects project the lepton asymmetry from $N_2$ decays
  onto a flavour direction that is protected against $N_1$ washout
  \cite{Barbieri:1999ma,Nielsen:2001fy,Strumia:2006qk,Engelhard:2006yg}.
\end{enumerate}
It is plausible that the role of $N_2$ leptogenesis cannot be
ignored also in the intermediate range for $N_1$ couplings, but the
analysis in this case is complicated and has not been carried out yet.

The $N_1$-decoupling scenario is simple to understand. It applies when
$N_1$ is weakly coupled to the lepton doublets (see eqn
(\ref{tildem}):
\beq
\tilde m_1\ll m_*,
\eeq
where $m_*=1.66g_*^{1/2}8\pi v_u^2/m_{\rm pl}\approx10^{-3}\ {\rm eV}$ and
\beq
\tilde m_i\equiv\frac{v_u^2(\lambda^\dagger\lambda)_{ii}}{M_i}.
\eeq
In this case, the asymmetry generated in thermal $N_1$ leptogenesis is
too small. Furthermore, the $N_1$ washout effects are negligible and,
consequently, the asymmetry generated in $N_2$ decays survives.

We next discuss $N_2$ leptogenesis in the strong $N_1$-coupling
limit. We are interested in the case that a sizeable asymmetry is
generated in $N_2$ decays, while $N_1$ leptogenesis is inefficient.
We thus assume that the $N_2$-related washout is not too strong, while
the $N_1$-related  washout is so strong that it makes $N_1$
leptogenesis fail:
\beq\label{eq:contms}
\tilde m_2\not\gg m_*,\ \ \ \ \tilde m_1\gg m_*.
\eeq
To further simplify the analysis, we impose two additional conditions
\cite{Engelhard:2006yg}: thermal leptogenesis, and strong hierarchy,
$M_2/M_1$. These two conditions together guarantee that
\beq
n_{N_1}(T\sim M_2)\approx 0,\ \ \ n_{N_2}(T\sim M_1)\approx 0.
\eeq
Thus, the dynamics of $N_2$ and $N_1$ are decoupled: there are neither
$N_1$-related washout effects during $N_2$ leptogenesis, nor
$N_2$-related washout effects during $N_1$ leptogenesis.

The $N_2$ decays into a combination of lepton doublets that we denote
by $\ell_2$:
\beq
|\ell_i\rangle=(\lambda^\dagger\lambda)_{ii}^{-1/2}
\sum_\alpha\lambda_{\alpha i}|\ell_\alpha\rangle.
\eeq
The second condition in (\ref{eq:contms}) implies that already at
$T\gsim M_1$ the $N_1$-Yukawa interactions are sufficiently fast to
quickly destroy the coherence of $\ell_2$. Then a statistical mixture
of $\ell_1$ and of the state orthogonal to $\ell_1$ builds up, and it
can be described by a suitable diagonal density matrix. On general
grounds, one expects that decoherence effects proceed faster than
washout. In the relevant range, $T\gsim M_1$, this is also ensured by
the fact that the dominant ${\cal O}(\lambda^2)$ washout process (the
inverse decay $\ell\phi\to N_1$) is blocked by thermal effects
\cite{Giudice:2003jh}, and only scatterings with top-quarks and gauge
bosons, that have additional suppression factors of $h_t^2$ and $g^2$,
contribute to the washout.

Let us consider the case where both $N_2$ and $N_1$ decay at
$T\gsim10^{12}\>$ GeV, so that flavour effects are irrelevant. We also
neglect the effects of $\Delta L = 2 $  interactions, which
are generically  in  equilibrium at $T \gsim$ few $\times 10^{12}$ GeV.
(This scenario is interesting to consider, even though
it may only occur in a narrow temperature range.)
A convenient choice for an orthogonal basis for the lepton doublets is
$(\ell_1,\ell_0,\ell_0^\prime)$ where, without loss of generality,
$\ell_0^\prime$ satisfies $\langle\ell_0^\prime|\ell_2\rangle=0$. Then
the asymmetry $\Delta Y_{\ell_2}$ produced in $N_2$ decays decomposes into
two components:
\beq
\Delta Y_{\ell_0}=c^2\,\Delta Y_{\ell_2},\qquad \ \Delta
Y_{\ell_1}=s^2\,\Delta Y_{\ell_2},
\eeq
where $c^2\equiv|\langle\ell_0|\ell_2\rangle|^2$ and $s^2=1-c^2$. The
crucial point here is that we expect, in general, $c^2\neq0$ and,
since $\langle\ell_0|\ell_1\rangle=0$, $\Delta Y_{\ell_0}$ is protected
against $N_1$ washout. Consequently, a finite part of the asymmetry
$\Delta Y_{\ell_2}$ from $N_2$ decays survives through $N_1$ leptogenesis. A
more detailed analysis \cite{Engelhard:2006yg} finds that $\Delta Y_{\ell_1}$
is not entirely washed out, and that the final lepton asymmetry is
given by $Y_{\Delta L}=(3/2)\Delta Y_{\ell_0}=(3/2)c^2\,\Delta Y_{\ell_2}$.

For $10^9\ GeV\lsim M_1\lsim10^{12}\ GeV$, flavour issues modify the
quantitative details, but the qualitative picture, and in particular
the survival of a finite part of $\Delta Y_{\ell_2}$, still hold. On the
other hand, for $M_1\lsim 10^9\ GeV$, the full flavour basis
$(\ell_e,\ell_\mu,\ell_\tau)$ is resolved and, in general, there are
no directions in flavour space where an asymmetry is protected and
$Y_{\ell_2}$ can be erased entirely.

The conclusion is that $N_2$ and $N_3$ leptogenesis cannot be ignored,
unless at least one of the following conditions applies:
\begin{enumerate}
\item The asymmetries and/or the washout factors vanish,
  $\epsilon_{N_2}\eta_2\approx0$ and $\epsilon_{N_3}\eta_3\approx0$.
\item $N_1$-related washout is still significant at $T\lsim10^9\ GeV$.
\item The reheat temperature is below $M_2$.
\end{enumerate}

In all other cases, the $N_{2,3}$-related parameters play a role in
determining the baryon asymmetry of the Universe. Consequently,
relations between these parameters and neutrino masses are
important. Such relations were obtained in
Ref. \cite{Endoh:2002wm,Engelhard:2007kf}, and they lead, for the case that the light
neutrino masses have normal hierarchy, to the following consequences:
\begin{itemize}
  \item In the framework with two singlet neutrinos, both $N_1$ and
    $N_2$ interactions are in the strong washout regime, with both
    $\tilde m_i\geq0.009$ eV \cite{Endoh:2002wm}, and at least one
    $\geq0.025$ eV.
    \item In the framework with three singlet neutrinos, at least two
      $N_i$'s have interactions in the strong washout regime, with
      $\tilde m_i\geq0.005$ eV and at least one $\geq0.02$ eV.
    \end{itemize}
The lower bounds are stronger for inverted hierarchy, and even more so
in the framework with three singlet neutrinos and quasi-degenerate
light neutrinos.

\subsubsection{Resonant leptogenesis}
\label{resonant}
A resonant enhancement of the CP asymmetry in $N_1$ decay occurs when
the mass difference between $N_1$ and $N_2$ is of the order of the
decay widths. Such a scenario has been termed `resonant leptogenesis',
and has benefited from many  studies in different formalisms 
\footnote{See \cite{Rangarajan:1999kt} for a comparison of
the different calculations.}
\cite{Flanz:1994yx,Covi:1996fm,Pilaftsis:1997jf,Pilaftsis:2003gt,Hambye:2004jf,Pilaftsis:2004xx,Albright:2004ws,Albright:2005bm,Pilaftsis:2005rv,Anisimov:2005hr,Xing:2006ms,West:2006fs,Cirigliano:2006nu,Branco:2006hz,DeSimone:2007rw,DeSimone:2007pa,Babu:2007zm,Cirigliano:2007hb}. 
As this review focuses on hierarchical $N_i$, 
we briefly list the idea and some references. 

The resonant effect is related to the self energy contribution to the
CP asymmetry. Consider, for simplicity, the case where only $N_2$ is
quasi-degenerate with $N_1$. Then, the self-energy contribution
involving the intermediate $N_2$, to the total
CP asymmetry (we neglect  important flavour
effects \cite {Pilaftsis:2005rv}) is given by
\beq
\epsilon_{N_1}({\rm
  self-energy})=-\frac{M_1}{M_2}\frac{\Gamma_2}{M_2}
\frac{M_2^2\Delta M^2_{21}}{(\Delta M^2_{21})^2+M_1^2\Gamma_2^2}
\frac{{\cal I}m[(\lambda^\dagger\lambda)_{12}^2]}
{(\lambda^\dagger\lambda)_{11}(\lambda^\dagger\lambda)_{22}}.
\eeq
The resonance condition reads
\beq
M_2-M_1=\Gamma_2/2.
\eeq
In this case
\beq
|\epsilon_{N_1}({\rm
  resonance})|\simeq\frac{1}{2}
\frac{|{\cal I}m[(\lambda^\dagger\lambda)_{12}^2]|}
{(\lambda^\dagger\lambda)_{11}(\lambda^\dagger\lambda)_{22}}.
\eeq
Thus, in the resonant case, the asymmetry is suppressed by neither the
smallness of the light neutrino masses, nor the smallness of their
mass splitting, nor small ratios between the singlet neutrino
masses. Actually, the CP asymmetry could be of order one. (More
accurately, $|\epsilon_{N_1}|\leq1/2$.)

With resonant leptogenesis, the Boltzmann Equations are different.
The densities of $N_1$ and $N_2$ are followed, since
the both contribute to the asymmetry, and the relevant
timescales are different.  
For instance,  the typical time-scale to build up
coherently the CP asymmetry is unusually long, of order $1/\Delta
M$. In particular, it can be larger than the time-scale for the change
of the abundance of the sterile neutrinos. This situation implies that
quantum effects in the Boltzmann equations can be significant,
 in the weak or mild washout
regime, for
resonant leptogenesis \cite{DeSimone:2007rw,DeSimone:2007pa}. 

The fact that the asymmetry could be large, independently of the
singlet neutrino masses, opens up the possibility of low scale
resonant leptogenesis. Models along these lines have been constructed
in Refs. \cite{Hambye:2004jf,Pilaftsis:2005rv,West:2006fs,West:2004me}.
It is  a  theoretical challenge to construct models
where a mass splitting as small as the decay width is naturally
achieved. For attempts that utilize approximate flavour symmetries see,
for example,
\cite{Pilaftsis:2004xx,Albright:2004ws,Xing:2006ms,Babu:2007zm,Branco:2005ye},
while
studies of this issue in the framework of minimal flavour violation can
be found in \cite{Cirigliano:2006nu,Branco:2006hz,Cirigliano:2007hb}.

\subsection{Soft leptogenesis}
\label{soft}
The modifications to standard leptogenesis due to supersymmetry have
been discussed in section \ref{sec:susylep}. The important parameters
there are the Yukawa couplings and the singlet neutrino parameters,
which are all superpotential terms (see eqn \ref{superpot}):
\beq\label{eq:nsuppot}
W^{(N)}=\lambda L H_u N^c+\frac12 MN^cN^c.
\eeq
Supersymmetry must, however, be
broken. In the framework of the supersymmetric standard model extended
to include singlet neutrinos (SSM+N), there are, in addition to the
soft supersymmetry breaking terms of the SSM, terms that involve the
singlet sneutrinos $\widetilde N$, in particular bilinear ($B$) and
trilinear ($A$) scalar couplings. These terms provide additional
sources of lepton number violation and of CP violation.  Scenarios
where these terms play a dominant role in leptogenesis have been
termed `soft leptogenesis'
\cite{Grossman:2003jv,D'Ambrosio:2003wy,Chun:2004eq,Boubekeur:2004ez,D'Ambrosio:2004fz,Chen:2004xy,Kashti:2004vj,Grossman:2005yi,Ellis:2005uk,Chun:2005ms,Medina:2006hi,Garayoa:2006xs,Chun:2007ny,BahatTreidel:2007ic}.

Soft leptogenesis would have taken place even with a single lepton
generation. To simplify things we work, therefore, in the framework
of a single generation SSM+N. The relevant soft supersymmetry
terms in the Lagrangian are given in eqn (\ref{eq:nsoft}):
\beq\label{eq:nsoft}
{\cal L}_{\rm soft}^{(N)}=-\left(
  A\lambda\tilde L\phi_u \widetilde N^c
  +\frac12 BM\widetilde N^c\widetilde N^c+{\rm h.c.}\right).
\eeq
In addition, the electroweak gaugino masses play a role:
\beq\label{eq:mwino}
{\cal L}_{\rm soft}^{(\lambda_2)}=-\left(
  m_2\lambda_2^a\lambda_2^a+{\rm h.c.}\right).
\eeq
Here $\lambda_2^a$ ($a=1,2,3$) are the $SU(2)$ gauginos. The effects
related to $\lambda_1$, the $U(1)_Y$ gaugino, are similar to (and
usually less important than) those of $\lambda_2$, so we do not
present them explicitly.

The Lagrangian derived from eqns (\ref{eq:nsuppot}), (\ref{eq:nsoft})
and (\ref{eq:mwino}) has two independent physical CP violating phases:
\bea
\phi_N&=&\arg(AB^*),\\
\phi_W&=&\arg(m_2 B^*).
\eea
These phases give the CP violation that is necessary to dynamically
generate a lepton asymmetry. If we set the lepton number of $N^c$ and
$\widetilde N^c$ to $-1$, so that $\lambda$ (and $A\lambda$) are lepton
number conserving, then $M$ (and $BM$) violate lepton number by two
units. Thus processes that involve $\lambda$ and $M$ give the lepton
number violation that is necessary for leptogenesis.

A crucial role in soft leptogenesis is played by the $\widetilde N-\widetilde
N^\dagger$ mixing amplitude,
\beq\label{eq:mixamp}
\langle\widetilde N|{\cal H}|\widetilde
N^\dagger\rangle=M_{12}-\frac{i}{2}\Gamma_{12}
\eeq
($M_{12}$ is the dispersive part of the mixing amplitude, while $\Gamma_{12}$
is the absorptive part), which induces mass and width differences,
\beq
x\equiv\frac{\Delta M}{\Gamma}\equiv\frac{M_H-M_L}{\Gamma},\ \ \
y\equiv\frac{\Delta\Gamma}{2\Gamma}\equiv\frac{\Gamma_H-\Gamma_L}{\Gamma},
\eeq
($\Gamma$ is the average width) between the two mass eigenstates, the
heavy $|\widetilde N_H\rangle$, and the light $|\widetilde N_L\rangle$,
\beq\label{eq:qp}
|\widetilde N_{L,H}\rangle=p|\widetilde N\rangle\pm q|\widetilde
N^\dagger\rangle.
\eeq
The ratio $q/p$ depends on the mixing amplitude ratio:
\beq
\left(\frac qp\right)^2=\frac{2M_{12}^*-i\Gamma_{12}^*}
{2M_{12}-i\Gamma_{12}}.
\eeq
Also of importance are the decay amplitudes of $\widetilde N$ or $\widetilde
N^\dagger$ into various final states $X$:
\beq
A_X=\langle X|{\cal H}|\widetilde N\rangle,\ \ \ \
\overline{A}_X=\langle X|{\cal H}|\widetilde N^\dagger\rangle.
\eeq

The decay width of the singlet sneutrino is given by (for
$|M|\gg|A|$)
\beq
\Gamma=\frac{|M\lambda^2|}{4\pi},
\eeq
while the mixing parameters are given by
\bea
x&=&\left|\frac{B}{M}\right|\frac{4\pi}{|\lambda|^2},\nonumber\\
y&=&\left|\frac{A}{M}\right|\cos\phi_N-\frac12\left|\frac{B}{M}\right|,\\
\left|\frac qp\right|&=&\left(1+\frac{2|A\lambda^2/(2\pi B)|\sin\phi_N}
  {1-|A\lambda^2/(2\pi B)|\sin\phi_N+\frac14|A\lambda^2/(2\pi
    B)|^2}\right)^{1/4}.\nonumber
\eea

The quantity of interest is the CP asymmetry in singlet sneutrino
decay,
\beq
\epsilon=\frac{\Gamma(\widetilde L)+\Gamma(L)-\Gamma(\widetilde
  L^\dagger)-\Gamma(\bar L)}{\Gamma(\widetilde L)+\Gamma(L)+\Gamma(\widetilde
  L^\dagger)+\Gamma(\bar L)},
\eeq
where $\Gamma(X)$ is the time-integrated decay rate into a final state
with leptonic content $X$. Here $L(\bar L)$ is the (anti)lepton
doublet and $\widetilde L(\widetilde L^\dagger)$ is the (anti)slepton
doublet.

Qualitatively, the most special feature of soft leptogenesis is that it
gets contributions that are related to CP violation in mixing. This is a
phenomenon that is analogous to the CP violation in $K-\bar K$ mixing, where
it leads to the $K_L\to\pi\pi$ decays, the process where CP violation
was first discovered. A relative CP violating phase between $M_{12}$
and $\Gamma_{12}$ of eqn (\ref{eq:mixamp}) gives [see
eqn (\ref{eq:qp})] $|q/p|\neq1$ which, in turn, leads to a situation
where the mass eigenstates $\widetilde{N}_{H,L}$ are not CP eigenstates.
One such
contribution is given by \cite{Grossman:2003jv,D'Ambrosio:2003wy}
\beq\label{eq:epsom}
\epsilon^{\rm mix}=\frac{x^2}{4(1+x^2)}\left(\left|\frac pq\right|^2
  -\left|\frac qp\right|^2\right)\Delta_{sf}={\cal O}\left(
  \frac{x}{1+x^2}\frac{m_{\rm susy}}{M}\Delta_{sf}\right).
\eeq
Here $m_{\rm susy}$ is the scale of the soft supersymmetry breaking
terms ($m_2\sim m_{\rm susy}$ and $A,B\lsim m_{\rm susy}$), and
\beq
\Delta_{sf}\equiv\frac{{\cal N}_s(|A_{\tilde
    L}|^2+|\overline{A}_{\tilde L^\dagger}|^2)-{\cal N}_f(|A_{\bar
    L}|^2+|\overline{A}_L|^2)}{{\cal N}_s(|A_{\tilde
    L}|^2+|\overline{A}_{\tilde L^\dagger}|^2)+{\cal N}_f(|A_{\bar
    L}|^2+|\overline{A}_L|^2)},
\eeq
where ${\cal N}_s$ (${\cal N}_f$) are phase space factors for final
states involving scalars (fermions). At zero temperature,
$\Delta_{sf}={\cal O}(m_{\rm susy}^2/M^2)$, but for temperature at the
time of decay that is comparable to the singlet neutrino mass we have
$\Delta_{sf}\approx({\cal N}_s-{\cal N}_f)/({\cal N}_s+{\cal
  N}_f)={\cal O}(1)$. The difference between ${\cal N}_f$ and ${\cal
  N}_s$ at finite temperature arises from the Pauli blocking of final
state fermions and Bose-Einstein stimulation of decays into scalars.

The contribution (\ref{eq:epsom}) stands out among the soft
leptogenesis contributions as the only one that is linear in the ratio
$m_{\rm susy}/M$. (All other contributions are quadratic in this
ratio.) Thus, for $M\gg10^2m_{\rm susy}$, this is the only
contribution that is potentially significant. Indeed, it could account
for the observed baryon asymmetry if the following conditions are all
fulfilled:
\begin{enumerate}
\item The lightest singlet sneutrino is light enough, $M\lsim10^9\>$
  GeV.
\item The Yukawa couplings are small enough, $\lambda\lsim10^{-4}$.
  The lighter is $M$, the smaller the Yukawa coupling must be.
\item The $B$ parameter is well below its naive value, $B/m_{\rm
    susy}\lsim10^{-3}$. The lighter is $M$, the more suppressed $B$
  must be.
      \end{enumerate}

For $M\lsim10^{2}m_{\rm susy}$, there are several other
contributions that can be significant \cite{Grossman:2003jv}.
All of these contributions
involve $\lambda_2$ and are therefore proportional to the weak
coupling $\alpha_2$. In addition, as mentioned above, they are
proportional to $(m_{\rm susy}/M)^2$. In particular, there are
contributions related to CP violation in decay (a phenomenon that is
analogous to the one giving ${\cal R}e(\epsilon^\prime)$ in
$K\to\pi\pi$ decays and to the vertex contribution in standard
leptogenesis), and to CP violation in the interference of decays with
and without mixing (analogous to $S_{\psi K}$ in $B$ decays). These
two contributions are suppressed by $\alpha_2(m_{\rm susy}/M)^2$ but
no other small factors. In particular, unlike the contribution
$\epsilon^{\rm mix}$ of eqn (\ref{eq:epsom}), these contributions do
not vanish when $x\gg1$ and therefore they allow $B\sim m_{\rm susy}$.

Soft leptogenesis is an interesting scenario in the framework of the
supersymmetric seesaw for several reasons. First, the relevant new
sources of CP violation and lepton number violation appear generically
in this framework. In this sense, soft leptogenesis is qualitatively
unavoidable in the SSM+N framework, and the question of its relevance
is a quantitative one. Second, if $M\lsim10^9$ GeV (in the
supersymmetric framework, this range is preferred by the gravitino
problem), then standard leptogenesis encounters problems, while soft
leptogenesis can be significant. Third, if $M\lsim10^2m_{\rm susy}$,
then it is almost unavoidable that soft leptogenesis plays an
important role.

\subsection{Dirac leptogenesis}
\label{dirac}
The extension of the Standard Model with singlet neutrinos allows two
different ways of giving the active neutrinos their very light
masses. First, one can invoke the seesaw mechanism which gives
Majorana masses to the light neutrinos. This extension has
at least three attractive features:
\begin{itemize}
\item No extra symmetries (and, in particular, no global symmetries)
  have to be imposed.
\item The extreme lightness of neutrino masses is linked to the
  existence of a high scale of new physics, which is well
  motivated for various other reasons ({\it e.g.} gauge unification).
\item Lepton number is violated, which opens the way for leptogenesis.
\end{itemize}
The second way is to impose lepton number and give Dirac masses to the
neutrinos. A-priori, one might think that all three attractive features of the
seesaw mechanism are lost. Indeed, one must usually impose additional
symmetries. But one can still construct natural models where the tiny Yukawa
couplings that are necessary for small Dirac masses are related to a small
breaking of a symmetry. What is perhaps most surprising is the fact that
leptogenesis could proceed successfully even if the neutrinos are Dirac
particles and lepton number is not broken \cite{Akhmedov:1998qx,Dick:1999je}
(except, of course, by the sphaleron interactions). Such scenarios have been
termed `Dirac leptogenesis'
\cite{Dick:1999je,Murayama:2002je,Boz:2004ga,Abel:2006hr,Cerdeno:2006ha,Thomas:2005rs,Thomas:2006gr}.
Actually, the success of Dirac leptogenesis is closely related to the extreme
smallness of the neutrino Yukawa couplings in this scenario.

An implementation of the idea is the following. A CP violating decay of a heavy
particle can result in a non-zero lepton number for left-handed
particles, and an equal and opposite non-zero lepton number for
right-handed particles, so that the total lepton number is zero. For
the charged fermions of the Standard Model, the Yukawa interactions
are fast enough that they quickly equilibrate the left-handed and the
right-handed particles, and the lepton number stored in each chirality
goes to zero. This is not true, however, for Dirac neutrinos. The size
of their Yukawa couplings is $\lambda\lsim10^{-11}$, which means that
equilibrium between the lepton numbers stored in left-handed and
right-handed neutrinos will not be reached until temperatures fall
well below the electroweak breaking scale. To see this, note that the
rate of the Yukawa interactions is given by
$\Gamma_\lambda\sim\lambda^2 T$. It becomes significant when it equals
the expansion rate of the Universe, $H\sim T^2/m_{\rm pl}$. Thus, the
temperature of equilibration between left-handed and right-handed
neutrinos can be estimated as $T\sim\lambda^2 m_{\rm pl}\sim
(\lambda/10^{-8}) T_{\rm EWPT}$. By this time, the left-handed lepton
number has been partially transformed into a net baryon number by the
sphaleron interactions.

More specifically, consider a situation where the CP violating decays
of some new, heavy particles have produced a negative lepton number in
left-handed neutrinos, and a positive lepton number, of equal
magnitude, in right-handed neutrinos. The sphalerons interact with the
left-handed neutrinos, violating $B+L$ and conserving $B-L$.
Consequently, part of the negative lepton number stored in the
left-handed neutrinos is converted to a positive baryon number. At
much lower temperatures, when sphaleron interactions (and $B$ and $L$
violation) are frozen, the remaining negative lepton number in
left-handed neutrinos equilibrates with the positive lepton number
stored in the right handed-neutrinos. Since, however, the negative
lepton number in left-handed neutrinos has become smaller in magnitude
than the positive lepton number in right-handed neutrinos, a net
lepton number remains. The final situation is then a Universe with
total positive baryon number, total positive lepton number, and
$B-L=0$.

A specific example of a natural, supersymmetric model with Dirac
neutrinos is presented in Ref. \cite{Murayama:2002je}. The Majorana
masses of the $N$-superfields are forbidden by $U(1)_L$ (lepton number
symmetry). The neutrino Yukawa couplings are forbidden by a $U(1)_N$
symmetry where, among all the SSM+N fields, only the $N$ superfields
are charged. The symmetry is spontaneously broken by the vacuum
expectation value of a scalar field $\chi$ that can naturally be at
the weak scale, $\langle\chi\rangle\sim v_u$. This breaking is
communicated to the SSM+N via extra, vector-like lepton doublet
fields, $\phi+\bar\phi$, that have masses $M_\phi$ much larger than
$v_u$.  Consequently, the neutrino Yukawa couplings are suppressed by
the small ratio $\langle\chi\rangle/M_\phi$.  The CP violation arises
in the decays of the vector-like leptons, whereby $\Gamma(\phi\to
NH_u^c)\neq\Gamma(\bar\phi\to N^cH_u)$ and $\Gamma(\phi\to
L\chi)\neq\Gamma(\bar\phi\to L^c\chi^c)$. The resulting asymmetries in
$N$ and in $L$ are equal in magnitude and opposite in sign.

The main phenomenological implications of Dirac leptogenesis is the
absence of any signal in neutrinoless double beta decays.

\subsection{Triplet scalar leptogenesis}
\label{triplet}
As explained in Subsection \ref{seesawii}, one can generate see-saw
masses for the light neutrinos by tree-level exchange of
$SU(2)$-triplet scalars. Since this mechanism necessarily involves
lepton number violation, and allows for new CP violating phases, it is
interesting to examine it as a possible source of leptogenesis
\cite{Ma:1998dx,Chun:2000dr,Hambye:2000ui,Joshipura:2001ya,Hambye:2003ka,%
D'Ambrosio:2004fz,Guo:2004mp,Antusch:2004xy,Antusch:2005tu,Chun:2005ms,%
Hambye:2005tk,Chun:2006sp,Gu:2006wj,Sahu:2007uh,McDonald:2007ka,%
Antusch:2007km,Chao:2007rm,Hallgren:2007nq}.
One obvious problem in this scenario is that, unlike singlet fermions,
the triplet scalars have gauge interactions that keep them close to
thermal equilibrium at temperatures $T\lsim10^{15}\>$ GeV. It turns
out, however, that successful leptogenesis is possible even at a much
lower temperature. This subsection is based in large part on
Ref. \cite{Hambye:2005tk}, where further details and, in particular,
an explicit presentation of the relevant Boltzmann equations can be
found.

The CP asymmetry that is induced by the triplet scalar decays is
defined as follows:
\beq
\epsilon_T\equiv2\frac{\Gamma(\bar T\to \ell\ell)-\Gamma(T\to
  \bar\ell\bar\ell)}{\Gamma_T+\Gamma_{\bar T}}.
\eeq
The overall factor of 2 comes because the triplet scalar decay
produces two (anti)leptons.

To calculate $\epsilon_T$, one should use the Lagrangian terms given
in eqn (\ref{ltrisca}). While a single triplet is enough to produce
three light massive neutrinos, there is a problem in leptogenesis if
indeed this is the only source of neutrinos masses: The asymmetry is
generated only at higher loops and in unacceptably small.

It is still possible to produce the required lepton asymmetry from a
single triplet scalar decays if there are additional sources for the
neutrino masses, such as type I, type III, or type II contributions
from additional triplet scalars. Define $m_{\rm II}$ ($m_{\rm I}$) as
the part of the light neutrino mass matrix that comes (does not come)
from the contributions of the triplet scalar that gives $\epsilon_T$:
\beq
m=m_{\rm II}+m_{\rm I}.
\eeq
Then, assuming that the particles that are exchanged to produce
$m_{\rm I}$ are all heavier than $T$, the CP asymmetry is given by
\beq
\epsilon_T=\frac{1}{4\pi}\frac{M_T}{v_u^2}\sqrt{B_LB_H}\,\frac{{\cal
    I}m[{\rm Tr}(m_{\rm II}^\dagger m_{\rm I})]}{{\rm Tr}(m_{\rm
    II}^\dagger m_{\rm II})},
\eeq
where $B_L$ ($B_H)$ is the tree-level branching ratio to leptons
(Higgs doublets). If these are the only decay modes, {\it i.e.}
$B_L+B_H=1$, then $B_L/B_H={\rm
  Tr}(\lambda_L\lambda_L^\dagger)/(\lambda_H\lambda_H^\dagger)$.
There is an upper bound on this asymmetry:
\beq\label{uppertri}
|\epsilon_T|\leq\frac{1}{4\pi}\frac{M_T}{v_u^2}\sqrt{B_LB_H\sum_i
  m_{\nu_i}^2}.
\eeq
Note that, unlike the singlet fermion case, $|\epsilon_T|$ increases
with larger $m_{\nu_i}$.

As concerns the efficiency factor, it can be close to maximal,
$\eta\sim1$, in spite of the fact that the gauge interactions tend to
maintain the triplet abundance very close to thermal
equilibrium. There are two  necessary conditions that have to be
fulfilled by the two decay rates, $T\to\bar\ell\bar\ell$ and
  $T\to\phi\phi$, in order that this will happen \cite{Hambye:2005tk}:
\begin{enumerate}
\item One of the two decay rates is faster than the $T\bar T$
  annihilation rate.
\item The other decay mode is slower than the expansion rate of the
    Universe.
\end{enumerate}
The first condition guarantees that gauge scatterings are ineffective:
the triplets decay before annihilating. The second condition
guarantees that the fast decays do not lead to a strong washout of the
lepton asymmetry: lepton number is violated only by the simultaneous
presence of $T\to\bar\ell\bar\ell$ and $T\to\phi\phi$.

Combining a calculation of $\eta$ and the upper bound on the CP
asymmetry (\ref{uppertri}), successful leptogenesis implies a lower
bound on the triplet mass $M_T$ varying between $10^9$ GeV and
$10^{12}$ GeV, depending on the relative weight of $m_{\rm II}$ and
$m_{\rm I}$ in the light neutrino mass.

Interestingly, in the supersymmetric framework, ``soft leptogenesis'',
namely one that is driven by the soft supersymmetry breaking terms,
can be successful even with the minimal set of extra fields -- a
single $T+\bar T$ -- generating both neutrino masses and the lepton
asymmetry \cite{D'Ambrosio:2004fz,Chun:2005ms}.

\subsection{Triplet fermion leptogenesis}
\label{sec:triplet}
As explained in Subsection \ref{seesawiii}, one can generate see-saw
masses for the light neutrinos by tree-level exchange of $SU(2)$-triplet
fermions. This mechanism necessarily involves lepton number violation,
and allows for new CP violating phases so we should examine it as a
possible source of leptogenesis
\cite{Brahmachari:2001bv,Hambye:2003rt,Hambye:2003ka}.  
This subsection is based in
large part on Ref. \cite{Hambye:2003ka}, where further details and, in
particular, an explicit presentation of the relevant Boltzmann
equations can be found.

As concerns neutrino masses, the formalism and qualitative features
are very similar to the singlet fermion case. As concerns
leptogenesis, there are, however, qualitative and quantitative
differences.

With regard to the CP asymmetry from the lightest triplet fermion decay,
the relative sign between the vertex loop contribution and the
self-energy loop contribution is opposite to that of the singlet
fermion case. Consequently, in the limit of strong hierarchy in the
heavy fermion masses, the asymmetry in triplet decay is 3 times
smaller than in the singlet decay (see the
discussion in Section \ref{esteff} for the singlet case and for
definitions).
On the other hand, since the triplet has three components, the ratio
between the final baryon asymmetry and $\epsilon\eta$ is 3 times
bigger. The decay rate of the heavy fermion is the same in both
cases. This, however, means that the thermally averaged decay rate is
3 times bigger for the triplet, as is the on-shell part of the $\Delta
L=2$ scattering rate.

A significant qualitative difference arises from the fact that the
triplet has gauge interactions. The effect on the washout factor
$\eta$ is particularly significant for $\tilde m\ll10^{-3}\>$eV, the
so-called ``weak washout regime'' (note that this name is
inappropriate for triplet fermions). The gauge interactions still
drive the triplet abundance close to thermal equilibrium. A relic
fraction of the triplet fermions survives. The decays of these relic
triplets produce a baryon asymmetry, with
\beq
\eta\approx M_1/10^{13}\ {\rm GeV}\ \ \ ({\rm for}\ \tilde
m\ll10^{-3}\ {\rm eV}).
\eeq
The strong dependence on $M_1$ is a result of the fact that the
expansion rate of the Universe is slower at lower temperatures.
On the other hand, for $\tilde m\gg10^{-3}\>$ eV, the Yukawa
interactions are responsible for keeping the heavy fermion abundance
close to thermal equilibrium, so the difference in $\eta$ between the
singlet and triplet case is only ${\cal O}(1)$.

Ignoring flavour effects, and assuming strong hierarchy between the
heavy fermions,  Ref. \cite{Hambye:2003ka} obtained 
a lower bound on the mass of the lightest
heavy triplet fermion:
\beq
M_1\gsim1.5\times10^{10}\ {\rm GeV}
\eeq

When the triplet fermion scenario is incorporated in a supersymmetric
framework, and the soft breaking terms do not play a significant role,
the modifications to the above analysis is by factors of
${\cal O}(1)$.

\newpage
\section{Conclusions}
\label{conclusions}

During the last few decades, a large set of experiments involving
solar, atmospheric, reactor and accelerator neutrinos have converged
to establish that the standard model neutrinos are massive. The
seesaw mechanism extends the standard model in a way that allows
neutrino masses. It provides a nice explanation of the suppression of
the neutrino masses with respect to the electroweak breaking scale,
which is the mass scale of all standard model charged fermions.
Furthermore, without any addition or modification, it can also account
for the observed baryon asymmetry of the Universe.  The possibility of
giving an explanation of two apparently unrelated experimental facts
-- neutrino masses and the baryon asymmetry -- within a single
framework that is a natural extension of the standard model, together
with the remarkable `coincidence' that the same neutrino mass scale
suggested by neutrino oscillation data is also optimal for
leptogenesis, makes the idea that baryogenesis occurs through
leptogenesis a very attractive one.

Leptogenesis can be quantitatively successful without any fine-tuning
of the seesaw parameters. Yet, in the non-supersymmetric seesaw
framework, a fine-tuning problem arises due to the large corrections
to the mass-squared parameter of the Higgs potential that are
proportional to the heavy Majorana neutrino masses 
(see Section~\ref{sec:susy}).  Supersymmetry can
cure this problem, avoiding the necessity of fine tuning. However, the
gravitino problem that arises in many supersymmetric models requires a
low reheat temperature after inflation, in conflict with generic
leptogenesis models (see Section~\ref{sec:susylep}).  Thus, constructing
a fully satisfactory theoretical framework that implements
leptogenesis within the seesaw framework is not a straightforward
task.

From the experimental side, the obvious question to ask is if it is
possible to test whether the baryon asymmetry has been really produced
through leptogenesis. Unfortunately it seems impossible that any
direct test can be performed.  To establish leptogenesis
experimentally, we need to produce the heavy Majorana neutrinos and
measure the CP asymmetry in their decays. However, in the most natural
seesaw scenarios, these states are simply too heavy to be produced,
while if they are light, then their Yukawa couplings must be very
tiny, again preventing any chance of direct measurements.

The possibility of indirect tests from measurements of the asymmetries
produced by leptogenesis is also ruled out. This is because there are too many
high energy parameters that are relevant to leptogenesis and, even if we adopt
the most optimistic point of view, there are at most four observables. These
are the value of the baryon asymmetry that is known with a good accuracy (see
Section~\ref{observations}), and the three cosmic neutrino flavour asymmetries
that we do not know how to measure. A measurement of the total lepton
asymmetry would be very valuable. If a value of the order of the baryon
asymmetry were found, it would provide strong evidence that electroweak
sphalerons have been at work in the early Universe.\footnote{The opposite
  result would not disprove sphaleron physics: see
  \cite{Shaposhnikov:2008pf,Laine:2008pg} for a model in which production of a
  large lepton asymmetry {\it after} sphalerons freeze-out can yield
  $Y_{\Delta L}$ orders of magnitude larger than $Y_{\Delta B}$.}  Even if we
have no reason to doubt that sphaleron processes occurred in the thermal bath
with in-equilibrium rates, they are the fundamental ingredient of the whole
leptogenesis idea, and thus the importance of an experimental test should not
be underappreciated.  Measurements of the single lepton flavour asymmetries
would provide some information on leptogenesis parameters (on the flavour CP
asymmetries if leptogenesis occurred in the unflavoured regime, or on a
combination of the flavour CP asymmetries and of the flavour-dependent
washouts otherwise). However, at present, we have not even observed the relic
neutrino background, and the possibility of revealing ${\cal O}(10^{-10})$
asymmetries in this background seems completely hopeless.

Lacking the possibility of a direct proof, experiments can still
provide circumstantial evidence in support of leptogenesis by
establishing that (some of) the Sakharov conditions for leptogenesis
(see Section~\ref{ingredients}) are realized in nature.

Planned neutrinoless double beta decay ($0\nu\beta\beta$) experiments
(GERDA~\cite{GERDA}, MAJORANA~\cite{Aalseth:2004yt},
CUORE~\cite{Ardito:2005ar}) aim at a sensitivity to the effective
$0\nu\beta\beta$ neutrino mass in the few $\times$ 10 meV range.
If they succeed in establishing the Majorana nature of the light
neutrinos,\footnote{This is likely to happen if neutrinos are
  quasi-degenerate or if the mass hierarchy is inverted} this will
strengthen our confidence that the seesaw mechanism is at the origin
of the neutrino masses and, most importantly, will establish that the
first Sakharov condition for the dynamical generation of a lepton
asymmetry, that is that lepton number is violated in nature, is
satisfied.

Proposed SuperBeam facilities~\cite{Autin:2000mn,Mezzetto:2003mm} and
second generation off-axis SuperBeams experiments
(T2HK~\cite{Itow:2001ee}, NO$\nu$A~\cite{Ayres:2004js}) can discover
CP violation in the leptonic sector. These experiments can only probe
the Dirac phase of the neutrino mixing matrix. They cannot probe
the Majorana low energy or the high energy phases, but the
important point is that they can establish that the second Sakharov
condition for the dynamical generation of a lepton asymmetry is
satisfied.\footnote{As discussed in~\cite{Anisimov:2007mw}, it is
  conceivable (but not natural) that the Dirac phase is solely
  responsible for the matter-antimatter asymmetry.}

In contrast to the previous two conditions, verifying that the decays
of the heavy neutrinos occurred out of thermal equilibrium (the third
condition) remains out of experimental reach, since it would require
measuring the heavy neutrino masses and the size of their couplings.

Given that we do not know how to prove that leptogenesis is the
correct theory, we might ask if there is any chance to falsify it.
Indeed, future neutrino experiments could weaken the case for
leptogenesis, or even falsify it, in two main ways: Establishing that
the seesaw mechanism is not responsible for the observed neutrino
masses, or finding evidence that the washout is too strong.

By itself, failure in revealing signals of $0\nu\beta\beta$ decays
will not disprove leptogenesis. Indeed, with normal neutrino mass
hierarchy one expects that the rates of lepton number violating
processes are below experimental sensitivity.  However, if neutrinos
masses are quasi-degenerate or inversely hierarchical, and future
measurements of the oscillation parameters will not fluctuate too much
away from the present best fit values, the most sensitive
$0\nu\beta\beta$ decay experiments scheduled for the near future
should be able to detect a signal
\cite{Strumia:2006db}.\footnote{$0\nu\beta\beta$ decay experiments are
  sensitive to the effective mass parameter
  $m_{\beta\beta}=\sum_iU^2_{ei}m_i$. In the inverted hierarchy
  scenario the $s^2_{13} m_3$ contribution to $m_{\beta\beta}$ can be
  neglected, and approximating for simplicity $c^2_{13}\simeq 1$ one
  gets $m_{\beta\beta}\approx
  m_1e^{2i\beta}c^2_{12}+m_2e^{2i\alpha}s^2_{12}$.  Since
  $\theta_{12}$ deviates from maximal mixing, we can then predict that
  $m_{\beta\beta}$ should be of the order of the atmospheric mass
  scale, independently of the values of the Majorana phases $\alpha$
  and $\beta$.}  If instead the limit on $|m_{\beta\beta}|$ is pushed
below $\sim 10\,$meV (a quite challenging task), this would suggest
that either the mass hierarchy is normal, or neutrinos are not
Majorana particles.  The latter possibility would disprove the seesaw
model and standard leptogenesis (see however
section~\ref{dirac}). Thus, determining the order of the neutrino mass
spectrum is extremely important to shed light on the connection
between $0\nu\beta\beta$ decay experiments and leptogenesis.

Long baseline neutrino experiments can achieve this
result~\cite{Hagiwara:2005pe,MenaRequejo:2005hn} by exploiting the
fact that oscillations in matter can determine the sign of the
atmospheric mass difference $\Delta m^2_{23}$. In fact, if the
oscillations of atmospheric neutrinos involve the heaviest (lightest)
state, corresponding to normal (inverted) hierarchy, then $\nu_\mu
\leftrightarrow \nu_e$ oscillations are enhanced (suppressed) while
$\bar\nu_\mu \leftrightarrow \bar\nu_e$, oscillations are suppressed
(enhanced).

Cosmology could also provide important information in establishing if
the neutrino mass hierarchy is inverted.  Cosmological observations
are sensitive to the sum of the three neutrino masses $m_{\rm
  cosmo}=m_1+m_2+m_3$.  In the near future, improved CMB data and more
precise large scale structure measurements can push the sensitivity on
$m_{\rm cosmo}$ down to the $0.1\,$eV level. For an inverted hierarchy
a signal should be detected around or above this value.

In summary, if it is established that the neutrino mass hierarchy is
inverted and at the same time no signal of $0\nu\beta\beta$ decays is
detected at a level $|m_{ee}|\lsim 10\,$meV, one could conclude that
the seesaw is not at the origin of the neutrino masses, and that
leptogenesis is not the correct explanation of the baryon asymmetry.

The neutrino mass scale has also more direct implications for
leptogenesis.  A quantitatively successful leptogenesis prefers a mass
scale for the light neutrinos not larger than a few times the
atmospheric neutrino mass scale (see Section~\ref{sec:upperbd}). In
particular, in the standard type I seesaw model (see
section~\ref{type1}), if leptogenesis occurs in the unflavoured regime
and the heavy Majorana neutrinos are sufficiently hierarchical,
leptogenesis would fail to produce enough baryon asymmetry when the
neutrino mass scale exceeds values as low as $\sim 0.1-0.2\,$eV.
Indeed, if past experiments had detected neutrino masses of order a
few eV, leptogenesis, if not completely ruled out, would have
certainly lost most of its theoretical appeal. In the future, a
discovery of a neutrino mass at the $\beta$ decay experiment KATRIN
will also put the type I seesaw scenario for leptogenesis in serious
doubt, since it will yield a mass value a bit too large (the claimed
discovery potential is for an effective $\beta$-decay mass parameter
$m^2_\beta = \sum_i |U_{ei}|^2 m^2_i$ of about 0.35 eV~\cite{KATRIN}).
However, cosmological observations have already reached a sensitivity
to $m_{\rm cosmo}$ somewhat below
$1\,$eV~\cite{Spergel:2006hy,Tegmark:2003ud,elgaroy-2005-7}, implying
that it is unlikely that leptogenesis could be put to doubt by a
direct detection of neutrino masses in laboratory experiments.

As concerns CP violation, a failure in detecting leptonic CP violation
will not weaken the case for leptogenesis in a significant way.
Instead, it would mean that the Dirac CP phase (or the $\theta_{13}$
mixing angle, or both) is small enough to render CP violating effects
unobservable.

Finally, the CERN LHC has the capability of providing information that
is relevant to leptogenesis. First, electroweak baryogenesis (see
Section~\ref{ingredients}) can be tested at the LHC. It will become
strongly disfavored (if not completely ruled out) if supersymmetry is
not found, or if supersymmetry is discovered but the stop and/or the
Higgs are too heavy. Eliminating various scenarios that are able to
explain the baryon asymmetry will strengthen the case for the
remaining viable possibilities, including leptogenesis. Conversely, if
electroweak baryogenesis is established by the LHC (and EDM)
experiments, the case for leptogenesis will become weaker.

Second, new physics discoveries at the LHC can play a fundamental role
in establishing that the origin of the neutrino masses is not due to
the seesaw mechanism, leaving no strong motivation for leptogenesis.
This may happen in several different ways. For example (assuming that
the related new physics is discovered), the LHC will be able to test
if the detailed phenomenology of any of the following models is
compatible with an explanation of the observed pattern of neutrino
masses and mixing angles: supersymmetric R-parity violating couplings and/or
L-violating bilinear terms~\cite{deCampos:2007bn,Allanach:2007vi};
leptoquarks~\cite{Mahanta:1999xd,AristizabalSierra:2007nf}; triplet
Higgses~\cite{Garayoa:2007fw,Kadastik:2007yd}; new scalar particles of
the type predicted in the Zee-Babu~\cite{Zee:1985id,Babu:1988ki} types
of models~\cite{AristizabalSierra:2006gb,Chen:2007dc,Nebot:2007bc}.
It is conceivable that such discoveries can eventually exclude the
seesaw mechanism and rule out leptogenesis.

To conclude, the seesaw framework provides the most natural and
straightforward explanation of the light neutrino masses and has, in
principle, all the ingredients that are necessary for successful
leptogenesis. This makes leptogenesis arguably the most attractive
explanation for the observed baryon asymmetry. This scenario has
limited predictive power for low energy observables, so it is
unlikely to be directly tested. Yet, future experiments have the
potential of strengthening, or weakening, or even falsifying
the case for leptogenesis.

\newpage
{\bf {\Large Acknowledgements}}

$~$

This review is based on collaborations with many people over many
years, and parts of it are copied from the resulting papers.  We thank
our various collaborators for the pleasure of working with
them, and for their assistance  in writing this review.  

SD is grateful to  many people
for  useful discussions,  and comments on the
manuscript;  in particular  
E Akhmedov, G Branco, L Covi, P Di Bari, J Garayoa, GF Giudice, F
Joaquim, FX Josse-Michaux, M Laine, S Lavignac,
M Losada,  S Petkov, G Raffelt, G Rebelo, V Rubakov  and A Santamaria. 
SD would like to thank all her
leptogenesis collaborators: A Abada, L Boubekeur, B Campbell, J
Garayoa, A Ibarra,R Kitano, M Losada, F-X Josse-Michaux, K Olive, F
Palorini, M Peloso, A Riotto, L Sorbo and N Rius, for productive
collaborations.  And
as usual, a  special thanks to A Strumia for comments, discussions and
contributions. The collaboration of   SD with M Losada
and A Abada  was partially  supported by ECOS.

EN acknowledges fruitful collaborations in leptogenesis with Diego
Aristizabal, Guy Engelhard, Yuval Grossman, Marta Losada, Luis Alfredo
Mu\~noz, Jorge Nore\~na, Juan Racker and Esteban Roulet.  The work of EN is supported
in part by Colciencias in Colombia under contract 1115-333-18739.

YN thanks his leptogenesis collaborators: Guy Engelhard, Yuval
Grossman, Tamar Kashti, Juan Racker and Esteban Roulet.
The research of YN is supported by the Israel Science Foundation
(ISF), by the United States-Israel Binational Science Foundation
(BSF), by the German-Israeli Foundation for Scientific
Research and Development (GIF), and by the Minerva Foundation.

\newpage
\section{Appendix: Notation}
\label{notn}

\begin{table}[h!]
\begin{tabular}{||c|c|c||}
\hline
&&\\[-10pt]
\hline
$z\phantom{^\big|}$ & dimensionless time variable $M_1/T$ & \ref{definez} \\
$\dot{Y}$  & $({sH_1}/{z}) ({dY}/{dz})$ & \ref{eq:dotY} \\
$H,\,H_1$ & Hubble rate, Hubble rate at $T= M_1$ & \ref{app6H}, \ref{defz}\\
$s$ & entropy density & \ref{defYX} \\
$Y_X$ & ratio of $X$ number density to entropy density &  \ref{defYX}\\
$n_X^{\rm eq}$ & number density for particle $X$ in kinetic
equilibrium  &  \ref{nMB}, \ref{nFB1} \\
$Y_X^{\rm eq}$ & $Y_X$ in kinetic equilibrium &\ref{eq:Ydef}, \ref{defYX} \\
$Y_\ell^{\rm eq}$ & ratio of $\nu_{\alpha}$ and $e_{\alpha}$
 density to entropy density ( $g_{\ell_\alpha} = 2)$ & after \ref{eq:Ydef}\\
$Y_{L}^{\alpha \alpha}$ &  density of doublets and singlets of flavour $\alpha$ & \ref{eq:La}\\
$y_X$ & $Y_X/Y_X^{\rm eq}$ & \ref{eq:Ydef}  \\
$\Delta y_X$ & $ y_{X}-y_{\bar X}$ & \ref{eq:Ydef}  \\
$ Y_{\Delta_\alpha}$ & ratio of $B/3 - L_\alpha$ asymmetry
to entropy density &\ref{eq:deltaalpha}, \ref{YDa} \\
$Y_{\Delta B}$ & ratio of baryon asymmetry to entropy density &\ref{YB}, \ref{approx}, \ref{with2g} \\
$ g_*,\,g_{*S}$ & degrees of freedom in the thermal bath &  \ref{g*}, \ref{g*S} \\
$g_i $ & internal degrees of freedom of  particle $i$ &  below \ref{K2} \\
$\eta_\alpha$ & efficiency parameter for flavour $\alpha$ & \ref{approx},
\ref{eq:tm>m*}, \ref{eq:tm<m*}\\
$\epsilon_{\alpha \alpha}$ & CP asymmetry in flavour $\alpha$ &\ref{epsaa},  \ref{flavour-CPasym} \\
$[m]$ & light neutrino mass matrix & \ref{mlight} \\
$m_{\rm max},\,m_{\rm min}$ & largest, smallest light neutrino mass &
\ref{epsaabd}, \ref{di2}
 \\
$\tilde{m},\tilde{m}_{\alpha \alpha}, m_*$ & (rescaled) $N$ decay rates, Hubble
expansion rate
&\ref{tildem} \\
$A$  & matrix relating   the  $Y_{\Delta \ell_\alpha}$
to the $Y_{\Delta_\alpha}$ &  \ref{A2gen} \ref{A3gen} \\
$\lambda$ & neutrino Yukawa coupling &\ref{L}  \\
$h_x$ & Yukawa coupling of particle $x$ (not a neutrino) & \ref{L}\\
$v_u$& Higgs vev ($v_u\simeq 174 \,$GeV) & \ref{L}\\
$\Gamma_{D} $ & total decay
 rate $\sum_\alpha \Gamma(N \rightarrow \phi \ell_\alpha,\,
\bar{\phi} \bar{\ell}_\alpha)$ & \ref{GammaD}, \ref{eq:app7Gamma} \\
$\Gamma_{ID}$ &  total inverse decay
 rate  & \ref{10} \\
$\Gamma_{\alpha \alpha}$ & partial decay rate  to flavour and
antiflavour $\alpha$: $ \Gamma(N \rightarrow \phi \ell_\alpha,\,
\bar{\phi} \bar{\ell}_\alpha)$ & \ref{10} \\
$B^X_{xz}$&  branching ratio $\Gamma(X \rightarrow y z )/\Gamma_{X}^{\rm tot}$ & \ref{epsaabd} \\
$\gamma^A_B$ & rate density for $A \rightarrow B$ & \ref{eq:intden} \\
$\gamma_{N\rightarrow 2}$ &  rate density for two body decays of $N_1$ & \ref{Nto2} \\
$\Gamma_\tau,\,\gamma_{\tau}$ & rate,  rate density  of $\tau$ Yukawa interaction  & \ref{app7gam} \\
$\Delta \gamma^A_B$ & CP difference: $ \gamma^A_B-\gamma^{\bar A}_{\bar B}$ &
 \ref{eq:DeltaAtoB}\\
$[A\leftrightarrow B]$ & difference between process and  its time reversed
& \ref{eq:tire} \\
$\gamma'{}^A_B$&  $\gamma^A_B$ with exchange of on-shell particle
removed & \ref{eq:gammap} \\
$K_1(z),\, K_2(z)$ & Bessel functions & \ref{K1}, \ref{K2} \\
${\cal M},\,{\cal A} $ & matrix element, amplitude & \ref{cetA} 
\\
$\tilde{\delta}$ & 4-momentum conservation:
$(2 \pi)^4 \delta^4(p_{a_1} + ... - p_{b_1} - ..)$ & \ref{tddPi}\\
$d \Pi$ & phase space $ d^3p/(16 E \pi^3)$ & \ref{tddPi}\\ [-5pt]
 &&\\[-2pt]
\hline
\hline
\end{tabular} 
\caption{Parameters and variables  introduced in the text, and
equations where they are defined. Simple BE for leptogenesis
that neglect scattering processes are given in eqns (\ref{BE-N.DID}).
  and (\ref{mess3}).  More complete  BE that include top-quark and 
gauge bosons scatterings  are given in eqns (\ref{eq:finalBEYN}), 
(\ref{eq:finalBEYDeltai}),  (\ref{eq:finalBEYDeltai-s})    and 
(\ref{eq:finalBEYDeltai-w}).}
\end{table}

\newpage
\section{Appendix: Kinetic Equilibrium}
\label{kinetic}
The interested reader can find a more complete set of formulae for
example in the Appendices of refs. \cite{Kolb:1979qa,Kolb:1988aj}.

\subsection{Number densities}
Of the particles relevant to leptogenesis, all but the $N_i$'s carry
charges under the standard model gauge group and are, therefore,
subject to gauge interactions. We take the distributions $f_i(p)$ of
the (non-singlet) particle species $i$ to have their kinetic
equilibrium form, because gauge interactions are fast in the relevant
temperature range, $f_i(p)=({n_i}/{n^{\rm eq}_{i}})f_i^{\rm eq}(p)$.
(The $N_i$'s may also be produced in an equilibrium distribution
because the initial state particles are in kinetic equilibrium, or
kinetic equilibrium may be obtained due to fast Yukawa interactions,
if $\lambda$ is in the strong washout regime.) With some additional
assumptions (see eqn \ref{appnotdYXdt}), this allows to solve the BE
for the total number density rather than mode by mode (as was done in
\cite{Basboll:2006yx}).  The number densities of particles and
anti-particles are considered separately. For some other applications,
it may be the convention to count the particles and anti-particles
together (for instance for $g_*,g_{*S}$, see
eqns~(\ref{g*}),(\ref{g*S})) so some care with factors of 2 is
required in matching.

Using Maxwell-Boltzmann statistics for $f_i$,
\beq
f^{\rm eq}_{i,{\rm MB}}(p)=e^{-(E_i-\mu_i)/T},
\eeq
the equilibrium number density of particles of species $i$ is given by
\beq\label{nMB}
n^{\rm eq}_{i,{\rm MB}}=\frac{g_i}{(2\pi)^3}\int d^3 p
f^{\rm eq}_{i,{\rm MB}}(p)=\left\{\begin{array}{cc}
\frac{g_i T^3 }{2 \pi^2} z_i^2 K_2(z_i) &
z_i=\frac{m_i}{T}\neq0,~\mu=0,   \\
\frac{g_i T^3}{ \pi^2} & m_i=0. \end{array} \right.
\eeq
Here $K_2(z)$ is a Bessel function (see ref. \cite{G+R} eqns 8.432.3
and 8.472.2, and ref. \cite{Buchmuller:2004nz} for useful analytic
approximations):
\beq
 z^2 K_2(z) =  \int_z^\infty xe^{-x} \sqrt{x^2 - z^2} dx
\rightarrow
\left\{ \begin{array}{cc}
2 & z \ll 1 \\
\left(\frac{15}{8} + z \right) \sqrt{\frac{\pi z}{2}} e ^{-z} & z \gg 1
\end{array}  \right.
\label{K2}
\eeq
and $g_i$ is the number of internal degrees of freedom of the
{\it particle}. For example, $g_N=2$ for the $N_i$'s, because they are
Majorana fermions, $g_{e_R}=1$ for the $SU(2)$-singlet leptons, and
$g_\ell=2$ for the $SU(2)$-doublet leptons and $g_\phi=2$ for the
Higgs. Antiparticle number densities have $g_{\bar{i}}=g_i$.

Using  Fermi-Dirac ($+$) or Bose-Einstein ($-$) statistics,
\beq
f_{i,\pm}^{\rm eq}(p) = \frac{1}{e^{(E_i - \mu_i)/T} \pm 1},
\eeq
gives the following equilibrium number densities:
\beq\label{nFB1}
n_{i,\pm}^{\rm eq}=\frac{g_i}{(2 \pi)^3}\int d^3pf_{i,\pm}^{\rm eq}(p)
\rightarrow\left\{\begin{array}{cc}
 \frac{g_i T^3}{ \pi^2}\times \left\{ \begin{array}{cc}
  \zeta(3) + \frac{\mu_i}{T} \zeta(2) + ... & ({\rm bosons})\\
\frac{3}{4} \zeta(3) + \frac{\mu_i}{T} \frac{\zeta(2)}{2}  + & ({\rm
  fermions})
\end{array}\right. & m_i \ll T, \\
\phantom{\Big|} n^{\rm eq}_{i,{\rm MB}} & \mu_i \ll T, m_i\gg T,
\end{array}\right.
\eeq
where $\zeta(2)=\pi^2/6$, $\zeta(3)=1.202$, and $\zeta(4)=\pi^4/90$.
Using Maxwell-Boltzmann instead of Fermi-Dirac statistics
makes a difference in $n^{\rm eq}$ of order $10\%$ at $T=m$.

\subsection{Rates}
Number densities decrease due to the expansion of the Universe. This
can be factored out of the BE by taking $Y_X$ to be the comoving
number density:
\beq
Y_X = \frac{n_X}{s}, ~~~~ ~~~~~s = \frac{g_{*S} 2 \pi^2}{45} T^3.
\label{defYX}
\eeq
The entropy density $s$ is conserved per {\it comoving} volume:
$$ \frac{ds}{dt} =  - 3 H s~~~~.$$
The expansion rate of the Universe is given by
\beq
H=\sqrt{\frac{8\pi\rho}{3m_{\rm pl}^2}}\simeq
\frac{1.66\sqrt{g_*}T^2}{m_{\rm pl}}.
\label{app6H}
\eeq
The density and the pressure of the gas of relativistic particles are
given by
$$\rho=\frac{\pi^2}{30}g_*T^4,~~~~~P=\frac{\pi^2}{90}g_* T^4.$$
The (temperature-dependent) parameters $g_{*} (g_{*S})$ are the
effective number of degrees of freedom contributing to the energy
(entropy) density:
\bea
\label{g*}
g_{*}& = & \frac{7}{8}\sum_{\rm fermions} g_f \left(\frac{T_f}{T_\gamma}\right)^4
+\sum_{\rm bosons} g_b \left(\frac{T_b}{T_\gamma}\right)^4  \\
 g_{*S} & = & \frac{7}{8}\sum_{\rm fermions} g_f \left(\frac{T_f}{T_\gamma}\right)^3
+\sum_{\rm bosons} g_b \left(\frac{T_b}{T_\gamma}\right)^3
\label{g*S}
\eea
where the particle and anti-particle should both appear in the sums
over species. (Recall that we defined $g_i$ as the internal degrees of
freedom for the particle only.) The SM Higgs and anti-Higgs contribute
$g_\phi+g_{\bar{\phi}}=4$. Each flavour of neutrino and its
anti-neutrino contribute $7/4$. At leptogenesis temperatures,
$g_{*S}=g_*$. The contribution to the present entropy density $s_0$
comes from the photons at $T_\gamma$, and from the neutrinos and
anti-neutrinos\footnote{The momentum distribution of the neutrinos
  freezes out when they are relativistic, so even though their mass
  may be greater than their temperature, they contribute to $s$ like a
  relativistic species.} at $T^3_\nu \simeq(4/11)
T^3_\gamma$. Thus $s_0\propto g_{*S} T^3_\gamma$, with
$g_{*S}=2+\frac{42}{8}\frac{4}{11}\simeq3.9$.

A convenient time variable for leptogenesis in $N_1$ decay is
\beq\label{definez}
z\equiv M_1/T,
\eeq
for which
\beq
\frac{dz}{dt}=-\frac{M_1}{T^2}\frac{dT}{dt}=zH(z)=\frac{H_1}{z},
\label{defz}
\eeq
where  $H_1\equiv H(T=M_1)$ is a constant.

Suppose the number density of $X$-particles can be changed by various
interactions. The time evolution of $Y_X$ is
\bea
\frac{d Y_X}{dt}&=&\sum_{\rm int}\left[-\int d \Pi_X  f_X d \Pi_a f_a\ldots
|{\cal M}(X+a+\ldots\to i+j+\ldots)|^2 \tilde{\delta}~ d\Pi_i(1\pm f^\pm_i)
d \Pi_j(1\pm f^{\pm}_j)\ldots\right. \nonumber \\
&+& \hspace{-3mm}\left. \int d\Pi_i f^\pm_i d\Pi_j f^{\pm}_j\ldots
|{\cal M}(i+j+\ldots\to X+a+\ldots)|^2
 \tilde{\delta}~(1\pm  f_X)d\Pi_X(1\pm  f_a)d\Pi_a\ldots\right],
\label{appnotdYXdt}
\eea
where $d\Pi_X$ and $\tilde{\delta}$ are defined in eqn (\ref{tddPi}),
and $|{\cal M}|^2$ is summed over the internal degrees of freedom of the
initial and final state. This equation simplifies if the particles are
assumed to be in kinetic equilibrium, and if the final state Bose
enhancement/Pauli blocking factors are ignored. Then one obtains:
\bea
\frac{d Y_X}{dt}&=&-\sum_{\rm int}
\left[ \frac{Y_X Y_a\ldots}{(s Y^{\rm eq}_X)(s Y^{\rm eq}_a)\ldots}
\gamma (X + a + \ldots \to i+j+\ldots)\right.\nonumber \\
&& \left.-\frac{ Y_i Y_j\ldots}{(s Y^{\rm eq}_i)(s Y^{\rm eq}_j)
 \ldots} \gamma (i +j + \ldots \rightarrow X + a +\ldots)\right].
\eea
Here $\gamma $ is an {\it interaction density}:
\bea\nonumber
\gamma^{X a \ldots}_{ij\ldots}  &\equiv&
\gamma (X + a + \ldots \rightarrow i+j+\ldots) \\
&=&
\int d \Pi_X f^{\rm eq}_X d \Pi_a    f^{\rm eq}_a \ldots| {\cal M}
(X + a + \ldots \rightarrow i+j+\ldots)|^2 \tilde{\delta}~ d \Pi_i
d \Pi_j ~~,
\label{eq:intden}
\eea
where we drop final state phase space factors. Recall that $|{\cal
  M}|^2$ is summed (not averaged) over the internal degrees of freedom
of the initial states.

The rates in this Appendix are obtained in $T=0$ field theory.
Calculating rates at finite temperature \cite{Giudice:2003jh}
gives $O(1)$ effects on the number densities as a function of time,
although it can change individual rates in a more significant way,
as discussed in Section \ref{sec:thermal}. It is useful, for the
purpose of calculating decay and scattering rates, to have
the expression for two body phase space:
 \bea
\int  ~\tilde{\delta}~
d\Pi_pd\Pi_q  &=& \int \frac{|\vec{p}_p|}
{16 \pi^2 \sqrt{s}} d \Omega_p \nonumber\\
& =&\frac{|\vec{p}_p - \vec{p}_q|}{8 \pi \sqrt{s}}
= \frac{\sqrt{({p}_p \cdot {p}_q)^2 - m_p^2 m_q^2}}
{4 \pi s},
\label{2bdyps}
\eea
where the first line has the familiar form of the center of mass
frame, while the second provides the Lorentz-invariant form.

Consider the two-body decay rate density,
\beq
\gamma^N_{\phi \ell_\beta}+\gamma^N_{\bar{\phi} \bar{\ell}_\beta}
\equiv \gamma(N_1 \to \ell_\beta \phi,\overline{\ell}_\beta \bar{\phi}).
\eeq
There are four possible final states, $\nu_\beta \phi^0,e_\beta \phi^+,
\overline{\nu}_\beta \bar{\phi}^0,\overline{e}_\beta{\phi}^- $.
The four rates are all equal:
$$|{\cal M}(N_1 \rightarrow \nu_\beta \phi^0 )|^2 =
2|\lambda_{\beta 1}|^2 p_N \cdot p_\ell =| \lambda_{\beta 1}|^2( M_1^2
+ m_\ell^2 - m_H^2).$$
We thus obtain for the decay rate density:
\bea
\gamma^N_{\phi \ell_\beta}+\gamma^N_{\bar{\phi} \bar{\ell}_\beta}
&=&\int \frac{ d^3p_N}{ 2E_N(2 \pi)^3}  e^{-E_N/T} ~
4 |\lambda_{\beta 1}|^2 M_1^2\int  ~\tilde{\delta}~
d\Pi_\phi d\Pi_\ell \nonumber \\
& = & 
\frac{1}{2}\frac{ T^2}{2 \pi^2}
\int_z^\infty   e^{-x} \sqrt{x^2 - z^2} dx ~
~\frac{ | \lambda_{\beta 1}|^2 M_1^2}{2 \pi} \nonumber \\
&=&s Y^{\rm eq} \frac{K_1(z)}{K_2(z)}
\Gamma(N_1 \to \ell_\beta \phi,\overline{\ell}_\beta \bar{\phi} )
\label{eq:app7.18}\\
&=&\frac{g_N T^3}{2 \pi^2} z^2 K_1(z)
\Gamma(N_1 \to \ell_\beta \phi,\overline{\ell}_\beta \bar{\phi} ),
\label{2bdps}
\eea
where $K_1$ is a Bessel function (see ref. \cite{G+R}, eqn 8.432.3):
\beq
z K_1(z) = \int_z^\infty   e^{-x} \sqrt{x^2 - z^2} dx  \rightarrow
\left\{ \begin{array}{cc}
1 & z \ll 1 \\
\sqrt{\frac{\pi z}{2}} e ^{-z} & z \gg 1
\end{array}  \right.
\label{K1}
\eeq
and  ($g_N = 2$)
\beq
\Gamma(N_1 \rightarrow \ell_\beta \phi,\overline{\ell}_\beta \bar{\phi} )
= \frac{| \lambda_{\beta 1}|^2 M_1}{4 \pi g_N}.
\label{eq:app7Gamma}
\eeq
The inclusive two-body decay rate density appears frequently in our
calculations:
\beq
\gamma_{N\rightarrow 2} =  \sum_\beta
(\gamma(N  \rightarrow \phi  \ell_\beta) +
\gamma(N  \rightarrow \overline{\phi} \overline{\ell}_\beta)) =
\frac{g_N T^3}{2 \pi^2} z^2 K_1(z) \Gamma_D.
\label{Nto2}
\eeq
In the literature it is frequently denoted by $\gamma_D$. It was
evaluated with quantum statistics in \cite{Giudice:2003jh}, in which
case it does not have a simple closed form, and differs by $\sim
10\%$.

Consider the scattering rate density,
\beq
\gamma^{ij}_{mn} \equiv \gamma( i+j \rightarrow m+n).
\eeq
It is given by
\bea
\gamma^{ij}_{mn}& =& \int d\Pi_i  d\Pi_j f^{eq}_i f^{eq}_j
\int |{\cal M}(i+j \rightarrow m+n )|^2 ~\tilde{\delta}~
d\Pi_m d\Pi_n \nonumber \\
& =&4g_ig_j \int d\Pi_i  d\Pi_j e^{-(E_i +E_j)/T}
\sqrt{(p_i \cdot p_j)^2 - m_i^2m_j^2}\  \sigma((p_i +p_j)^2),
\eea
where the $g_i g_j$-factor appears because cross-sections are usually
averaged over initial state internal degrees of freedom.  The square
root cancels the initial particle flux factor of the cross-section. It
is convenient to separate the center of mass from initial state phase
space integrals, by multiplying by $1=\int d^4Q\delta^4(Q-p_i-p_j)$.
Using eqn (\ref{2bdyps}) gives:
\bea
\gamma(i+j \rightarrow m+n)&=&g_ig_j \int \frac{dQ_0d^3Q}{(2 \pi)^4}
\,  \frac{e^{-Q_0/T}}{\pi s}\, [(p_i\cdot p_j)^2-m_i^2m_j^2]\,
\sigma(Q^2) \nonumber \\
&=&\frac{g_ig_j}{32\pi^5}\int s ds \, d\Omega\, \int_{\sqrt{s}}dQ_0
e^{-Q_0/T}\sqrt{Q_0^2-s}~
\lambda\left(1,\frac{m_i^2}{s},\frac{m_j^2}{s}\right) \sigma(s)
\nonumber \\
&=&\frac{g_ig_jT}{32\pi^4}\int ds s^{3/2} \,
K_1\left(\frac{\sqrt{s}}{T}\right)
 \lambda\left(1,\frac{m_i^2}{s},\frac{m_j^2}{s}\right)\sigma(s),
\label{scatapp}
\eea
where $\lambda(a,b,c) = (a - b - c)^2 - 4 bc$,  $zK_1(z)$ is given in
eqn (\ref{K1}), and we used $d^3 Q = \sqrt{Q_0^2 - s}~
ds \,  d\Omega/2$ to obtain the second equality.

In the massless limit, $\lambda(1,x,y) \to 1$ and $s$ can be integrated
from 0 to $\infty$, so the definite integral
\beq
\int_0^\infty x^n K_1(x) = 2^{n-1} \Gamma(1 + n/2) \Gamma(n/2)
\eeq
is useful for obtaining $\gamma$. For interaction terms in the
Lagrangian of dimension 4 and 5, one obtains
\beq
\gamma=\left\{ \begin{array}{cc}
\frac{g_i g_j T^4}{8 \pi^4}& \sigma =\frac{1}{s},\\
\frac{g_i g_j T^6}{\pi^4 M^2} & \sigma =\frac{1}{M}.\end{array}\right.
\label{scatapp4d}
\eeq

\newpage
\section{Appendix: Chemical Equilibrium}
\label{chempot}
In this section, the conditions of chemical equilibrium are used to
derive the $A$-matrix and the factor of 12/37 relating (within the
Standard Model) $\Delta Y_B$ to $\Delta Y_{B-L}$.

Various Standard Model interactions can change the number of particles
of different species. For instance, the Lagrangian term $h_\alpha
\overline{\ell}_\alpha \phi^c e_{R\alpha}$ changes a charged
$SU(2)$-singlet lepton into a Higgs and an $SU(2)$-doublet lepton. If
such interactions are fast, compared to the expansion rate $H$, they
lead to an equilibrium state where the comoving number densities of
the participating particles remain constant.  This is described by
conditions of chemical equilibrium (see chapter 10 of ref.
\cite{Landau:1989}): The sum of the chemical potentials, over all
particles entering the interaction, should be zero.  For example, if
the charged lepton Yukawa interaction is fast, we have
\beq
\mu_{e_{\alpha}} - \mu_{{\ell}_\alpha} + \mu_{\phi}   = 0.
\eeq
Thus, the set of reactions that are in chemical equilibrium enforce
algebraic relations between various chemical potentials
\cite{Khlebnikov:1988sr,Harvey:1990qw}.

The chemical potentials can be  related  to the asymmetries
in the particle number densities, by expanding the distribution
functions of eqn (\ref{nFB1}) for small $\mu/T$:
\bea\label{eq:cpf}
n_f - n_{\bar{f}} &= & \frac{g_i T^3}{6}\ \mu~~~~~{\rm fermions}\\
n_b - n_{\bar{b}} &= & \frac{g_i T^3}{3}\ \mu~~~~~{\rm bosons}
\label{eq:cpb}
\eea
where $g_i$ is the number of degrees of freedom of the particle.

At the temperatures where the lepton asymmetry is generated, the
interactions mediated by the top-quark Yukawa coupling $h_t$, and by
the $ SU(3)\times SU(2)\times U(1)$ gauge interactions, are always in
equilibrium. This situation has the following consequences:
\begin{itemize}
\item All components of a gauge multiplet share the same chemical
  potential, and the chemical potential of gauge bosons is zero.
\item Hypercharge neutrality implies
\begin{equation} \label{hyper}
\sum_{i=1,2,3}\left( \mu_{q_i}+2\mu_{u_i}-\mu_{d_i}\right) -
\sum_{\alpha=e,\mu,\tau}\left(\mu_{\ell_\alpha}-\mu_{e_\alpha}\right)+
2\mu_\phi =0\,.
\end{equation}
\item The equilibrium condition for the Yukawa interactions
of the top-quark $\mu_t = \mu_{q_3}+\mu_\phi$ yields:
\begin{equation}
\label{tophiggs}
\Delta y_t-\frac{\Delta y_{q_3}}{2}=\frac{\Delta y_\phi}{4}\,,
\end{equation}
where the relative factor 1/2 between $\Delta y_{q_3}$ and $\Delta
y_\phi$ can be traced back to eqns~(\ref{eq:cpf})-(\ref{eq:cpb}): the
relation of the number asymmetry to the chemical potential is
different by a factor of 2 between bosons and fermions.
\end{itemize}

Using these relations, the asymmetries in $B_i$ and $L_\alpha$ can
also be expressed in terms of the chemical potentials, for example,
\bea
Y_{\Delta B_3}&=& \frac{T^3}{6s} (2\mu_{q_3}+
\mu_{t_R} + \mu_{b_R}) ~~~,\nonumber \\
Y_{\Delta L_\tau}&=& \frac{T^3}{6s}(2\mu_{\ell_\tau}+\mu_{\tau_R}),
\eea
and similarly for the lighter generations.  We use the notation $
Y_{\Delta X}$ for the {\it asymmetry} $Y_X-Y_{\bar{X}}$, because we
think it is clearer, though not standard: for instance $Y_B$ is
usually the baryon asymmetry. However, this should not cause
confusion.

The baryon flavour asymmetries can always be taken to be equal:
\beq\label{rules}
Y_{ \Delta B_3} = Y_{ \Delta B_2} = Y_{\Delta B_1},
\eeq
because they are conserved before flavour-changing quark interactions
come into equilibrium, and we assume them to be equal as an initial
condition. (Once flavour-changing quark interactions enter equilibrium,
these asymmetries are driven to be equal.) The conservation rules
(\ref{rules}) impose two conditions on the sixteen chemical potentials
of the SM fields, and hypercharge neutrality, eqn (\ref{hyper}),
imposes a third.

We work in the approximation that each of the SM interactions is
either negligible, in which case there are additional conserved
quantum numbers, or in chemical equilibrium. Either case yields a
restriction on the chemical potentials. The interaction rate densities
for processes involving SM Yukawa couplings can be obtained from eqn
(\ref{scatapp}), assuming the Yukawa coupling at one vertex, and a
gauge or top Yukawa coupling at the other. Then the interaction rate
for a Yukawa coupling that brings into equilibrium an $SU(2)$-singlet
$r$ can be estimated as $\Gamma_r\simeq\sum_{i,j,n} \gamma(ij\to
rn)/Y_j^{\rm eq}$.

The rate of the charged lepton Yukawa interactions is important for
leptogenesis. A careful study \cite{Cline:1993bd} of the relevant
processes, which takes into account thermal masses, finds that the
Higgs decay gives the fastest rate $\sim 10^{-2}h_\alpha^2 T$. Using
the results obtained in \cite{CDEO,Cline:1993bd} for $\alpha=e$,
we find that the $h_\tau,h_\mu,h_e$ Yukawa interactions are in
equilibrium for $T\lsim10^{12},\ 3\times10^9,\ 10^5$ GeV, respectively.
Here, as an example, we estimate the (subdominant) $h_\tau$-related
scattering rate at zero temperature.  It can be calculated
from interactions such as $q_3+\ell_\tau\to\tau_R +t_R$ and $W+
\ell_\tau \rightarrow \tau_R +H$.  In the massless limit, the
cross-sections for $\tau_R$ production satisfy
\beq
\sigma(\nu_{\tau}\,t_R^a\to\overline{\tau_R} b_L^a) =
\sigma(\nu_{\tau}\, \overline{b_L}^a\to\overline{\tau_R}\overline{t_R}^a) =
\sigma({b_L}^a\, \overline{t_R}^a\to\overline{\nu_{\tau}}\tau_R ) =
\frac{h^2_t h^2_\tau}{16 \pi s},
\eeq
where $a$ is a colour index, and
\beq
\sigma (\nu_{\tau}  \, W^0 \to \tau_R  \phi_d^+) =
\sigma (\nu_\tau  \, W^- \to \tau_R  \phi_d^0) =
 \frac{g^2 h^2_\tau}{16 \pi s} ~~.
\eeq
Including all these processes, summing over color and $SU(2)$ indices,
and using eqn (\ref{scatapp4d}), we obtain
\beq
\Gamma_\tau = \frac{\gamma_\tau}{\frac{1}{2}\, n_{\ell}^{\rm eq}} =
\frac{(9 h_t^2 + 2 g^2) h_\tau^2 T^4/128 \pi^4}{T^3/\pi^2}
\simeq 5\times 10^{-3} h_\tau^2\,T.
\label{app7gam}
\eeq

The temperature ranges in which the various SM interactions are faster
than $H$ are given in table \ref{tab:spectator-results}.

Consider the temperature range $10^{12}$ GeV $>T>10^{9}$ GeV,
with SM particle content. 
The strong and electroweak sphalerons, and the top,
bottom, tau and charm Yukawa interactions are taken to be in
chemical equilibrium. This imposes six conditions, in addition to the
three conditions of eqns (\ref{rules}) and (\ref{hyper}). Moreover,
the asymmetries in $\mu_R$, $e_R$, $u_R-d_R$, and $d_R-s_R$,
are conserved and presumably zero. So the system of equations can be
solved for the three asymmetries
\beq
Y_{\Delta_\alpha} = \frac{1}{3} Y_{\Delta B} -Y_{\Delta L_\alpha}
\label{YDa}
\eeq
which are conserved in the SM, as a function of the
$\mu_{\ell_\alpha}$:
\beq
\left(
\begin{array}{c}
 Y_{\Delta_e} \\  Y_{\Delta_\mu} \\  Y_{\Delta_\tau}
\end{array}
\right) = \frac{T^3}{6s}
\left[
\begin{array}{ccc}
-22/9 & -4/9 & -4/9 \\
 -4/9 &-22/9 & -4/9 \\
-2/9 &-2/9 & - 139/45
\end{array}
\right]
\left(
\begin{array}{c}
\mu_{\ell_e} \\ \mu_{\ell_\mu} \\ \mu_{\ell_\tau}
\end{array}
\right)
\eeq
However, in the range of temperatures that we consider, $\ell_\mu$ and
$\ell_e$ are indistinguishable, so their chemical potentials should be
summed, $\mu_o = \mu_{\ell_e} + \mu_{\ell_\mu}$, giving
\beq
\left( \begin{array}{c}
 Y_{\Delta_e} + Y_{\Delta_\mu} \\
 Y_{\Delta_\tau}
\end{array}
\right) =\frac{T^3}{6s}
\left[
\begin{array}{ccc}
-26/9 & -8/9 \\
-2/9 & -139/45
\end{array}
\right]
\left(
\begin{array}{c}
\mu_{\ell_o} \\ \mu_{\ell_\tau}
\end{array}
\right)
\label{2genAinv}
\eeq

The $\lambda$-Yukawa interactions can generate an asymmetry in the
lepton doublets. However, SM interactions rapidly change the doublet
densities, so it is better to write Boltzmann Equations for the
asymmetries $Y_{\Delta_\alpha}$, which are conserved by all the SM
interactions. The $N$ interactions can add a lepton to the asymmetry,
or remove a doublet lepton $\ell_\alpha$. Since the washout
interactions can only destroy the part of the $B/3 - L_\alpha$
asymmetry stored in $\ell_\alpha$, we need to express the $ Y_{\Delta
  \ell_\alpha} \sim 2\mu_{\ell_\alpha}T^3/(6s)$ asymmetry as a
function of the charges $Y_{\Delta_\beta}$ eqn~(\ref{YDa}). Inverting
eqn (\ref{2genAinv}) gives
\beq\label{A2gen}
\left(\begin{array}{c} Y_{\Delta \ell_o} \\
    Y_{\Delta \ell_\tau}\end{array}\right) =
\left[\begin{array}{ccc}
-417/589 & 120/589 \\
30/589 & -390/589 \end{array}\right]
\left( \begin{array}{c}
 Y_{\Delta_e} + Y_{\Delta_\mu} \\
 Y_{\Delta_\tau}\end{array}\right)
\eeq
and the matrix is called ``the $A$-matrix'' \cite{Barbieri:1999ma}.

At temperatures below $ \sim 10^9$ GeV, the strange and muon Yukawa
interactions are in equilibrium, so the asymmetries in $d_R-s_R$, and in
$\mu_R$ are no longer conserved. There are three distinguishable flavours, and
the $A$-matrix is:
\beq\label{A3gen}
\left(\begin{array}{c}
 Y_{\Delta \ell_e} \\  Y_{\Delta \ell_\mu} \\  Y_{\Delta \ell_\tau}
\end{array}\right)=
\left[\begin{array}{ccc}
-151/179 & 20/179 & 20/179 \\
 25/358 &-344/537 & 14/537 \\
25/358 & 14/537 & - 344/537
\end{array}\right]
\left(\begin{array}{c}
 Y_{\Delta_e} \\  Y_{\Delta_\mu} \\  Y_{\Delta_\tau}
\end{array}\right).
\eeq

These expression for the $A$-matrix agree with refs.
\cite{Abada:2006ea,Nardi:2006fx}. Ref. \cite{Nardi:2006fx} defines
$Y_{\Delta_\ell}$ per gauge degree of freedom, that is 1/2 of what we
use here. Therefore the $A$-matrices they quote differ by a factor of
2. Our expressions differ from ref. \cite{Abada:2006fw}, where the $u$
and $d$ Yukawa interactions are taken in equilibrium, and from ref.
\cite{Barbieri:1999ma}, where the strong sphalerons are not included.

The coefficient relating $Y_{\Delta B}$ to the $Y_{\Delta_\alpha}$
should be calculated at the temperature when the sphalerons go out of
equilibrium. The reason is that, after the period of leptogenesis, the
$\Delta_\alpha$'s are conserved. Baryon number $B$ is, however, not
conserved in the presence of the EW sphalerons (see eqn
\ref{eq:Lsphal}), and the relation of $Y_{\Delta B}$ to the
$Y_{\Delta_\alpha}$ depends on which other interactions are in
equilibrium. Using the chemical equilibrium and charge conservation
equations, one obtains the following relation, depending on whether
the sphalerons go out of equilibrium above or below the electroweak
phase transition \cite{Harvey:1990qw}:
\beq
\label{eq:app6C}
Y_{\Delta B}=\sum_\alpha  Y_{\Delta_\alpha}\times \left\{
  \begin{array}{cc}
\frac{24+4m}{66+ 13m} & T > T_{\rm EWPT},\\ [6pt]
\frac{32+4m}{98 + 13m}& T < T_{\rm EWPT}, \end{array}\right.
\eeq
where $m$ is the number of Higgs doublets. For the Standard Model
(where $m=1$), one obtains \cite{Khlebnikov:1988sr,Harvey:1990qw}
\beq
\label{eq:app6CSM}
Y_{\Delta B}^{\rm SM}=\sum_\alpha Y_{\Delta_\alpha}\times \left\{
 \begin{array}{cc}
\frac{28}{79} & T > T_{\rm EWPT},\\  [4pt]
\frac{12}{37} & T < T_{\rm EWPT}.
\end{array}\right.
\eeq
In ref. \cite{Laine:1999wv}, the relation is given for any value
of the Higgs VEV, valid through the EWPT.
In the minimal supersymmetric Standard Model (where $m=2$), one
obtains
\beq\label{eq:app6CMSSM}
Y_{\Delta B}^{\rm MSSM}=\sum_\alpha  Y_{\Delta_\alpha}\times \left\{
  \begin{array}{cc}
\frac{8}{23} & T > T_{\rm EWPT},\\  [4pt]
\frac{10}{31}& T < T_{\rm EWPT}. \end{array}\right.
\eeq

\newpage
\section{Appendix: Evolution of  Flavoured Number Operators}
\label{apenrho}
The aim of this appendix is twofold. First, to understand which flavour
basis is appropriate for the Boltzmann Equations. Second, we
investigate whether the physics that defines the preferred flavour
basis introduces additional significant flavour-related effects
on the baryon asymmetry.

To address the first question, one would like to have a formalism
which is covariant under flavour transformations, where it is possible
to ``derive'' the Boltzmann Equations. One possible approach is to
study Schwinger-Dyson equations for the non-equilibrium Green's
functions of the particles involved in leptogenesis.  The resulting
Kadanoff-Baym \cite{KadanoffBaym} equations for $N_1$ two-point functions
were first discussed in \cite{Buchmuller:2000nd}. Since the $N_1$
population must be out of equilibrium for leptogenesis,
the aim of this paper was to study quantum non-equilibrium
effects. The Boltzmann Equations were found to be
a good approximation  for non-relativistic $N_1$.\footnote{See however,
\cite{Lindner:2007am} for numerical comparisons
of the Kadanoff-Baym and Boltzmann Equations in a 
related model.}  
Non-equilibrium  Schwinger-Dyson equations were also
used in \cite{DeSimone:2007rw,DeSimone:2007pa} to show the importance
of time-dependent CP violation in resonant leptogenesis.
An elegant approach to flavour effects in leptogenesis
would  use the finite temperature Schwinger-Dyson equations for
flavoured fermionic propagators, with a flavour off-diagonal
$\epsilon_{\alpha \beta}$ as a source for flavour off-diagonal
propagators. An alternative approach, which should
in principle be equivalent, is to study
the  equations of motion, in field theory, for a flavoured number
operator (following, for instance, chapter 17 of
\cite{Bjorken:1979dk} . For a pedagogical introduction,
see {\it e.g.} \cite{Cardall:2007zw}.) In this framework, however, neither the
notation nor the formalism are transparent (similar to perturbative
calculations without Feynman diagrams and rules).
 Here, we simplify this formalism and study a toy model of Simple Harmonic
Oscillators (SHOs), then guess an extrapolation to field theory.
 We show that, in this toy model, fast interactions choose a basis where
the equations of motion have the form of Boltzmann Equations. Thus,
the flavour basis is determined by which interactions are fast at the
time of leptogenesis.

The Boltzmann equations for number densities treat interactions as a
series of quantum processes of a classical particle. They neglect
quantum effects, such as oscillations. In field theory, effects of
quantum mechanical evolution appear in the equations of motion
for  the (flavour-dependent) number operator
\bea
\label{fdelta}
\hat{f}_\Delta^{\alpha \beta} (\vec{p})
=a_+^{\alpha \dagger}(\vec{p})  a^\beta_+ (\vec{p})-
 a_-^{\beta \dagger}(\vec{p})  a^\alpha_-(\vec{p}),
\eea
which counts  the asymmetry in lepton doublets. (We denote the
operator $\hat{f}$ with a hat, to distinguish it from its expectation
value.) The operator $a_+^{\alpha\dagger}$ ($a_-^{\alpha \dagger}$)
creates lepton doublets (anti-lepton doublets) of flavour
$\alpha$. Notice the inverted flavour order between the particle and
antiparticle number operators \cite{Sigl:1992fn}. The operator is a
matrix in flavour space, analogous to the density matrix of quantum
mechanics, and is sometimes called a density matrix. The diagonal
elements are the flavour asymmetries stored in the lepton doublets.
The trace, which is flavour-basis-independent, is the total lepton
asymmetry. The off-diagonals elements encode (quantum) correlations
between the different flavour asymmetries.

Variants on such an operator have been studied in  the context of
neutrino oscillations in the early Universe
\cite{Sigl:1992fn,Stodolsky:1986dx,McKellar:1992ja,Raffelt:1992uj},
and, in particular, the generation of a lepton asymmetry by
active-sterile oscillations
\cite{Enqvist:1990ad,Foot:1996qc,Abazajian:2004aj,Asaka:2005pn,Asaka:2006rw}. 
The first discussion of
such an operator in the context of thermal leptogenesis was made in
ref. \cite{Barbieri:1999ma}.

For simplicity, we replace $\hat{f}_\Delta^{\alpha \beta}(\vec{p})$
by the number operator of "flavoured" harmonic oscillators. Our aim is
to separate the (possibly quantum) flavour effects from the (assumed
classical) particle dynamics. That is, we extract the flavour structure
from a toy model of harmonic oscillators, and input the particle
dynamics and the Universe expansion, by analogy with the Boltzmann
equations. The  system contains harmonic oscillators of  two
``flavours'', coupled to match the interactions of the Lagrangian in
eqn.~(\ref{L}). By taking expectation values of the number operator in
a ``thermal'' state,  this model can be reduced to a two-state quantum
system. We then extrapolate the flavour structure of these equations of
motion to obtain ``flavoured'' equations of motion for the lepton
asymmetry number operator in the early Universe.

\subsection{A toy model}
Consider a system of two simple harmonic oscillators with ``flavour''
labels 3 and 2, which have different frequencies in the free
Hamiltonian: $H_0=(\omega_2 a^\dagger_2 a_2+\omega_3 a^\dagger_3
a_3)I$. Usually, the number operator is introduced as $a^\dagger_2
a_2+a^\dagger_3 a_3$. However, since we are interested in changing
the basis of the SHOs, we introduce the number operator as a matrix
$\hat{f}$:
\beq
\hat{f}  = \left[  \begin{array}{cc}
a^\dagger_2 a_2  &  a^\dagger_2 a_3  \\
 a^\dagger_3 a_2  & a^\dagger_3 a_3
\end{array} \right],
\eeq
whose trace is the usual number operator. The commutation
relations\footnote{In studying flavour, we consider ``bosonic''
leptons. We anticipate no difficulties in  generalizing to fermionic
leptons.} are $[a_2,a_2^\dagger]=1$, $[a_3,a_3^\dagger]=1$. The number
operator evolves according to the Hamiltonian equations of motion:
$\frac{d\hat{f}}{dt}=+i[H,\hat{f}]$. Since our Hamiltonian conserves
particle number, we can take  expectation values -- for instance, in a
thermal bath -- and reduce our toy model to a two-state system
described by the amplitude of the particles to be $3$'s, or $2$'s. The
density matrix of this two-state system,  is
$$f\equiv\left[\begin{array}{cc}
 \langle a^\dagger_2 a_2 \rangle & \langle a^\dagger_2 a_3  \rangle\\
 \langle a^\dagger_3 a_2 \rangle & \langle a^\dagger_3 a_3  \rangle
\end{array} \right]$$
satisfying
\bea\label{dry}
\frac{df}{dt}&=&-i\left[\begin{array}{ccc}
 0 & (\omega_2-\omega_3)f^{23} \\
 (\omega_3-\omega_2)f^{32} & 0
\end{array} \right].
\eea
If at $t=0$ the system is created in some state, then the probability
to be found in this same state at a later time $t$ is
${\rm Tr}\{f(0)f(t)\}$, which can oscillate.

It is useful to make an analogy to the use of this formalism in the
context of neutrino oscillations \cite{Raffelt:1996wa}. We take the
expectation value of the neutrino number operator in a beam of
neutrinos of momentum $\vec{p}$. This gives a density matrix $f$
representing the flavour of the beam. For two neutrino species, this
is a two-state quantum system. For an initial state
$|\nu(t=0)\rangle=|\nu_\mu\rangle=c_{23}|\nu_2\rangle+s_{23}|\nu_3\rangle$,
(so that $\hat f(t=0)=|\nu_\mu \rangle \langle  \nu_\mu |$),
the density matrix at $t=0$ is given by
\beq\label{initnumu}
f(t=0)=\left[  \begin{array}{cc}
s^2_{23}& c_{23}s_{23} \\ c_{23}s_{23} & c^2_{23}\end{array}\right].
\eeq
The solution to eqn.~(\ref{dry}) with the initial condition
(\ref{initnumu}) is
\beq\label{app7Delta}
f(t) =    \left[  \begin{array}{cc}
 s^2_{23}   &   c_{23}s_{23}  e^{-i \Delta t}\\
  c_{23}s_{23} e^{+i \Delta t}  &   c^2_{23}
\end{array} \right]  ~ ~ ~, ~~ ~~\Delta  = \frac{m_3^2 - m_2^2}{2E}.
\eeq
We obtain the $\nu_\mu$ survival probability:
\beq
P_{\mu \rightarrow \mu}(t) = {\rm Tr} \{f  ( 0)f  (t ) \}
= 1 - \sin^2 2 \theta_{23} \sin^2 \Delta  t
\eeq

Returning to lepton flavour in the early Universe,  consider doublets
$\ell_\alpha$ of momentum $\vec{p}$, in a temperature range where the
rate associated to the $\tau$ Yukawa coupling is fast
($\Gamma_\tau\gsim H$), while the muon and electron Yukawa
interactions can be neglected. The $\ell_\alpha$'s have thermal masses
\cite{Weldon:1982bn,Davidson:1994gn}:
\beq\label{mtherm}
m_\tau^2(T) \simeq \left( \frac{3 g^2 }{32}+ \frac{g'^{2} }{32} +
 \frac{h_\tau^2 }{16} \right)T^2 ~~,~~~~~~
m_{\beta\neq\tau}^2(T)\simeq\left(\frac{3
    g^2}{32}+\frac{g'^{2}}{32}\right)T^2~.
\eeq
Notice the difference with respect to neutrino oscillations in the
lab: In the early Universe, the mass basis is the flavour basis. The
contribution of the neutrino Yukawa coupling to the thermal mass has
been neglected in eqn.~(\ref{mtherm}), because we assume that
$h_\tau\gg|\lambda_{\alpha 1}|$. For $h_\tau<|\lambda_{\alpha1}|$, an
additional term $\propto \lambda_{\alpha 1} \lambda^*_{\beta 1} f_{N_1}$
should be included. This case is discussed in section \ref{lambda}.

The two harmonic oscillators in the early Universe can be labeled
$\tau$ and $o$, where $o$ is the projection onto $e$ and $\mu$ of the
direction into which $N_1$ decays:
$\vec{o}\propto\lambda_{e1}\hat{e}+\lambda_{\mu 1}\hat{\mu}$.
The eigenvalues of the free Hamiltonian are the particle energies
$\omega(\vec{p})$, and the time evolution of the number operator is
determined by the energy differences:
\beq\label{thdiff}
\omega^2_\tau - \omega^2_o =  \frac{h_\tau^2 T^2}{16}.
\eeq

By analogy with the neutrino oscillation example, we anticipate that
``flavour oscillations'' might affect the lepton asymmetry. Imagine
the excitations of the oscillators to be the lepton {\it asymmetry},
carried by particles of energies $\omega_\tau \sim\omega_o\sim E$.
Then $\omega_\tau  - \omega_o \simeq  h_\tau^2 T^2/(32E)$. We treat
the production and washout of the lepton asymmetry as, respectively,
initial condition and subsequent measurement on the system. In other
words, we consider the production of an asymmetry in some linear
combination of $o$ and $\tau$, we allow it to evolve, and later turn
on the inverse decays. Then the inverse decay rate can be suppressed
or enhanced because the asymmetry changed flavour during time evolution
(analogous to a survival probability in neutrino oscillations)
\cite{Abada:2006fw}. Notice that the oscillation timescale is of order
$(h_\tau^2 T)^{-1}$, which is parametrically the same as the timescale
of the $h_\tau$-mediated scattering rates that cause decoherence.
This is different from, {\it e.g.} the MSW effect in matter, where the
rate for decohering scattering can be neglected on the oscillation
timescale.

The flavour-blind gauge contribution is neglected in 
eqn.~(\ref{thdiff}). This is subtle, because the early Universe version 
of $\Delta$ (see eqn.~(\ref{app7Delta})) depends on the energy of the
lepton, and within the oscillation timescale, a lepton participates in
many energy-changing gauge interactions. Ref. \cite{Abada:2006fw} used
the thermally averaged energy $\langle E\rangle\simeq 3T$ to estimate
$\Delta$ \cite{Enqvist:1990ad}. In path integral language, this means
approximating the integral $(i\int Ed\tau)$ along the path from one
lepton-number violating interaction to the next, to be $i\langle
E\rangle\int d\tau$. This is in agreement with the analytic and
numerical analysis of \cite{Bell:2000kq}, which indicates that fast
gauge interactions do not affect the coherence of flavour
oscillations: if the timescale for transitions between different
energy levels is much shorter than the oscillation timescale, then a
particle spends time at many different energies during an oscillation
timescale. The probability to have an energy $E$ is proportional to
the distribution function $f$ of eqn.~(\ref{nFB1}). Therefore, on the
oscillation timescale, the particles all have the thermal average
energy, and oscillate coherently.  In any case, this does not affect
the principal claim of this appendix, that charged lepton Yukawa
interactions choose the flavour basis for the Boltzmann Equations.

We now include interactions among the SHOs. This allows one to take
into account the production and annihilation of particles, and can
lead to decoherence. Recall that in ordinary neutrino oscillations,
decoherence happens anyway after a few oscillation lengths, due to
``spreading of the wave packet'' (see {\it e.g.}
\cite{Akhmedov:2007fk}). Here we are looking for decohering
interactions, which, in the quantum mechanical analogy, collapse the
wavefunctions onto an eigenbasis. We will see that in this basis, the
equations of motion look like Boltzmann Equations.

The decays and inverse decays due to the Yukawa interactions can be
included by perturbing in an interaction Hamiltonian, $H_I$.  We first
consider the production and washout of leptons, due to $\lambda$, in a
model of four harmonic oscillators: $N,\phi,$ and two leptons flavours
$\{\ell_\alpha\}$. Although we use the index $\alpha$, this discussion
is covariant in flavour space, so any basis choice for doublet flavour
space is valid.  We are looking for behaviors analogous to the lepton
number production and washout in the decay/inverse decay $N
\leftrightarrow \phi \ell_\alpha$. We take $\phi$ and $\ell_\alpha$ to
be massless and $N$ massive, and ignore the free Hamiltonian
(considered previously) which does not change the particle
numbers. The interaction Hamiltonian  is
\beq\label{HIY}
H_{I}=\lambda_{\alpha 1} a_N^\dagger a_\phi
a_{\ell_\alpha}+\lambda^*_{\alpha 1} a_N a_\phi^\dagger
a_{\ell_\alpha}^{\dagger}.
\eeq
We perturbatively expand the Heisenberg equations of motion
and obtain
\bea
\frac{\partial }{\partial t} f_\ell^{\alpha \beta}
&=& -  [ H_{I}, [H_{I} ,f_\ell^{\alpha \beta} ]] =-
\lambda^*_{\rho 1} \lambda_{\alpha 1}
[a_N a_\phi^\dagger a_{\ell_\rho}^{ \dagger}, a_N^\dagger
a_\phi   a_{\ell_\beta}]+\lambda_{\rho 1} \lambda^*_{\beta 1}
[a_N^\dagger  a_\phi a_{\ell_\rho}  ,
a_N   a_\phi^\dagger  a_{\ell_\alpha}^{ \dagger}]  \nonumber\\
&=&-\lambda_{\alpha 1}\lambda^*_{\rho 1}(f_\phi f_\ell^{\rho\beta} -
f_N f_\phi\delta ^{\rho\beta}-f_N f_\ell^{\beta\rho}-f_N\delta^{\beta\rho})
+ (  f_N \delta ^{\alpha \rho}+ f_N f_\phi \delta ^{\alpha \rho}
+ f_N f_\ell^{\alpha \rho}- f_\phi f_\ell^{\alpha \rho} )
\lambda_{\rho 1} \lambda^*_{ \beta 1} ~
\nonumber \\
& =  &  -\lambda_{\alpha 1}  \lambda^*_{\rho 1}
\Big[ f_\phi f_\ell^{\rho  \beta} (1 + f_N) -
  f_N (1+f_\phi)( \delta ^{ \rho \beta } +  f_\ell^{\rho  \beta})  \Big]
\nonumber \\
&&+\Big[f_N(1+f_\phi)(\delta^{\alpha\rho}+f_\ell^{\alpha \rho})
-f_\phi f_\ell^{\alpha \rho}(1+f_N)\Big]\lambda_{\rho 1}\lambda^*_
{\beta 1}~.
\label{longdamp}
\eea
This has the form of inverse decays and decays of $N$, with Bose
enhancement factors (recall that the toy model leptons are bosons) for
the final states. For simplicity, we now drop the Bose enhancement
factors. A similar calculation can be performed for the tau Yukawa
coupling $\vec{h}_\tau$ ($|\vec{h}_\tau|=h_\tau$). The interaction
Hamiltonian is
\beq
H_{I,h}=h^\rho_\tau a_\phi^\dagger a_{\tau^c}a_{\ell_\rho}+
h^{*\rho}_\tau a_\phi a_{\tau^c}^\dagger a_{\ell_\rho}^\dagger.
\eeq
We assume that the processes $\phi\leftrightarrow\ell_\tau\tau^c$ are
allowed, because the thermal mass of Higgs gets contributions from the
top Yukawa interaction. (To avoid appealing to thermal masses,
one could go to higher orders in $H_I$, and consider scattering such
as $W\ell\to\tau^c\phi$.) We neglect Bose enhancement factors and take
expectation values. We obtain, in the charged lepton mass eigenstate
basis,
\bea
\frac{\partial }{\partial t} \left[\begin{array}{cc}
f_\ell^{oo} & f_\ell^{o \tau} \\
f_\ell^{\tau o} &  f_\ell^{\tau \tau}\end{array} \right]
& =&   -  f_{\tau^c} \left[\begin{array}{cc}
0 & 0 \\ 0 &  |h_\tau|^2\end{array} \right]
\left[\begin{array}{cc} f_\ell^{oo} & f_\ell^{o \tau} \\
f_\ell^{\tau o} &  f_\ell^{\tau \tau}\end{array} \right]
 - f_{\tau^c}\left[\begin{array}{cc}
f_\ell^{oo} & f_\ell^{o \tau} \\
f_\ell^{\tau o} &  f_\ell^{\tau \tau}\end{array} \right]
\left[\begin{array}{cc}0 & 0 \\ 0 &  |h_\tau|^2
\end{array} \right] \nonumber \\ & &
 + 2 f_\phi  \left[\begin{array}{cc}
0 & 0 \\ 0 &  |h_\tau|^2\end{array} \right].
\label{decohere}
\eea
It is reassuring that the $t$-independent equilibrium solution
of this equation imposes $f_{\tau^c}f_\ell^{\tau \tau}-f_\phi=0$.
Combined with eqn.~(\ref{dry}), this gives damped oscillations for the
off-diagonal elements:
\beq
 \partial f^{o \tau}_\ell/\partial t = -i ( \omega_o - \omega_\tau -
i |h_\tau|^2f_{\tau^c} ) f^{o \tau}_\ell,
\label{app7decoffd}
\eeq
We conclude that exchanging doublet with singlet leptons destroys the
coherence between different doublet flavours. In the ``flavour'' basis,
the off-diagonal elements of the $\ell_\alpha$ number operator vanish
when the charged lepton Yukawa interactions are fast. The charged
lepton Yukawa interactions are therefore the desired
``basis-choosing'' physics, that collapses the equations of motion of
the density matrix to Boltzmann Equations for the number densities on
the diagonal.

It is convenient to write these equations in the notation of
projectors, introduced in eqn.~(\ref{eq:Pab}). In an arbitrary basis,
we define, at tree level, the projectors onto the decay
direction of $N_1$, and onto the $\tau$ flavour:
\beq
\overline{P}^{ab}=P^{ab}=\frac{\lambda_{a1}\lambda^*_{b1}}
{\sum_d|\lambda_{d1}|^2}, ~~~~~~ P^{ab}_\tau =
 \frac{h_\tau^a h^{* b}_\tau}{\sum_d|h^d_\tau|^2}.
\eeq
When loop contributions are included in the decays of $N_1$, a CP
asymmetry can arise. It can be represented by a difference between $P$
and $\overline{P}$, where $\overline{P}$ is the projector onto the
anti-lepton direction into which $N_1$ decays.

Equations for the anti-lepton density can be derived in a similar
way. For simplicity, we take $\phi$ and $N$ to be their own
anti-particles, so there are now eight coupled oscillators:
$N,\phi,\ell^{\alpha},\bar{\ell}^\alpha,\tau^c$ and $\bar{\tau}^c$.
The expectation value of the  difference between the two equations,
again in the charged lepton mass eigenstate basis, is
\bea\label{damp}
\frac{\partial }{\partial t} f_{\Delta }^{\alpha \beta}
&=&-|\lambda_1|^2 P^{\alpha \rho}f_\phi f_{\Delta }^{\rho \beta}
- |\lambda_1|^2 f_\phi f_{\Delta }^{ \alpha \rho} P^{ \rho \beta}
+ 2 (f_N - f_N^{\rm eq}) \epsilon^{\alpha\beta}\nonumber \\
&& - i (\omega_\alpha - \omega_\beta) f_{\Delta }^{\alpha\beta}
 - \frac{|h_\tau|^2}{2} (f_{\tau^c}
+ f_{\bar{\tau}^c}) ( f_{\Delta }^{ \alpha \tau}  \delta_{\tau \beta}
+ \delta_{\tau \alpha}   f_{\Delta }^{\tau \beta} )\nonumber \\
&&- \frac{|h_\tau|^2}{2}(f_{\tau^c}-f_{\bar{\tau}^c})
[(f_{\ell }+f_{\bar{\ell} }^T)^{\alpha\tau}\delta_{\tau \beta}
+\delta_{\tau\alpha}(f_{\ell}+f_{\bar{\ell}}^T)^{\tau\beta}],
\eea
where $\rho,\alpha,\beta=o,\tau$. The purpose of this equation is to
motivate the flavour index structure we use in the Boltzmann
equations. We now briefly discuss the various terms.

${\bf\bullet}$ The first two terms describe washout by inverse decays.
By construction, there is no washout by $\Delta L = 2$ scattering
(we will introduce the resonant part by hand). Notice that the usual
partial decay rates $\Gamma(N\to\phi\ell_\alpha)$ become the diagonal
elements of a matrix. The total decay rate, which is flavour-basis
invariant, is the trace of this matrix.

${\bf\bullet}$ The third  term, which is $\propto \epsilon$, has been
obtained artificially. It is well-known  in field theory that no
asymmetry is created at second order in $\lambda$ (see eqn.~\ref{CPloop}). 
A CP asymmetry is proportional to the imaginary part of
the loop amplitude times the imaginary part of the product of tree and
loop coupling constants (see eqn.~\ref{hateps}). We include the loop as
an effective (non-unitary) interaction in the Hamiltonian:
$(\delta \lambda)_\alpha\Omega~a_N^\dagger a_\phi a_{\ell_\alpha}+
(\delta\lambda)^*_\alpha\Omega~a_N a_\phi^\dagger
a_{\ell_\alpha}^\dagger$. Here $\Omega$ encodes the imaginary part of
the loop amplitude; it is flavour independent, of little direct
interest here, and can be obtained by matching to the calculation of
$\epsilon_{\alpha \alpha}$ in section \ref{ssCP}. Just as the decay
rate $\Gamma$ becomes a matrix in flavour space, the CP asymmetry
$\epsilon$ also becomes a matrix in flavour space, proportional to
$[(\delta \lambda)_\alpha\lambda^*_\beta-
\lambda_\alpha(\delta\lambda)^*_\beta]
\propto P^{\alpha \beta} -\overline{P}^{\alpha \beta}$.
This gives
\beq\label{guess1}
\epsilon_{  \alpha \beta} \propto {\rm Im} \{  \lambda_{\alpha}
(\delta \lambda)^*_{\beta} -
 (\delta \lambda)_{\alpha} \lambda_{\beta}^* \}
=\frac{1}{v_u^2}{\rm Im}\{\lambda_{\alpha}m^*_{ \beta\rho}\lambda_\rho
-m_{\alpha \rho}\lambda^{* \rho}\lambda^*_\beta\}.
\eeq
Using the equilibrium condition,
$$2f_N^{\rm eq} [\epsilon]=
\frac{f_\phi}{2}([\epsilon][f_\ell+f_{\bar\ell}^T]
+[f_\ell+f_{\bar\ell}^T ][\epsilon])$$
(where square brackets are matrices in flavour space), we obtain a
production term $2(f_N+f_N^{\rm eq})\epsilon^{\alpha\beta}$. This
cannot be complete, because it gives an asymmetry in thermal
equilibrium. The resonant part of $\Delta L=2$ scattering should be
subtracted \cite{Kolb:1979qa}, which ensures that Tr $f_\Delta$
vanishes in thermal equilibrium. Working in the charged lepton mass
basis, it is straightforward to obtain  eqn.~(\ref{damp}) by following
section \ref{simplestBE}, or ref. \cite{Kolb:1979qa}.

${\bf\bullet}$ The remaining terms of eqn.~(\ref{damp}) describe the
effects of the tau Yukawa coupling. The asymmetry oscillates in flavour
space, as described at the beginning of the section, because the
off-diagonal flavour matrix elements have time-dependent phases. We
drop this oscillatory term from the equation of motion, when
extrapolating to the early Universe. (See \cite{Abada:2006fw} for a
discussion including these oscillations).

The tau Yukawa coupling also causes the off-diagonal elements to decay
away, via the last two terms, This happens separately for the lepton
and anti-lepton densities, as shown in eqn.~(\ref{app7decoffd}).
However, the equation is not so simple to solve for the asymmetry,
because the last term depends also on the asymmetry in the
$SU(2)$-singlet $\tau^c$s. The coupled equations for the singlet and
doublet asymmetries should be solved. Instead, we attempt to include
the singlet asymmetry via the $A$-matrix, as is done in the Boltzmann
Equations, and argue in the following that the singlet asymmetry has
little effect on the decay of the flavour-off-diagonal terms.

The last two terms of  eqn.~(\ref{damp}) drive the diagonal asymmetry
$Y_{\Delta \ell_\tau}$ to the correct relation with $Y_{\Delta_\tau}$,
when the tau Yukawa coupling is in equilibrium. (In our toy model,
where $\phi$ is a real scalar, this means
$(f_{\tau^c}-f_{\bar{\tau}^c})=f_{\Delta }^{\tau\tau}$.
In the SM, there is a Higgs asymmetry
\cite{Harvey:1990qw,Khlebnikov:1988sr,Barbieri:1999ma}.)
This can be clarified by studying the evolution of
\beq\label{app7F}
F^{\alpha\beta}=f_\Delta^{\alpha\beta}
-P_\tau^{\alpha\beta}f_{\Delta\tau^c},
\eeq
where $P_\tau$ is the projector onto the $\tau$ direction in some
arbitrary basis for lepton doublets, and
$f_{\Delta\tau^c}=f_{\tau^c}-f_{\overline{\tau}^c}$. The trace of $F$
is the total lepton number stored in $o_L$, $\tau_L$ and $\tau_R$, so
studying $F$ in the toy model is analogous to replacing  the asymmetry
in  lepton doublets with the asymmetry in $B/3-L_\alpha$
\cite{Barbieri:1999ma} in  usual leptogenesis calculations.
For the remainder of this subsection, we focus on the decohering
effects of the $\tau$ Yukawa. This means we neglect the oscillation
term and the neutrino Yukawa $\lambda$. The equation for the doublet
asymmetry is
\bea
\frac{\partial }{\partial t} [f_\Delta] &=&-
|h_\tau|^2 {\Big \{ } [ P_\tau], [f_\Delta]  {\Big \} }
\frac{(f_{\tau^c} + f_{\overline{\tau}^c} )}{2}
-|h_\tau|^2 {\Big \{ } [ P_\tau], [f_\ell + {f}_{\overline{\ell}}^T]
{\Big \} }  \frac{(f_{\tau^c} - f_{\overline{\tau}^c} )}{2} ~,
\eea
where  square brackets denote
matrices in doublet space. The equation for the singlet asymmetry is
\bea
\frac{\partial }{\partial t}f_{\Delta \tau^c}
&=&-|h_\tau|^2 (f_{\tau^c}
+f_{\bar{\tau}^c})  f_{\Delta }^{ \tau \tau}
-|h_\tau|^2 f_{\Delta\tau^c}(f_{\ell}+f_{\bar{\ell}}^T) ^{\tau\tau},
\eea
where
$f_{\Delta}^{\tau\tau}=\hat{h}^\dagger\cdot[f_\Delta]\cdot\hat{h}$.
The equation for the lepton asymmetry $F^{\alpha \beta}$  is
\bea\label{50}
\frac{\partial }{\partial t}[F]
&=&-|h_\tau|^2 \left( {\Big \{ }[P_\tau],[f_\Delta]{\Big\}}-
2 [ P_\tau] {\rm Tr}[ P_\tau f_\Delta] \right)
\frac{(f_{\tau^c} + f_{\overline{\tau}^c} )}{2}\nonumber \\
&&-|h_\tau|^2\left({\Big\{ }[ P_\tau],[f_\ell+{f}_{\overline{\ell}}^T]
{\Big \} }-2[P_\tau]{\rm Tr}[ P_\tau(f_{\ell}+f_{\bar{\ell}
}^T)]\right)\frac{(f_{\tau^c} - f_{\overline{\tau}^c} )}{2} ~,
\eea
where the second line is of the same order as the first (see
eqn.~\ref{damp}). The trace of the right hand side is zero as expected,
since Tr($F$) is conserved in the absence of $\lambda$. In the flavour
basis, eqn.~(\ref{50}) can be rewritten as
\beq\nonumber
\frac{\partial }{\partial t} \left[
\begin{array}{cc}f_\Delta^{oo} & f_\Delta^{o \tau} \\
f_\Delta^{\tau o} &  f_\Delta^{\tau \tau} -  f_{\Delta \tau^c}
\end{array} \right] =
-\frac{ |h_\tau|^2}{2 } \left[\begin{array}{cc}
0 &   f_\Delta^{o \tau} \\
 f_\Delta^{\tau o} &  0   \end{array} \right]
(f_{\tau^c} + f_{\overline{\tau}^c} )
- \frac{ |h_\tau|^2}{2 }  \left[\begin{array}{cc}
0  &  f_\ell^{o \tau}
+  f_{\overline{\ell}}^{ \tau o} \\
 f_\ell^{ \tau o}+  f_{\overline{\ell}}^{ o \tau}
 &   0 \end{array} \right] f_{\Delta \tau^c}.
\eeq
The first term causes $f^{o\tau}_\Delta$ to decay. The second term is
of the same order, and can be of either sign. To see that,
nevertheless, $f^{o\tau}_\Delta$ and $f^{\tau o}_\Delta$ are driven to
zero as $h_\tau$ comes into equilibrium, notice that 
eqn.~(\ref{app7decoffd}) implies that $f_\ell^{\tau o}
+f_{\overline{\ell}}^{o\tau}$ vanishes exponentially fast. So the
problematic second term vanishes as $h_\tau$ comes into equilibrium,
and the equilibrium (time-independent) solution is
$f^{o\tau}_\Delta=0$.

In summary, the Yukawa coupling $h$ causes the flavour off-diagonal
asymmetries to decay away, as it did for the off-diagonal number
densities in eqn.~(\ref{decohere}).  It has no effect on the asymmetry
in either flavour. To obtain eqn.~(\ref{osc}), we simply drop the second
line (so we neglect the asymmetry in $SU(2)$-singlets).

\subsection{Extrapolating to the early Universe}
A matrix equation describing the asymmetry $\Delta Y_\ell$ in the
early Universe can be guessed by matching the flavour structure of
eqn.~(\ref{damp}) to the Boltzmann Equations:
\bea
\frac{d}{dz}
\left[\begin{array}{cc}
 Y_{\Delta}^{oo} & Y_{\Delta}^{o \tau}  \\
Y_{\Delta}^{ \tau o}& Y_{\Delta}^{ \tau \tau}
\end{array}\right]
& = & \frac{z}{s H_1}\left(\gamma_{N\rightarrow 2}
\left( \frac{Y_{N_{1}}}{Y_{N_{1}}^{\rm eq}} -1 \right)
\left[\begin{array}{cc}
\epsilon^{oo} &\epsilon^{o\tau}  \\
\epsilon^{ \tau o}&\epsilon^{\tau \tau}
\end{array}\right]
\right. \nonumber \\
& & - \left. \frac{1}{2 Y_{\ell}^{\rm eq}}\left\{
\left[\begin{array}{cc}
\gamma ^{oo}_{N\rightarrow 2} &\gamma ^{o\tau}_{N\rightarrow 2}  \\
\gamma ^{\tau o}_{N\rightarrow 2}&\gamma ^{\tau \tau }_{N\rightarrow 2}
\end{array}\right],
\left[\begin{array}{cc}
 Y_{\Delta \ell}^{oo} & Y_{\Delta \ell}^{o \tau}  \\
 Y_{\Delta \ell}^{ \tau o}&  Y_{\Delta \ell}^{ \tau \tau}
\end{array}\right]
 \right\}\right.\nonumber \\
&& \left.- \,\frac{1}{2 Y_{\ell}^{\rm eq}}\left\{[\gamma_\tau],
\left[\begin{array}{cc}
 Y_{\Delta}^{oo} & Y_{\Delta}^{o \tau} \\
Y_{\Delta}^{ \tau o}& Y_{\Delta}^{ \tau \tau}
\end{array}\right]
\right\}+\frac{{\rm Tr} [\gamma_\tau Y_{\Delta}]}{ Y_{\ell}^{\rm eq}}
\delta_{\tau\tau} \right)
\label{osc}
\eea
where $\left\{\cdot,\cdot\right\}$ stands for anti-commutator. We now
discuss each of the three terms on the right hand side of eqn.~(\ref{osc}) 
in turn:
\begin{enumerate}
\item The first term describes the creation of the asymmetry. The CP
asymmetry in the $\alpha$-flavour is $\epsilon^{\alpha\alpha}$, and
$\epsilon=\sum_\alpha\epsilon^{\alpha\alpha}$. Notice that
$\epsilon^{\alpha \beta}$ is normalized by the total decay rate (so
that $[\epsilon]$ transforms as a tensor under $\ell$ basis
rotations). The matrix $[\epsilon]$  is now defined as
\bea\label{asym}
\epsilon_{\alpha\beta}&=&\frac{1}{(16\pi)}
\frac{1}{[\lambda\lambda^{\dagger}]_{11}}
\sum_J {\rm Im}\left\{(\lambda_{1 \alpha})
[\lambda\lambda^{\dagger}]_{1j}\lambda^*_{j\beta}
-(\lambda^*_{1\beta})[\lambda^*\lambda^{T}]_{1j}
\lambda_{j \alpha} \right\}
g\left(\frac{M_{j}^2}{M_{1}^{2}}\right)
\nonumber \\
&\simeq&\frac{3}{(16\pi v^2)}
\frac{M_1}{[\lambda\lambda^{\dagger}]_{11}}
{\rm Im}\{\lambda_{\alpha}\lambda_\sigma[m_\nu^*]_{\sigma\beta}
-\lambda^*_\sigma[m_\nu]_{\sigma \alpha}\lambda^*_\beta\}\,\,
~~~~~~( {\rm for}~M_1 \ll M_{2,3}),
\eea
where $g(x)$ is the loop function of eqn.~(\ref{gSM}).
\item
The second term is the washout of the asymmetry by decays and inverse
decays. The matrix $[\gamma_{N\to 2}]$ is defined as
\beq
\gamma^{\alpha \beta }_{N\rightarrow 2}  =
 \gamma_{N\rightarrow 2} P^{\alpha \beta}
= \gamma_{N\rightarrow 2} \frac{\lambda_{1\alpha}
\lambda_{1 \beta}^*}{\sum_\rho|\lambda_{1 \rho}|^2} , ~~~
\label{gij}
\eeq
where $\gamma_{N\to2}=\sum_\rho\gamma^{\rho\rho}_{N\to2}$ is the total
(thermally averaged) decay rate. For $\alpha=\beta$,
$\gamma^{\alpha\alpha}_{N\rightarrow 2}$ is the decay rate of $N_1$ to
the  $\alpha$-flavour. This term is written as a function of the
asymmetry in doublets, $\Delta Y_{\ell}$, rather than the asymmetry in
$B-L$: $Y_{\Delta}-\hat{h}\hat{h}^\dagger Y_{\Delta \tau^c}$. One way
to proceed is to solve simultaneously this equation and the equation
for $Y_{\Delta\tau^c}$. Alternatively, one can include the singlet
asymmetry in an approximate way via an $A$-matrix, as one does in the
Boltzmann equations. Since the $h_\tau$-interaction is entering
equilibrium, the $A$-matrix is time-dependent:
\beq
A^{\alpha\beta}(t)=A^{\alpha\beta}_{T>}e^{-\Gamma_\tau t}+
A^{\alpha \beta}_{T<}(1-e^{-\Gamma_\tau t}),
\eeq
where $A^{\alpha \beta}_{T>}$ ($A^{\alpha \beta}_{T<}$) are the matrix
elements before (after) $h_\tau$ comes into equilibrium, and
$\Gamma_\tau=Y_{\ell}^{\rm eq}\gamma(\ell\leftrightarrow \tau_R)$ is
the interaction rate associated to the tau Yukawa coupling, see 
eqn.~(\ref{app7gam}). We use this to obtain eqn.~(\ref{compact}).

\item The last term describes the decay of the quantum correlations
between $o$ and $\tau$, that happens as the $h_\tau$ interaction comes
into equilibrium and makes the $\tau$'s distinguishable.  In the
flavour basis, we have
\beq
[\gamma_\tau] =
\left[\begin{array}{cc} 0 & 0 \\ 0 &\gamma_\tau \end{array}\right],
\eeq
where $\gamma_\tau$ is the interaction density for the  $\tau$
Yukawa coupling (see discussion around eqn.~\ref{app7gam}):
\beq
 \gamma_\tau \simeq 5\times 10^{-3} h_\tau^2\,T  n_{\ell}^{\rm eq}.
\eeq
These interactions collapse the density matrix onto its
diagonal elements in the flavour basis.
\end{enumerate}

Following Stodolsky \cite{Stodolsky:1986dx,Raffelt:1996wa}, we write
eqn.~(\ref{osc}) in a more transparent way by expanding the various
matrices on the basis provided by the $\sigma$-matrices,
$\sigma^\mu=(I,\vec{\sigma})$. A hermitian $2\times2$ matrix $Y$ can
be written as
\beq
Y= Y^\mu \sigma_\mu , ~~~~ Y^\mu = \frac{1}{2}
{\rm Tr} \{Y \sigma^\mu\}.
\label{sigmaS}
\eeq
The asymmetry matrix in the flavour basis is then replaced by a
four-vector with components $Y^0_\Delta=(1/2) {\rm Tr}\,[ Y_\Delta]$,
which represents half of the total lepton asymmetry, $Y^z_\Delta=(1/2)
\left(Y_\Delta^{oo}-Y_\Delta^{\tau\tau}\right)$, which is half of the
difference of the asymmetries in the $o$ and $\tau$ flavour densities,
and $Y_\Delta^{x,y}$, which account for the quantum correlations
between the lepton asymmetries.

In this notation, the Boltzmann equation (\ref{osc}) becomes a system
of equations of the form
\begin{eqnarray}
\frac{dY^0_\Delta}{dz}&=&\frac{z}{sH_1}\left(\gamma_{N\to 2}
  \left(\frac{Y_{N_{1}}}{Y_{N_{1}}^{\rm eq}}- 1\right)\epsilon^0
  -\frac{1}{Y_\ell^{\rm eq}}\gamma^0_{N\to 2} [A Y_\Delta]^0
-\frac{1}{Y_\ell^{\rm eq}}\vec{\gamma}_{N\rightarrow 2}
\cdot \stackrel{\longrightarrow}{[AY_\Delta]} \right),\nonumber\\
\frac{d\vec{Y}_\Delta}{dz}&=&\frac{z}{sH_1}\left( \gamma_{N\to 2}
\left( \frac{Y_{N_{1}}}{Y_{N_{1}}^{\rm eq}} -1 \right)
\vec{\epsilon}-\vec{\gamma}_D \frac{ [AY_\Delta]^0}{Y_\ell^{\rm eq}}
-\frac{1}{Y_\ell^{\rm eq}}\gamma^0_{N\to 2}
\stackrel{\longrightarrow}{[AY_\Delta]}
+\hat{\gamma}_\tau\times\vec{\gamma}_\tau\times\vec{Y}_\Delta\right).
\label{compact}
\end{eqnarray}

Had we kept the thermal masses in our equations, we would find that
the components $Y_\Delta^{x}$ and $Y_\Delta^{y}$ precess around the
$z$-direction with an angular velocity set by the thermal mass
\cite{Abada:2006fw}. At the same time, such a precession is damped by
the tau-Yukawa interactions at a rate $\sim\gamma_{\tau}$, as
described by the term
$\hat{\gamma}_\tau\times\vec{\gamma}_\tau\times\vec{Y}_\Delta$.

For $\gamma_\tau\gg\gamma_{N\to2}$, the asymmetry $\vec{Y}_\Delta$ is
projected onto $\vec{\gamma}_\tau$ before the washout interactions
have time to act. So the last cross product term involving $
\vec{\gamma}_\tau$ can be dropped from the equations, provided that
$\vec{Y}_\Delta \parallel\vec{\gamma}_\tau$ is imposed. To summarize,
we find that for $\gamma_\tau>\gamma_{N\to2}$, the baryon asymmetry
can be obtained by solving the Boltzmann Equations in the flavour
basis.

\subsection{Decoherence due to  $\lambda$ }
\label{lambda}
In this section we consider the case where the $N_1$ decay rate is
faster than, or of order of, the rate of $\tau$ Yukawa interactions
\cite{DeSimone:2006,Blanchet:2006ch}. The evolution of the asymmetry
density operator in this transition region was studied in
\cite{DeSimone:2006}, including the oscillations due to the thermal
mass (\ref{mtherm}).

For $\gamma_{N\to2}>\gamma_\tau$, the distinguished direction in the
doublet-lepton flavour space is $\ell_{N_1}$, the direction into which
$N_1$ decays. Therefore the ``single flavour'' leptogenesis calculation
applies, even if $\Gamma_\tau>H$. To see this effect analytically, we
use the orthonormal basis $\{ \hat{\ell}_{N_1}, \hat{\ell}_p \}$ of
the plane spanned by ${\ell}_{N_1}$ and $\hat{\tau}$ ($p$ stands for
perpendicular: $\hat{\ell}_{N_1}\cdot\hat{\ell}_p=0$). In this basis
for the two-dimensional flavour space,  the four-vector components are
$\epsilon^0=\epsilon^z=\epsilon/2$,
$\gamma^0_{N\to2}=\gamma^z_{N\to2}=\gamma_{N\to2}/2$, and
$\gamma^x_{N\to2}=\gamma^y_{N\to2}=0$. The lepton asymmetry matrix
$Y_{\Delta \ell}^{ij}$  can be written as the four-vector
$Y_\ell^{\rm eq}(\Delta y_\ell^0,\Delta y_\ell^x,\Delta y_\ell^y,
\Delta y_\ell^z)$. The equations (\ref{compact}) become
\begin{eqnarray}
\frac{2 sH_1}{\tilde{z}}\frac{dY^0_\Delta}{d\tilde{z}}&=& 
\gamma_{N\rightarrow 2}
\left( y_{N_1} 
- 1\right)\epsilon -\gamma_{N\rightarrow 2}
(\Delta y_{ \ell}^0
 +
 \Delta y^z _\ell ) \label{eq:comp0}\\
\frac{2 sH_1}{\tilde{z}}
\frac{d{Y}^z_{\Delta \ell}}{d\tilde{z}}&=&  \gamma_{N\rightarrow 2}
\left(y_{N_1} 
-1 \right)
\epsilon - \gamma_{N\rightarrow 2}  (\Delta y_\ell^0
+
\Delta y_\ell^z)
+ 2 (\hat{\gamma}_\tau \times  \vec{\gamma}_\tau
\times \Delta \vec{y}_\ell ) \cdot \hat{z} \label{eq:comp1} \\
\frac{sH_1}{\tilde{z}}
\frac{d{Y}^y_\Delta}{d\tilde{z}}&=&  \gamma_{N\rightarrow 2}
\left(y_{N_1} 
-1 \right)
\epsilon^y
-
\frac{\gamma_{N\rightarrow 2}}{2}
\Delta y_\ell^y
+  (\hat{\gamma}_\tau \times  \vec{\gamma}_\tau
\times \Delta\vec{y}_\ell) \cdot \hat{y}
\\
\frac{sH_1}{\tilde{z}}
\frac{d{Y}^x_\Delta}{d\tilde{z}}&=&  \gamma_{N\rightarrow 2}
\left(y_{N_1} 
-1 \right)
\epsilon^x
-
\frac{\gamma_{N\rightarrow 2}}{2}
\Delta y_\ell^x
+  (\hat{\gamma}_\tau \times  \vec{\gamma}_\tau
\times \Delta\vec{y}_\ell) \cdot \hat{x}.
\label{compactBBR}
\end{eqnarray}
where the time variable $z=M/T$ has exceptionally been represented as
$\tilde{z}$, to avoid confusion with the four-vector index.  In the
above equations, the $A$-matrix has been neglected for two reasons.
First, it includes the effects of other interactions, in the
approximation that they are fast compared to $N_1$ decays, while we
are interested in the case where the interactions rates of $N_1$,
$\tau_R$ and the sphalerons are of the same order. Second, it is
simpler to consider the equations for the doublet asymmetry $Y_{\Delta
  \ell}$ rather than the $B/3-L_\alpha$ asymmetries
$Y_{\Delta_\alpha}$. The $A$-matrix is included in
\cite{DeSimone:2006}.

The final baryon asymmetry is proportional to the total asymmetry in
the doublets, which is the trace $Y_{\Delta \ell}^0=Y^{\rm eq}_\ell
\Delta y_\ell^0$. The second  washout term in eqn.~(\ref{eq:comp0})
depletes the asymmetry less effectively for $\Delta y_\ell^0+\Delta
y_\ell^z\simeq0$. When  $\Delta y_\ell^0+\Delta y_\ell^z<0$, it
contributes to generating the total asymmetry by reprocessing a flavour
asymmetry. This change of sign must be accomplished by $\gamma_\tau$,
before the washout interactions have time to act.

Consider an asymmetry produced in $N_1$ decay. In a short timescale,
before the $\ell$ have time to interact via $\lambda$ or $h_\tau$,
$\Delta y _\ell^a\propto\epsilon^a $ (where $a=0,x,y,z$). Flavour
effects can modify the evolution of the asymmetry if $\Delta
y_\ell^z$ is determined by the $h_\tau$ interactions (recall that
$\hat{z}=\hat{\gamma}_{N\to2}$):
\beq\label{eq:condbasis}
(\hat{\gamma}_\tau \times  \vec{\gamma}_\tau
\times \vec{\epsilon}) \cdot \hat{\gamma}_{N\rightarrow 2}=
(\hat{\gamma}_\tau \cdot   \vec{\epsilon} ~
\hat{\gamma}_\tau  \cdot \hat{\gamma}_{N\rightarrow 2}
- \epsilon^0 )|\gamma_\tau|
\gg \gamma_{N\rightarrow 2}\epsilon^0.
\eeq
Notice that $|\vec{\epsilon}|\gg\epsilon^0$ is possible, corresponding
to $\epsilon_{oo}\simeq-\epsilon_{\tau \tau}\gg\epsilon$. 
Eqn.~(\ref{eq:condbasis}) is satisfied for $|\vec{\epsilon}|\gamma_\tau\gg
\gamma_{N\to2}\epsilon$, provided that certain misalignment
conditions hold. For the purpose of rough estimates, this condition
can be taken to be $\gamma_\tau\gg\gamma_{N\to2}$, because
$\vec{\epsilon}$ is washed out by $\gamma_{N \to2}$, so the
approximation that $\Delta y_\ell^j \propto \epsilon^j$ is only valid
in a time-step shorter than the $N_1$ inverse decay time. The
misalignment conditions are  expected: if for instance
$N_1$ decays only to $\tau$s ($\hat{\tau} \cdot \hat{\ell}_{N_1} = 1$),
the single-flavour analysis is correct. The misalignment
is probably more transparent in flavour space
than in 4-vector space.

\subsection{Summary}
\label{sec:appenrhosum}
The charged lepton Yukawa couplings are relevant for leptogenesis,
because they can determine the initial state of washout interactions,
and washout is crucial for thermal leptogenesis.

In this appendix, we explicitly saw that the equations of motion of
the ``asymmetry number operator'', which are covariant in flavour
space, are projected onto the flavour basis by the interactions of
the charged lepton Yukawa couplings. In this flavour basis, the
equations for the diagonal elements of the asymmetry operator become
the familiar Boltzmann Equations, in which the charged lepton Yukawa
interactions do not appear. We expect this projection to occur if the
charged lepton Yukawa interactions are fast compared to leptogenesis
timescales.  Indeed, the flavour off-diagonal elements are negligible
in the equations when
\beq\label{eq:cdn1}
\gamma_\tau \gg \gamma_{N\to2} ~~~.
\eeq
Intuitively, this makes sense: if a lepton $\ell$ interacts many times
via the $h_\tau$ Yukawa coupling before interacting with $N_1$, it
will participate in the washout process as a flavour eigenstate
($\hat{\ell}_\tau$ or $\hat{\ell}_o$).

The temperature ranges where  flavour effects should be included
can be estimated by comparing rates, rather than rate densities.  The
timescale for leptogenesis is $H^{-1}$,  because the
``non-equilibrium'' is provided by the Universe expansion, so the
Yukawa couplings can be neglected when they are out-of-equilibrium,
that is, slower than $H$. Comparing $H$ to the rates for $h_\tau$- or
$h_\mu$-mediated interactions, such as $q_L\overline{t_R}
\to\ell_\tau\overline{\tau_R}$, one finds
%
\beq
\Gamma_\alpha 
\simeq10^{-2}h_\alpha^2 T > H {\rm ~  for ~} T<\left\{
  \begin{array}{cc}
    10^{12} ~ {\rm GeV} & \alpha=\tau\\
    10^9 ~ {\rm GeV} & \alpha=\mu. \end{array}\right.
\eeq
Below $T\sim10^{12}$ GeV, $h_\tau$ is in equilibrium, and
there can be two distinguishable flavours down to $T\sim10^{9}$ GeV.
Below $T \sim 10^{9} $ GeV, $h_\mu$ is also in  equilibrium, and
there can be three.

To impose the flavour basis in the washout processes, the charged
lepton Yukawa interactions should also be faster than the inverse
decays of $N_1$. (Notice that the Boltzmann factor $e^{-M_1/T}$,
heuristically included in $\Gamma_{ID}$, appears in the rate density
$\gamma_{N\to2}$ of eqn.~(\ref{eq:cdn1}).) We thus define a temperature
$T_{\rm fl}$, at which $\Gamma_\tau= \Gamma_{ID}$. Below $T_{\rm fl}$,
the lepton mass eigenstates are the flavour states. In flavoured
leptogenesis, one can estimate a temperature $T_B$, after which the
baryon asymmetry is generated. If $T_B<T_{\rm fl}$, then it is
consistent to calculate the baryon asymmetry with flavoured Boltzmann
Equations.

With strong washout for all flavours, it is more convenient to keep
track of the fraction of the $N_1$ population remaining, $\propto
e^{-M_1/T}$. At $T_{\rm fl}$, when the thermal mass eigenstates become
the flavour states, this fraction is $\sim\Gamma_{\tau}/\Gamma_D$. The
fraction remaining when a flavoured lepton asymmetry can survive is
$\sim H/\Gamma_{\alpha\alpha}$, where $\Gamma_{\alpha\alpha}={\rm
  min}\{\Gamma_{oo},\Gamma_{\tau \tau}\}$. ($\Gamma_\tau$ is the
$h_\tau$-mediated scattering rate, and
$\Gamma_{\tau\tau}=\Gamma(N_1\to\tau\phi)$.) So leptogenesis can be
calculated using BE written the flavour basis, when
\beq
\frac{\Gamma_\tau}{\Gamma_D} \gg \frac{H}{\Gamma_{\alpha \alpha}}.
\label{eq:app7dernier}
\eeq
(Recall that we are discussing the case of $\Gamma_{\alpha \alpha}>H$.)
For instance, for $M_1\sim10^{10}$ GeV, and $\tilde{m}\sim m_{\rm atm}$,
one finds that $\Gamma_\tau (M_1)\gsim\Gamma_{D}(M_1)$, so the baryon
asymmetry is generated after $T_{\rm fl}$ (for strong washout in all
flavours).

There are, however, regions of parameter space where $T_B>T_{\rm fl}$
\cite{Blanchet:2006ch}, such as the strong washout regime at large
$M_1$. In this case, the baryon asymmetry could be divided into three
parts: \\
{\it i)} An unflavoured contribution from the decays taking place
before $\Gamma_{ID} \sim \Gamma_\tau$. In  strong washout and assuming
$\Gamma_\tau > H$, these are negligible. \\
{\it ii)} A contribution from the transition region where
$\Gamma_{ID}\sim\Gamma_\tau$, to be obtained by solving the equations
of the covariant asymmetry density operator. \\
{\it iii)} A flavoured contribution from the decays taking place after
$\Gamma_{ID}\sim\Gamma_\tau$.

In the limit where $ii)$ is neglected, the final baryon asymmetry can
be estimated analytically: It is the flavoured asymmetry that is
produced in the decays of the $N_1$'s that remained at $T_{\rm fl}$.
See \cite{DeSimone:2006} for a detailed discussion based on solving
the equations for the asymmetry density operator.

\newpage
\section{Appendix: Analytic Approximations to $Y_{\Delta B}$}
\label{recipes}
In this Appendix, we reproduce analytic estimates of the baryon
asymmetry, derived in ref. \cite{Abada:2006ea}. These are approximate
solutions of the BE which take into account decays, inverse decays,
and $\Delta L=1$ scatterings with $N_1$ on an external leg. CP
violation in all these processes is included.  The off-diagonal
elements of the $A$-matrix are neglected (they are included in Section
\ref{sec:flavormix}), as is $\Delta L = 2$ scattering, so the
equations for the $B/3-L_\alpha$ asymmetries are decoupled. This
simplifies the solutions.  Ref.~\cite{Abada:2006ea} estimates the
overall uncertainty related to the various approximations to be of
order $30 \%$.

A detailed discussion of analytic approximations (as well as numerical
evaluations) can be found in ref. \cite{Buchmuller:2004nz}, for
single-flavour leptogenesis and without CP violation in $\Delta L=1$
scatterings. We here modify the formulae of \cite{Buchmuller:2004nz}
to include flavour and CP violation in scattering, providing simple and
useful approximations.

The  Boltzmann equations  can be written as follows:
\begin{eqnarray}
\label{eq:app9YN}
\frac{d}{dz}Y_{N_1}&=& - S(z), \\
\frac{d}{dz} Y_{\Delta \alpha}&=&\epsilon_{\alpha \alpha}\, S(z)
- W_{\alpha \alpha}(z)  Y_{\Delta \alpha} ~~, ~~
\label{eq:app9Yda}
\end{eqnarray}
where the source and washout functions can be read from eqns
(\ref{eq:finalBEYDeltai-s}) and (\ref{eq:finalBEYDeltai-w}).  Their
functional form can be found in ref. \cite{Buchmuller:2004nz}, or,
with finite temperature effects, in ref. \cite{Giudice:2003jh}.

As discussed earlier, the values of the light neutrino mass-squared
differences suggest that $\tilde{m}>m_*$. In this strong washout
regime, there is a useful approximation to $Y_{N_1}(z) - Y_{N_1}^{\rm
  eq}(z)$ at large $z$ (see chapter 6.4 of ref. \cite{Kolb:1988aj}).
Indeed, the asymmetry is produced when the inverse decays go out of
equilibrium, which happens at $z\gg1$. The relevant equation is then
(\ref{BE-N.DID}):
\beq
\frac{d }{dz} \Big[ Y_{N_1} - Y_{N_1}^{\rm eq} \Big]  =
- \frac{z}{sH_1}\Big[ { Y_{N_1} } - { Y^{\rm eq}_{N_1} }\Big]
\frac{\gamma_{N  \rightarrow 2}}{Y_{N_1}^{\rm eq}}
-  \frac{d }{dz}  Y_{N_1}^{\rm eq} ~~.
\eeq
When the interaction rate $\gamma_{N \to 2}/(sY_{N_1}^{\rm eq})$ is
fast compared to $H$, the differential equation is ``stiff'' and can
be solved approximately by setting the right-hand-side to zero, and
taking $dY_{N_1}^{\rm eq}/dz\simeq-Y_{N_1}^{\rm eq}$:
\beq
\Big[ { Y_{N_1} }- { Y^{\rm eq}_{N_1}}\Big]  \simeq
\frac{s Y_{N_1}^{\rm eq} }{\gamma_{N \rightarrow 2} }
\frac{H_1}{z} Y_{N_1}^{\rm eq}
\simeq \frac{z K_2(z)H_1}{4 g_*  \Gamma_D}.
\label{swYN}
\eeq
The last equality uses eqn (\ref{nMB}) and, from eqn
(\ref{eq:app7.18}), $\gamma_{N\to 2}/(s Y_{N_1}^{\rm eq}) \simeq
\Gamma_D$.

Using the trick of rewriting  eqn (\ref{eq:app9Yda}) as
$(f Y)' = fS\epsilon_{\alpha\alpha}$, with
$f' =W_{\alpha\alpha} f$, it has the formal solution
\beq
 Y_{\Delta \alpha} (z) = \epsilon_{\alpha \alpha}
\int_0^z dx S(x) e^{-\int_x^z dy  W_{\alpha \alpha}(y)}.
\label{eq:app9soln}
\eeq
We are interested in  $Y_{\Delta \alpha} (z\rightarrow \infty)$.
This can be approximated  via the ``steepest descent'', or
``saddle-point'' evaluation  of  an integral
along the contour $C$ \cite{MorseFeshbach}:
\beq
J(x) = \int_C e^{ x F(t)} dt  \simeq
\sqrt{\frac{2 \pi}{- x F''(t_0)}}e^{x F(t_0)},
\label{saddlept}
\eeq
where $t_0$ is an extremum of the integrand.  In the case of eqn
(\ref{eq:app9Yda}), the extremum is at $ W_{\alpha \alpha} = d/dz (\ln
|S|)$.  Using the saddle point approximation, and simple forms for the
functions $S$ and $W_{\alpha\alpha}$,
%
 ref. \cite{Abada:2006ea}
obtained the following approximations for the efficiency parameter :
\bea
\eta_\alpha & = &
\left[ \left(\frac{|A_{\alpha \alpha}| \tilde{m}_{\alpha \alpha}}{2.1
      m_*}\right)^{-1}+
\left(\frac{m_*}{ 2 |A_{\alpha \alpha}| \tilde{m}_{\alpha \alpha}}
\right)^{-1.16}     \right]^{-1}   ~~~~~~~~ ( \tilde{m} > m_*)
\label{eq:tm>m*} \\
&=&  \frac{|A_{\alpha \alpha}| \tilde{m}_{\alpha \alpha}}{ 2.5  m_*}
\frac{ \tilde{m}}{ m_*} 
~~~~~~~~~~~~~~~~~~~~~~~~~~~~~~~~~~~~~
( \tilde{m} < m_*)
\label{eq:tm<m*}
\eea
where $m_*$ and $\tilde{m}_{\alpha \alpha}$ are defined in eqn
(\ref{tildem}), and the $A_{\alpha \alpha}$ are defined in eqn
(\ref{eq:CC}). 
The $B/3 - L_\alpha$ asymmetries are given by
\bea
Y_{\Delta \alpha} & = & 4 \times 10^{-3} \epsilon_{\alpha \alpha} \eta_\alpha
\eea
and finally  the baryon asymmetry (in the SM)  is 
\beq
Y_{\Delta B} \simeq     ~~~ \frac{12}{37}  \sum_{\alpha}
Y_{\Delta \alpha}.
\label{with2g}
\eeq

The temperature range where tau Yukawa interactions are faster than the
expansion rate $H$ is $T<(1+\tan^2\beta)\times 10^{12}$ GeV ($\tan\beta=1$ in
the SM). If the tau Yukawa interactions are also faster than $\Gamma_{ID}$
when the asymmetries are generated (see Appendix \ref{sec:appenrhosum}), then
there are two relevant CP asymmetries --
$\epsilon_{oo}=\epsilon_{ee}+\epsilon_{\mu \mu}$ and $\epsilon_{\tau\tau}$ --
and two relevant partial decay rates --
$\tilde{m}_{oo}=\tilde{m}_{ee}+\tilde{m}_{\mu \mu}$ and $\tilde{m}_{\tau
  \tau}$.  So the sum in eqn (\ref{with2g}) is over $\alpha=o,\tau$ and one
should use eqn (\ref{A2gen}) for the $A_{\alpha\alpha}$. For $
T<(1+\tan^2\beta)\times 10^9$ GeV, the muon Yukawa interactions are also
faster than $H$ and there are three distinguishable flavours. The sum in eqn
(\ref{with2g}) is over $\alpha=e,\mu,\tau$, and one should use eqn
(\ref{A3gen}) for the $A_{\alpha \alpha}$.

\bibliography{PRbiblio}{}

\bibliographystyle{unsrt}

\end{document}